%% file: susybhs_arXiv2.tex
\documentclass[12pt,a4paper]{article}
\pdfoutput=1
\usepackage{jheppub}

\usepackage{placeins}
\usepackage{epsfig}
\usepackage{latexsym}
\usepackage[bbgreekl]{mathbbol}
\usepackage{mathtools}
\usepackage{amsmath,amssymb,amsthm,amsbsy,amsfonts,mathrsfs}
\usepackage{multirow}
\usepackage{slashed}
\usepackage{float}
\usepackage[font=footnotesize,labelfont=bf,width=.94\textwidth]{caption}
\usepackage{esint}
\usepackage{epsfig}
\usepackage{lscape}
\usepackage{graphicx}
\usepackage{psfrag}
\usepackage[utf8]{inputenc}
\usepackage{subfigure}
\usepackage{nicefrac}
\usepackage[stable]{footmisc}
\usepackage{rotating}
\usepackage{afterpage}
\usepackage{bbm}
\usepackage{color,xcolor}
\usepackage{cancel}
\usepackage{comment}

\usepackage{cancel}
\usepackage{comment}
\usepackage[framemethod=tikz]{mdframed}
\usepackage{empheq}
\usepackage{tikz}
\usepackage{tikz-feynman}
\usepackage{tcolorbox}
\tcbuselibrary{theorems}

\newmdenv[%
middlelinecolor=gray!30!,
middlelinewidth=1pt,
backgroundcolor=gray!10!,
roundcorner=3pt
]{identity}

\newmdenv[%
middlelinecolor=black,
middlelinewidth=1pt,
backgroundcolor=blue!20!,
roundcorner=3pt
]{titlebox}

\newmdenv[%
middlelinecolor=gray!80!white,
middlelinewidth=1pt,
backgroundcolor=white,
roundcorner=10pt,
subtitlebelowline=true,
frametitle={Calculation},
frametitlefont={\normalfont\bfseries\sffamily\color{gray!90!white}},
]{calculation}

\mdfdefinestyle{mystyle}{%
middlelinecolor=gray!80!white,
middlelinewidth=1pt,
backgroundcolor=blue!5!white,
roundcorner=10pt,
subtitlebelowline=true,
frametitlefont={\normalfont\bfseries\sffamily\color{gray!90!white}},
innertopmargin=.1cm,
innerbottommargin=.3cm,
}

\def\bebx#1\eebx{\begin{empheq}[box={\tcbhighmath[colframe=gray!90!white,colback=white]}]{align} #1 \end{empheq}}

\def\bbxd#1\ebxd{\begin{identity} \vskip -.4cm #1 \end{identity}\vskip-.2cm}

\def\btbox#1\etbox{\begin{titlebox} \vskip -.0cm #1 \end{titlebox}\vskip-.0cm}

\def\bframe#1{\begin{mdframed}[style=mystyle,frametitle={#1}]}
\def\eframe{\end{mdframed}}

\input{macros}

\title{\LARGE Black hole superpotential as a unifying entropy function and BPS thermodynamics \vskip-.8cm}

\author[a]{Praxitelis Ntokos}
\author[b]{Ioannis Papadimitriou} 

\affiliation[a]{School of Mathematics and Maxwell Institute for Mathematical Sciences,
University of Edinburgh, King’s Buildings, Edinburgh, EH9 3JZ, UK}

\affiliation[b]{Beijing Institute of Mathematical Sciences and Applications, Huairou District, Beijing 101408, China}

\emailAdd{praxitelis.ntokos@ed.ac.uk}
\emailAdd{ioannis@bimsa.cn}

\abstract{We develop an effective superpotential formalism for the SU(2)$\times$U(1) invariant sector of $\cn=2$ gauged supergravity in five dimensions with a U(1)$^3$ Fayet-Iliopoulos gauging, and determine the exact superpotential that describes all 1/4 BPS solutions in this sector. This includes the Gutowski-Reall black holes, but also a much broader class of solutions with a squashed $S^3$, magnetic flux and vector multiplet sources, as well as complex Euclidean BPS saddles. Some of these solutions are known only numerically, but the exact superpotential allows us to analytically evaluate the on-shell action, holographic one-point functions and conserved charges of all BPS solutions and to study their thermodynamics. In particular, by examining the supersymmetry Ward identities we show that solutions with supersymmetric vector multiplet sources break supersymmetry spontaneously. We also demonstrate the first law for black holes in the SU(2)$\times$U(1) invariant sector and show that the conserved charges of supersymmetric solutions satisfy the generalized BPS relation derived in \cite{Papadimitriou:2017kzw}, which includes the supersymmetric Casimir energy as a consequence of the anomalous supersymmetry transformation of the $\cn=1$ supercurrent at the boundary. Finally, we show that the effective superpotential provides a unifying entropy extremization principle, reproducing Sen's entropy function for near extremal black holes and the Hosseini-Hristov-Zaffaroni functional for complex Euclidean BPS saddles.
}

\keywords{BPS black holes, effective superpotential, black hole entropy, conserved charges, supersymmetric Casimir energy, AdS/CFT}

\begin{document}  
	\maketitle


\addtocontents{toc}{\protect\setcounter{tocdepth}{2}}

\section{Introduction and summary}
\label{intro}
\setcounter{equation}{0}

Supersymmetric black holes provide a rich terrain for exploring quantum gravity, since their microstates can often be identified either within string theory
\cite{Strominger:1996sh} or in a holographic dual quantum field theory \cite{Benini:2015eyy,Benini:2016rke} (see \cite{Zaffaroni:2019dhb} for a comprehensive review and extensive list of references). While much progress has been made in the microscopic description of non-supersymmetric near extremal black holes through near horizon holography \cite{Strominger:1997eq,Sen:2007qy,Sen:2008vm,Maldacena:2016upp}, the field theory duals of supersymmetric Anti de Sitter (AdS) black holes are better understood and more accessible.   

AdS$_5$ BPS black holes are holographically dual to four dimensional superconformal field theories, such as $\cn=4$ super Yang-Mills, which are among the best understood. Yet, the microscopic description of regular AdS$_5$ BPS black holes is complicated by the fact that they are necessarily rotating. As a result, early attempts to count the  microstates responsible for the macroscopic entropy of large 1/16 BPS AdS$_5$ black holes through the superconformal index of $\cn=4$ super Yang-Mills theory were unsuccessful \cite{Kinney:2005ej,Romelsberger:2005eg,Grant:2008sk,Chang:2013fba}. This problem was resolved recently in \cite{Cabo-Bizet:2018ehj,Choi:2018hmj,Benini:2018ywd} by realizing the necessity of complex fugacities or chemical potentials. Further progress was made in \cite{Honda:2019cio,ArabiArdehali:2019tdm,Kim:2019yrz,Cabo-Bizet:2019osg,Hosseini:2019lkt,Hosseini:2020mut}.

A related observation that applies to supersymmetric AdS black holes in various dimensions was that their Bekenstein-Hawking entropy can be obtained from an extremization principle  
\cite{Benini:2015eyy,Benini:2016rke,Hosseini:2017mds,Hosseini:2018dob,Choi:2018fdc,Hosseini:2019ddy}. The extremization functional is given by the Legendre transform of the free energy with respect to the R-symmetry and rotational chemical potentials, respectively $\D^I$ and $\o^\a$, which are subject to a linear constraint. For AdS$_5$ black holes the free energy takes the form \cite{Hosseini:2017mds}
\bbxd
\vskip.4cm
\be\label{partition-intro}
\log Z = -\frac12 N^2 \frac{\D^1\D^2\D^3}{\o^1\o^2}\,,
\ee
\ebxd
and the chemical potentials satisfy the complex constraint
\bbxd
\vskip.5cm
\be\label{constraint-intro}
\sum_\a \o^\a+ \sum_I \D^I=2\p i\,.
\ee
\ebxd
These results were obtained holographically from the on-shell supergravity action in 
\cite{Cabo-Bizet:2018ehj}. 

The free energy \eqref{partition-intro} is intimately related with the Casimir energy obtained from supersymmetric partition functions on $S^1\times S^3$ \cite{Assel:2014paa,Lorenzen:2014pna,Assel:2015nca,Bobev:2015kza,Martelli:2015kuk,BenettiGenolini:2016qwm,Brunner:2016nyk}. Nevertheless, the role of the Casimir energy in the BPS relation among the conserved charges is often overlooked. The general form of the BPS relation that supersymmetric AdS$_5$ black holes satisfy was obtained in \cite{Papadimitriou:2017kzw} from the anomalous supersymmetry transformation of the supercurrent in the dual field theory. The conserved mass, angular momenta and R-charges satisfy
\bbxd
\vskip.2cm
\be\label{BPS-intro}
M-\O^\a J_\a-\F^I Q_{I}=M_{\text{Casimir}}\,,
\ee
\ebxd
where the BPS chemical potentials $\O^\a$ and $\F^I$ take specific values and the supersymmetric Casimir energy $M_{\text{Casimir}}$ is given by
\bbxd
\bal\label{Casimir-intro}
M_{\text{Casimir}}\propto &\;\int_{\cm_3} d\s_{i}\Big[i(\bar\e^{\,-}\G_{j}\e^-)\Big(R^{ij}-\frac12 Rg^{ij}+\frac{\sqrt{3}}{2}\e^{ijkl}F_{kl}\Big)\NO\\
&\hspace{4.cm}-\frac{1}{2}(\bar\e^{\,+}\e^-+\bar\e^{\,-}\e^+)\e^{ijkl}\Big(F_{kl} A_{j}-\frac12\wt F^a_{kl}\wt A^a_{j}\Big)\Big]\,.
\eal
\ebxd
Here $\e^+$, $\e^-$ are components of the conformal supergravity Killing spinor on the AdS$_5$ boundary, $R_{ij}$ is the Ricci tensor of the boundary metric $g_{ij}$, while $A_i$ and $\wt A_i^a$ with $a=1,2$ are the gauge fields that couple to the R-symmetry currents.\footnote{We parameterize the spectrum in terms of an $\cn=1$ gravity multiplet and two Abelian vector multiplets.} This expression for the Casimir energy applies specifically to the case of equal anomaly coefficients, i.e. $a=c$. The supersymmetric Casimir energy for arbitrary $a$ and $c$ was determined for the conformal and R-multiplet of $\cn=1$ currents from the corresponding background supergravities in    \cite{Papadimitriou:2019gel,Papadimitriou:2019yug}. 

One of the main motivations for the present work is to demonstrate through concrete examples that supersymmetric AdS$_5$ black holes satisfy the BPS relation \eqref{BPS-intro} and to clarify its connection with the partition function \eqref{partition-intro} and BPS thermodynamics. The second motivation is to derive the Hosseini-Hristov-Zaffaroni (HHZ) entropy functional for supersymmetric AdS$_5$ black holes from first principles and identify any relation with Sen's entropy function for near extremal (both supersymmetric and non-supersymmetric) black holes \cite{Sen:2007qy,Sen:2008vm}.     

To address these questions we focus on AdS$_5$ black holes with two equal angular momenta that possess an enhanced SU(2)$\times$U(1) isometry. This is the minimal setup where we can explore the consequences of the supersymmetric Casimir energy and at the same time obtain a simple unifying entropy extremization function. The main character in our analysis is the effective black hole superpotential $\cf$ that allows us to formulate the dynamics of both supersymmetric and non-supersymmetric SU(2)$\times$U(1) invariant solutions in terms of a set of first order gradient flow equations. The effective superpotential is {\em not} the supergravity superpotential but it is related to it in case of BPS black holes. The effective superpotential depends on the conserved charges and is a solution of the radial Hamilton-Jacobi equation. 

\begin{figure}[h]
\begin{center}
\captionsetup{
singlelinecheck=false
}
\scalebox{.55}{\includegraphics{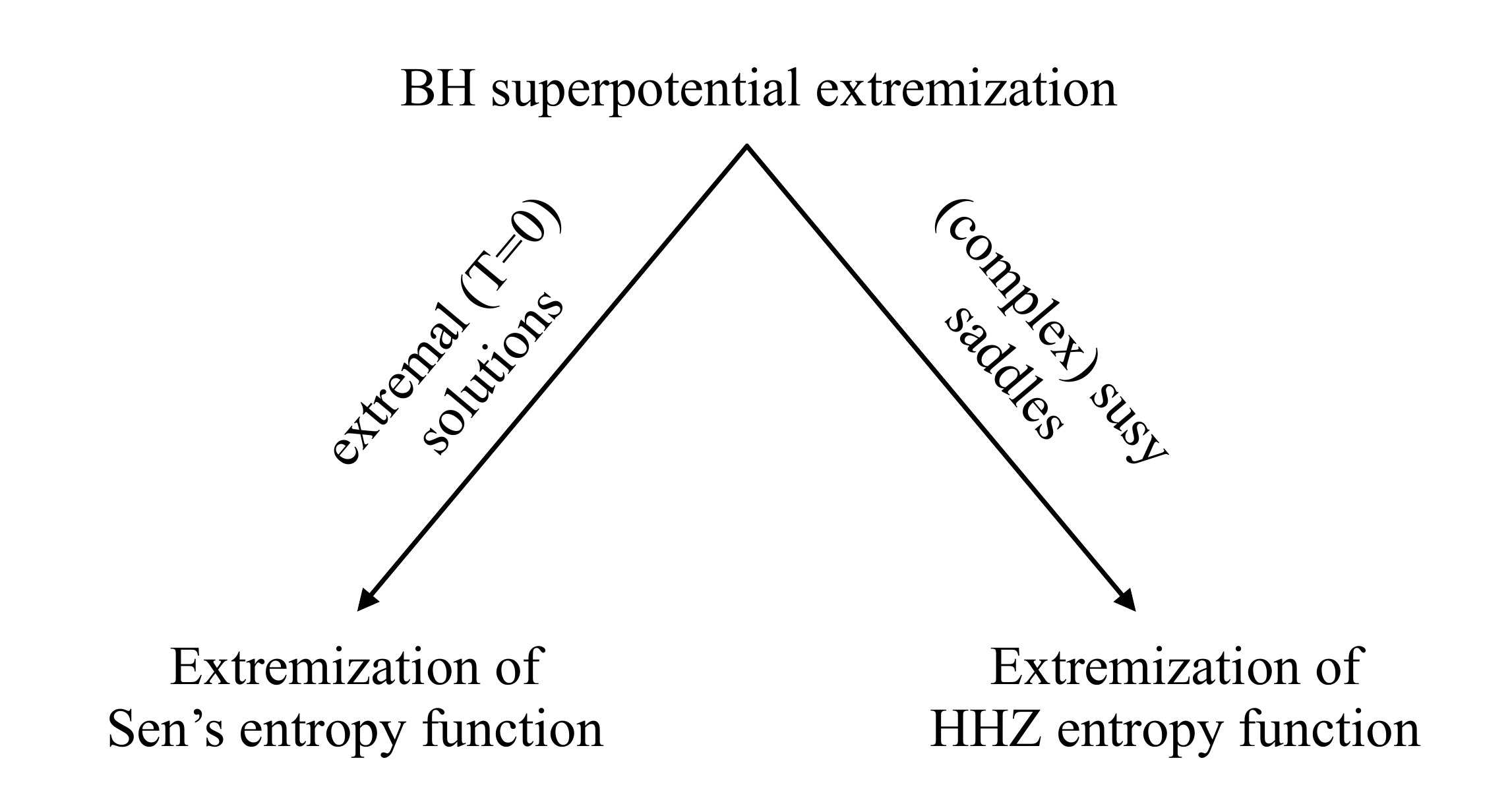}}
\vskip-.2cm
\caption{The effective black hole superpotential evaluated on the horizon provides a universal entropy extremization functional. For near extremal black holes it determines Sen's entropy function \cite{Sen:2007qy,Sen:2008vm}, while for complex BPS black saddles it reproduces the Hosseini-Hristov-Zaffaroni entropy extremization functional \cite{Hosseini:2017mds}.}
\label{fig1}
\end{center}
\end{figure}

The ultimate reason why the effective superpotential is the correct object to consider in the context of black hole entropy extremization is that Hamilton-Jacobi theory relates it directly to the on-shell action of the solutions, which is in turn identified with the free energy of the holographic dual field theory. Both Sen's and HHZ's entropy functions are obtainable from the on-shell action of the black hole solutions. Moreover, the gradient flow equations associated with the effective superpotential coincide with the attractor equations \cite{Ferrara:1996dd,Larsen:2006xm,LopesCardoso:2007qid,Cacciatori:2009iz,DallAgata:2010ejj,Cabo-Bizet:2017xdr}. As a consequence, the effective superpotential provides an overarching entropy extremization principle, as is schematically indicated in figure \ref{fig1}.

However, the utility of the effective superpotential goes beyond entropy extremization. It permits the evaluation of the on-shell action without use of the explicit form of the solution. In particular, one need not perform any integral over the bulk geometry -- Hamilton-Jacobi theory expresses the on-shell action as a boundary term, given by the effective superpotential. Moreover, the effective superpotential admits a direct holographic interpretation as a (bare) quantum effective potential in the dual field theory. As a result, the holographic one-point functions of the dual operators can be read off the effective superpotential.   

A central result of the present article is the exact effective superpotential for 1/4 BPS (1/16 BPS in $\cn=8$ supergravity) solutions of the SU(2)$\times$U(1) invariant sector of the five dimensional STU model, namely the U(1)$^3$ truncation of $\cn=8$ gauged supergravity. This effective superpotential is given in \eqref{BPS-superpotential-F}-\eqref{c1&2-main} and, to the best of our knowledge, was not known previously. It describes all 1/4 BPS SU(2)$\times$U(1) invariant solutions -- both  regular and singular -- with different solutions arising from different values of the integration constants of the corresponding BPS gradient flow equations.    

The exact 1/4 BPS superpotential allows us to compute the holographic one-point functions associated with any supersymmetric SU(2)$\times$U(1) invariant solution and to examine the Ward identities these satisfy. In particular, we obtain concrete criteria for explicit and spontaneous supersymmetry breaking in this sector. Moreover, we show that the 1/4 BPS superpotential can be expressed as the modulus of a complex holomorphic superpotential (see \eqref{BPS-superpotential-F-complex}), in direct analogy with the effective superpotential for static AdS$_4$ BPS black holes \cite{DallAgata:2010ejj}. This enables us to consider complex field configurations corresponding to Euclidean black saddles, again analogous to those obtained in four dimensions in  \cite{Bobev:2020pjk}. The near horizon analysis of these black saddles provides a direct connection between the 1/4 BPS superpotential and the HHZ entropy extremization functional.      

The rest of the paper is organized as follows. In section \ref{sec:5D-STU} we summarize some basic aspects of five dimensional $\cn=2$ gauged supergravity with a U(1)$^3$ Fayet-Iliopoulos gauging and review known supersymmetric and non-supersymmetric cohomogeneity one solutions, i.e. solutions that possess an enhanced SU(2)$\times$U(1) isometry. In section \ref{sec:superpotential} we consistently truncate $\cn=2$ gauged supergravity to its SU(2)$\times$U(1) invariant sector and derive the corresponding one dimensional effective Lagrangian. Formulating the point particle dynamics in Hamilton-Jacobi language we obtain a universal description of the SU(2)$\times$U(1) invariant sector in terms of first order flow equations and an effective superpotential that satisfies a reduced Hamilton-Jacobi equation. 

We present three exact effective superpotentials that describe respectively the general Reissner-Nordstr\"om-AdS black hole, the near horizon region of non-supersymmetric extremal black holes, and all SU(2)$\times$U(1) invariant 1/4 BPS solutions. We show that the effective superpotential describing the near horizon region of extremal black holes is directly related to Sen's entropy function \cite{Sen:2007qy,Sen:2008vm}. The 1/4 BPS superpotential is one of our main results and constitutes the basis for most of the subsequent analysis in sections \ref{sec:superpotential} and \ref{sec:thermodynamics}. In section \ref{sec:superpotential} we use the 1/4 BPS superpotential in order to determine the general form of the dual field theory background compatible with supersymmetry, as well as the holographic current and vector multiplet one-point functions. Moreover, we show how some of the known black holes reviewed in section \ref{sec:5D-STU} can be obtained from the 1/4 BPS superpotential by appropriate choice of the integration constants in the corresponding first order flow equations.

The thermodynamics of both supersymmetric and non-supersymmetric black holes of $\cn=2$ gauged supergravity is studied extensively in section \ref{sec:thermodynamics}. Following a brief overview of the holographic charges \cite{Papadimitriou:2005ii} and their generalization in the presence of Chern-Simons terms \cite{Papadimitriou:2017kzw}, we derive a gauge invariant version of the conserved charges by introducing suitable gauge compensators. We then focus on the SU(2)$\times$U(1) invariant sector and show that both the conserved charges and the on-shell action of generic black holes within this sector are determined by the effective superpotential. This leads to a simple proof of the first law of thermodynamics for all SU(2)$\times$U(1) invariant black holes.

We then turn to supersymmetric black holes described by the 1/4 BPS superpotential obtained in section \ref{sec:superpotential} and show that the conserved charges satisfy the generalized BPS relation derived in \cite{Papadimitriou:2017kzw}. Moreover, we show that the first order flow equations following from the 1/4 BPS superpotential correspond to a supersymmetric attractor flow that determines the values of the fields at the horizon (and hence the entropy), as well as a nonlinear constraint among the conserved charges. Finally, extending the analysis to complex configurations, we demonstrate that the 1/4 BPS superpotential provides a first principles derivation of the Hosseini-Hristov-Zaffaroni entropy extremization functional \cite{Hosseini:2017mds}.

We conclude with a brief discussion of future directions in
section \ref{sec:discussion}. Several technical results are presented in three appendices. In appendix \ref{sec:holography} we implement the holographic renormalization of the bosonic sector of $\cn=2$ Fayet-Iliopoulos gauged supergravity, derive the anomalous holographic Ward identities, and obtain general expressions for the renormalized one-point functions of all SU(2)$\times$U(1) invariant solutions -- both supersymmetric and non-supersymmetric -- in terms of the effective superpotential. Appendix \ref{sec:boundary} is a self contained discussion of the field theory backgrounds induced at the AdS boundary by SU(2)$\times$U(1) invariant supergravity solutions in the bulk. Using boundary supersymmetry, we determine the general form of SU(2)$\times$U(1) invariant backgrounds that admit Killing spinors, as well as the relations between the current and vector multiplet operator vacuum expectation values imposed by $\cn=1$ rigid supersymmetry. Finally, in appendix \ref{sec:KS} we solve the bulk Killing spinor equations for the SU(2)$\times$U(1) invariant sector and determine the general form of the corresponding Killing spinor.

\section{5D $\mathbf{\cn = 2}$ FI gauged supergravity and the STU model}
\label{sec:5D-STU}

The black holes we study in this paper are solutions of the gauged STU model in five dimensions, namely an $\cn=2$ gauged supergravity with a U(1)$^3$ Fayet-Iliopoulos (FI) gauging. This theory is a consistent truncation of five dimensional $\cn=8$ gauged supergravity and, therefore, can be holographically identified with a sector of the $\cn=4$ super-Yang-Mills (SYM) theory in four dimensions. In this section, we briefly review the STU model and summarize certain known black hole solutions that are relevant to our subsequent analysis. 

\subsection{Action and supersymmetry transformations}
\label{sec:model}

Five dimensional $\cn=2$ FI gauged supergravity coupled to $n_V$ vector multiplets is most efficiently described in terms of `very special real geometry' \cite{Gunaydin:1983bi} (see section 20.3.2 of \cite{Freedman:2012zz} for an overview). The STU model corresponds to the case $n_V=2$. In this framework, the $n_V$ gauge fields, $\wt A^a$, $a=1,\cdots,n_V$, in the vector multiplets and the graviphoton, $A$, correspond to specific linear combinations of the $n_V+1$ gauge fields $A^I$, $I=1,\cdots,n_V+1$. In particular, for the STU model these linear combinations are   
\be\label{gauge-field-decomposition}
A=-\frac{1}{\sqrt{3}}(A^1+A^2+A^3)\,,\qquad \wt A^1=\frac{1}{\sqrt{6}}(A^1+A^2-2A^3)\,,\qquad \wt A^2=\frac{1}{\sqrt{2}}(A^1-A^2)\,,
\ee
where the normalization of the graviphoton is chosen such that it possesses a canonically normalized kinetic term (cf. eq.~(4) in \cite{Chong:2005hr}). We now outline some basic aspects of $\cn=2$ FI supergravity, focusing specifically on the STU model (see also appendix A of \cite{Hosseini:2017mds} and \cite{Looyestijn:2010pb}).

In order to utilize the properties of very special real geometry, the $n_V$ independent scalars in the vector multiplets, $\vf^a$, with $a=1,\cdots,n_V$, are parameterized in terms of sections, $L^I=L^I (\vf^a)$, satisfying the constraint 
\be\label{LI-constraint}
L_IL^I=1 \, ,
\ee
where 
\be\label{LIdown}
L_I\equiv \frac16 C_{IJK}L^JL^K \, ,
\ee
and $C_{IJK}$ is a totally symmetric constant tensor. These relations imply that
\be\label{special-geometry-id}
L_I\pa_a L^I=0\,.
\ee
The scalar potential of the theory is  
\be\label{scalar-potential}
V=- 9 \x_I \x_J \big(L^I L^J - G^{IJ}\big) \, ,
\ee
where $\x_I$ are FI parameters and $G^{IJ}$ is the inverse of the metric on the scalar manifold 
\be\label{scalar-metric}
G_{IJ}= 9 L_I L_J - C_{IJK} L^K \, .
\ee

The bosonic part of the $\cn=2$ supergravity action takes the form\footnote{In order to simplify the radial Hamiltonian formalism, we choose the orientation of the bulk manifold such that the Levi-Civita symbol, $\ve_{\m\n\r\cdots}=\pm 1$, satisfies $\ve_{rt\cdots}=1$. The corresponding Levi-Civita tensor is defined as usual by $\e_{\m\n\r\cdots}=\sqrt{-g}\,\ve_{\m\n\r\cdots}$. This orientation convention differs from the more common one, where $\ve_{tr\cdots}=1$, used e.g. in \cite{Hosseini:2017mds}. However, our gauge fields also differ from \cite{Hosseini:2017mds} by an overall sign so that the sign of the Chern-Simons term is the same.\label{foot-orientation}}
\bbxd
\be\label{5Daction}
\actn_{\text{B}}=\frac{1}{2\k_5^2}\int d^5x\sqrt{-g}\Big(R-\frac12 G_{IJ}\pa_\m L^I\pa^\m L^J-\frac14 G_{IJ} F^I_{\m\n}F^{J\m\n}-V-\frac{1}{24} C_{IJK}\e^{\m\nu\rho\s\l}F^I_{\m\nu}F^J_{\rho\s}A^K_\l\Big),
\ee
\ebxd
where $\k^2_5=8\p G_5$ is the gravitational constant in five dimensions. As usual, the bulk action should be supplemented with the Gibbons-Hawking term
\be\label{GH}
\actn\sbtx{GH}=\frac{1}{2\k_5^2}\int d^4x\sqrt{-\g}\,2K \, ,
\ee
where $\g_{ij}$ is the induced metric on the regularized boundary and $K=\g^{ij} K_{ij}$ denotes the trace of its extrinsic curvature. However, the Gibbons-Hawking term alone does not render the variational problem on asymptotically locally AdS (AlAdS) manifolds well posed. This is achieved by adding further boundary terms -- the counterterms $\actn_{\text{ct}}$ -- to the action \cite{Papadimitriou:2005ii}
\be\label{full-action}
\actn=\actn_{\text{B}} + \actn\sbtx{GH} \, .
\ee
These additional boundary terms are discussed in detail in appendix \ref{sec:holography}.

Focusing on the STU model, we have $n_V=2$ and the constant symmetric tensor $C_{IJK}$ satisfies $C_{IJK}=1$ if $I,J,K$ are all different and zero otherwise.
Consequently, the scalar metric \eqref{scalar-metric} becomes (no index summation)
\be\label{scalar-metric-STU}
G_{IJ}=\left(L^I\right)^{-2}\d_{IJ} \, .
\ee

Demanding a canonical kinetic term and a diagonal mass matrix for the vector multiplet scalars, $\vf^a$,  leads to the following parameterization of the sections, $L^I$, in terms of the vector multiplet scalars \cite{Chong:2005da}\footnote{In the context of the STU model, the sections $L^I$ are often denoted by $X_I$, as in e.g. \cite{Chong:2005da}.\label{foot-XI}}
\be
L^1=e^{-\frac{1}{\sqrt{6}}\vf^1-\frac{1}{\sqrt{2}}\vf^2},\qquad L^2=e^{-\frac{1}{\sqrt{6}}\vf^1+\frac{1}{\sqrt{2}}\vf^2},\qquad L^3=e^{\frac{2}{\sqrt{6}}\vf^1}\,.
\ee
Notice that with this parameterization the definition \eqref{LIdown} implies that
\be\label{LIdownSTU}
L_I=\frac13(L^I)^{-1}\,.
\ee
Moreover, one may easily verify that 
\be
G_{IJ}\pa_\m L^I \pa^\m L^J=\d_{ab}\pa_\m\vf^a\pa^\m\vf^b\equiv\pa_\m\vec\vf\cdot \pa^\m\vec\vf \, ,
\ee
which allows us to use the Cartesian metric, $\d^{ab}$, to freely raise or lower the vector multiplet indices. In particular, we will often write $\vf_1$, $\vf_2$ for the vector multiplet scalars. Finally, the inverse metric on the scalar manifold takes the form
\be
\label{inverse-GIJ}
G^{IJ}=(L^I)^2\d^{IJ}=\pa_{a}L^{I}\pa_{a}L^{J}+\frac{1}{3}L^{I}L^{J}\,.
\ee

The FI parameters for the STU model are related with the AdS radius, $\ell$, as $\x_I = \frac{\sqrt{2}}{3\ell}$, and as a result the scalar potential \eqref{scalar-potential} simplifies to
\be\label{scalar-potential-STU}
V=-\frac{4}{\ell^2}\sum_{I=1}^3 \frac{1}{L^I} \, .
\ee
The scalar potential can be alternatively expressed in terms of the STU superpotential 
\be\label{superpotential-STU}
W=\frac{1}{\ell}\sum_{I=1}^3 L^I \, ,
\ee
through the relation
\be\label{V-W-eq}
\frac12V=(\pa_aW)^2-\frac{2}{3}W^2\, .
\ee
The superpotential, $W$, enters in the supersymmetry transformations of the gravitino and gauginos, which take the form \cite{Klemm:2000nj,Azzola:2018sld}
\bbxd
\bal\label{susy-transformations}
\d\j_\m=&\;\Big(\pa_\m+\frac14\o_{\m\r\s}\G^{\r\s}+\frac{i}{2\ell}\sum_{I=1}^3A^I_\m+\frac{i}{8}L_I(\G_\m{}^{\r\s}-4\d_\m^\r\G^\s)F^I_{\r\s}-\frac{1}{6}W\G_\m \Big)\e \, ,\NO\\
\d\l_a=&\;-\frac{i}{4}\Big(\G^\m\pa_\m \vf^a+\frac{3i}{2}\G^{\m\n}F^I_{\m\n}\pa_a L_I+2\pa_a W\Big)\e \, ,
\eal
\ebxd
where $\G^{\m\n\r\cdots}$ denotes totally antisymmetrized products of gamma matrices.\footnote{We use the spinor conventions of \cite{Freedman:2012zz} and the radial decomposition described in appendix A of \cite{Papadimitriou:2017kzw}.}

Finally, for the STU model, the field equations following from the action \eqref{5Daction} are
\bal\label{eoms}
&R_{\m\n}-\frac12 R\; g_{\m\n}
-\frac12\sum_{I=1}^3 \left(L^I\right)^{-2}
\Big((F^I)_{\m\r}(F^I)_\n{}^\r-\frac14(F^I)^2g_{\m\n}\Big)\NO\\
&\hskip1.2in-\frac12\Big(\pa_\m\vec\vf\cdot\pa_\n\vec\vf-\frac12(\pa\vec\vf)^2g_{\m\n}\Big)-\frac{2}{\ell^2}\sum_{I=1}^3\frac{1}{L^I}g_{\m\n}=0\, ,\NO\\
&\square_g\vf_1-\frac{1}{2\sqrt{6}}
\Big(\left(L^1\right)^{-2}(F^1)^2+\left(L^2\right)^{-2}(F^2)^2
-2\left(L^3\right)^{-2}(F^3)^2\Big)+\frac{4}{\ell^2\sqrt{6}}
\left(\frac{1}{L^1}+\frac{1}{L^2}-\frac{2}{L^3}\right)=0\, ,\NO\\
&\square_g\vf_2-\frac{1}{2\sqrt{2}}
\Big(\left(L^1\right)^{-2}(F^1)^2-\left(L^2\right)^{-2}(F^2)^2\Big)+\frac{4}{\ell^2\sqrt{2}}\left(\frac{1}{L^1}-\frac{1}{L^2}\right)=0\,,\NO\\
&\nabla_\m\Big(\left(L^K\right)^{-2}(F^K)^{\m\l}\Big)
-\frac18 \sum_{I,J=1}^3 C_{IJK}\e^{\m\n\r\s\l}F_{\m\n}^IF_{\r\s}^J=0 \, ,
\eal
where, for the sake of clarity, we explicitly indicate all summations over the indices $I,J,K$.

\subsubsection{Consistent truncation to minimal gauged supergravity}

The minimal supergravity in five dimensions that admits supersymmetric AdS black holes can be obtained as a consistent truncation of the STU model by setting all vector multiplet fields to zero, i.e. $\wt A^1=\wt A^2=0$ and $\vf_1=\vf_2=0$. From the relations \eqref{gauge-field-decomposition} follows that setting to zero the vector multiplet gauge fields amounts to setting
\be\label{minimal-sugra-As}
A^1=A^2=A^3=-\frac{1}{\sqrt{3}}A\,.
\ee

\subsection{Black hole solutions of the STU model}

The STU model in five dimensions admits asymptotically locally AdS$_{5}$ black hole solutions. Such solutions are characterized by their mass, $M$, two angular momenta, $J_{1,2}$, associated with the Cartan generators of $\text{SO}(4)\subset\text{SO}(4,2)$ and three electric charges, $Q_{I}$, corresponding to the $\text{U}(1)^{3}$ gauge symmetry of \eqref{5Daction}. When embedded in type IIB supergravity \cite{Cvetic:1999xp}, the latter corresponds to the Cartan subgroup of the isometries of the internal $S^{5}$.

In this subsection we briefly review known black hole solutions of the STU model that have equal angular momenta, i.e. $J_1=J_2$, and therefore possess an enhanced SU(2)$\times$U(1) isometry. This is the class of solutions we focus on in this paper, solely on technical grounds. However, our analysis goes beyond the analytically known solutions, whose boundary is conformal to $\bb R\times S^3$ (or the corresponding rotating Einstein universe), and applies also to solutions with a squashed boundary, magnetic flux and/or scalar sources, some of which have been found recently and in most cases only numerically.       

Our main interest is on supersymmetric black hole solutions of the STU model, but the techniques we develop, as well as several of our results, apply equally to non-supersymmetric solutions. We therefore find it instructive to summarize both the supersymmetric and non-supersymmetric solutions with an SU(2)$\times$U(1) isometry known in the literature. We begin by reviewing the most general non-BPS solution with two equal angular momenta, three independent charges and a round $S^3$ at the boundary. We then discuss the supersymmetric counterparts of these black holes, before summarizing some of the recent developments on mostly numerical solutions with a squashed $S^3$ at the boundary.

\subsubsection{Non-supersymmetric black holes}
\label{sec:Non BPS black holes}

The earlier solutions of the STU model go back to the Reissner-Nordstr\"om-AdS solutions of Einstein-Maxwell theory \cite{Romans:1991nq,London:1995ib}. These were generalized to non-extremal solutions of the STU model in \cite{Behrndt:1998jd}. Soon after, the first rotating solution was obtained in \cite{Hawking:1998kw}, namely the Kerr-AdS$_5$ black hole of pure Einstein gravity with a negative cosmological constant. All static solutions that admit a BPS limit turned out to be singular in that limit. Non-extremal black holes in AdS$_{5}$ with a regular BPS limit are necessarily rotating and were first found in \cite{Chong:2005hr} for minimal supergravity and in \cite{Cvetic:2004hs,Cvetic:2004ny,Chong:2005da} for the STU model. The most general analytically known non-supersymmetric black hole of the STU model with all six charges $M$, $J_{1,2}$, $Q_{I}$ independent was obtained in \cite{Wu:2011gq}.

The most general analytically known solution of the STU model that has three independent electric charges and an enhanced SU(2)$\times$U(1) isometry is the Cveti\v c-L\"u-Pope (CLP) black hole obtained in \cite{Cvetic:2004ny}. In its simplified form presented in \cite{Cvetic:2005zi} this solution depends on five physical parameters, $m$, $a$, $\d_{I}$ with $I=1,2,3$, which correspond to the five charges $M$, $J_{1}=J_{2}$ and $Q_{I}$. Writing
\be
c_{I}\equiv\cosh\d_{I}\,,\qquad s_{I}\equiv\sinh\d_{I}\,,
\ee
the solution takes the form 
\bal\label{sol:CLP}
ds^{2} = &\; -\frac{RY}{f_{1}}dt^{2}+\frac{r^{2}R}{Y}dr^{2}+\frac{1}{4}R\,d\Omega_{2}^{2}+\frac{f_{1}}{4R^{2}}\Big(\s_{3}-\frac{2f_{2}}{f_{1}}dt\Big)^{2}\,,\NO \\
A^{I} = &\; \frac{2m}{r^{2}H_{I}}\Big(s_{I}c_{I}dt+\frac{a}{2}(c_{I}s_{J}s_{K}-s_{I}c_{J}c_{K})\s_{3}\Big)\,,\qquad I\neq J\neq K\,,\NO \\
L^{I} = &\; \frac{R}{r^{2}H_{I}}\,,
\eal
where 
\be\label{sigma3}
\s_{3} = d\j+\cos\th\,d\f\,,
\ee
is one of the SU(2) left-invariant one-forms (see appendix \ref{sec:boundary}) and 
\bal
\label{nonBPSsolution}
f_{1} = &\; R^{3}+2m\, a^{2}r^{2}+4m^{2}a^{2}\big(2(c_1c_2c_3-s_1s_2s_3)s_1s_2s_3-s_1^2s_2^2-s_1^2s_3^2-s_2^2s_3^2\big)\,,\NO \\
f_{2} = &\; 2m\, a(c_1c_2c_3-s_1s_2s_3)r^2+4m^{2}a\,s_1s_2s_3\,,\NO \\
f_{3} = &\;2m\, a^{2}(1+r^2/\ell^2)+4\ell^{-2}m^{2}a^{2}\big(2(c_1c_2c_3-s_1s_2s_3)s_1s_2s_3-s_1^2s_2^2-s_1^2s_3^2-s_2^2s_3^2\big)\,,\NO \\
Y = &\; f_{3}+\ell^{-2}R^{3}+r^{4}-2m\, r^{2}\,,
\eal
with
\be
R=r^{2}(H_{1}H_{2}H_{3})^{1/3}\,,\qquad H_{I}=1+\frac{2m s_{I}^{2}}{r^{2}}\,.
\ee
Our notation differs relative to \cite{Cvetic:2005zi} in that $g_{\text{there}}=\ell_{\text{here}}^{-1}$.

The BPS limit of this solution was studied extensively in \cite{Cvetic:2005zi}. In particular, setting 
\be\label{CLP-BPS1}
\frac{a}{\ell}=\exp\Big(-\sum_I\d_I\Big)\,,
\ee
results in a supersymmetric solution that has naked closed timelike curves. Regular BPS black holes are obtained if one imposes in addition
\be\label{CLP-BPS2}
m=\frac{\ell^2}{2\sinh(\d_1+\d_2)\sinh(\d_1+\d_3)\sinh(\d_2+\d_3)}\,.
\ee
The resulting BPS black holes depend only on the three independent parameters $\d_I$ and coincide with those found by Gutowski and Reall (GR) a few months earlier \cite{Gutowski:2004yv}. These are the solutions we review next. As we will discuss in section \ref{sec:superpotential}, however, the entire family of singular BPS solutions is described by an exact superpotential, with the regularity condition \eqref{CLP-BPS2} arising from the extremization of this superpotential on the horizon.

\subsubsection{1/4 BPS black holes}
\label{sec:BPS black holes}

Early examples of supersymmetric asymptotically AdS$_5$ black hole solutions suffered from naked singularities \cite{Romans:1991nq,London:1995ib,Behrndt:1998ns} and closed timelike curves \cite{Klemm:2000gh}. Regular BPS domain walls and black string solutions of $\cn=2$ gauged supergravity coupled to vector multiplets were obtained in \cite{Cacciatori:2003kv}, while \cite{Gauntlett:2003fk} provided a complete classification of all supersymmetric solutions of minimal supergravity. The first regular BPS black hole solutions in minimal supergravity were constructed in \cite{Gutowski:2004ez}, and in \cite{Gutowski:2004yv} they were generalized to $\cn=2$ supergravity with an arbitrary number of vector multiplets. These solutions possess two equal angular momenta and, as we mentioned above, in the case of the STU model correspond to the BPS limit of the CLP solution \eqref{sol:CLP}. A more general class of regular BPS solutions with two independent angular momenta was obtained in \cite{Kunduri:2006ek}, while supersymmetric near horizon solutions with topology other than $S^3$ were found in \cite{Kunduri:2007qy,Kunduri:2006uh}.\footnote{Supersymmetric AdS$_5$ black holes of a different truncation of $\cn=8$ gauged supergravity than the one we consider here include \cite{Bhattacharyya:2010yg,Markeviciute:2018yal,Markeviciute:2018cqs}.}

Supersymmetric asymptotically AdS$_{5}$ black holes generically preserve two real supercharges, i.e. 1/4 of the supersymmetry parameterized by a Dirac spinor in five dimensions. The solutions we are interested in here are the Gutowski-Reall black holes obtained in \cite{Gutowski:2004yv}, which possess an enhanced SU(2)$\times$U(1) isometry. These black holes are also obtained from those in \cite{Kunduri:2006ek} by setting the two angular parameters equal. In discussing these solutions we mostly adopt the notation of \cite{Kunduri:2006ek}, except that
\be
A^I_{\text{there}}=-A^I_{\text{here}}\,, \qquad
g_{\text{there}}=\ell_{\text{here}}^{-1}\,,
\ee
and we use Euler instead of Hopf coordinates (see appendix \ref{sec:boundary}). Some additional minor differences in conventions are pointed out in footnotes \footref{foot-orientation} and \footref{foot-XI}.

Setting the two rotation parameters in the solution of \cite{Kunduri:2006ek} equal,
the Gutowski-Reall black holes \cite{Gutowski:2004yv} can be expressed in Euler coordinates in the form 
\bal\label{GR}
ds^{2}=&\;-f^{2}\Big(dt+\frac{\ell}{2}h\s_{3}\Big)^{2}+\frac{1}{f\X}\Big(\frac{1}{\D_{r}}dr^{2}+\frac{r^{2}}{4}\big(d\O_{2}^{2}+\D_r\s_{3}^{2}\big)\Big)\,,\NO\\
A^{I}=&\;-H_{I}^{-1}dt+\frac{\ell}{2}\big(1-\D_r+\ell^{-2}\m_{I}-H_I^{-1}h\big)\s_{3}\,,\NO\\
L^I=&\;\frac{1}{f H_I}\,,
\eal
where $\s_{3}$ is again the SU(2) left-invariant form given in \eqref{sigma3} and  
\bal\label{GR-functions-1}
\D_r =&\; \frac{r^2+(2a+\ell)^2}{\ell^2\X}\,,\NO\\
h = &\; -\frac{1}{ r^{2}\X}\Big((r^{2}+a^{2})(\X\D_r-1)+\frac{\X^2(\m_1\m_2+\m_1\m_3+\m_2\m_3)-8a^3(a+\ell)}{2\ell^2}\Big)\,,
\eal
with
\be\label{GR-functions-2}
f=(H_{1}H_{2}H_{3})^{-1/3}\,,\qquad H_{I}=1+\frac{\X\m_{I}}{r^{2}}\,,\qquad\X=1-a^{2}/\ell^{2}\,.
\ee
This solution depends on the four real constants $a$ and $\m_I$ that satisfy the relation
\be\label{constraintKLR}
\sum_I\m_I=\frac{(2a+\ell)^2}{\X}-\ell^2\,,\qquad  \m_I > 0 \, .
\ee
Unless this condition is satisfied, \eqref{GR} is not a solution of the field equations \eqref{eoms}.  

The Gutowski-Reall black hole \eqref{GR} is the canonical example for our discussion of BPS black holes of the STU model with an SU(2)$\times$U(1) isometry, but our analysis encompasses a much wider class of solutions, as we explain next.

\subsubsection{Solutions with a squashed $S^3$ and vector multiplet sources}
\label{sec:general black holes}

The black holes we reviewed in the previous subsections are not the most general solutions of the STU model with an SU(2)$\times$U(1) isometry. As we will see in section \ref{sec:superpotential}, these solutions have a round $S^3$ at the boundary and no scalar sources or magnetic flux. Supersymmetry imposes certain constraints on the form of the boundary sources (see eq.~\eqref{susy-from-the-bulk} and appendix \ref{sec:boundary}), but there exist nontrivial choices for the sources that preserve the SU(2)$\times$U(1) isometry, with a subset thereof preserving supersymmetry as well. One may therefore wonder whether regular black hole solutions with such more general asymptotics exist. The recent uniqueness result obtained in \cite{Lucietti:2021bbh} may help address this question for regular supersymmetric black holes. Moreover, in section \ref{sec:superpotential} we show that solutions with supersymmetric vector multiplet sources necessarily break supersymmetry spontaneously. 

Nevertheless, a number of supersymmetric and non-supersymmetric solutions with an SU(2)$\times$U(1) isometry and more general asymptotic form have been found in recent years, most only numerically. Analytic SU(2)$\times$U(1) invariant gravitational solitons with magnetic flux were obtained recently in \cite{Durgut:2021rma}. Both supersymmetric and non-supersymmetric black holes of minimal supergravity with more general asymptotics were obtained numerically in \cite{Blazquez-Salcedo:2017kig,Blazquez-Salcedo:2017ghg}. Further supersymmetric solutions with a squashed $S^3$ at the boundary as well as nontrivial sources for the vector multiplet scalars were presented in \cite{Cassani:2018mlh,Bombini:2019jhp}. 

Analyzing any of these solutions specifically is beyond the scope of the present manuscript, but several of our results are applicable to these solutions as well. In particular, the dynamics of all these solutions can be described in terms of an effective superpotential using the first order formulation of the dynamics we develop in section \ref{sec:superpotential}. Most notably, all 1/4 BPS solutions of the STU model with an SU(2)$\times$U(1) isometry are governed by the exact superpotential in eq.~\eqref{BPS-superpotential-F}, which therefore provides a valuable new tool for exploring the physics of such solutions, especially those that are only known numerically.

\section{A superpotential for the SU(2)$\times$U(1) sector of the STU model}
\label{sec:superpotential}

We are interested in solutions of the STU model \eqref{5Daction} that possess an SU(2)$\times$U(1) isometry. Our main focus are BPS solutions with this symmetry, but most of our analysis applies to non-supersymmetric solutions as well, such as the CLP black hole \eqref{sol:CLP} or the more general solutions reviewed in section \ref{sec:general black holes}. In this section we present a general ansatz for solutions with an SU(2)$\times$U(1) isometry and obtain first order flow equations that describe any solution within the ansatz in terms of an effective superpotential. Moreover, we determine the exact superpotential for all 1/4 BPS black holes of the STU model and study their asymptotic structure and holographic observables. The role of the effective superpotential in the black hole thermodynamics is discussed in section \ref{sec:thermodynamics}.

\subsection{SU(2)$\times$U(1) ansatz and minisuperspace Lagrangian}
\label{sec:ansatz}

Stationary solutions of the STU model \eqref{5Daction} with an SU(2)$\times$U(1) isometry are of the form
\bbxd
\bal
\label{ansatz}
ds^2=&\;N(r)^2dr^2+e^{-u_1(r)+u_2(r)}\Big(-e^{-4u_2(r)}dt^2+e^{-u_3(r)}d\O^2_2+e^{2u_3(r)}\big(\s_3+u_4(r)dt\big)^2\Big),\NO\\
A^I=&\;a^I(r)dt+v^I(r)\s_3,\qquad \vf^a=\vf^a(r),
\eal
\ebxd
where $d\O_{2}^{2}$ is the canonical metric on $S^2$
\be
d\O_{2}^{2} = \s_{1}^{2}+\s_{2}^{2}=d\th^2+\sin^2\th d\f^2\,,
\ee
and $\s_{1,2,3}$ are the left invariant 1-forms on SU(2), which in the Euler parameterization are 
\be\label{lifs}
\s_{1} = -\sin\j\,d\th+\cos\j\,\sin\th\,d\f\,,\quad
\s_{2} = \cos\j\,d\th+\sin\j\,\sin\th\,d\f\,,\quad
\s_{3} = d\j+\cos\th\,d\f\,.
\ee
The range of the Euler angles is $0\leq \th\leq \p$, $0\leq \f\leq 2\p$, $0\leq \j\leq 4\p$, and the metric on the round $S^3$ is given by
\be\label{S3-metric}
d\O_3^2=\frac14(\s_1^2+\s_2^2+\s_3^2)\,.
\ee
As we discuss in appendix \ref{sec:boundary}, a nonzero value of the ansatz function $u_3$ corresponds to a squashing of $S^3$.

Besides the $\bb R_t\times$SU(2)$\times$U(1) isometries, which not only preserve the form of the ansatz \eqref{ansatz}, but also the value of the ansatz functions, there are certain transformations that preserve the ansatz \eqref{ansatz}, but change the value of the ansatz functions. An example of such transformations is shifts $a^I\to a^I+a_0^I$ with constant $a_0^I$. Locally, this shift symmetry corresponds to U(1)$^{n_V+1}$ gauge transformations with time-dependent gauge parameters of the form $\a^I=a_0^I t$. However, there is no residual gauge symmetry along $\s_3$ since $d\s_3=\s_1\wedge\s_2\neq 0$ and so there is no shift symmetry associated with the functions $v^I(r)$. A second transformation that preserves the form of the ansatz but changes the values of the ansatz functions is the coordinate transformation $\j\to\j+c t$, for a constant $c$. Under this transformation the ansatz is preserved, but $u_4\to u_4+c$ and $a^I \to a^I +c v^I$. This amounts to changing the rotation frame and can be used in order to select a non-rotating frame at the boundary. Finally, the non-compactness of the time isometry results in a third symmetry of this type, namely constant rescalings of the time coordinate, $t\to \l t$. This is equivalent to transforming the ansatz functions as $u_1\to u_1-\frac12\log\l$, $u_2\to u_2-\frac12\log\l$, $u_4\to \l u_4$, $a^I\to \l a^I$. As we will see momentarily, the last two transformations leave the 1D effective action invariant. However, the invariance of the 1D effective action under the shift transformations $a^I\to a^I+a_0^I$ is broken by the Chern-Simons term.

Evaluating the STU model action \eqref{5Daction} on the ansatz \eqref{ansatz} we obtain
\be\label{S_B-ansatz}
\actn_{\text{B}}=\int dr\int dt \Big[L_{\rm 1D}+\frac{4\p^2}{\k_5^2}\pa_r\Big(\frac{8}{N}e^{-2u_1}\dot u_1-\frac13C_{IJK}a^Iv^Jv^K\Big)\Big],
\ee
where the one dimensional (point particle) effective Lagrangian $L_{\rm 1D}$ takes the form
\bbxd
\bal\label{1DLagrangian}
L_{\rm 1D}=&\;\frac{8\p^2}{\k_5^2}Ne^{-2u_1}\Big(-\frac{1}{2N^2}\big(-6\dot u_1^2+6\dot u_2^2+3\dot u_3^2-e^{4u_2+2u_3}\dot u_4^2+\dot\vf_1^2+\dot\vf_2^2\NO\\
&\;-e^{u_1+3u_2}G_{IJ}(\dot a^I-u_4\dot v^I)(\dot a^J-u_4\dot v^J)+e^{u_1-u_2-2u_3}G_{IJ}\dot v^I\dot v^J\big)+\frac{1}{2N}e^{2u_1}C_{IJK}\dot a^Iv^Jv^K\NO\\
&\;-V-\frac12e^{2u_1-2u_2+2u_3}G_{IJ}v^Iv^J-\frac12e^{u_1-u_2+u_3}(e^{3u_3}-4)\Big),
\eal
\ebxd
and the total derivative terms in \eqref{S_B-ansatz} are not only necessary to obtain the correct value of the on-shell action, but also, as we will see momentarily, play an important role in maintaining the symmetries of the problem. Solutions of the STU model \eqref{5Daction} within the ansatz \eqref{ansatz} are described by the 1D effective Lagrangian \eqref{1DLagrangian}. In particular, this is a consistent truncation of the STU model: the equations of motion following from the 1D Lagrangian \eqref{1DLagrangian} (see \eqref{1Deoms-metric}, \eqref{1Deoms-maxwell} and \eqref{1Deoms-scalars} below) imply the 5D equations of motion \eqref{eoms}, and so any solution of the 1D theory uplifts to a solution of the STU model. 

Let us now examine how the 1D action \eqref{S_B-ansatz} behaves under the transformations $u_4\to u_4+c$, $a^I\to a^I+a_0^I + c v^I$ that preserve the form of the ansatz \eqref{ansatz}. The shifts $a^I\to a^I+a_0^I$ leave the effective Lagrangian \eqref{1DLagrangian} invariant, but the full action \eqref{S_B-ansatz} is not invariant under such transformations due to the total derivative terms originating in the Chern-Simons form. This non-invariance of the action reflects the fact that the Chern-Simons term is not invariant under large gauge transformations. In particular, for closed gauge field variations $\d A^I$, i.e. $d(\d A^I)=0$, the variation of the Chern-Simons form is     
\be\label{BZ-current} 
\d\int_\cm C_{IJK} A^I\wedge F^J\wedge F^K=\int_\cm C_{IJK} \d A^I\wedge F^J\wedge F^K= \int_\cm d(\d A^I\wedge\cx_I)\,, 
\ee
where $\cx_I\equiv -C_{IJK} A^J\wedge F^K$ is (up to a normalization factor) the Bardeen-Zumino 3-form \cite{Bardeen:1984pm}. For backgrounds described by the ansatz \eqref{ansatz}, the pullback of $\d A^I\wedge\cx_I$ on a surface of constant radial coordinate is proportional to $C_{IJK}a_0^Iv^Jv^K$, in agreement with the transformation of the 1D action \eqref{S_B-ansatz} under the constant shift transformations $a^I\to a^I+a_0^I$.\footnote{Viewing the shifts $a^I\to a^I+a_0^I$ as gauge transformations $A^I\to A^I+d\a^I$ with gauge parameters $\a^I=t a_0^I$, one may naively conclude that the Chern-Simons term is invariant under such transformations, since the pullback of $\a^I C_{IJK}F^J\wedge F^K=-\a^Id\cx_I$ on a surface of constant radial coordinate vanishes within the ansatz \eqref{ansatz}. However, this conclusion is incorrect because $\d A^I=a_0^Idt$ are only locally exact, since the gauge function $\a_0t$ does not die off at past or future infinity. Equivalently, it does not satisfy the necessary periodic boundary condition around the Euclidean time circle.}

In contrast to the shifts $a^I\to a^I+a_0^I$, the transformation $u_4\to u_4+c$, $a^I\to a^I + c v^I$ preserves the effective Lagrangian \eqref{1DLagrangian} only up to a total derivative, but the full action \eqref{S_B-ansatz} is invariant under such transformations. This is expected since we have seen that these transformations are equivalent to a diffeomorphism, under which the total action is invariant. Finally, under the third transformation that leaves the form of the ansatz invariant, namely $u_1\to u_1-\frac12\log\l$, $u_2\to u_2-\frac12\log\l$, $u_4\to \l u_4$, $a^I\to \l a^I$, both the effective Lagrangian \eqref{1DLagrangian} and the total derivative terms in \eqref{S_B-ansatz} transform homogeneously by an overall factor, which can be absorbed into a rescaling of the time coordinate.

\subsubsection{Equations of motion}

In terms of the radial derivative 
\be\label{prime}
{}'=\pa_\r=N^{-1}\pa_r\,,
\ee
the equations of motion following from the 1D effective Lagrangian \eqref{1DLagrangian} take the form
\bal\label{1Deoms-metric}
&\big(6e^{-2u_1}u_1'\big)'+e^{-2u_1}\Big(6 u_1'^2-6u_2'^2-3 u_3'^2+e^{4u_2+2u_3}u_4'^2-\vf_1'^2-\vf_2'^2-\frac{1}{2}e^{u_1-u_2-2u_3}G_{IJ} v'^I v'^J\NO\\
&\;+\frac{1}{2}e^{u_1+3u_2}G_{IJ}(a'^I-u_4v'^I)(a'^J-u_4v'^J)\Big)-2e^{-2u_1}V-\frac12e^{-u_1-u_2+u_3}(e^{3u_3}-4)=0\,,\NO\\
\rule{.0cm}{.8cm}&\big(6e^{-2u_1} u'_2\big)'+e^{-2u_2+2u_3}G_{IJ}v^Iv^J+\frac12e^{-u_1-u_2+u_3}(e^{3u_3}-4)\NO\\
&\;+\frac{1}{2}e^{-2u_1}\Big(4e^{4u_2+2u_3}u_4'^2+3e^{u_1+3u_2}G_{IJ}(a'^I-u_4 v'^I)(a'^J-u_4v'^J)+e^{u_1-u_2-2u_3}G_{IJ}v'^Iv'^J\Big)=0\,,\NO\\
\rule{.0cm}{.8cm}&\big(3e^{-2u_1} u'_3\big)'+e^{-2u_1+4u_2+2u_3} u_4'^2+e^{-u_1-u_2-2u_3}G_{IJ}v'^Iv'^J\NO\\
&\;-e^{-2u_2+2u_3}G_{IJ}v^Iv^J-2e^{-u_1-u_2+u_3}(e^{3u_3}-1)=0\,,\NO\\
\rule{.0cm}{.8cm}&\big(e^{-2u_1+4u_2+2u_3} u'_4\big)'+e^{-u_1+3u_2}G_{IJ}v'^I( a'^J-u_4v'^J)=0\,,
\eal
\bal\label{1Deoms-maxwell}
&\Big(e^{-u_1+3u_2}G_{IJ}(a'^J-u_4v'^J)+\frac{1}{2}C_{IJK}v^Jv^K\Big)'=0,\\
\rule{.0cm}{.8cm}&\Big(e^{-u_1-u_2-2u_3}G_{IJ} v'^J+e^{-u_1+3u_2}u_4G_{IJ}( a'^J-u_4v'^J)\Big)'+C_{IJK}a'^Jv^K-e^{-2u_2+2u_3}G_{IJ}v^J=0\,.\NO
\eal
\bal\label{1Deoms-scalars}
&\big(e^{-2u_1} \vf'_a\big)'-e^{-2u_1}\pa_{\vf_a}V-\frac12e^{-2u_2+2u_3}\pa_{\vf_a}G_{IJ}v^Iv^J\NO\\
&\;+\frac{1}{2}e^{-u_1+3u_2}\pa_{\vf_a}G_{IJ}(a'^I-u_4v'^I)( a'^J-u_4v'^J)-\frac{1}{2}e^{-u_1-u_2-2u_3}\pa_{\vf_a}G_{IJ}v'^Iv'^J=0\,.
\eal
Moreover, the equation of motion for the non-dynamical field $N$ is the Hamiltonian constraint
\bal\label{1Deoms-Hamiltonian}
&\frac{1}{2}\big(-6u_1'^2+6u_2'^2+3u_3'^2-e^{4u_2+2u_3} u_4'^2+\vf_1'^2+\vf_2'^2-e^{u_1+3u_2}G_{IJ}(a'^I-u_4v'^I)(a'^J-u_4v'^J)\NO\\
&\;+e^{u_1-u_2-2u_3}G_{IJ}v'^I v'^J\big)-V-\frac12e^{2u_1-2u_2+2u_3}G_{IJ}v^Iv^J-\frac12e^{u_1-u_2+u_3}(e^{3u_3}-4)=0\,.
\eal
In fact, the equations of motion \eqref{1Deoms-metric}, \eqref{1Deoms-maxwell} and \eqref{1Deoms-scalars} imply that the l.h.s. of \eqref{1Deoms-Hamiltonian} is constant. The Lagrange multiplier $N$ sets this constant to zero, as a consequence of the 1D reparameterization invariance of the Lagrangian \eqref{1DLagrangian}.

\subsubsection{Canonical momenta and conservation equations}

As a first step towards deriving a system of first order equations for the 1D effective theory, we obtain the radial canonical momenta following from the Lagrangian \eqref{1DLagrangian}, namely 
\bal
\label{1dmomenta}
\p_{u_1}=&\;\frac{48\p^2}{\k_5^2}e^{-2u_1}u'_1\,,\quad
\p_{u_2}=-\frac{48\p^2}{\k_5^2}e^{-2u_1}u'_2\,,\quad
\p_{u_3}=-\frac{24\p^2}{\k_5^2}e^{-2u_1}u'_3\,,\NO\\
\p_{u_4}=&\;\frac{8\p^2}{\k_5^2}e^{-2u_1+4u_2+2u_3}u'_4\,,\quad
\p_{\vf_1}=-\frac{8\p^2}{\k_5^2}e^{-2u_1}\vf'_1\,,\quad
\p_{\vf_2}=-\frac{8\p^2}{\k_5^2}e^{-2u_1}\vf'_2\,,\NO\\
\p^a_{I}=&\;\frac{8\p^2}{\k_5^2}\Big(e^{-u_1+3u_2}G_{IJ}(a'^J-u_4 v'^J)+\frac12C_{IJK}v^Jv^K\Big)\,,\NO\\
\p^v_{I}=&\;-\frac{8\p^2}{\k_5^2}\Big(e^{-u_1-u_2-2u_3}G_{IJ} v'^J+e^{-u_1+3u_2}u_4G_{IJ}(a'^J-u_4v'^J)\Big)\,.
\eal

The constraint \eqref{1Deoms-Hamiltonian} is equivalent to the vanishing of the 1D Hamiltonian 
\be\label{1DHamiltonian}
H_{\rm 1D}=\dot u_1\p_{u_1}+\dot u_2\p_{u_2}+\dot u_3\p_{u_3}+\dot u_4\p_{u_4}+\dot \vf_a\p_{\vf_a}+\dot a^I\p^a_I+\dot v^I\p^v_I-L_{\rm 1D}\,,
\ee
which, in terms of the canonical momenta \eqref{1dmomenta}, takes the form
\bal
\label{Hamiltonian}
&-\frac{1}{24}\p_{u_1}^2+\frac{1}{24}\p_{u_2}^2+\frac{1}{12}\p_{u_3}^2-\frac{1}{4}e^{-4u_2-2u_3}\p_{u_4}^2+\frac{1}{4}\p_{\vf_1}^2+\frac{1}{4}\p_{\vf_2}^2\NO\\
&-\frac{1}{4}e^{-u_1-3u_2}G^{IJ}\Big(\p^a_I-\frac{4\p^2}{\k_5^2}C_{IKL}v^Kv^L\Big)\Big(\p^a_J-\frac{4\p^2}{\k_5^2}C_{JPQ}v^Pv^Q\Big)\NO\\
&+\frac14e^{-u_1+u_2+2u_3}G^{IJ}\Big(\p^v_I+u_4\Big(\p^a_I-\frac{4\p^2}{\k_5^2}C_{IKL}v^Kv^L\Big)\Big)\Big(\p^v_J+u_4\Big(\p^a_J-\frac{4\p^2}{\k_5^2}C_{JPQ}v^Pv^Q\Big)\Big)\NO\\
&-\frac12\Big(\frac{8\p^2}{\k_5^2}\Big)^2e^{-4u_1}\Big( V+\frac12e^{2u_1-2u_2+2u_3}G_{IJ}v^Iv^J+\frac12e^{u_1-u_2+u_3}(e^{3u_3}-4)\Big)=0\,.
\eal

This is the first of three conservation laws the 1D effective theory admits. The other two follow from the first equation in \eqref{1Deoms-maxwell} and the last one in \eqref{1Deoms-metric}, which can be expressed in terms of the canonical momenta as
\be
\p'^a_I=0\,,\qquad \Big(\p_{u_4}+v^I\Big(\p^a_I-\frac{4\p^2}{3\k_5^2}C_{IJK}v^Jv^K\Big)\Big)'=0\,.
\ee
Integrating these we determine that 
\bbxd
\vskip.4cm
\be\label{charges}
\p^a_I=\frac{4\p^2}{\k_5^2}q_I\,,\qquad \p_{u_4}+\frac{4\p^2}{\k_5^2}v^I\Big(q_I-\frac13C_{IJK}v^Jv^K\Big)=\frac{4\p^2}{\k_5^2}j\,,
\ee
\ebxd
where $q_I$ and $j$ are integration constants. We will see in section \ref{sec:thermodynamics} that these correspond respectively to the electric charges and angular momentum of the 5D solutions.

\subsection{Effective superpotential from Hamilton-Jacobi theory}
\label{sec:flow}

Instead of solving directly the second order equations \eqref{1Deoms-metric}, \eqref{1Deoms-maxwell} and \eqref{1Deoms-scalars}, it is often useful to reformulate these in terms of first order flow equations. For supersymmetric solutions, such first order equations are typically provided by the BPS equations. However, first order equations can also be obtained for non-supersymmetric solutions, as was first pointed out in the context of `fake supergravity' \cite{Freedman:2003ax}. Hamilton-Jacobi theory provides a general and systematic way for deriving genuine first order flow equations for both supersymmetric and non-supersymmetric systems, and has been applied not only to domain wall type solutions in different settings \cite{Papadimitriou:2006dr,Skenderis:2006rr,Papadimitriou:2007sj,Janssen:2007rc,HoyosBadajoz:2008fw,Lindgren:2015lia,Cremonini:2020rdx,Kim:2020dqx}, but also to black holes \cite{Andrianopoli:2009je,Hyun:2012bc,Lindgren:2015lia,Dorronsoro:2016pin,Klemm:2016kxw,Klemm:2017pxv,Cabo-Bizet:2017xdr}. While the Hamilton-Jacobi approach is applicable to any system, alternative approaches to deriving first order flow equations may exist in certain special cases   
\cite{Andrianopoli:2007gt,LopesCardoso:2007qid,Perz:2008kh,Ceresole:2009iy,DallAgata:2010ejj,Kiritsis:2012ma,Gnecchi:2012kb}.

In the Hamilton-Jacobi approach, the first order flow equations follow from the fact that the canonical momenta can be expressed as gradients of an `effective superpotential' $\cs_{\rm 1D}(u_1,u_2,u_3,\vf^a,v^I,u_4,a^I)$ -- the so called Hamilton's principal function -- that is a function of all generalized coordinates, but not of the velocities or of the momenta. Namely, 
\bal
\label{HJmomenta}
&\p_{u_1}=\frac{\pa\cs_{\rm 1D}}{\pa u_1},\quad \p_{u_2}=\frac{\pa\cs_{\rm 1D}}{\pa u_2},\quad \p_{u_3}=\frac{\pa\cs_{\rm 1D}}{\pa u_3},\quad \p_{u_4}=\frac{\pa\cs_{\rm 1D}}{\pa u_4}\,,\NO\\
&\qquad\qquad \p_{\vf_a}=\frac{\pa\cs_{\rm 1D}}{\pa \vf_a},\quad \p^a_{I}=\frac{\pa\cs_{\rm 1D}}{\pa a^I},\quad \p^v_{I}=\frac{\pa\cs_{\rm 1D}}{\pa v^I}\,. 
\eal
The dependence of $\cs_{\rm 1D}$ on $a^I$ and $u_4$ is determined by the conservation equations \eqref{charges}:\footnote{Notice that, were it not for the Chern-Simons term, the transformations $u_4\to u_4+c$, $a^I \to a^I +a_0^I+c v^I$ would shift $\cs_{\rm 1D}$ by a constant, reflecting the invariance of the effective Lagrangian \eqref{1DLagrangian}. However, the Chern-Simons term implies that $\cs_{\rm 1D}$  picks up a non-constant term under $u_4\to u_4+c$, $a^I \to a^I +c v^I$, which cancels against the transformation of the boundary term in \eqref{S_B-ansatz}.}  
\vskip-.0cm
\bbxd
\vskip.4cm
\be\label{HJ1d}
\cs_{\rm 1D}=\frac{4\p^2}{\k_5^2}\Big[\cu(u_1,u_2,u_3,\vf^a,v^I)+\Big(j-v^Iq_I+\frac13C_{IJK}v^Iv^Jv^K\Big)u_4+q_Ia^I\Big]\,,
\ee
\ebxd
while the Hamiltonian constraint \eqref{Hamiltonian} requires that the function $\cu(u_1,u_2,u_3,\vf^a,v^I)$ satisfies the partial differential equation
\bal
\label{HJeq}
&-\frac{1}{24}(\pa_{u_1}\cu)^2+\frac{1}{24}(\pa_{u_2}\cu)^2+\frac{1}{12}(\pa_{u_3}\cu)^2+\frac{1}{4}(\pa_{a}\cu)^2+\frac14e^{-u_1+u_2+2u_3}G^{IJ}\pa_{v^I}\cu\pa_{v^J}\cu\NO\\
&-\frac{1}{4}e^{-4u_2-2u_3}\Big(j-v^Iq_I+\frac13C_{IJK}v^Iv^Jv^K\Big)^2-\frac{1}{4}e^{-u_1-3u_2}G^{IJ}\big(q_I-C_{IKL}v^Kv^L\big)\big(q_J-C_{JPQ}v^Pv^Q\big)\NO\\
&-2e^{-4u_1}\Big( V+\frac12e^{2u_1-2u_2+2u_3}G_{IJ}v^Iv^J+\frac12e^{u_1-u_2+u_3}(e^{3u_3}-4)\Big)=0\,.
\eal

Given a solution $\cu(u_1,u_2,u_3,\vf^a,v^I)$ of this equation, inserting the corresponding principal function \eqref{HJ1d} and the canonical momenta \eqref{1dmomenta} in the relations \eqref{HJmomenta} results in the following first order equations for the generalized coordinates $u_1$, $u_2$, $u_3$, $u_4$, $\vf^a$, $a^I$, $v^I$: 
\bbxd
\bal\label{flow-eqs-U}
&e^{-u_1+3u_2}G_{IJ}(a'^J-u_4v'^J)=\frac12\big(q_I-C_{IJK}v^Jv^K\big)\,,\\ &e^{-2u_1+4u_2+2u_3}u'_4=\frac12 \Big(j-v^Iq_I+\frac13C_{IJK}v^Iv^Jv^K\Big)\,,\quad e^{-u_1-u_2-2u_3}G_{IJ}v'^J=-\frac12\frac{\pa\cu}{\pa v^I}\,,\NO\\
&e^{-2u_1}u'_1=\frac{1}{12}\frac{\pa\cu}{\pa u_1}\,,\quad
-e^{-2u_1}u'_2=\frac{1}{12}\frac{\pa\cu}{\pa u_2}\,,\quad
-e^{-2u_1}u'_3=\frac{1}{6}\frac{\pa\cu}{\pa u_3}\,,\quad
-e^{-2u_1}\vf'^a=\frac12\frac{\pa\cu}{\pa\vf^a}\,.\NO
\eal
\ebxd
As can be explicitly verified with a bit of algebra, Hamilton-Jacobi theory ensures that, given any solution $\cu(u_1,u_2,u_3,\vf^a,v^I)$ of \eqref{HJeq}, integrating the corresponding first order equations \eqref{flow-eqs-U} results in a solution of the second order equations \eqref{1Deoms-metric}, \eqref{1Deoms-maxwell} and \eqref{1Deoms-scalars}. In the remaining of this section we will determine the general asymptotic form of the superpotential $\cu$ corresponding to AlAdS solutions, and provide a number of exact solutions of the Hamilton-Jacobi equation \eqref{HJeq} -- most notably the exact superpotential for all 1/4 BPS solutions in the SU(2)$\times$U(1) invariant sector of the STU model.

\subsubsection{Solutions with non-compact timelike isometry}

We saw above that there exists a set of transformations that preserve the form of the ansatz but shift the values of the ansatz functions. These are $u_4\to u_4+c$, $a^I\to a^I+a_0^I + c v^I$, as well as $u_1\to u_1-\frac12\log\l$, $u_2\to u_2-\frac12\log\l$, $u_4\to \l u_4$, $a^I\to \l a^I$, for constant $c$, $a_0^I$ and $\l$. The first two of these transformations do not affect the superpotential, but the last one, which is a consequence of the non-compactness of the timelike isometry, imposes a stringent constraint on the form of $\cu(u_1,u_2,u_3,\vf^a,v^I)$. Namely, it requires that -- apart from an additive constant that does not affect the dynamics -- $\cu$ transforms homogeneously with weight 1 under this transformation. Without loss of generality, therefore, physical solutions correspond to a superpotential of the form 
\bbxd
\vskip.25cm
\be\label{separable-U}
\cu(u_1,u_2,u_3,\vf^a,v^I)=e^{-2u_2}\cf(u,u_3,\vf^a,v^I)+\cu_0\,,\qquad u\equiv u_1-u_2\,, 
\ee 
\ebxd
where the reduced superpotential, $\cf$, depends only on the difference, $u$, of the generalized coordinates $u_1$ and $u_2$, and $\cu_0$ is a constant. This is a general result for all solutions with a non-compact timelike isometry, but we will verify it explicitly in the examples we will discuss later or, including all 1/4 BPS black holes.  

Introducing the variables 
\bbxd
\vskip.4cm
\be\label{F0-def}
\cf_\pm \equiv \cf\pm\cf_0\,,\qquad \cf_0\equiv e^{-u_3}\Big(j-v^Iq_I+\frac13C_{IJK}v^Iv^Jv^K\Big)\,,
\ee
\ebxd
the Hamilton-Jacobi equation \eqref{HJeq} reduces to
\bbxd
\bal
\label{HJ-F}
&\frac{1}{12}(\pa_u+2)(\cf_+\cf_-)+\frac{1}{12}\pa_{u_3}\cf_+\pa_{u_3}\cf_-+\frac{1}{4}\pa_{a}\cf_+\pa_{a}\cf_-\\
&+\frac14e^{-u+2u_3}G^{IJ}\pa_{v^I}\cf_+\pa_{v^J}\cf_--2e^{-4u}\Big( V+\frac12e^{2u+2u_3}G_{IJ}v^Iv^J+\frac12e^{u+u_3}(e^{3u_3}-4)\Big)=0\,,\NO
\eal
\ebxd
while the first order equations \eqref{flow-eqs-U} become
\bbxd
\bal\label{flow-eqs-F}
&e^{-2u+2u_2+u_3}u'_4=\frac12\cf_0,\qquad e^{-u+2u_2-u_3}G_{IJ}(a'^J-u_4 v'^J)=-\frac12\frac{\pa\cf_0}{\pa v^I},\NO\\
&e^{-u-2u_3}G_{IJ}v'^J=-\frac12\frac{\pa\cf}{\pa v^I},\qquad
u_{2}'= \frac{1}{12}\pa_u(e^{2u}\cf),\qquad
e^{-2u}u'=-\frac{1}{6}\cf,\NO\\
&e^{-2u}u'_3=-\frac{1}{6}\frac{\pa\cf}{\pa u_3},\qquad
e^{-2u}\vf'^a=-\frac12\frac{\pa\cf}{\pa\vf^a}\,.
\eal
\ebxd

\subsubsection{Reissner-Nordstr\"om-AdS black hole}

It is instructive to illustrate the superpotential formalism with a simple example. Besides global AdS$_5$ and the AdS-Schwarzschild black hole, the simplest solution of the STU model is the Reissner-Nordstr\"om black hole 
\cite{London:1995ib,Chamblin:1999tk}. Setting $j=0$, it is straightforward to confirm that it is consistent to set $u_3=\vf_a=v^I=0$ in the second order equations \eqref{1Deoms-metric}, \eqref{1Deoms-maxwell} and \eqref{1Deoms-scalars}, provided the electric charge parameters satisfy the constraints  
\be
\pa_{\vf_a}G_{IJ}\big|_{\vf_a=0}q_Iq_J=0\,,
\ee
or equivalently
\be\label{no-scalar-charge-conditions}
q_1^2+q_2^2-2q_3^2=0\,,\qquad q_1^2-q_2^2=0\,. 
\ee
These correspond to setting to zero the two vector multiplet gauge fields in \eqref{gauge-field-decomposition} so that only the minimal supergravity gauge field $A$ in \eqref{minimal-sugra-As} survives, with 
\be\label{minimal-sugra-q}
q_1=q_2=q_3=-\frac{1}{\sqrt{3}}q\,.
\ee 

With these conditions, the superpotential $\cu$ depends only on $u_2$ and $u_1$. Using the form \eqref{separable-U} of the superpotential, equation \eqref{HJ-F} takes the trivially solvable form
\be
\pa_u\Big(e^{2u}\cf^2-\frac{144}{\ell^2}e^{-2u}-36e^{-u}-3q^2e^{u}\Big)=0\,,
\ee
and therefore
\bbxd
\vskip.4cm
\be\label{F_RN}
\cf=\frac{12}{\ell}e^{-2u}\sqrt{1+\frac{\ell^2}{4}e^{u}-\frac{m\ell^2}{16}e^{2u}+\frac{\ell^2}{48}e^{3u}q^2}\,,
\ee
\ebxd
where $m$ is an integration constant that will be identified with the mass parameter and the overall sign of $\cf$ is fixed by the asymptotic conditions.

Inserting \eqref{F_RN} in the first order equations \eqref{flow-eqs-F} we can determine the corresponding form of the metric and gauge potentials. The flow equation for $u$ can be written as 
\be
Ndr=-6e^{-2u}\cf^{-1}du\,,
\ee
which allows us to replace the radial coordinate $r$ with the variable $u$. In terms of $u$, the general solution of the flow equations for $u_1$ and $a$ is
\be
e^{-2u_1}=k_1\cf\,,\qquad a=-\frac{1}{\sqrt{3}}(a^1+a^2+a^3) = -3qk_1 e^u+k_2\,,
\ee
where $k_1$ and $k_2$ are integration constants. Inserting these expressions in the ansatz \eqref{ansatz} and setting $4e^{-u}=r^2$ (this amounts to choosing a function $N(r)$), and $k_1=1/6$ we obtain 
\be
ds^2=\frac{d r^2}{\Big(\frac{r^2}{\ell^2}+1-\frac{m}{r^2}+\frac{4q^2}{3r^4}\Big)}-\Big(\frac{r^2}{\ell^2}+1-\frac{m}{r^2}+\frac{4q^2}{3r^4}\Big)dt^2+r^2d\O_3^2\,,\qquad a=-\frac{2q}{r^2}+k_2\,,
\ee
where the metric on the round $S^3$ is given in \eqref{S3-metric}. This shows that the Reissner-Nordstr\"om-AdS$_5$ black hole is described by the superpotential \eqref{F_RN}.

\subsubsection{AdS$_2\times$S$^3(b)$ solutions and Sen's entropy function}

Another important example is the general attractor solution that arises in the near horizon limit of extremal black holes \cite{Larsen:2006xm}. It corresponds to AdS$_2$ of radius $\ell_2$ times a squashed 3-sphere with squashing parameter $b$ (see \eqref{squashedS3} in appendix \ref{sec:boundary}) and takes the form
\bbxd
\bal\label{attractor}
&N=\ell_2 r^{-1}\,,\quad e^{-2u_1}=\ell_2 e^{-3\bar u/2} r\,,\quad e^{-2u_2}=\ell_2 e^{\bar u/2} r\,,\quad a^I=e^I r\,,\quad u_4= \o r\,, \NO\\
&u=\bar u\,,\quad u_3=\bar u_3\,,\quad\vf_a=\bar\vf_a\,,\quad v^I=\bar v^I\,,
\eal
\ebxd
where $e^I$, $\o$, $\bar u$, $\bar u_3$, $\bar v^I$ and $\bar \vf_a$ are constants that are constrained by the equations of motion. The squashing parameter is determined by $\bar u_3$ through the relation $b^2=e^{3\bar u_3}$. In fact, the equations of motion that determine the values of these constants can be cast in the form of an extremization problem for a specific function -- Sen's entropy function \cite{Sen:2007qy,Sen:2008vm}.

Sen's entropy function is closely related with the 1D Hamiltonian \eqref{1DHamiltonian}. Like $H_{\rm 1D}$, the entropy function $\ce$ corresponds to the Legendre transform of the 1D Lagrangian \eqref{1DLagrangian}, but includes the first total derivative term in \eqref{S_B-ansatz}, namely\footnote{The fact that the total derivative from the Chern-Simons term should not be included is a well known subtlety, discussed extensively in \cite{Sen:2007qy,Choi:2008he}.}
\bal\label{EF0}
\ce=&\;\dot u_1\p_{u_1}+\dot u_2\p_{u_2}+\dot u_3\p_{u_3}+\dot u_4\p_{u_4}+\dot \vf_a\p_{\vf_a}+\dot a^I\p^a_I+\dot v^I\p^v_I-L_{\rm 1D}-\frac{4\p^2}{\k_5^2}\pa_r\Big(\frac{8}{N}e^{-2u_1}\dot u_1\Big)\NO\\
=&\;H_{\rm 1D}-\frac{4\p^2}{\k_5^2}\pa_r\Big(\frac{8}{N}e^{-2u_1}\dot u_1\Big)\,.
\eal
Using the 1D Lagrangian \eqref{1DLagrangian} and replacing the momenta $\p_{u_4}$ and $\p^a_I$ with the conserved charges \eqref{charges}, the entropy function \eqref{EF0} evaluated on the background \eqref{attractor} becomes
\bbxd
\bal\label{EF}
\ce=&\;\frac{4\p^2}{\k_5^2}\Big[4e^{-\frac32\bar u}+e^I\big(q_I-C_{IJK}\bar v^J\bar v^K\big)+\o\Big(j-\bar v^Iq_I+\frac13C_{IJK}\bar v^I\bar v^J\bar v^K\Big)\NO\\
&-\frac{1}{\ell_2^{2}}e^{-\frac52\bar u}\big(e^{2\bar u_3}\o^2+e^{\bar u}G_{IJ}(\bar \vf)e^I e^J\big)\NO\\
&+2\ell_2^2 e^{-\frac32\bar u}V(\bar \vf)+\ell_2^2e^{\frac12\bar u+2\bar u_3}G_{IJ}(\bar\vf)\bar v^I\bar v^J+\ell_2^2e^{-\frac12\bar u+\bar u_3}(e^{3\bar u_3}-4)\Big]\,.
\eal
\ebxd
This is the entropy function for general SU(2)$\times$U(1) invariant extremal black holes \cite{Morales:2006gm,Choi:2008he}.

Extremizing $\ce$ with respect to $\o$ and $e^I$, i.e. requiring that 
\be
\frac{\pa \ce}{\pa \o}=\frac{\pa \ce}{\pa e^I}=0\,,
\ee
reproduces the conservation laws \eqref{charges} in the form
\bbxd
\bal\label{velocities}
\o=&\;\frac{\ell_2^2}{2}e^{5\bar u/2-2\bar u_3}\Big(j-\bar v^Iq_I+\frac13C_{IJK}\bar v^I\bar v^J\bar v^K\Big)\,,\NO\\
e^I=&\;\frac{\ell_2^2}{2}e^{3\bar u/2}G^{IJ}(\bar\vf)(q_J-C_{JKL}\bar v^K\bar v^L)\,.
\eal
\ebxd

Inserting these expressions back in \eqref{EF} and extremizing with respect to the remaining variables, namely setting
\be\label{EF-extrema}
\frac{\pa\ce}{\pa \bar u}=\frac{\pa\ce}{\pa \ell_2}=\frac{\pa\ce}{\pa \bar u_3}=\frac{\pa\ce}{\pa \bar v^I}=\frac{\pa\ce}{\pa \bar \vf^a}=0\,,
\ee
leads to the attractor equations
\bbxd
\bal\label{attractor-eqs}
&e^{\bar u+\bar u_3}-\ell_2^{-2}-V(\bar\vf)=0\,,\NO\\
\rule{.0cm}{.6cm}&e^{2\bar u+2\bar u_3}G_{IJ}(\bar\vf)\bar v^I\bar v^J-\ell_2^{-4}e^{-\bar u+2\bar u_3}\o^2+2e^{\bar u+\bar u_3}(e^{3\bar u_3}-1)=0\,,\NO\\
\rule{.0cm}{.6cm}&\ell_2^{-4}G_{IJ}(\bar\vf)e^Ie^J+2\ell_2^{-4}e^{-\bar u+2\bar u_3}\o^2-e^{\bar u+4\bar u_3}-2\ell_2^{-2}=0\,,\NO\\
\rule{.0cm}{.6cm}&G_{IJ}(\bar\vf)\big(e^{2\bar u+2\bar u_3}\bar v^J-\ell_2^{-4}e^J\o\big)-\ell_2^{-2}e^{\frac32 \bar u}C_{IJK}e^J\bar v^K=0\,,\NO\\
\rule{.0cm}{.6cm}&\pa_{\vf^a}G_{IJ}(\bar\vf)\big(e^{2\bar u+2\bar u_3}\bar v^I\bar v^J-\ell_2^{-4}\pa_{\vf^a}e^Ie^J\big)+2\pa_{\vf^a}V(\bar\vf)=0\,,
\eal
\ebxd
where $\o$ and $e^I$ are evaluated using \eqref{velocities}. Approximate solutions to these algebraic equations are discussed in \cite{Morales:2006gm,Choi:2008he}. We will determine the general solution for supersymmetric black holes in section \ref{sec:thermodynamics}.

Here we instead highlight a remarkable connection between the entropy function $\ce$ and the superpotential $\cf$. From the expression \eqref{EF0} for the entropy function follows that
\be
\int dr\, \ce
=\int dr\Big[H_{\rm 1D}-\frac{4\p^2}{\k_5^2}\pa_r\Big(\frac{8}{N}e^{-2u_1}\dot u_1\Big)\Big]\,.
\ee
Moreover, Hamilton-Jacobi theory implies that (up to a constant)
\be
\int^{r}drL_{\text{1D}}=\cs_{\text{1D}}\,,
\ee
where $\cs_{\text{1D}}$ is Hamilton's principal function \eqref{HJ1d}. Hence, using the form of the attractor background \eqref{attractor} and the conservation equations \eqref{charges}, we obtain
\bal\label{H-F}
\int dr\, H_{\rm 1D}=&\;\int dr \big(\dot u_4\p_{u_4}+\dot a^I\p^a_I-L_{\rm 1D}\big) =u_4\p_{u_4}+a^I\p^a_I- \int dr\,L_{\rm 1D}\NO\\
=&\;-\frac{4\p^2}{\k_5^2}e^{-2u_2}\cf=-\frac{4\p^2}{\k_5^2}\ell_2 e^{u/2} r \cf\,,
\eal
together with
\be
-\frac{4\p^2}{\k_5^2}\int dr\pa_r\Big(\frac{8}{N}e^{-2u_1}\dot u_1\Big)=\frac{16\p^2}{\k_5^2}re^{-3u/2}\,.
\ee

Since also 
\be
\int dr\,\ce(\bar u,\ell_2,\bar u_3,\bar v^I,\bar \vf^a) =r\, \ce(\bar u,\ell_2,\bar u_3,\bar v^I,\bar \vf^a)\,,
\ee
we conclude that the entropy function $\ce$ and the superpotential $\cf$ are related as
\bbxd
\vskip.4cm
\be
\label{EF-F}
\ce=\frac{4\p^2}{\k_5^2}\big(4e^{-3u/2}-\ell_2 e^{u/2}\cf\big)\,.
\ee
\ebxd

Together with the expression \eqref{EF} for the entropy function, this relation allows us to read off the superpotential for the attractor solution \eqref{attractor}, namely
\bbxd
\bal\label{F-attractor}
\cf(u,u_3,\vf^a,v^I)=&\;-\ell_2\Big(\frac{1}{4}e^{u+2u_3}G^{IJ}(\vf)\pa_I\cf_0\pa_J\cf_0+\frac{1}{4}e^{2u}\cf_0^2\NO\\
&\;+2 e^{-2u}V(\vf)+e^{2u_3}G_{IJ}(\vf)v^Iv^J+e^{-u+u_3}(e^{3u_3}-4)\Big)\,,
\eal
\ebxd
where $\cf_0$ is defined in \eqref{F0-def}. It is straightforward to verify that this superpotential satisfies all flow equations \eqref{flow-eqs-F} for the attractor solution \eqref{attractor}, and hence solves the Hamilton-Jacobi equation \eqref{HJ-F}. The flow equations for $u'_3$, $v'^I$ and $\vf'^a$ follow immediately from the identification \eqref{EF-F} and the extremization conditions \eqref{EF-extrema}. Moreover, we see from \eqref{H-F} that the superpotential for the attractor solution is proportional to the 1D Hamiltonian $H_{\rm 1D}$, and hence vanishes at the extremum, i.e. $\cf=0$, which verifies the flow equation for $u'$. Finally, \eqref{EF-F} and the extremization conditions \eqref{EF-extrema} imply that at the extremum
\be
0=\pa_u\big(4e^{-3u/2}-\ell_2 e^{u/2}\cf\big)=-6e^{-3u/2}-\ell_2 e^{u/2}\pa_u\cf\,,
\ee
or
\be
\label{F-u-der-extremal}
\pa_u\cf=-6\ell_{2}^{-1}e^{-2u}\,,
\ee
and so at the extremum (where $\cf=0$) the flow equation for $u_2'$ becomes
\be
u_2'=\frac{1}{N}\dot u_2=\frac{1}{12}\pa_u(e^{2u}\cf)=-\frac{1}{2\ell_{2}}\,,
\ee
in agreement with the attractor solution \eqref{attractor}.

\subsection{Asymptotic solutions and the mass perturbation}
\label{sec:asymptotics}

The asymptotic form of the superpotential $\cu(u_1,u_2,u_3,\vf^a,v^I)$ plays a key role in understanding the physics of SU(2)$\times$U(1) invariant asymptotically locally AdS solutions. Asymptotic solutions of the superpotential equation \eqref{HJeq} or \eqref{HJ-F} are associated with a boundary condition, specified by the leading asymptotic behavior of the ansatz functions $u_1,u_2,u_3,\vf^a,v^I$. 

We will consider three distinct classes of asymptotic solutions. The first one corresponds to the most general asymptotic form of the fields consistent with AlAdS asymptotics and the SU(2)$\times$U(1) ansatz. The leading asymptotic form of the ansatz functions in this case is given in \eqref{leading-asymptotics}, with all sources arbitrary but restricted to the SU(2)$\times$U(1) ansatz. The second class of asymptotic superpotentials corresponds to solutions for which the sources for the scalars $\vf^a$ vanish, allowing only for nonzero expectation values. Finally, supersymmetry imposes certain relations among the sources on the boundary, which we derive in  appendix \ref{sec:boundary}. These relations provide a third boundary condition for the asymptotic solution of the superpotential equation \eqref{HJ-F}.

In this subsection we focus on the structure of the first two classes of asymptotic superpotentials and their normalizable deformations. The asymptotic form of the superpotential for supersymmetric solutions will be discussed separately, since it can be deduced from the exact supersymmetric superpotential we present in the next subsection.        

\subsubsection{Superpotential for asymptotically locally AdS solutions}

In order to specify what we mean by an asymptotic solution of the superpotential equation \eqref{HJ-F} we need to examine the leading asymptotic form of the fields for AlAdS solutions within the SU(2)$\times$U(1) ansatz. As we discuss in detail in appendix \ref{sec:holography}, the asymptotic form of the most general AlAdS solutions of the $\cn=2$ supergravity theory \eqref{5Daction} is specified by the set of arbitrary functions of the boundary coordinates, $g_{(0)ij}(x), A_{(0)i}^I(x),\vf^a_{(0)}(x)$, according to the leading asymptotic expansions \eqref{leading-asymptotics}. For solutions within the ansatz \eqref{ansatz}, the form of these sources is constrained such that they are parameterized by twelve constants $u_{1,2,3,4(0)}$, $a^I_{(0)}$, $v^I_{(0)}$, $\vf^a_{(0)}$. In particular, in the gauge $N=1$, the leading asymptotic form of the variables that parameterize the SU(2)$\times$U(1) ansatz \eqref{ansatz} is
\bbxd
\bal\label{ansatz-asymptotics-u}
&e^{-u_{1}} \sim e^{-u_{1(0)}}e^{2r/\ell}\,,\qquad 
e^{-u_{2}} \sim e^{-u_{2(0)}}\,,\qquad
e^{-u_{3}} \sim e^{-u_{3(0)}}\,,\qquad u_4\sim u_{4(0)}\NO\\
&a^I \sim a^I_{(0)}\,,\qquad v^I \sim v^I_{(0)}\,,\qquad \vf^{a} \sim \vf_{(0)}^{a}\frac{r}{\ell}e^{-2r/\ell}\,.
\eal
\ebxd

It is convenient to reparameterize the three functions $u_{1,2,3}$ in the metric ansatz so that the dual field theory sources become more apparent. Namely, we write 
\be\label{u-to-h}
e^{-u_{1}} = 2^{-3/2}(\mtrc_{1}\mtrc_{3})^{1/4}\mtrc_{2}^{1/2}\,,\qquad 
e^{-u_{2}} = \sqrt{2}\mtrc_{1}^{1/4}\mtrc_{2}^{-1/6}\mtrc_{3}^{-1/12}\,,\qquad
e^{-u_{3}} = (\mtrc_{2}/\mtrc_{3})^{1/3}\,,
\ee 
so that the functions 
\be
\label{h-to-u}
\mtrc_{1}=e^{-u_{1}-3u_{2}}\,,\qquad\mtrc_{2}=4e^{-u_{1}+u_{2}-u_{3}}\,,\qquad\mtrc_{3}=4e^{-u_{1}+u_{2}+2u_{3}}\,,
\ee
admit standard asymptotic expansions of the form 
\bbxd
\vskip.0cm
\be\label{ansatz-asymptotics-h}
\mtrc_{1,2,3} \sim \mtrc_{1,2,3(0)}e^{2r/\ell}\,,\quad u_4\sim u_{4(0)}\,,\quad a^I \sim a^I_{(0)}\,,\quad v^I \sim v^I_{(0)}\,,\quad \vf^{a} \sim \vf_{(0)}^{a}\frac{r}{\ell}e^{-2r/\ell}\,,
\ee
\ebxd
where the constants $h_{1,2,3(0)}$ parameterize components of the boundary metric $g_{(0)_{ij}}$.

The leading asymptotic form of the ansatz functions suggests that the corresponding asymptotic solution of the superpotential equation \eqref{HJ-F} can be sought in the form of an asymptotic expansion in $u$.\footnote{This can be viewed as a special case of the expansion in eigenfunctions of the operator $\d_\g$ in \eqref{gamma-derivative}, which was used to obtain the asymptotic solution of the general functional Hamilton-Jacobi equation \eqref{HJ-eqn}.}  The generic form of such an expansion turns out to be
\bbxd
\vskip.2cm
\be
\label{F-expansion}
\cf=e^{-2u}w_{(0)}+e^{-u}w_{(2)}+u\,\wt w_{(4)}+w_{(4)}+\co(e^{u})\,,
\ee
\ebxd
where all coefficients are independent of $u$. Notice that terms beyond the order shown in \eqref{F-expansion} vanish asymptotically and hence do not affect the physical observables, such as the conserved charges, free energy, or the holographic 1-point functions. Inserting the leading term in this expansion in the flow equations \eqref{flow-eqs-F} determines that $w_{(0)}=w_{(0)}(\vf)$. Moreover, inserting the expansion \eqref{F-expansion} in the superpotential equation \eqref{HJ-F} and collecting terms of the same order in $u$ leads to the following set of equations for the coefficients  
\bbxd
\bal\label{F-exp-eqs}
&(\pa_{a}w_{(0)})^2-\frac{2}{3}w_{(0)}^2-8V=0\,,\NO\\
\rule{.0cm}{.9cm}&\pa_{a}w_{(0)}\pa_a w_{(2)}
-\frac{1}{3}w_{(0)}w_{(2)}+\frac12e^{2u_3}G^{IJ}\pa_{v^I}w_{(2)}\pa_{v^J}w_{(2)}- 2e^{u_3}(e^{3u_3}-4)=0\,,\NO\\
\rule{.0cm}{.9cm}&\pa_a w_{(0)}\pa_a\wt w_{(4)}+e^{2u_3}G^{IJ}\pa_{v^I}w_{(2)}\pa_{v^J}\wt w_{(4)}=0\,,\NO\\
\rule{.0cm}{.9cm}&\pa_a w_{(0)}\pa_a w_{(4)}+e^{2u_3}G^{IJ}\pa_{v^I} w_{(2)}\pa_{v^J} w_{(4)}\NO\\
&+\frac{1}{3}w_{(0)}\wt w_{(4)}+\frac{1}{6}(\pa_{u_3}w_{(2)})^2+\frac{1}{2}(\pa_{a}w_{(2)})^2-2e^{2u_3}G_{IJ}v^Iv^J=0\,.
\eal
\ebxd

These equations govern all possible classes of asymptotic solutions we mentioned above. Starting from the first equation that determines $w_{(0)}(\vf)$, it is straightforward to show that there exist two distinct asymptotic branches (see 
appendix \ref{sec:counterterms} for a derivation), namely
\be\label{F-exp-BF}
w_{(0)}^{\rm source}(\vf)=\frac{12}{\ell}+\frac{2}{\ell}\vec\vf^2\Big(1+\frac{2}{\log \vec\vf^2}\Big)+\cdots\,,\qquad w_{(0)}^{\rm vev}(\vf)=\frac{12}{\ell}+\frac{2}{\ell}\vec\vf^2+\cdots\,.
\ee
Inserting these expressions in the flow equation for the scalars $\vf^a$ in \eqref{flow-eqs-F} we deduce that the $w_{(0)}^{\rm source}$ branch corresponds to asymptotic solutions with nonzero scalar sources $\vf^a_{(0)}$, while $w_{(0)}^{\rm vev}$ describes solutions with zero scalar sources. Notice that the ellipses in \eqref{F-exp-BF} stand for terms that are at least of the same order as $e^u$ asymptotically and so, as the $\co(e^u)$ terms in \eqref{F-expansion}, do not contribute to physical observables.  

The rest of the equations in \eqref{F-exp-eqs} can be solved similarly, again keeping only terms that are asymptotically nonzero. As we pointed out above and saw explicitly in the solution for the leading coefficient $w_{(0)}$, the form of the asymptotic solutions depends on the choice of boundary conditions, i.e. on the boundary sources. For example, noting that $e^{-u}\vec\vf^2=\co(u^2 e^{u})$ and demanding that the sources for $v^I$ are unconstrained, the flow equations \eqref{flow-eqs-F} together with the second equation in \eqref{F-exp-eqs} determine that $w_{(2)}$ is of the form 
\be
w_{(2)}=-\frac{\ell}{2}e^{u_3}(e^{3u_3}-4)+k_1 e^{-2u_3}G_{IJ}v^Iv^J+\cdots\,,
\ee
where $k_1$ takes the two possible values
\be
k_1=0\,,\qquad k_1=\frac{2}{\ell}\,.
\ee
Inserting this form of $w_{(2)}$ in the flow equation for $v^I$ in \eqref{flow-eqs-F} gives
\be
v'^I=-k_1v^I+\cdots\,,
\ee
and so the two values of $k_1$ correspond to the two independent modes of $v^I$. This is analogous to the two branches for $w_{(0)}$ in \eqref{F-exp-BF} that correspond to the two independent scalar modes. In this case, the value $k_1=0$ corresponds to a nonzero source $v_{(0)}^I$, i.e. $v^I\sim v_{(0)}^I$, while $k_1=2/\ell$ gives $v^I=\co(e^{u})$. Therefore, the term proportional to $k_1$ in $w_{(2)}$ is either identically zero ($k_1=0$) or asymptotically zero ($k_1=2/\ell$), and so in either case it can be dropped. 

Using analogous arguments we find that the third equation in \eqref{F-exp-BF} does not impose any constraint on the relevant asymptotic form of $\wt w_{(4)}$. Similarly the last equation in \eqref{F-exp-BF} does not determine the asymptotic form of $w_{(4)}$, but it does determine that of $\wt w_{(4)}$. Namely, 
\be
\wt w_{(4)}=-\frac{\ell^3}{6}e^{2u_3}(e^{3u_3}-1)^2+\frac{\ell}{2}e^{2u_3}G_{IJ}v^Iv^J+\cdots\,,
\ee   
where again the ellipses denote terms that asymptote to zero. 

Collecting the above results, we find that for generic sources within the SU(2)$\times$U(1) invariant ansatz, the asymptotic form of the reduced superpotential $\cf$ is 
\bal\label{F-asymptotics-general}
\cf =&\; 
\frac{12}{\ell}e^{-2u}\bigg[1+\frac{1}{6}\vec{\vf}^{2}\Big(1+\frac{2}{\log \vec\vf^2}\Big)
-\frac{\ell^2}{24}e^{u+u_{3}}(e^{3u_{3}}-4)\NO\\
&\hspace{3.0cm}-\frac{\ell^4}{24} u\,e^{2u+2u_3}\Big(\frac13 (e^{3u_3}-1)^2-\frac{1}{\ell^2}v^Iv^I\Big)\bigg]+\Hat w_{(4)}+\cdots\,,
\eal
where $\Hat w_{(4)}$ is asymptotically constant but otherwise undetermined by the asymptotic analysis alone. We have distinguished this from $w_{(4)}$ in \eqref{F-expansion} since in general the asymptotic expansion of the coefficients $w_{(0)}$, $w_{(2)}$ and $\wt w_{(4)}$ may contribute additional finite terms. In appendix \ref{sec:counterterms} we derive the boundary counterterms required in order to holographically renormalize the full $\cn=2$ theory \eqref{5Daction}. For solutions within the SU(2)$\times$U(1) invariant ansatz these reduce to the expression in \eqref{counterterms-ansatz-F}, which can alternatively be deduced directly from the asymptotic form of the superpotential for generic boundary sources in \eqref{F-asymptotics-general}. Moreover, inserting this asymptotic superpotential in the flow equations \eqref{flow-eqs-F} one can derive the full asymptotic expansions of the ansatz variables.

\subsubsection{Superpotential mass perturbation}

Given a solution of the superpotential equation \eqref{HJeq} or \eqref{HJ-F}, it is often important to consider perturbative deformations of this superpotential. For example, this is necessary for studying the thermodynamics of extremal black holes through the corresponding near extremal solutions \cite{Castro:2018ffi}. A key property of the superpotential is that such perturbations are necessarily normalizable \cite{Papadimitriou:2006dr,Papadimitriou:2007sj}. This follows from Hamilton-Jacobi theory: normalizable perturbations correspond to the generalized initial momenta that parameterize a complete integral, while non-normalizable perturbations correspond to the initial positions and arise from integrating the flow equations.  

We can see this explicitly in the present context as follows. Let $\cf$ be a given solution of the superpotential equation \eqref{HJ-F}. A small perturbation $\D\cf$ around the superpotential $\cf$ is governed by the linearized version of the superpotential equation \eqref{HJ-F}, namely
\be\label{JHeq-linear}
\frac{1}{3}(\pa_u+2)(\cf\D\cf)+\frac{1}{3}\pa_{u_3}\cf\pa_{u_3}\D\cf+\pa_{a}\cf\pa_{a}\D\cf+e^{-u+2u_3}G^{IJ}\pa_{v^I}\cf\pa_{v^J}\D\cf=0\,.
\ee
Using the asymptotic form of $\cf$ in \eqref{F-expansion}, we see that the first term in this equation dominates asymptotically, leading to an asymptotic solution of the form 
\bbxd
\vskip.4cm
\be\label{mass-pert-F}
\D\cf=-\frac{9\D m}{2} \big(e^{-2u}\cf^{-1}+\cdots\big)\sim -\frac{3\ell\D m}{8}\,,
\ee
\ebxd
where $\D m$ is an arbitrary integration constant. The overall factor has been chosen judicially such that $\D m$ coincides with the variation of the mass parameter $m$ in the superpotential \eqref{F_RN} that describes the Reissner-Nordstr\"om-AdS$_5$ black hole. The perturbation \eqref{mass-pert-F} is therefore normalizable and corresponds to a perturbation of the coefficient $w_{(4)}$ in \eqref{F-expansion}.   

The expression \eqref{mass-pert-F} for the mass perturbation of the superpotential is in general only valid asymptotically near the conformal boundary. For example, as we show in section \ref{sec:thermodynamics}, $\cf$ diverges on non-extremal black hole horizons and so the linearized approximation for $\D\cf$ breaks down there. Determining the superpotential perturbation $\D\cf$ in a form that is uniformly valid in the bulk -- as would be required e.g. for relating the mass perturbation $\D m$ to the black hole temperature -- is a nontrivial problem that is particularly relevant for the study of near extremal black holes. Addressing this problem for solutions of the STU model is not required for our current purposes, but an example of such an analysis for the near horizon region of the near extremal Kerr-AdS$_5$ black hole can be found in \cite{Castro:2018ffi}.

Let us conclude this subsection with the observation that the mass parameter $\D m$ in \eqref{mass-pert-F} is analogous to the angular momentum parameter $j$ and the electric charge parameters $q^I$. All these parameters enter in the Hamilton principal function \eqref{HJ1d} together with the variable that transforms under the corresponding symmetry. In particular, $j$ is associated with constant shifts in $u_4$, $q^I$ with constant shifts in $a^I$, and $\D m$ with constant shifts in $u_2$. Moreover, as we show in appendix \ref{sec:holography} (see eq.~\eqref{T-vevs}), the time-time component of the holographic stress tensor is conjugate to $u_2$ (at fixed $u$), and hence proportional to $\D m$.

\subsection{The 1/4 BPS superpotential and supersymmetric black holes}
\label{sec:exactBPSsuperpotential}

We are now ready to present our main result, namely an exact superpotential for all 1/4 BPS solutions in the SU(2)$\times$U(1) invariant sector of the STU model \eqref{5Daction}. This superpotential describes a much wider class of supersymmetric solutions than those previously known in the literature and provides a tool for analytically studying black holes that are presently only known numerically. As we will see, the thermodynamics of these black holes and their holographic observables can be directly obtained from the exact superpotential, without explicit knowledge of the solution.   

The 1/4 BPS superpotential is an exact solution of the HJ equation \eqref{HJ-F} given by
\bbxd
\vskip.4cm
\be\label{BPS-superpotential-F}
\cf_{\rm BPS}=8\sqrt{c_1^2+c_2^2}\,,
\ee
\ebxd
where $c_1$ and $c_2$ are the following functions of the SU(2)$\times$U(1) invariant ansatz variables:
\vskip-.05cm
\bbxd
\bal\label{c1&2-main}
c_1\equiv&\;\frac12e^{-2u}W-\frac{\projsp}{8}\big(\cf_0-6e^{-u+u_3}L_Iv^I\big)\,,\NO\\
c_2\equiv&\;\frac{1}{4}e^{-\frac32u-u_3}\Big(e^{3u_3}+2-2\projsp\ell^{-1}\sum_Iv^I\Big)-\frac{1}{8}e^{-\frac12u+u_3}L^I\pa_{v^I}\cf_0\,.
\eal
\ebxd
In these expressions, $\projsp=\pm 1$ corresponds to a choice of spinor projection -- see eq.~\eqref{eigenspinor} and $\cf_0$ is the quantity defined in \eqref{F0-def}. Although \eqref{BPS-superpotential-F} could be obtained by solving the superpotential equation \eqref{HJ-F}, we derived it through a systematic analysis of the bulk Killing spinor equations, which we present in appendix \ref{sec:KS}. This analysis also allows us to determine the bulk Killing spinor for all 1/4 BPS solutions, and helps understand their structure.    

A key property of all 1/4 BPS solutions that follows from the exact superpotential \eqref{BPS-superpotential-F} is that they admit four additional integrals of motion, besides those corresponding to the angular momentum and electric charges in  \eqref{charges}. These extra constants of motion are
\bbxd
\bal\label{constants-of-motion-main}
\mathscr C_{4} \equiv &\; u_{4}-\projsp e^{-2u_{2}-u_{3}}\cos2g\,,\NO \\
\mathscr C^{I} \equiv &\; a^{I}-u_{4}v^{I}+\projsp e^{-2u_{2}}(e^{-u_{3}}\cos2g\,v^{I}-e^{-\frac{1}{2}u}\sin2g\,L^{I})\,,
\eal
\ebxd
where 
\be\label{g-def}
g\equiv-\frac{\projsp}{2}\arcsin\Big(\frac{c_2}{\sqrt{c_1^2+c_2^2}}\Big)\,.
\ee
As we explain in appendix \ref{sec:KS}, using the BPS superpotential  in \eqref{BPS-superpotential-F} in the flow equations \eqref{flow-eqs-F}, one can verify that the quantities \eqref{constants-of-motion-main} are constants of motion. These supersymmetric conservation laws will play a crucial role in the discussion of BPS black hole thermodynamics in section \ref{sec:thermodynamics}. The constants $\mathscr C^{I}$ in \eqref{constants-of-motion-main} determine also the time-dependence of the bulk Killing spinor, which takes the form 
\bbxd
\vskip.2cm
\be\label{KSbulk}
\e=e^{-\frac14 u-u_2+ig\G_3}e^{i\o_0 t}\e_0\,,\qquad \o_{0}=-\frac{1}{2\ell}\sum_{I}\mathscr C^{I}\,, \qquad \e_0=\text{constant spinor}\,.
\ee
\ebxd

The superpotential \eqref{BPS-superpotential-F} may be parameterized in an alternative way that will prove very useful for our analysis in section \ref{sec:thermodynamics}. We start by introducing the complex variables
\bbxd
\vskip.2cm
\be\label{c-complex}
\cz_\pm=c_1\pm ic_2\,,
\ee
\ebxd
in terms of which the superpotential \eqref{BPS-superpotential-F} becomes
\bbxd
\vskip.3cm
\be\label{BPS-superpotential-F-complex}
\cf_{\rm BPS}=8\sqrt{\cz_+\cz_-}\,.
\ee
\ebxd
Moreover, the function $g$ in \eqref{g-def} can be expressed as
\bbxd
\vskip.3cm
\be\label{g-def-complex}
e^{-2i\projsp g}= \sqrt{\cz_+/\cz_-}\,.
\ee
\ebxd
Notice that when the ansatz functions (and hence $c_1$ and $c_2$) are real the variables $\cz_\pm$ are complex conjugates. In that case we can write
\be\label{complex-real}
\cz_+\equiv \cz\,,\qquad \cz_-=\lbar \cz\,,\qquad \cz=|\cz|e^{-2i\projsp g}\,,\qquad \text{real solutions only}\,.
\ee
However, the more general parameterization \eqref{c-complex} allows us to accommodate complexified solutions, which we consider in section \ref{sec:thermodynamics}.

Using the form of $c_1$, $c_2$ in \eqref{c1&2-main} we can express $\cz_\pm$ in terms of the complex variables
\bbxd
\vskip.2cm
\be\label{V-variables}
V_{\pm}^{I}=v^{I}\pm i\projsp e^{-\frac{1}{2}u+u_{3}}L^{I}\,,
\ee
\ebxd
as
\bbxd
\vskip.1cm 
\be\label{Zpm}
\cz_{\pm}=\frac{\projsp}{8}e^{-u_{3}}\Big(\sum_{I}V_{\pm}^{I}\big(q_{I}\mp4i\ell^{-1}e^{-\frac{3}{2}u}\big)-\frac{1}{3}C_{IJK}V_{\pm}^{I}V_{\pm}^{J}V_{\pm}^{K}-\big(j\mp4i\projsp e^{-\frac{3}{2}u}\big)\Big)\,.
\ee
\ebxd
Like $\cz_\pm$, the variables $V_\pm^I$ are complex conjugates when the ansatz functions are real, so in that case we may define 
\be\label{V-variables-real}
V^I\equiv V^I_+\,,\qquad V^I_-=\lbar V_+\,,\qquad \text{real solutions only}\,.
\ee

The expressions \eqref{Zpm} for $\cz_\pm$ are remarkable for two reasons. Firstly, $\cz_+$ and $\cz_-$ are holomorphic in respectively $V^I_+$ and $V^I_-$. Secondly, all dependence of $\cz_\pm$ on $v^I$ and on the scalars $\vf^a$ is absorbed in the new variables $V^I_\pm$. This allows us to eliminate the five variables $v^I$ and $\vf^a$ in favor of the six new variables $V^I_\pm$, subject to the condition \eqref{LI-constraint} that constrains the imaginary parts of $V^I_\pm$. In particular, the derivatives in the two bases are related as
\bbxd
\begin{eqnarray}\label{V-derivatives}
\pa_{v^{I}} & = & \pa_{V_{+}^{I}}+\pa_{V_{-}^{I}}\,,\NO \\
\pa_{\vf^{a}} & = &\projsp e^{-\frac{1}{2}u+u_{3}}(\pa_{\vf^{a}}L^{I})i(\pa_{V_{+}^{I}}-\pa_{V_{-}^{I}})\,,\NO\\
\pa_{u}|_{v^{I},\vf^{a}} & = & \pa_{u}|_{V_{+}^{I},V_{-}^{I}}-\frac{\projsp}{2}e^{-\frac{1}{2}u+u_{3}}L^{I}i(\pa_{V_{+}^{I}}-\pa_{V_{-}^{I}})\,,\NO \\
\pa_{u_{3}}|_{v^{I},\vf^{a}} & = & \pa_{u_{3}}|_{V_{+}^{I},V_{-}^{I}}+\projsp e^{-\frac{1}{2}u+u_{3}}L^{I}i(\pa_{V_{+}^{I}}-\pa_{V_{-}^{I}})\,.
\end{eqnarray}
\ebxd
In the following we will use both sets of variables depending on the context.

\subsubsection{Asymptotic expansions and supersymmetric sources}

The superpotential \eqref{BPS-superpotential-F} allows us to determine the asymptotic expansions for the ansatz variables directly from the flow equations \eqref{flow-eqs-F}. To this end, we expand the superpotential \eqref{BPS-superpotential-F} as in \eqref{F-expansion} and read off the coefficients
\bbxd
\bal\label{F-exp-BPS}
w_{(0)} = &\; 4W\,,\NO \\
w_{(2)} =&\; \frac12 e^{-2u_3}Wx^2+6\projsp e^{u_{3}}L_{I}v^{I}\,,\NO \\
\wt{w}_{(4)} =&\; 0\,,\NO \\
w_{(4)} =&\; -\frac{1}{32}e^{-4u_{3}}Wx^{4}-\frac{3\projsp}{4}e^{-u_{3}}L_{I}v^{I}x^{2}-\frac{1}{2}xL^{I}\pa_{v^{I}}\cf_{0}-\projsp cf_{0}\,,
\eal
\ebxd
where
\be
x\equiv \frac{e^{3u_{3}}+2-2\projsp \ell^{-1}\sum_Iv^I}{W}\,.
\ee

One may verify that the expressions \eqref{F-exp-BPS} satisfy equations \eqref{F-exp-eqs}, as is guaranteed by the fact that \eqref{BPS-superpotential-F} is an exact solution of the superpotential equation \eqref{HJ-F}. However, notice that they do not have the general form \eqref{F-asymptotics-general} that ensures that all sources are arbitrary and independent. We therefore expect that the field theory sources for supersymmetric solutions described by the superpotential \eqref{F-exp-eqs} are not all arbitrary. As we will see momentarily, the constrains these sources satisfy can be deduced by looking at the leading asymptotic form of the flow equations \eqref{flow-eqs-F}. This reproduces the result obtained by solving the conformal supergravity Killing spinor equations on the boundary, which we present in appendix \ref{sec:boundary}. 

Since the reduced superpotential $\cf$ depends only on the variables $u$, $u_{3}$, $v^{I}$ and $\vf^{a}$, the flow equations for these variables form a closed set of coupled equations. The flow equations \eqref{flow-eqs-F} then determine the asymptotic expansion for $u_2$, $u_4$ and $a^I$ by direct integration:   
\bal\label{integral-relations}
u_{2}=&\; \frac{1}{12}\int d\r\; \pa_u(e^{2u}\cf)\,,\NO\\
u_4 = &\; \frac12 \int d\r\;e^{2u-2u_2-2u_3}\Big(j-v^Iq_I+\frac13C_{IJK}v^Iv^Jv^K\Big)\,,\NO\\
a^I=&\; \int d\r\;\Big(u_4v'^I+\frac12e^{u-2u_2}G^{IJ}(q_J-C_{JKL}v^Kv^L)\Big)\,,
\eal
where the radial coordinate $\r$ was introduced in \eqref{prime}. Alternatively, the expansions of $u_4$ and $a^I$ can be derived from the integrals of motion \eqref{constants-of-motion-main}. In the gauge $N=1$, in which case $\r=r$, the asymptotic expansions of the variables $u$, $u_{3}$, $v^{I}$ and $\vf^{a}$ take the form \cite{Bianchi:2001kw,Sahoo:2010sp,Genolini:2016ecx,Papadimitriou:2017kzw,Cassani:2018mlh}
\bal\label{FG-expansions-BPS}
e^{-u} = &\; e^{-u_{(0)}}e^{2r/\ell}+e^{-u}_{(2)}+\big(e^{-u}_{(4,2)}r^2/\ell^{2}+e^{-u}_{(4,1)}r/\ell+e^{-u}_{(4,0)}\big)e^{-2r/\ell}+\cdots\,,\NO \\
\rule{.0cm}{.5cm}e^{u_{3}} = &\; e^{u_{3(0)}}+e^{u_{3}}_{(2)}e^{-2r/\ell}+\big(e^{u_{3}}_{(4,2)}r^2/\ell^{2}+e^{u_{3}}_{(4,1)}r/\ell+e^{u_{3}}_{(4,0)}\big)e^{-4r/\ell}+\cdots\,,\NO \\
\rule{.0cm}{.5cm}v^{I} = &\; v_{(0)}^{I}+\big(v_{(2,1)}^{I}r/\ell+v_{(2,0)}^{I}\big)e^{-2r/\ell}+\cdots\,,\NO \\
\rule{.0cm}{.5cm}\vf^{a} = &\; \big(\vf_{(0)}^{a}r/\ell+\Hat\vf_{(0)}^{a}\big)e^{-2r/\ell}+\cdots\,,
\eal
where all coefficients are independent of the radial coordinate.

Inserting these expansions in the flow equations \eqref{flow-eqs-F} and expanding $\cf$ as in \eqref{F-expansion} with coefficients \eqref{F-exp-BPS}, we determine the expansion coefficients in \eqref{FG-expansions-BPS} up to the normalizable modes. We find that the leading order coefficients are arbitrary, except for the constraints
\bbxd
\vskip.4cm
\be\label{susy-from-the-bulk}
v_{(0)}=\frac{\projsp \ell}{\sqrt{3}}(e^{3u_{3(0)}}-1)\,,\qquad\vf_{(0)}^{a}=-\projsp \ell e^{u_{(0)}+u_{3(0)}}\wt{v}_{(0)}^{a}\,,
\ee
\ebxd
where we have separated the gravity and vector multiplet variables $v^I$ according to \eqref{gauge-field-decomposition} as 
\be\label{v-bases}
v=-\frac{1}{\sqrt{3}}(v^1+v^2+v^3)\,,\qquad \wt v^1=\frac{1}{\sqrt{6}}(v^1+v^2-2v^3)\,,\qquad \wt v^2=\frac{1}{\sqrt{2}}(v^1-v^2)\,.
\ee
As we demonstrate in appendix \ref{sec:boundary} (see eqs.~\eqref{eq:v-squashing} and \eqref{b-susy-vector}), these constraints ensure that the field theory background on the boundary is supersymmetric. 

At higher orders, all coefficients before the normalizable modes in the expansions \eqref{FG-expansions-BPS} are uniquely determined and are given by
\bal
&v_{(2,1)}=-\frac{\projsp \ell^{3}}{2\sqrt{3}}e^{u_{(0)}+4u_{3(0)}}(e^{3u_{3(0)}}-1)\,,\qquad\wt{v}_{(2,1)}^{a}=\frac{\projsp \ell}{2}e^{3u_{3(0)}}\vf_{(0)}^{a}\,,\\
\rule{.0cm}{.7cm}
&e^{-u}_{(2)}=\frac{\ell^{2}}{24}e^{u_{3(0)}}(e^{3u_{3(0)}}-4)\,,\quad e^{-u}_{(4,2)}=-\frac{1}{12}e^{-u_{(0)}}\vec{\vf}_{(0)}^{2}\,,\quad e^{-u}_{(4,1)}=\frac{1}{24}e^{-u_{(0)}}\vec{\vf}_{(0)}\cdot(\vec{\vf}_{(0)}-4\vec{\Hat\vf}_{(0)})\,,\NO\\
\rule{.0cm}{.7cm}
&e^{u_{3}}_{(2)}=\frac{\ell^{2}}{6}e^{u_{(0)}+2u_{3(0)}}(1-e^{3u_{3(0)}})\,,\quad e^{u_{3}}_{(4,2)}=0\,,\quad
e^{u_{3}}_{(4,1)}  =  \frac{1}{12}e^{u_{3(0)}}\big(\ell^{4}e^{2u_{(0)}+5u_{3(0)}}(e^{3u_{3(0)}}-1)-\vec{\vf}_{(0)}^{2}\big)\,.\NO
\eal
However, the normalizable modes $\Hat\vf_{(0)}^{a}$, $v_{(2,0)}$, $\wt{v}_{(2,0)}^{a}$, $e^{u_{3}}_{(4,0)}$, $e^{-u}_{(4,0)}$ are not completely determined, except for the following relations among them
\bbxd
\bal\label{BPS-modes}
&\Hat\vf_{(0)}^{a}-\frac{2\projsp}{\ell}e^{-3u_{3(0)}}\wt{v}_{(2,0)}^{a}=-\frac{3}{2}e^{u_{(0)}-2u_{3(0)}}\big(q_I-C_{IJK}v_{(0)}^Jv_{(0)}^K\big)\pa_aL_I|_{\vf=0}\,,\NO\\
\rule{.0cm}{.9cm}
&e^{u_{3}}_{(4,0)}+\frac{\projsp\ell}{2\sqrt{3}}e^{u_{(0)}+2u_{3(0)}}\Big(v_{(2,0)}-\frac{1}{2\sqrt{3}}\Hat\vf_{(0)}^{a}\wt{v}_{(0)}^{a}\Big) = \frac{\ell^{4}}{48}e^{2u_{(0)}+3u_{3(0)}}(e^{6u_{3(0)}}-3e^{3u_{3(0)}}+2)\NO\\
&+\frac{\projsp\ell}{24}e^{2u_{(0)}}\Big(j-v_{(0)}^{I}q_{I}+\frac{1}{3}C_{IJK}v_{(0)}^{I}v_{(0)}^{J}v_{(0)}^{K}\Big)- \frac{1}{48}e^{u_{3(0)}}\vec{\vf}_{(0)}^{2}\,,\hspace{-.2cm}\NO\\
\rule{.0cm}{.9cm}&e^{-u}_{(4,0)}+\frac{1}{96}e^{-u_{(0)}}\Big(\vec{\vf}_{(0)}^{2}-4\vec{\vf}_{(0)}\cdot\vec{\Hat\vf}_{(0)}+8\vec{\Hat\vf}_{(0)}^{2}\Big) = \frac{\ell^{4}}{96}e^{u_{(0)}+5u_{3(0)}}\Big(1-\frac{5}{8}e^{3u_{3(0)}}\Big)\NO \\
&+\frac{\projsp\ell}{24}e^{u_{(0)}-u_{3(0)}}\Big(j-v_{(0)}^{I}q_{I}+\frac{1}{3}C_{IJK}v_{(0)}^{I}v_{(0)}^{J}v_{(0)}^{K}\Big)-\frac{\ell^{2}}{48}e^{u_{(0)}+2u_{3(0)}}\sum_{I}q_{I}\,.
\eal
\ebxd
It follows that the normalizable modes $\wt{v}_{(2,0)}^{a}$, $e^{u_{3}}_{(4,0)}$ and $e^{-u}_{(4,0)}$ can be expressed in terms of the two unconstrained coefficients $\Hat\vf_{(0)}^{a}$ and $v_{(2,0)}$. These are the only modes not determined by supersymmetry alone and fixing their value requires input from the interior of the bulk geometry that is not accessible through the asymptotic solutions. To obtain their values it is necessary to know the solution of the flow equations \eqref{flow-eqs-F} associated with the BPS superpotential \eqref{BPS-superpotential-F} in a form that is uniformly valid for all values of the radial coordinate. As an example, we will determine these coefficients in the Gutowski-Reall black hole below. 

Finally, inserting the expansions \eqref{FG-expansions-BPS} in the relations \eqref{integral-relations}, we obtain the expansions of the variables $u_2$, $u_4$ and $a^I$, namely
\bal\label{FG-expansions-aux}
e^{-u_{2}}= &\; e^{-u_{2(0)}}\Big[1-\frac{\ell^{2}}{48}e^{u_{(0)}+u_{3(0)}}(e^{3u_{3(0)}}-4)e^{-2r/\ell}\NO\\
&\hskip-1.3cm-\frac{1}{12}\Big(\frac{r}{2\ell}\vec{\vf}_{(0)}^{2}+12e^{u_{(0)}}e^{-u}_{(4,0)}+\vec{\Hat\vf}_{(0)}^{2}-\frac{\ell^{4}}{128}e^{2u_{(0)}+2u_{3(0)}}\big((e^{3u_{3(0)}}-4)^2-32\ell^{-2}v_{(0)}^2\big)\Big)e^{-4r/\ell}+\cdots\Big]\,,\NO\\
\rule{.0cm}{.7cm}u_{4}=&\; u_{4(0)}-\frac{\ell}{8}e^{2u_{(0)}-2u_{2(0)}-2u_{3(0)}}\Big(j-v_{(0)}^{I}q_{I}+\frac{1}{3}C_{IJK}v_{(0)}^{I}v_{(0)}^{J}v_{(0)}^{K}\Big)e^{-4r/\ell}+\cdots\,,\\
\rule{.0cm}{.7cm}a^{I}=&\;a_{(0)}^{I}+\Big(u_{4(0)}v_{(2,1)}^Ir/\ell+u_{4(0)}v^I_{(2,0)}-\frac{\ell}{4}e^{u_{(0)}-2u_{2(0)}}(q_{I}-C_{IJK}v_{(0)}^{J}v_{(0)}^{K})\Big)e^{-2r/\ell}+\cdots\,.\NO
\eal

\subsubsection{Supersymmetric one-point functions}

The supersymmetric boundary sources \eqref{susy-from-the-bulk} ensure that the dual field theory is supersymmetric, but provide no information on whether the state, which is specified by the expectation values of the dual operators, preserves supersymmetry or not. Requiring that the state be supersymmetric results in a set of BPS relations among the 1-point functions of the gravity and vector multiplet operators, which we derive in appendix \ref{sec:boundary} using purely field theoretic arguments. We now evaluate the holographic 1-point functions of the conformal current and vector multiplet operators on the boundary following from the 1/4 BPS superpotential \eqref{BPS-superpotential-F} and examine whether they satisfy the BPS relations derived in appendix \ref{sec:boundary}. In section \ref{sec:thermodynamics}, we will utilize these 1-point functions in order to determine the conserved charges of BPS black holes and the BPS relations they satisfy. 

The computation of the holographic 1-point functions for generic AlAdS solutions of the STU model is discussed in detail in appendix \ref{sec:holography}. In particular, holographic renormalization is summarized in appendix \ref{sec:counterterms} and the renormalized 1-point functions are defined in \eqref{1pt-generic}. These results are then specialized to SU(2)$\times$U(1) invariant solutions, for which the local counterterms can be expressed in terms of the `counterterm superpotential' given in \eqref{counterterms-ansatz-F}, while all 1-point functions are determined in terms of the renormalized superpotential and its derivatives through the expressions \eqref{T-vevs}, \eqref{J-vevs} and \eqref{scalar-vevs}. These results apply to any SU(2)$\times$U(1) invariant solutions, both supersymmetric and non-supersymmetric. 

When applying the results of appendix \ref{sec:holography} to supersymmetric solutions described by the exact superpotential \eqref{BPS-superpotential-F}, it is important to keep in mind that the counterterms \eqref{counterterms-ansatz-F} are not identical to the asymptotic form of the supersymmetric superpotential \eqref{BPS-superpotential-F}. This is because, as we pointed out above, the counterterms follow from the asymptotic superpotential for arbitrary boundary sources, which we obtained above in \eqref{F-asymptotics-general}, while the asymptotic form of the supersymmetric superpotential is dictated by the constraints \eqref{susy-from-the-bulk} on the sources. Nevertheless, as we now discuss, the two asymptotic superpotentials do coincide, precisely due to these constraints.        

The renormalized superpotential is obtained by adding the counterterms \eqref{counterterms-ansatz-F} to the asymptotic expansion of the superpotential that describes the solutions we are interested in. The result is again of the form \eqref{F-expansion}, but with the renormalized coefficients
\bbxd
\bal\label{F-exp-ren}
w_{(0)}^{\rm ren}=&\;w_{(0)}-\frac{12}{\ell}-\frac{2}{\ell}\vec{\vf}^{2}\Big(1-\frac{\ell}{2r_0^2}(r_0-\a_3)\Big)\,,\NO\\
w_{(2)}^{\rm ren}=&\;w_{(2)}+\frac{\ell}{2}e^{u_3}(e^{3u_3}-4)\,,\NO\\
\wt w_{(4)}^{\rm ren}=&\;\wt w_{(4)}+\frac{\ell}{2}e^{2u_3}\Big(\frac{\ell^2}{3} (e^{3u_3}-1)^2-v^Iv^I\Big)\,,\NO\\
w_{(4)}^{\rm ren}=&\;w_{(4)}-\a_2\frac{\ell^2}{3}e^{2u_3} (e^{3u_3}-1)^2\NO\\
&+\frac{1}{3}e^{2u_3}\Big(\Big(\frac{3\ell}{2}u_{(0)}+\a_2+2\a_3\Big)\d_{JK}+(\a_2-
\a_3)\sum_IC_{IJK}\Big)v^Jv^K\,,
\eal
\ebxd
where $w_{(0)}$, $w_{(2)}$, $\wt w_{(4)}$ and $w_{(4)}$ are the coefficients in the asymptotic expansion \eqref{F-expansion} of the superpotential describing the solution. For supersymmetric solutions described by the BPS superpotential \eqref{BPS-superpotential-F} these are given in \eqref{F-exp-BPS}. Notice that the cancellation of asymptotic divergences need not -- and in general does not -- take place order by order in this expansion. As we show in the case of supersymmetric solutions, if the boundary sources are constrained, the cancellation of asymptotic diverges occurs between terms of different orders in the asymptotic expansion in $u$. 

For supersymmetric solutions described by the BPS superpotential \eqref{BPS-superpotential-F}, the coefficients \eqref{F-exp-ren} can be evaluated explicitly using \eqref{F-exp-BPS}. Expanding further in the scalars and keeping only relevant terms, we obtain
\bal\label{F-exp-ren-BPS}
w_{(0)}^{\rm ren}=&\;\frac{1}{r_0^2}\vec{\vf}^{2}(r_0-\a_3)+\cdots\,,\NO\\
w_{(2)}^{\rm ren}=&\;\frac{2\ell}{3} e^{-2u_3}\Big(e^{3u_{3}}-1+\projsp\ell^{-1}\sum_Iv^I\Big)^2+6\projsp e^{u_{3}}\vf^a\pa_aL_I v^I+\cdots\,,\NO\\
\wt w_{(4)}^{\rm ren}=&\;\frac{\ell}{2}e^{2u_3}\Big(\frac{\ell^2}{3} (e^{3u_3}-1)^2-v^Iv^I\Big)\,,\NO\\
w_{(4)}^{\rm ren}=&\;-\frac{\ell^3}{32\times 27}e^{-4u_{3}}\Big(e^{3u_{3}}+2-2\projsp\ell^{-1}\sum_Iv^I\Big)^{4}\NO\\
&-\frac{\projsp\ell^2}{36}e^{-u_{3}}\Big(e^{3u_{3}}+2-2\projsp\ell^{-1}\sum_Iv^I\Big)^{2}\sum_{J}v^{J}\NO\\
&+\frac{\ell}{6}e^{-u_3}\Big(e^{3u_{3}}+2-2\projsp\ell^{-1}\sum_Iv^I\Big)\sum_J(q_J-C_{JKL}v^Kv^L)\NO\\
&-\projsp e^{-u_3}\Big(j-v^Iq_I+\frac13C_{IJK}v^Iv^Jv^K\Big)-\a_2\frac{\ell^2}{3}e^{2u_3} (e^{3u_3}-1)^2\NO\\
&+\frac{1}{3}e^{2u_3}\Big(\Big(\frac{3\ell}{2}u_{(0)}+\a_2+2\a_3\Big)\d_{JK}+(\a_2-
\a_3)\sum_IC_{IJK}\Big)v^Jv^K+\cdots\,.
\eal

As we emphasized, the cancellation of divergences among these coefficients is nontrivial and relies on the constraints \eqref{susy-from-the-bulk} that supersymmetric sources satisfy. These may be expressed in covariant form in terms of fields on the radial cutoff as 
\bbxd
\vskip.2cm
\be\label{susy-from-the-bulk-cov}
e^{3u_{3}}-1+\projsp\ell^{-1}\sum_Iv^I=\co(e^{u})\,,\quad \frac{1}{r_0}e^{-u}\vf^a+\projsp e^{u_3}\,\wt v^a=\frac{1}{r_0}e^{-u_{(0)}}\Hat\vf_{(0)}^a+\co(e^u)\,.
\ee
\ebxd
In fact, in order to evaluate the derivatives $\pa_{u_3}\cf_{\rm ren}$ and $\pa_{v^I}\cf_{\rm ren}$ we will need the subleading terms in the gravity multiplet constraint in \eqref{susy-from-the-bulk-cov} that can be deduced from the asymptotic expansions \eqref{FG-expansions-BPS}. These allow us to show that
\bal\label{constraint-1}
&\;4\ell e^{-u+u_3}\Big(e^{3u_{3}}-1+\projsp \ell^{-1}\sum_Jv^J\Big)-2r_0\ell^2e^{5u_3} (e^{3u_3}-1)\NO\\
=&\;4\projsp e^{-u_{(0)}+u_{3(0)}}\sum_{J}v_{(2,0)}^J-2\ell^3e^{5u_{3(0)}}(e^{3u_{3(0)}}-1)+\cdots\,.
\eal

In addition, using the identities
\bal\label{identities}
&v^I=3\wt v^a\pa_a L_I|_{\vf=0}+\frac{1}{3} \sum_J v^J=3\wt v^a\pa_a L_I|_{\vf=0}-\frac{\projsp\ell}{3} (e^{3u_3}-1)+\co(e^u)\,,\NO\\
&v^Iv^I=\frac13\sum_{I,J}v^Iv^J+\wt v^a\wt v^a\,,\qquad 3\vf^av^I\pa_aL_I|_{\vf=0} =\vf^a\wt v^a+\co(\vf^2)\,,
\eal
the constraints \eqref{susy-from-the-bulk-cov} lead to the following set of asymptotic relations 
\bal\label{constraint-ids}
&e^{2u_3}\wt v^a\wt v^a=\frac{1}{r_0^2}e^{-2u}\vec\vf^2-\frac{2}{r_0}e^{-2u_{(0)}}\vec\vf_{(0)}\cdot\vec{\Hat\vf}_{(0)}\ell^{-1}+\co(u^{-2})\,,\NO\\
&3\projsp e^{-u+u_3}\vf^a\pa_aL_I v^I=-\frac{1}{r_0}e^{-2u}\vec\vf^2+e^{-2u_{(0)}}\vec\vf_{(0)}\cdot\vec{\Hat\vf}_{(0)}\ell^{-1}+\co(u^{-1})\,,\NO\\
&e^{2u_3}\Big(\frac{\ell^2}{3} (e^{3u_3}-1)^2-v^Iv^I\Big)=-\frac{1}{r_0^2}e^{-2u}\vec\vf^2+\frac{2}{r_0}e^{-2u_{(0)}}\vec\vf_{(0)}\cdot\vec{\Hat\vf}_{(0)}\ell^{-1}+\co(u^{-2})\,.
\eal
These relations facilitate the cancellation of asymptotic divergences among the coefficients \eqref{F-exp-ren-BPS}, and help us simplify their asymptotic form. However, they should only be used {\em after} applying any derivatives to the renormalized superpotential.

The holographic 1-point functions for general SU(2)$\times$U(1) invariant solutions are given in \eqref{T-vevs}, \eqref{J-vevs} and \eqref{scalar-vevs} in appendix \ref{sec:holography} in terms of the renormalized superpotential $\cf_{\rm ren}$ and its derivatives. In order to evaluate these 1-point functions for BPS solutions described by the superpotential \eqref{BPS-superpotential-F}, we start with the expansion of the renormalized superpotential specified by the coefficients \eqref{F-exp-ren-BPS}, evaluate its derivatives and then simplify the result using the relations \eqref{susy-from-the-bulk-cov}, \eqref{constraint-1}, \eqref{identities}, \eqref{constraint-ids}, as well as  
\be\label{decID}
\sum_IC_{IJK}v_{(0)}^{J}v_{(0)}^{K}=2\big(v_{(0)}^{2}v_{(0)}^{3}+v_{(0)}^{1}v_{(0)}^{3}+v_{(0)}^{1}v_{(0)}^{2}\big)
=2v_{(0)}^2-\wt v_{(0)}^a \wt v_{(0)}^a\,.
\ee
In particular, these imply the relations
\bal
&e^{2u_{3(0)}}\sum_IC_{IJK}v_{(0)}^Jv_{(0)}^K+e^{-2u_{(0)}}\ell^{-2}\vec\vf_{(0)}^2-\frac{2\ell^2}{3}e^{2u_{3(0)}}(e^{3u_{3(0)}}-1)^2=0\,,\\
&e^{2u_{3(0)}}\sum_JC_{IJK}v_{(0)}^K+\frac{2\ell \projsp}{3}e^{2u_{3(0)}}(e^{3u_{3(0)}}-1)-3\projsp e^{-u_{(0)}+u_{3(0)}}\ell^{-1}\vf_{(0)}^a\pa_aL_I|_{\vf=0}=0\,,\NO
\eal
which we use to simplify the contribution of the finite counterterms. Eventually we obtain
\bbxd
\bal\label{F-ren-BPS-derivatives}
\cf_{\rm BPS}^{\rm ren}=&\;\frac{\ell^3}{32}(5e^{8u_3}-8e^{5u_{3}})+\frac{\ell}{2}e^{2u_3}\sum_J(q_J-C_{JKL}v^Kv^L)\NO\\
&\hspace{-0.cm}-\projsp e^{-u_3}\Big(j-v^Iq_I+\frac13C_{IJK}v^Iv^Jv^K\Big)+\cdots\,,\NO\\
\rule{.0cm}{.8cm}\pa_u\cf_{\rm BPS}^{\rm ren}=&\;-\frac{2}{\ell}e^{-2u_{(0)}}\vec\vf_{(0)}\cdot\big(\vec{\Hat\vf}_{(0)}-\a_3\ell^{-1}\vec\vf_{(0)}\big)+\cdots\,,\NO\\
\rule{.0cm}{.8cm}\pa_{\vf^a}\cf_{\rm BPS}^{\rm ren}=&\;\frac{2}{r_0}e^{-2u_{(0)}+2r_0/\ell}\big(\Hat\vf^a_{(0)}-\a_3\ell^{-1}\vf^a_{(0)}\big)+\co(1/r_0^2)\,,\NO\\
\rule{.0cm}{.8cm}\pa_{u_3}\cf_{\rm BPS}^{\rm ren}=&\;4\projsp e^{-u_{(0)}+u_{3(0)}}\sum_{J}v_{(2,0)}^J-\frac{\ell^3}{4}\Big(7+\frac{8\a_2}{\ell}\Big)e^{5u_{3(0)}}(e^{3u_{3(0)}}-1)\NO\\
&\hspace{-0.8cm}+\projsp e^{-u_3}\Big(j-v^Iq_I+\frac13C_{IJK}v^Iv^Jv^K\Big)-\frac{2}{\ell}e^{-2u_{(0)}}\vec\vf_{(0)}\cdot\big(\vec{\Hat\vf}_{(0)}-\a_3\ell^{-1}\vec\vf_{(0)}\big)+\cdots\,,\NO\\
\rule{.0cm}{.8cm}\pa_{v^I}\cf_{\rm BPS}^{\rm ren}=&\;\frac{\projsp}{3\ell}\Big[4\projsp e^{-u_{(0)}-2u_{3(0)}}\sum_{J}v_{(2,0)}^J-\ell^3\Big(1+\frac{2\a_2}{\ell}\Big)e^{2u_{3(0)}}(e^{3u_{3(0)}}-1)\Big]\NO\\
&\hspace{-0.cm}+\projsp e^{-u_3}\sum_J\Big(\d_{IJ}-\frac13\Big)\big(q_J-C_{JKL}v^Kv^L\big)\NO\\
&\hspace{-0.cm}+6\projsp e^{-u_{(0)}+u_{3(0)}}\Big(\Hat\vf_{(0)}^a-\Big(\frac{\ell}{2}+\a_3\Big)\ell^{-1}\vf_{(0)}^a\Big)\pa_aL_I|_{\vf=0}+\cdots.
\eal
\ebxd

As we discuss in appendix \ref{sec:holography}, the 1-point functions \eqref{T-vevs}, \eqref{J-vevs} and \eqref{scalar-vevs} for generic  SU(2)$\times$U(1) invariant solutions trivially satisfy all Ward identities \eqref{WardIDs}, except for the trace Ward identity, which within the SU(2)$\times$U(1) sector reduces to the condition \eqref{trace-ansatz} on the renormalized superpotential $\cf_{\rm ren}$. An immediate consequence of the expressions \eqref{F-ren-BPS-derivatives} is that this condition is satisfied for all BPS solutions, due to the fact that $\pa_u\cf_{\rm BPS}^{\rm ren}=-\vf^a\pa_{\vf^a}\cf_{\rm BPS}^{\rm ren}$ and the constraints \eqref{susy-from-the-bulk} on the supersymmetric sources imply that the Weyl anomaly vanishes on supersymmetric backgrounds.  

The expressions \eqref{F-ren-BPS-derivatives} for the renormalized BPS superpotential enable us to evaluate all supersymmetric 1-point functions. Those of certain components of the energy-momentum tensor and of the U(1) currents, given respectively in \eqref{T-vevs} and \eqref{J-vevs}, are independent of the superpotential $\cf$ and so have universal form, applicable to both supersymmetric and non-supersymmetric solutions. Namely, for all solutions
\bbxd
\bal\label{T-vevs-BPS-universal}
\<\ct^{t}{}_{\j}\> = &\;  \sec\th\,\<\ct^t{}_\f\>= -\frac{1}{4\k_{5}^{2}}e^{2u_{1(0)}}\Big(j-v_{(0)}^Iq_I+\frac13C_{IJK}v_{(0)}^Iv_{(0)}^Jv_{(0)}^K\Big)\,,\NO \\
\rule{.0cm}{.7cm}\<\cj_{I}^{t}\> = &\; \frac{1}{4\k_{5}^{2}}e^{2u_{1(0)}}\Big(q_{I}-\frac{1}{3}C_{IJK}v_{(0)}^{J}v_{(0)}^{K}\Big)\,.
\eal
\ebxd

For the components of the energy-momentum tensor that depend on the superpotential, inserting the results \eqref{F-ren-BPS-derivatives} in \eqref{T-vevs} we determine that for supersymmetric solutions
\bbxd
\bal\label{T-vevs-BPS}
\<\ct^{t}{}_{t}\>_{\rm BPS} = &\;\big(u_{4(0)}-\projsp e^{-2u_{2(0)}-u_{3(0)}}\big)\<\ct^{t}{}_{\j}\>_{\rm BPS}\NO\\
&\hspace{-0.cm}-\frac{1}{4\k_{5}^{2}}e^{2u_{(0)}}\Big(\frac{\ell^3}{32}(5e^{8u_{3(0)}}-8e^{5u_{3(0)}})+\frac{\ell}{2}e^{2u_{3(0)}}\sum_J(q_J-C_{JKL}v_{(0)}^Kv_{(0)}^L)\Big)\,,\NO \\
\rule{.0cm}{.7cm}\<\ct^{\j}{}_{\f}\>_{\rm BPS} = &\; -\cos\th\Big[\big(u_{4(0)}-\projsp e^{-2u_{2(0)}-u_{3(0)}}\big)\<\ct^{t}{}_{\j}\>_{\rm BPS}-\vf_{(0)}^a\<\co_{a}\>_{\rm BPS}\NO\\
&\hspace{-.0cm}+\frac{1}{4\k_{5}^{2}}e^{2u_{(0)}}\Big(4\projsp e^{-u_{(0)}+u_{3(0)}}\sum_{J}v_{(2,0)}^J-\frac{1}{4}(7\ell+8\a_2)\ell^2e^{5u_{3(0)}}(e^{3u_{3(0)}}-1)\Big)\Big]\,,\NO\\
\rule{.0cm}{.7cm}\<\ct^{\th}{}_{\th}\>_{\rm BPS} = &\;\rule{.0cm}{.7cm}\<\ct^{\f}{}_{\f}\>_{\rm BPS}= -\frac13\big(\<\ct^{t}{}_{t}\>_{\rm BPS}+\sec\th\<\ct^{\j}{}_{\f}\>_{\rm BPS}+2\vf_{(0)}^a\<\co_{a}\>_{\rm BPS}\big)\,.
\eal
\ebxd
The 1-point functions of the remaining nonzero components of the energy-momentum tensor are then obtained algebraically using the general relations in \eqref{T-vevs}.

The $\j$-component of the U(1) currents is similarly determined by inserting the expression for $\pa_{v^I}\cf^{\rm ren}_{\rm BPS}$ given in \eqref{F-ren-BPS-derivatives} in the corresponding 1-point function in \eqref{J-vevs}. This gives
\bbxd
\bal\label{J-vevs-BPS}
\<\cj_{\text{cov}\,I}^{\j}\>_{\rm BPS} = &\;-\big(u_{4(0)}-\projsp e^{-2u_{2(0)}-u_{3(0)}}\big)\<\cj_{\text{cov}\,I}^{t}\>+3\projsp\ell e^{u_{(0)}+u_{3(0)}}\<\co_a\>_{\rm BPS}\pa_aL_I|_{\vf=0}\NO\\
&\hspace{-.5cm}+\frac{\projsp}{12\k_{5}^{2}\ell}e^{2u_{(0)}-3u_{3(0)}}\Big(4\projsp e^{-u_{(0)}+u_{3(0)}}\sum_{J}v_{(2,0)}^J-(\ell+2\a_2)\ell^2e^{5u_{3(0)}}(e^{3u_{3(0)}}-1)\NO\\
&\hspace{-.5cm}-\ell e^{2u_{3(0)}}\sum_J\big(q_J-C_{JKL}v_{(0)}^Kv_{(0)}^L\big)\Big)-\frac{3\projsp}{4\k_{5}^{2}}e^{u_{(0)}+u_{3(0)}}\vf^a_{(0)}\pa_a L_I|_{\vf=0}\,,
\eal
\ebxd
where the covariant form of the U(1) currents \eqref{cov-currents} for SU(2)$\times$U(1) invariant solutions is given in \eqref{cov-currents-STU} and we have used the first identity in \eqref{identities} together with the fact that 
\be
\sum_JC_{IJK}\pa_a L_K|_{\vf=0}=-\pa_a L_I|_{\vf=0}\,,\quad \Leftrightarrow\quad  \pa_a L_1|_{\vf=0}+\pa_a L_2|_{\vf=0}+\pa_a L_3|_{\vf=0}=0\,.
\ee

Finally, from \eqref{scalar-vevs} and \eqref{F-ren-BPS-derivatives} follows that the supersymmetric 1-point functions of the vector multiplet scalar operators are given by
\bbxd
\vskip.4cm
\be\label{scalar-vevs-BPS}
\<\co_{a}\>_{\rm BPS} =  \frac{1}{2\k_{5}^{2}\ell}\big(\Hat\vf^a_{(0)}-\a_3\ell^{-1}\vf^a_{(0)}\big)\,.
\ee
\ebxd

As we pointed out above, supersymmetry imposes restrictions not only on the form of the dual field theory background -- see \eqref{susy-from-the-bulk} -- but also on the 1-point functions of the current and vector multiplet operators. The relations on the 1-point functions following from supersymmetry hold on supersymmetric states of the theory and their violation indicates spontaneous breaking of supersymmetry. In appendix \ref{sec:boundary} we derive the general form of the BPS relations that supersymmetric 1-point functions must satisfy, using a purely field-theoretic analysis. These relations follow from the rigid supersymmetry transformations of the fermionic current operators in the theory and are given in \eqref{BPS-relations-VEVs}.

In section \ref{sec:thermodynamics} we will confirm that the holographic 1-point functions obtained above satisfy an integrated version of the BPS relation following from the supersymmetry variation of the $\cn=1$ supercurrent on the boundary. Here, we focus on the supersymmetry variation of the fermionic currents in the vector multiplets. The 1-point function \eqref{J-vevs-BPS} can be written as 
\be\label{J-vevs-BPS-compact}
\frac{1}{\ell}\ck_{(0)i}\<\wt\cj^i_{a\,{\rm cov}}\>\sbtx{BPS}-e^{-2u_{2(0)}+2u_{3(0)}}\<\co_a\>\sbtx{BPS}=-\frac{1}{4\k_5^2\ell}e^{-2u_{2(0)}+2u_{3(0)}}\vf^a_{(0)}\,,
\ee
where $\ck_{(0)i}$ is the Killing vector given in \eqref{KV} in appendix \ref{sec:boundary} and satisfies 
\bal
\ck_{(0)i}\<\wt\cj^i_{a\,{\rm cov}}\>\sbtx{BPS}=&\;g_{(0)ij}\ck_{(0)}^i\<\wt\cj^j_{a\,{\rm cov}}\>\sbtx{BPS}\\
=&\;\big(g_{(0)tt}\ck_{(0)}^t+g_{(0)t\j}\ck_{(0)}^\j\big)\<\wt\cj^t_{a\,{\rm cov}}\>\sbtx{BPS}+\big(g_{(0)t\j}\ck_{(0)}^t+g_{(0)\j\j}\ck_{(0)}^\j\big)\<\wt\cj^\j_{a\,{\rm cov}}\>\sbtx{BPS}\NO\\
=&\;\projsp e^{-u_{1(0)}-u_{2(0)}+u_{3(0)}}\big(\<\wt\cj^\j_{a\,{\rm cov}}\>\sbtx{BPS}+(u_{4(0)}-\projsp e^{-2u_{2(0)}-u_{3(0)}})\<\wt\cj^t_{a\,{\rm cov}}\>\sbtx{BPS}\big)\,.\NO
\eal

Inserting the expression \eqref{J-vevs-BPS-compact} in the supersymmetry transformation of the fermionic currents in the vector multiplets given in \eqref{BPS-relations-VEVs} we determine that
\bbxd
\bal\label{SSB}
i\bar\e_{(0)}^{\,+}\d_\e\<j^\l_a\>\sbtx{BPS}
=&\;\frac{1}{\ell}\ck_{(0)i}\<\wt\cj^i_{a\,{\rm cov}}\>\sbtx{BPS}-e^{-2u_{2(0)}+2u_{3(0)}}\<\co_a\>\sbtx{BPS}\NO\\
=&\;-\frac{1}{4\k_5^2\ell}e^{-2u_{2(0)}+2u_{3(0)}}\vf^a_{(0)}\,.
\eal
\ebxd
In particular, the BPS relation \eqref{BPS-scalar-VEVs} is violated in the presence of a nonzero scalar source for the vector multiplets. This result shows that any supergravity solution with nonzero supersymmetric vector multiplet sources spontaneously breaks supersymmetry. Notice that nonzero scalar sources result also in a nonzero trace for the stress tensor, as follows from the trace Ward identity in \eqref{WardIDs}. Recall that the Weyl anomaly is numerically zero on supersymmetric backgrounds. We should emphasize that the result \eqref{SSB} is independent of the choice of supersymmetric scheme, since the holographic 1-point functions have been determined with an arbitrary choice of supersymmetric finite local counterterms.

\subsubsection{Integrating the BPS flow equations}

The asymptotic analysis of supersymmetric solutions described by the BPS superpotential \eqref{BPS-superpotential-F} shows that it encompasses a much wider class of solutions than the known black holes reviewed in section \ref{sec:5D-STU}. However, it is instructive to demonstrate with a couple of examples how these specific solutions are recovered by integrating the flow equations \eqref{flow-eqs-F} following from the BPS superpotential \eqref{BPS-superpotential-F}.

Starting with the Gutowski-Reall black hole \eqref{GR}, we can write it in terms of the ansatz variables \eqref{ansatz} and the functions $h_{1,2,3}$ introduced in \eqref{h-to-u} as
\bbxd
\bal\label{GR-ansatz}
&N^{2} = \frac{1}{f\X\D_{r}}\,,\quad
\mtrc_{1} = f^2+\frac14u_4^2 h_3\,,\quad
\mtrc_{2} = \frac{r^{2}}{f\X}\,,\quad
\mtrc_{3} = \frac{r^2\D_r}{f\X}-\ell^2h^2f^2\,,\quad
u_{4} = -\frac{2\ell hf^2}{h_3}\,,\NO\\
&a^{I} = -H_{I}^{-1}\,,\quad
v^{I} = \frac{\ell}{2}\big(1-\D_r+\ell^{-2}\m_{I}-H_I^{-1}h\big)\,,\quad
L^{I} = \frac{1}{f H_{I}}\,,
\eal
\ebxd
where the quantities $h$, $\D_r$, $f$, $H_I$ and $\X$ are defined in \eqref{GR-functions-1} and \eqref{GR-functions-2}. It can be readily checked that these solve the flow equations \eqref{flow-eqs-F} obtained from the BPS superpotential \eqref{BPS-superpotential-F} with $\projsp=1$ provided the charge parameters $q_I$ and $j$ \eqref{charges} are given by\footnote{Alternatively, the Gutowski-Reall solution \eqref{GR-ansatz} can be parameterized in terms of $j$ and $q_I$ by writing 
\be
\m_I=\frac{\ell j-\ell^2q_{I}}{q_{I}-\sum_Jq_J+\ell^2/2}\,,
\ee
and expressing $a$ in terms of $j$ and $q_I$ through \eqref{constraintKLR-gamma}. In this parameterization, the regularity constraint \eqref{nonlinear-constraint} must also be imposed on the charge parameters $j$ and $q_I$. This is the parameterization of the Gutowski-Reall solution obtained by integrating the flow equations \eqref{flow-eqs-F} following from the BPS superpotential \eqref{BPS-superpotential-F}.}
\bbxd
\vskip.4cm
\be
q_I=\frac{\g_3}{2\ell^2\m_I}-\frac{\g_2}{4\ell^2}-\frac{\m_I}{2}\,,\qquad j=\frac{\g_2}{4\ell}+\frac{\g_3}{2\ell^3}\,,
\ee
\ebxd
where 
\be
\g_1=\m_1+\m_2+\m_3\,,\qquad \g_2=\m_2\m_3+\m_1\m_3+\m_1\m_2\,,\qquad \g_3=\m_1\m_2\m_3\,,
\ee
and the parameters $a$, $\m_I$ satisfy the relation \eqref{constraintKLR}, i.e.
\be\label{constraintKLR-gamma}
\g_1+\ell^2=\frac{(2a+\ell)^2}{\X}\,.
\ee
Notice that $\m_I$ correspond to the three roots of the cubic equation
\be
\m_I^3-\g_1\m^2_I+\g_2\m_I-\g_3=0\,,\qquad I=1,2,3\,.
\ee

The nonzero coefficients in the asymptotic expansions \eqref{FG-expansions-BPS} and \eqref{FG-expansions-aux} of the Gutowski-Reall solution take the form
\bbxd
\bal\label{GR-FG-coeffs}
&e^{-u_{(0)}} = \frac{\ell^{2}}{4\X}\,,\quad e_{(2)}^{-u} = -\frac{\ell^{2}}{8}\,,\quad
e_{(4,0)}^{-u} =  \frac{\X}{6\ell}\Big(\frac{5\g_2}{8\ell}+\frac{\g_3}{2\ell^3}+\frac{\ell \g_1}{4}+\frac{3\ell^{3}}{32}-\frac{\g_{1}^{2}}{12\ell}\Big)\,,\NO \\
&e^{u_{3(0)}} = 1\,,\quad
e_{(4,0)}^{u_{3}} = \frac{\X^{2}\g_{3}}{3\ell^{6}}\,,\quad v_{(2,0)}^{I} = \frac{\X}{2\ell^3}\Big(\frac{\g_3}{\m_I}-\frac{\g_2}{2}\Big)\,,\quad \Hat{\vf}_{(0)}^{a} = \frac{3\X}{\ell^{2}}\m_{I}\pa_{a}L_{I}|_{\vf=0}\,,\NO\\
&e^{-2u_{2(0)}}=\frac{2}{\ell}\,,\quad u_{4(0)}=\frac{2}{\ell}\,,\quad a_{(0)}^{I}=-1\,.
\eal
\ebxd
In particular, using \eqref{v-bases} and the identity $\sum_I\m_I^{-1}=\g_2/\g_3$ we find that the three coefficients $v_{(2,0)}$ and $\Hat\vf^a_{(0)}$ that remain undetermined by the asymptotic analysis take the values 
\bbxd
\vskip.4cm
\be
v_{(2,0)}=\frac{1}{\sqrt{3}}\frac{\X\g_2}{4\ell^3}\,, \qquad \Hat\vf^a_{(0)}=\frac{3\X}{\ell^{2}}\m_{I}\pa_{a}L_{I}|_{\vf=0}\,.
\ee
\ebxd

It is straightforward to verify that the expressions \eqref{GR-FG-coeffs} satisfy the relations among the expansion coefficients required by supersymmetry, namely the constraints \eqref{susy-from-the-bulk} among the sources (trivially) and those on the normalizable modes in \eqref{BPS-modes}, which now simplify to
\bal
\Hat{\vf}_{(0)}^{a}-2\ell^{-1}\wt{v}_{(2,0)}^{a} = &\; -\frac{3}{2}e^{u_{(0)}}q_{I}\pa_{a}L_{I}|_{\vf=0}\,,\NO \\
e_{(4,0)}^{u_{3}}+\frac{\ell}{2\sqrt{3}}e^{u_{(0)}}v_{(2,0)} = &\; \frac{\ell}{24}e^{2u_{(0)}}j\,,\NO \\
e_{(4,0)}^{-u}+\frac{1}{12}e^{-u_{(0)}}\vec{\Hat{\vf}}_{(0)}^{2} = &\; \frac{\ell}{24}e^{u_{(0)}}\Big(\frac{3\ell^3}{32}+j-\frac{\ell}{2}\sum_I q_I\Big)\,.
\eal

Finally, using the expansion coefficients \eqref{GR-FG-coeffs} we evaluate the BPS 1-point functions \eqref{T-vevs-BPS-universal}, \eqref{T-vevs-BPS}, \eqref{J-vevs-BPS} and \eqref{scalar-vevs-BPS} corresponding to the Gutowski-Reall black hole:
\bbxd
\bal
&\<\ct^{t}{}_{\j}\>_{\rm GR} = \sec\th\,\<\ct^t{}_\f\>_{\rm GR}= -\frac{\X^2}{\k_{5}^{2}\ell^4}\Big(\frac{\g_2}{2}+\frac{\g_3}{\ell^2}\Big)\,,\qquad \<\ct^{\j}{}_{t}\>_{\rm GR} = -\frac{8\X^2}{\k_{5}^{2}\ell^2}\Big(\frac{1}{8}+\frac{\g_1}{3\ell^2}\Big)\,,\NO\\
\rule{.0cm}{.7cm}&\<\ct^{t}{}_{t}\>_{\rm GR} = \frac{\X^2}{\k_{5}^{2}\ell}\Big(\frac{3}{8}+\frac{\g_1}{\ell^2}+\frac{\g_2}{2\ell^4}\Big)\,,\qquad \<\ct^{\th}{}_{\th}\>_{\rm GR} = \<\ct^{\f}{}_{\f}\>_{\rm GR}= -\frac{\X^2}{\k_{5}^{2}\ell}\Big(\frac{1}{8}+\frac{\g_1}{3\ell^2}+\frac{\g_2}{2\ell^4}\Big)\,,\NO \\
\rule{.0cm}{.7cm}&\<\ct^{\j}{}_{\j}\>_{\rm GR} =  -\frac{\X^2}{\k_{5}^{2}\ell}\Big(\frac{1}{8}+\frac{\g_1}{3\ell^2}-\frac{\g_2}{2\ell^4}\Big)\,,\qquad \<\ct^{\j}{}_{\f}\>_{\rm GR} =  \frac{\X^2\g_2}{\k_{5}^{2}\ell^5}\cos\th\,,
\eal
\ebxd
\bbxd
\vskip.4cm
\be
\<\cj_{I}^{t}\>_{\rm GR} =  \frac{\X^2}{\k_{5}^{2}\ell^3}\Big(\frac{\g_3}{\ell^2\m_I}-\frac{\g_2}{2\ell^2}-\m_I\Big)\,,\qquad \<\cj_{I}^{\j}\>_{\rm GR} = \frac{2\X^2}{\k_{5}^{2}\ell^4}\m_I\,,
\ee
\ebxd
\bbxd
\vskip.4cm
\be
\<\co_{a}\>_{\rm GR} =  \frac{3\X}{2\k_{5}^{2}\ell^3}\m_{I}\pa_{a}L_{I}|_{\vf=0}\,.
\ee
\ebxd
It can be easily verified that these satisfy the Ward identities \eqref{WardIDs}, as well as the BPS relations \eqref{BPS-relations-VEVs} required of a supersymmetric state.

As a second example let us consider the general non-supersymmetric CLP solution \eqref{sol:CLP}. In terms of the ansatz \eqref{ansatz} it takes the form
\bbxd
\bal\label{CLP-ansatz}
&N^{2}=\frac{r^{2}R}{Y}\,,\qquad\mtrc_{1}=\frac{RY}{f_{1}}\,,\qquad\mtrc_{2}=R\,,\qquad\mtrc_{3}=\frac{f_{1}}{R^{2}}\,,\qquad u_{4}=-2\frac{f_{2}}{f_{1}}\,,\NO\\
&a^{I}=\frac{2m s_{I}c_{I}}{r^{2}H_{I}}\,,\qquad v^{I}=\frac{m a}{r^{2}H_{I}}(c_{I}s_{J}s_{K}-s_{I}c_{J}c_{K})\,,\qquad I\neq J\neq K\,,
\eal
\ebxd
where the charge parameters $j$ and $q_I$ are given by
\bbxd
\vskip.2cm
\be\label{CLP-charge-parameters}
q_{I} = -m\,c_{I}s_{I}\,,\qquad j = ma\,\big(c_{1}c_{2}c_{3}-s_{1}s_{2}s_{3}\big)\,.
\ee
\ebxd

As we saw in section \ref{sec:5D-STU}, a four-parameter family of BPS solutions that contain closed timelike curves is obtained by setting 
\be\label{CLP-BPS1-ansatz}
\frac{a}{\ell}=\projsp \exp\Big(-\sum_I\d_I\Big)\,,
\ee
where now we allow for the two possible spinor projections corresponding to $\projsp=\pm1$. It can be checked that this solution satisfies the flow equations \eqref{flow-eqs-F} following from the 1/4 BPS superpotential \eqref{BPS-superpotential-F}. Regular BPS solutions -- namely the Gutowski-Reall black holes discussed above -- are obtained by imposing also the condition \eqref{CLP-BPS2}, which ensures that the charge parameters \eqref{CLP-charge-parameters} satisfy the constraint \eqref{nonlinear-constraint}. In section \ref{sec:thermodynamics} we show that this constraint follows from the extremization of the BPS superpotential \eqref{BPS-superpotential-F} on the horizon.

\section{Black hole thermodynamics}
\label{sec:thermodynamics}

We now turn to a discussion of black hole thermodynamics. AdS black hole thermodynamics is equivalent to that of the dual field theory \cite{Papadimitriou:2005ii}. However, Chern-Simons terms and extremal or BPS black holes present certain subtleties. After a brief review of some basic aspects of black hole thermodynamics for asymptotically locally AdS black holes, we focus on black holes with an SU(2)$\times$ U(1) isometry, discussing  non-supersymmetric and BPS black holes separately. As we will show, in both cases the superpotential $\cf$ is directly related with the free energy and an entropy extremization principle.

\subsection{General STU black holes}
\label{sec:STUthermo}

The holographic dictionary, which we discuss in detail in appendix \ref{sec:holography},  allows us to express most thermodynamic variables directly in field theory language, without explicit reference to the bulk. In particular, all conserved charges follow from field theory Ward identities and can be expressed in terms of the corresponding Noether currents on the boundary. Of course, the expectation values of these currents -- which determine the charges --   do depend on the bulk geometry. The same holds for the free energy, which corresponds to a field theory partition function, as well as the temperature and other chemical potentials that can be identified with certain holonomies on the boundary. However, the black hole entropy is given by the area of the horizon and does not seem directly expressible in terms of boundary data alone. One of the main goals of this section is to demonstrate a connection between the entropy and the superpotential $\cf$, which admits a direct field theory interpretation.    

\subsubsection{Electric charges}

The electric charges can be expressed in terms of the U(1) currents  \eqref{1pt-generic} dual to the bulk gauge fields \cite{Papadimitriou:2005ii}. However, this identification becomes more subtle in the presence of Chern-Simons terms \cite{Marolf:2000cb}. An important consequence of such terms is that there exists no choice of a U(1) current that is simultaneously gauge invariant and conserved on a general field theory background. Of course, this should not come as a surprise given that bulk Chern-Simons terms are holographically dual to axial anomalies \cite{Witten:1998qj}.      

In order to incorporate the effects of Chern-Simons terms in the definition of the electric charges, let us consider the one-parameter family of currents \cite{Papadimitriou:2017kzw}  
\be\label{U(1)-current-family}
\<\cj_{I}^{i\anom}\>=\<\cj_{I}^{i}\>-\frac{\anom}{24\k_{5}^{2}}C_{IJK}\epsilon_{(0)}^{ijkl}A_{(0)j}^{J}F_{(0)kl}^{K}\,,
\ee
where the constant $\anom$ is arbitrary for now. These currents are gauge invariant only for $\anom = -2$, in which case they coincide with the covariant currents  \eqref{cov-currents}. However, the U(1) Ward identities in \eqref{WardIDs} imply that 
\be\label{R-current-div}
D_{(0)i}\<\cj_{I}^{i\anom}\>=(1-\anom)\ca_{I(0)}\,,
\ee
where $\ca_{I(0)}$ are the R-symmetry and flavor anomalies defined in \eqref{anomalies}\,,
and so $\<\cj_{I}^{i\anom}\>$ is conserved identically only for $\anom =1$. Hence, only the currents $\<\cj_{I}^{i\anom=1}\>$ lead to conserved electric charges in the presence of non-vanishing R-symmetry and/or flavor anomalies. These are the so called Page charges \cite{Page:1984qv,Marolf:2000cb,Copsey:2005se,Benini:2007gx}
\bbxd
\vskip.0cm
\be\label{el-charges-page}
\hspace{-.2cm}Q_{I}^{\rm Page}
=\int_{\cc_{(0)}}d\s_{i}\<\cj_{I}^{i\anom=1}\>
=\int_{\cc_{(0)}}d\s_{i}\Big(\<\cj_{I}^{i}\>-\frac{1}{24\k_{5}^{2}}C_{IJK}\epsilon_{(0)}^{ijkl}A_{(0)j}^{J}F_{(0)kl}^{K}\Big)\,,\quad \hspace{-.1cm}\ca_{I(0)}\neq 0\,,
\ee
\ebxd
where the integral is over any Cauchy surface $\cc_{(0)}$ on the boundary. Although the currents $\<\cj_{I}^{i\anom=1}\>$ are not gauge invariant, the Page charges \eqref{el-charges-page} are gauge invariant under small gauge transformations provided the gauge fields $A_{(0)j}^{I}$ are globally well defined on the boundary, i.e. in the absence of Dirac string singularities \cite{Marolf:2000cb}. However, they are never invariant under large gauge transformations.

For backgrounds with numerically vanishing R-symmetry and/or flavor anomalies, the currents \eqref{U(1)-current-family} are conserved for any value of the parameter $\anom$. In such cases there exists a corresponding family of conserved electric charges given by
\bbxd
\vskip.0cm
\be\label{el-charges-omega}
Q_{I}^{\anom}
=\int_{\cc_{(0)}}d\s_{i}\<\cj_{I}^{i\anom}\>
=\int_{\cc_{(0)}}d\s_{i}\Big(\<\cj_{I}^{i}\>-\frac{\anom}{24\k_{5}^{2}}C_{IJK}\epsilon_{(0)}^{ijkl}A_{(0)j}^{J}F_{(0)kl}^{K}\Big)\,,\qquad \ca_{I(0)}=0\,.
\ee
\ebxd
Besides the Page charges \eqref{el-charges-page}, this family includes the gauge invariant Maxwell charges corresponding to $\anom=-2$, which are determined by the covariant currents \eqref{cov-currents}.

\subsubsection{Magnetic dipole charges} 

Another type of charges that solutions of the STU model in five dimensions can possess is magnetic dipole charges, which create a Dirac string singularity. These are given by an integral of the field strengths over a closed 2-cycle $\S_2$ at the boundary, namely  
\bbxd
\vskip.4cm
\be\label{mag-charges}
P^{I}
=\int_{\S_2}F^I_{(0)}\,.
\ee
\ebxd
The interplay of the magnetic dipole charges with the Chern-Simons terms is somewhat subtle \cite{Copsey:2005se}. For example, as we mentioned above, the presence of Dirac string singularities that magnetic dipole charges create cause the Page charges \eqref{el-charges-page} not to be gauge invariant under small gauge transformations. However, dipole charges exist only in the presence of a nontrivial 2-cycle at the boundary, which is absent if the boundary has the topology $\bb R\times S^3$, but do exist, for example, for lens space boundaries of the form $\bb R \times L(p,q)$, or in the black string geometry \cite{Klemm:2000nj,Azzola:2018sld}.

\subsubsection{Killing charges}

The last type of conserved charges we need to discuss are those associated with isometries. Bulk Killing vectors in asymptotically locally AdS spaces are in one-to-one correspondence with conformal Killing vectors at the boundary. The holographic dictionary, therefore, allows one to express all bulk Killing charges in terms of the corresponding boundary conformal Killing vector and the field theory Ward identities \cite{Papadimitriou:2005ii}. Of course, in the presence of a nonzero conformal anomaly, only the subset of bulk isometries that map to boundary isometries -- not conformal isometries -- result in a conserved charge. Additional complications arise in the presence of bulk Chern-Simons terms \cite{Papadimitriou:2017kzw}. We now  briefly review the construction of the holographic Killing charges in the context of the STU model.  

Let us consider a boundary conformal Killing vector $\kbdy^i$, corresponding to a bulk Killing vector $\x^\m$ (see \cite{Papadimitriou:2005ii} for the explicit map between bulk Killing vectors and boundary conformal Killing vectors). The Lie derivative along $\kbdy^i$ acts on the field theory background as 
\be\label{conformal-Killing}
\hskip-.1cm\cl_{\kbdy}g_{(0)ij}=\frac{1}{2} \big(D_{(0)k} \kbdy^k\big) g_{(0)ij} ,\quad 
\cl_{\kbdy}A^I_{(0)i}= \pa_i \L_{(0)\kbdy}^{I} ,\quad 
\cl_{\kbdy}\vf^a_{(0)}=-\frac{1}{2}\big(D_{(0)k} \kbdy^k\big)  \vf^a_{(0)},
\ee
where the arbitrary functions $\L_{(0)\kbdy}^{I} (x)$ arise from the solution of the gauge invariant conformal Killing condition on the gauge field background,  $\cl_{\kbdy}F^I_{(0)ij}=0$. These functions transform under boundary gauge transformations $A^{I}_{(0)}\to A^{I}_{(0)}+d\alpha^{I}_{(0)}$ according to  
\be\label{gauge-boundary}
\L_{(0)\kbdy}^{I}\to\L_{(0)\kbdy}^{I}+\iota_{\kbdy}d\alpha^{I}_{(0)}\,,
\ee
and, as we review below, can be used in order to render the conserved charges and chemical potentials gauge invariant \cite{Kunduri:2013vka,Elgood:2020nls}.

The diffeomorphism Ward identity in \eqref{WardIDs} can be expressed in terms of the family of U(1) currents $\<\cj_{I}^{i\anom}\>$ introduced in \eqref{U(1)-current-family} as \cite{Jensen:2012kj,Papadimitriou:2017kzw,Minasian:2021png}  
\be\label{DiffWID-parameter}
D_{(0)j}\<\ct^{j}{}_{i}\>+\<\co_{a}\> D_{(0)i}\vf_{(0)}^{a}+F_{(0)ij}^{I}\<\cj_{I}^{j\anom}\>=\frac{\anom+2}{2}\,A_{(0)i}^{I}\ca_{I(0)}\,,
\ee
where we have used the Schouten identity $0=\d_{p}^{[q}\epsilon_{(0)}^{ijkl]}$ in order to write 
\be
F_{(0)pi}^{I}C_{IJK}\epsilon^{ijkl}F_{(0)jk}^{J}A_{(0)l}^{K}=-12\k_{5}^{2}A_{(0)p}^{I}\ca_{I(0)}\,.
\ee
Notice that the r.h.s. of \eqref{DiffWID-parameter} vanishes when $\<\cj_{I}^{j\anom}\>$ are the covariant currents, i.e. for $\anom=-2$. Contracting \eqref{DiffWID-parameter} with the boundary conformal Killing vector $\kbdy^i$ and rearranging the result using \eqref{conformal-Killing}, we obtain 
\bal\label{EM-current-div}
D_{(0)j}\Big(\<\ct^{j}{}_{i}\>\kbdy^{i}
-\<\cj_{I}^{j\anom}\>\big(\kbdy^{i}A_{(0)i}^{I}-\L_{(0)\kbdy}^{I}\big)\Big)
=&\;
\frac{1}{4} (D_{(0)k} \kbdy^k)\ca_{(0)}^{\text{W}}\NO\\
&\hskip-3.5cm+\Big(\frac{\anom+2}{2}\kbdy^{i}A_{(0)i}^{I}-(1-\anom)\big(\kbdy^{i}A_{(0)i}^{I}-\L_{(0)\kbdy}^{I}\big)\Big)\ca_{I(0)}\,,
\eal
where the conformal and U(1) anomalies, respectively $\ca_{(0)}^{\text{W}}$ and $\ca_{I(0)}$, are given in \eqref{anomalies}.

This result implies that a conserved charge associated with a boundary conformal Killing vector $\x_{(0)}^i$ exists if the r.h.s. of \eqref{EM-current-div} vanishes. The first term vanishes if either the conformal anomaly is zero or $\x_{(0)}^i$ is a boundary Killing vector, instead of a conformal Killing vector \cite{Papadimitriou:2005ii}. The second term vanishes for all values of $\anom$ if the U(1) anomalies $\ca_{I(0)}$ are numerically zero, or if $\ca_{I(0)}\neq 0$ (which, as we have seen, forces us to take $\anom=1$) and $\kbdy^{i}A_{(0)i}^{I}=0$. When any of the conditions that set the r.h.s. of \eqref{EM-current-div} to zero are met, the conserved charge associated with the conformal Killing vector $\kbdy^{i}$ is given by 
\bbxd
\vskip.4cm
\be\label{conformal-Killing-charges}
Q^{\anom}[\kbdy]=-\int_{\cc_{(0)}}d\s_{i}\Big(\<\ct^{i}{}_{j}\>\kbdy^{j}
-\<\cj_{I}^{i\anom}\>\big(\kbdy^{j}A_{(0)j}^{I}-\L_{(0)\kbdy}^{I}\big)\Big)\,.
\ee
\ebxd
This generalizes the Killing charges derived in \cite{Papadimitriou:2017kzw} by including the gauge compensators $\L_{(0)\kbdy}^{I}$, which ensure that $\kbdy^{j}A_{(0)j}^{I}-\L_{(0)\kbdy}^{I}$ is gauge invariant. Provided this quantity is also covariantly constant, i.e. $\x_{(0)}^iF_{(0)ij}=0$, the charges \eqref{conformal-Killing-charges} are invariant under small gauge transformations for any value of $\anom$ (once again, for $\anom\neq 1$ this requires $\ca_{I(0)}=0$ at least numerically, or else there exists no conserved charge) due to the Bianchi identity that the background gauge fields $A^I_{(0)i}$ satisfy. However for $\anom=-2$ these charges are gauge invariant even under large gauge transformations.

Although the choice of the parameter $\anom$ in the electric \eqref{el-charges-omega} and Killing charges \eqref{conformal-Killing-charges} may be different in principle, we will show below that a gauge invariant free energy can only be defined if they are the same and set equal to $1$.

\subsubsection{Hawking temperature and extremality}

Next, let us consider thermodynamic variables that depend on the existence of a Killing horizon. This horizon can be either degenerate or non-degenerate, corresponding respectively to extremal and non-extremal black holes. Let us discuss the non-extremal case first. The Hawking temperature is given by the surface gravity, $\hat \k$, through the relation
\be\label{Hawking-T-kappa}
T=\frac{\hat{\k}}{2\pi}\,.
\ee
Recall that the surface gravity is constant on the horizon and may be expressed in terms of the null generator of the horizon, $\c$, as 
\be
\hat{\k}^{2}=-\frac{1}{2}(\nabla^{\m}\c^{\nu})(\nabla_{\m}\c_{\nu})\,.
\ee

In concrete calculations, it is often useful to have a more explicit expression for the temperature. Focusing on spacetimes with a globally defined timelike Killing vector, $\pa_t$, and two commuting spacelike Killing vectors, $\pa_{\f^\a}$, $\a=1,2$, the metric can be written in the form 
\be\label{BH-metric}
ds^{2}=N^2dr^{2}+\g_{\th\th}d\th^{2}+\g_{tt}dt^{2}+2\g_{t\a}dtd\f^{\a}+\t_{\a\b}d\f^{\a}d\f^{\b}\,,
\ee
where $\t_{\a\b}\equiv \g_{\a\b}$ is the metric on the torus of spatial isometries, and all metric components are functions of $r$ and $\th$ only. A non-degenerate Killing horizon at $r=r_H$ is characterized by the fact that the timelike component of the metric in a non-rotating frame has a zero. 

Introducing static coordinates on the horizon  
\be\label{static-hor}
\hat{t}=t\,,\qquad \hat{\f}^{\a} = \f^{\a} -\O_{H}^{\a}t\,,
\ee
where $\O_{H}^{\a}\equiv-\tau^{\a\b}\g_{t\b}|_H$, existence of a Killing horizon at $r=r_H$ implies that 
\be\label{blackening-zero-general}
\g_{\hat{t}\hat{t}}|_H=(\g_{tt}-\tau^{\a\b}\g_{t\a}\g_{t\b})|_H
=0\,.
\ee
More specifically, for a non-degenerate Killing horizon, there exist (Schwarzschild type) coordinates such that the $r\hat t$-plane part of the metric admits the near horizon expansions
\be\label{Rindler-expansions}
N^{-2}=(r-r_H)\pa_{r}(N^{-2})|_H+\co(r-r_H)^2\,,\qquad
\g_{\hat{t}\hat{t}}=(r-r_H)\,\dot\g_{\hat{t}\hat{t}}|_H+\co(r-r_H)^2\,,
\ee
with $\pa_{r}(N^{-2})|_H\neq 0$ and $\dot\g_{\hat{t}\hat{t}}|_H\neq 0$.  It follows that, introducing the radial coordinate
\be
r-r_H=\frac14\pa_{r}(N^{-2})|_H\r^2+\co(\r^3)\,,
\ee
the near horizon geometry admits a Rindler metric of the form  
\be
\label{eq:near-horizon-Rindler}
ds^{2}= d\r^{2}+\frac14\pa_{r}(N^{-2})|_H\dot\g_{\hat{t}\hat{t}}|_H\,\r^2d\hat{t}^{2}+\cdots\,.
\ee

Demanding absence of a conical deficit in the Euclidean section of the Rindler metric \eqref{eq:near-horizon-Rindler} under\footnote{\label{periodicity}In the original (rotating) coordinates this periodicity condition reads $(t,\f^\a)\sim (t-i\b,\f^\a-i\b\O^\a_{H})$.} $\hat{t}\sim\hat{t}-i\b$, we arrive at the alternative formula for the Hawking temperature
\bbxd
\vskip.0cm
\be\label{Hawking-general}
T=\b^{-1}=\frac{1}{4\pi}(-\pa_{r}(N^{-2})\dot{\g}_{\hat{t}\hat{t}}|_H)^{1/2}
=\frac{1}{4\pi}\big(-
\pa_{r}(N^{-2})\pa_r(\g_{tt}-\tau^{\a\b}\g_{t\a}\g_{t\b})|_H\big)^{1/2}
\,.
\ee
\ebxd

Let us now consider the case of extremal horizons. Extremal black holes arise from non-extremal ones in the limit that at least two Killing horizons, located say at $r_\pm$ with $r_{-}<r_{+}$, coalesce, i.e. $r_{-}\to r_{+}=r_{H}$, which generally results in a vanishing Hawking temperature. As we saw above, this means that the functions $N^{-2}$ and $\g_{\hat{t}\hat{t}}$ parameterizing the metric \eqref{BH-metric} have simple zeros at $r_\pm$, resulting in a double zero in the limit $r_{-}\to r_{+}=r_{H}$. In particular, at an extremal horizon the derivatives of these functions vanish as well,
\be
\pa_{r}(N^{-2})|_{H}=0\,,\qquad\dot{\g}_{\hat{t}\hat{t}}|_{H}=0\,,
\ee
which reflects the fact that the Hawking temperature \eqref{Hawking-general} is zero.

For an extremal horizon, the near horizon expansions \eqref{Rindler-expansions} are therefore replaced by 
\be\label{extremal-expansions}
N^{-2}=\frac12(r-r_H)^2\pa_{r}^2(N^{-2})|_H+\co(r-r_H)^3\,,\qquad
\g_{\hat{t}\hat{t}}=\frac12(r-r_H)^2\,\ddot\g_{\hat{t}\hat{t}}|_H+\co(r-r_H)^3\,.
\ee
Introducing a suitable radial coordinate $\wt\r$ through 
\be
r-r_{H}=2(\pa_{r}^{2}(N^{-2})|_{H}\ddot{\g}_{\hat{t}\hat{t}}|_{H})^{-1/2}\wt\r+\co({\wt\r}^{2})\,,
\ee
the $r\hat{t}$-plane metric obtained from \eqref{BH-metric} in the near horizon limit is
the AdS$_{2}$ metric 
\be
ds^{2}=\ell_{2}^{2}\Big(\frac{1}{\wt\r^2}d{\wt\r}^{2}-\wt\r^2d\hat{t}^{2}\Big)+\cdots\,,
\ee
with AdS$_2$ radius
\be\label{L-IR}
\ell_{2}^{2}=2(\pa_{r}^{2}(N^{-2})|_{H})^{-1}\,.
\ee

\subsubsection{Bekenstein-Hawking entropy}

The entropy is given by the area of the horizon through the Bekenstein-Hawking formula
\be\label{BH-entropy}
S=\frac{\text{Area}_{H}}{4G_{5}}=\frac{1}{4G_{5}}\int_{H}d^{3}x\sqrt{\g_{\th\th}\,\det(\tau_{\a\b})}\,.
\ee
Again, it is useful to provide an alternative expression for the entropy that, as we shall see, arises naturally in the subsequent analysis.  

To this end, observe that the volume element induced on a radial slice and the trace of its extrinsic curvature admit the near horizon expansions 
\bal
\sqrt{-\g} = &\; \Big(-\dot{\g}_{\hat{t}\hat{t}}\g_{\th\th}\,\det(\tau_{\a\b})|_H(r-r_{H})+\co(r-r_H)^2\Big)^{1/2}\,,\NO \\
K  = &\; \frac{1}{2N}\g^{ij}\dot\g_{ij}\approx \frac12 N^{-1}\g^{\hat t\hat t}\dot\g_{\hat t\hat t}\approx\frac{1}{2}(r-r_{H})^{-1/2}\big(\pa_{r}(N^{-2})|_H\big)^{1/2}\,.
\eal
It follows that the product of these quantities is finite on the horizon and proportional to its area. Using the Euclidean time $t_E=it$ and the temperature \eqref{Hawking-general} we obtain \cite{Cabo-Bizet:2017xdr,Castro:2018ffi}
\bbxd
\vskip.4cm
\be\label{entropy-GH-at-hor}
S=\frac{i}{\k_{5}^{2}}\int_{0}^{-i\b}d\hat{t}d^{3}x\sqrt{-\g}K|_{H}
=\frac{1}{\k_{5}^{2}}
\int_{t_E=0}^{t_E=\b}\!\!\!\!d^{4}x_E\,\,\sqrt{-\g}K|_{H}\,.
\ee
\ebxd
To help interpret this expression for the entropy it is useful to rewrite it in the form
\bbxd
\vskip.4cm
\be\label{entropy-RG-flow}
S=\frac{\b}{\k_5^2}\int_H d^3x\,\pa_\r\sqrt{-\g}|_H\,.
\ee
\ebxd
As we will see in the next subsection, it is this expression for the entropy that provides a direct link with the superpotential $\cu$.

\subsubsection{Chemical potentials}

Another set of thermodynamic variables that depend on the presence of a horizon consists of various chemical potentials. These include the angular velocities conjugate to the conserved angular momenta, as well as electric chemical potentials conjugate to electric charges. Magnetic chemical potentials (magnetizations) may also be introduced, but we will not discuss them here since we focus on solutions without 2-cycles that could support magnetic charges.    

One may expect that the angular velocities and electric chemical potentials may be treated in a unified way, as would be natural, for example, from the point of view of a lower dimensional effective description. However, this is not possible in general due to a number of reasons. Firstly, a straightforward comparison is only possible in the absence of magnetic fluxes, although a unified treatment may persist in the presence of magnetic charges once NUT charges are included. Another difference between angular momenta and electric charges in the present context is the absence of gravitational Chern-Simons terms, corresponding to gravitational or mixed anomalies. Finally, our definition of Killing symmetries results in a third difference, due to the fact that bulk isometries, $\x$, are required to leave the metric invariant, while for the gauge fields we demand that $\cl_{\kbulk}F^{I}=0$, so that $\cl_{\kbulk}A^{I}=d\L_{\kbulk}^{I}$ for some unspecified $\L_{\kbulk}^{I}$ \cite{Kunduri:2013vka,Elgood:2020nls}, in analogy to the way we defined conformal Killing isometries on the boundary in the previous subsection. However, the stronger condition $\cl_{\kbulk}A^{I}=0$ can be imposed through partial gauge fixing.

Given the general form of the black hole metric in \eqref{BH-metric}, the physical angular velocities associated with the two isometries $\pa_{\f^\a}$ are defined as \cite{Papadimitriou:2005ii}
\bbxd
\vskip.4cm
\be\label{angular-velocities-general}
\O^\a\equiv\O_{H}^\a-\O_{\infty}^\a\,,
\ee
\ebxd
where $\O_{H}^{\a}\equiv-\tau^{\a\b}\g_{t\b}|_H$ and $\O_{\infty}^{\a}\equiv-\tau^{\a\b}\g_{t\b}|_\infty$.

The electric chemical potentials could be defined similarly, but the gauge compensators $\L^I_\x$ allow us to define them more generally. Under a gauge transformation $A^{I}\to A^{I}+d\alpha^{I}$, the compensators transform as $\L_{\kbulk}^{I}\to\L_{\kbulk}^{I}+\iota_{\kbulk}d\alpha^{I}$ and hence, the quantities  
\be
\F_{\kbulk}^{I}\equiv -\iota_{\kbulk}A^{I}+\L_{\kbulk}^{I}\,,
\ee
are gauge invariant. Moreover, they satisfy $d\F_{\kbulk}^{I}=\iota_{\kbulk}F^{I}$. This implies that $\F^I_\c$, where $\c$ is the null generator of the horizon, are not only gauge invariant, but also constant on the horizon. In particular, for any vector field $\z$ tangent to the horizon,  
\begin{eqnarray}
\z^\m\pa_\m\F_{\c}^{I} =&\; \iota_{\z}d\F_{\c}^{I}=\iota_{\z}\iota_{\c}F^{I}\,.
\end{eqnarray}
Since $\z\propto\c$ on the horizon, it follows that $\F_{\c}^{I}$
are constant on the horizon. 

Given that $\F_\c^I=-\c^{\m}A_{\m}^{I}+\L^I_{\c}$ are gauge invariant and constant on the horizon, we define the physical electric chemical potentials as
\bbxd
\vskip.4cm
\be\label{el-potential-general}
\F^{I}\equiv \F_\c^I|_H=\big(-\c^{\m}A_{\m}^{I}+\L^I_{\c}\big)_{H}\,,
\ee
\ebxd
This definition is consistent with that in \cite{Papadimitriou:2005ii} which used  the gauge $\cl_{\c}A^I=0$ so that $\c^{\m}A_{\m}^{I}$ are constant on the horizon. However, the more general definition \eqref{el-potential-general} allows us to alternatively demand that the gauge fields be regular on the horizon, which amounts to the condition that the holonomies around the contractible thermal circle vanish, i.e. $A^I_{\hat t}|_H=\c^\m A_\m^I|_H=0$. In section \ref{sec:thermodynamics} we will make yet another choice for $\L_\c$ that is natural for SU(2)$\times$U(1) invariant solutions, especially in the presence of supersymmetry. 

We conclude this subsection with the observation that the electric chemical potentials \eqref{el-potential-general} are only defined up to an arbitrary constant. This ambiguity is inherited from the gauge compensators $\L^I_{\c}$, which are also defined up to a constant, and shifts the conserved Killing charges \eqref{conformal-Killing-charges} by a linear combination of the electric charges \eqref{el-charges-omega}. As we will see in the next subsection, the free energy is invariant under such a change of charge basis.

\subsection{STU black holes with SU(2)$\times$U(1) isometry}
\label{STU-thermodynamics}

We now turn our attention to the special case of black holes described by the SU(2)$\times$U(1) invariant ansatz \eqref{ansatz}. The description of such black holes in terms of the superpotential $\cf$ permits a more detailed understanding of the thermodynamics, without explicit knowledge of the black hole solutions. For example, as we showed in section \ref{sec:superpotential}, the superpotential is closely related with Sen's entropy function for extremal black holes \cite{Sen:2008vm,Sen:2007qy} (see eq.~\eqref{EF-F}). After a brief discussion of the near horizon behavior of the superpotential for extremal and non-extremal black holes, we evaluate the thermodynamic variables introduced in the previous subsection. Subsequently, we show that the superpotential determines the free energy in terms of the conserved charges and their chemical potentials, and provides a particularly elegant derivation of the first law of thermodynamics for solutions within the ansatz.

\subsubsection{Near horizon behavior of the superpotential}

Considering first non-extremal black holes, the near horizon expansions \eqref{Rindler-expansions} determine 
\be\label{NHE-ansatz}
e^{-2u_{2}}=\co\big((r-r_{H})^{1/2}\big)\,,\qquad N^{-2}=\co(r-r_{H})\,,
\ee
and so the flow equations \eqref{flow-eqs-U} imply that the superpotential $\cu$ satisfies
\be
\pa_{u_3}\cu|_H=\pa_{\vf^a}\cu|_H=\pa_{v^I}\cu|_H=0\,,\qquad \pa_{u_1}\cu|_H=-\pa_{u_2}\cu|_H=-12\p T e^{-3u_H/2}\,.
\ee
These relations become more transparent when viewing $\cu$ as a function of the variables $u,u_2,u_3,v^I,\vf^a$ instead of $u_1,u_2,u_3,v^I,\vf^a$. Using the transformation of the partial derivatives
\be\label{change_of-variables}
\pa_{u_1}|_{u_2}=\pa_{u}|_{u_2}\,,\qquad \pa_{u_1}|_{u_2}+\pa_{u_2}|_{u_1}=\pa_{u_2}|_u\,,
\ee
we conclude that $\cu(u,u_2,u_3,v^I,\vf^a)$ satisfies
\bbxd
\bal\label{U-horizon-extremization}
&\pa_{u_2}\cu|_H=\pa_{u_3}\cu|_H=\pa_{\vf^a}\cu|_H=\pa_{v^I}\cu|_H=0\,,\NO\\
&\rule{.0cm}{.8cm}\pa_{u}\cu|_H=-12\p T e^{-3u_H/2}=-\frac{3\k_5^2}{8\p^2}ST\,.
\eal
\ebxd
It follows that non-extremal horizons extremize the superpotential $\cu$ with respect to all variables except $u$. We will see momentarily that these relations hold for extremal horizons as well, in which case $T=0$ and so $\cu$  is extremized with respect to all variables. 

Near non-extremal horizons the superpotential $\cf$ satisfies 
\be\label{non-degenerate-scaling}
\cf=\pa_{u_{3}}\cf=\pa_{\vf^{a}}\cf=\pa_{v^{I}}\cf=\co((r-r_{H})^{1/2})\,,\qquad \pa_{u}\cf=\co((r-r_{H})^{-1/2})\,.
\ee
Hence, $\pa_u\cf$ diverges at a non-degenerate horizon, while
\bbxd
\vskip.2cm
\be\label{eq:regularity-generic}
\cf|_{H}=\pa_{u_{3}}\cf|_{H}=\pa_{\vf^{a}}\cf|_{H}=\pa_{v^{I}}\cf|_{H}=0\,.
\ee
\ebxd

For near extremal black holes, the expansions \eqref{extremal-expansions} imply instead that near the horizon
\be
e^{-2u_{2}}=\co(r-r_{H})\,,\qquad N^{-2}=\co((r-r_{H})^{2})\,.
\ee
The flow equations \eqref{flow-eqs-F} then determine that
\be\label{degenerate-scaling}
\cf=\pa_{u_{3}}\cf=\pa_{\vf^{a}}\cf=\pa_{v^{I}}\cf=\co(r-r_{H})\,,
\ee
which shows that the relations \eqref{eq:regularity-generic} -- and hence \eqref{U-horizon-extremization} -- hold for extremal horizons as well. An important difference, however, is that while $\pa_u\cf$ diverges on non-extremal horizons (see \eqref{non-degenerate-scaling}), it remains finite on extremal horizons and is related to the entropy via (see \eqref{F-u-der-extremal})
\be\label{extremal-entropy-generic}
\pa_{u}\cf|_{H}=-6\ell_{2}^{-1}e^{-2u_{H}}=-6\ell_2^{-1}\Big(\frac{\k_{5}^{2}}{32\pi^{3}}\Big)^{4/3}S^{4/3}\,.
\ee
Moreover, while $\pa_u\cu|_H$ is finite and proportional to the Hawking temperature for non-extremal solutions, in the extremal case $\pa_u\cu|_H=0$. In that case, the superpotential equations \eqref{U-horizon-extremization} are equivalent to the attractor equations \eqref{attractor-eqs}, thanks to the relation \eqref{EF-F} between the entropy function and the superpotential $\cu$.

\subsubsection{Conserved charges}

The conserved charges can be evaluated generically for any solution within the SU(2)$\times$U(1) ansatz \eqref{ansatz} using the 1-point functions of the stress tensor and of the U(1) currents, given respectively in \eqref{T-vevs} and \eqref{J-vevs}. Since the U(1) anomalies \eqref{anomalies} vanish numerically for all solutions in the SU(2)$\times$U(1) invariant sector, any value of the parameter $\anom$ in the definition of the charges \eqref{el-charges-omega} and \eqref{conformal-Killing-charges} leads to a conserved charge. As should be expected, we find that only the mass depends on the superpotential $\cu$, while the remaining charges are determined in terms of the integration constants $j$ and $q_I$. 

Using the expectation value of the time component of the U(1) currents within the SU(2)$\times$U(1) invariant ansatz in \eqref{J-vevs}, we find that the electric charges \eqref{el-charges-omega} become  
\bbxd 
\vskip.4cm
\be\label{el-charges-ansatz}
Q_{I}^{\anom}=\frac{4\pi^{2}}{\k_{5}^{2}}\Big(q_{I}+\frac{\anom-1}{3}C_{IJK}v_{(0)}^{J}v_{(0)}^{K}\Big)\,.
\ee
\ebxd
As we just pointed out, the fact that the U(1) anomalies \eqref{anomalies} vanish numerically for all solutions within the SU(2)$\times$U(1) ansatz, these electric charges are conserved for any value of the parameter $\anom$. Recall that the value 
$\anom=1$ corresponds to the Page charges, which are identified with the integration constants $q_I$ within the SU(2)$\times$U(1) invariant ansatz. 

A generic solution within the SU(2)$\times$U(1) invariant sector possesses the three commuting Killing vectors $\pa_t$, $\pa_\j$ and $\pa_\f$. In fact, these are also isometries on the conformal boundary, and so lead to a conserved charge even in the presence of a nonzero conformal anomaly. The mass is the conserved charge associated with the timelike Killing vector in a static frame on the boundary, which is obtained through the coordinate transformation
\be\label{boundary-static-frame}
t=\bar{t}\,,\qquad \j=\bar\j+\O_{\infty}\bar t=\bar\j-u_{4(0)}\bar t\,,\qquad \f=\bar\f\,.
\ee
Hence, the Killing vectors in the static and rotating frame on the boundary are related as 
\be
\pa_{\bar t}= \pa_t+\O_\infty\pa_\j\,,\qquad \pa_{\bar\j} = \pa_\j\,,\qquad \pa_{\bar\f}=\pa_\f\,.
\ee 

Using the timelike Killing vector in the static frame and the expectation values of the stress tensor and U(1) currents in \eqref{T-vevs} and \eqref{J-vevs}, one finds that the mass is given by
\bbxd
\vskip.4cm
\be\label{mass-ansatz}
M^{\anom} \equiv Q^{\anom}[-\pa_{\bar{t}}]=- \frac{4\pi^{2}}{\k_{5}^{2}}e^{-2u_{2(0)}}\lim_{r_{0}\to\infty}\cf_{\text{ren}}-(a_{(0)}^{I}+\O_{\infty}v_{(0)}^{I}-\L_{\pa_{\bar{t}}}^{I})Q_{I}^{\anom}\,.
\ee
\ebxd
Notice that the asymptotic form of the mass deformation of the superpotential $\cf$
in \eqref{mass-pert-F} implies that the corresponding change in conserved mass is 
\bbxd
\vskip.4cm
\be\label{mass-pert-M}
\D M^{\anom}=\frac{3\pi^{2}\ell}{2\k_{5}^{2}}e^{-2u_{2(0)}}\D m\,.
\ee
\ebxd

The angular momentum along $\bar\j$ is independent of the superpotential and is given by
\bbxd
\vskip.4cm
\be\label{ang-mom-ansatz}
J^{\anom}\equiv Q^{\anom}[\pa_{\bar{\j}}] = \frac{4\pi^{2}}{\k_{5}^{2}}\Big(
j+\frac{\anom}{3}C_{IJK}v_{(0)}^{I}v_{(0)}^{J}v_{(0)}^{K}\Big)-\L_{\pa_{\bar{\j}}}^{I}Q_{I}^{\anom}\,,
\ee
\ebxd
while that along $\bar\f$ takes the form
\be
Q^{\anom}[\pa_{\bar{\f}}]=-\frac{4\pi^{2}}{\k_{5}^{2}}\L_{\pa_{\bar{\f}}}^{I}Q_{I}^{\anom}\,.
\ee
As we will see shortly, the angular velocity along the $\f$-direction vanishes for solutions in the SU(2)$\times$U(1) invariant sector and hence $Q^{\anom}[\pa_{\bar{\f}}]$ does not contribute to the thermodynamics. Moreover, the gauge compensators $\L^I_{\pa_{\bar t}}$, $\L^I_{\pa_{\bar\j}}$ and $\L^I_{\pa_{\bar\f}}$ are constant throughout the bulk for any solution in this sector, which is why we do not distinguish between these and the corresponding compensators on the boundary. Although their introduction may look unnecessary for solutions described by the SU(2)$\times$U(1) invariant ansatz, they are important in order to ensure that the quantities $\x^\m A_\m^I-\L_\x$ are invariant under the constant shift transformations $a^I\to a^I+a_0^I$ discussed in section \ref{sec:ansatz}. 

Finally, we should emphasize that the conserved charges derived above include nontrivial Casimir contributions from the background fields on the boundary. The Casimir contribution to the electric charges and the angular momentum is shown explicitly and originates exclusively in the background gauge fields. The Casimir contribution to the mass \cite{Balasubramanian:1999re,Awad:1999xx,Papadimitriou:2005ii} is hidden in the renormalized superpotential $\cu\sbtx{ren}$. We will determine its form explicitly in the next subsection, using the supersymmetric superpotential $\cf\sbtx{BPS}$ in  \eqref{BPS-superpotential-F}.

\subsubsection{Chemical potentials}

From the SU(2)$\times$U(1) invariant ansatz \eqref{ansatz} we see that the local angular velocities associated with the spacelike isometries at the boundary and on the horizon are respectively
\be\label{ang-vel-hor-ansatz}
\O_{\infty}\equiv\O_{\j\infty}=-u_{4(0)}\,,\qquad\O_{H}\equiv\O_{\j H}=-u_{4H}\,,\qquad\O_{\f\infty}=\O_{\f H}=0\,.
\ee
Hence, only the angular velocity along the $\j$-direction is nontrivial and takes the form 
\bbxd
\vskip.2cm
\be\label{eq:ang-vel-explicit}
\O\equiv\O_{H}-\O_{\infty}=u_{4(0)}-u_{4H}\,.
\ee
\ebxd

Next, in order to evaluate the electric chemical potentials \eqref{el-potential-general}, we note that the null generator of the Killing horizon is
\be\label{Killing-hor-general}
\c =\pa_{\hat{t}}=\pa_t + \O_{H} \pa_{\j} 
=\pa_{\bar{t}} + \O \pa_{\bar{\j}} \,.
\ee
It follows that the electric chemical potentials \eqref{el-potential-general} are given by
\bbxd
\vskip.3cm
\be\label{eq:el-potential-explicit}
\F^{I}=-a_{H}^{I}+u_{4H}v_{H}^{I}+\L^I_{\c}
\,,
\ee
\ebxd
where the gauge compensators $\L^I_{\c}$ are constant throughout the bulk for all solutions in the SU(2)$\times$U(1) invariant sector.

\subsubsection{Entropy}

Applying the Bekenstein-Hawking entropy formula \eqref{BH-entropy} to the ansatz \eqref{ansatz} gives
\bbxd
\vskip.4cm
\be\label{entropy-on-ansatz}
S=\frac{32\pi^{3}}{\k_{5}^{2}}e^{-3u_H/2}\,.
\ee
\ebxd
However, the alternative expression for the entropy we derived in \eqref{entropy-RG-flow} gives instead
\be\label{entropy-on-ansatz-U}
S=\frac{\b}{\k_5^2}\int_H d^3x\,\pa_\r\sqrt{-\g}|_H=\frac{16\p^2\b}{\k_5^2}\pa_\r (e^{-2u_1})|_H=-\frac{8\p^2\b}{3\k_5^2}\pa_{u_1}\cu|_H\,,
\ee
where the flow equations \eqref{flow-eqs-U} were used in the last step. Of course,   these two results for the entropy coincide, but \eqref{entropy-on-ansatz-U} provides the advertised connection between the entropy and the superpotential $\cu$. As we discuss next, equating the two expressions for the entropy leads to an alternative expression for the temperature in terms of $\cu$.

\subsubsection{Temperature}

From the ansatz \eqref{ansatz} follows that near a non-extremal horizon  
\be
\g_{\hat{t}\hat{t}}=-\mtrc_{1}+\frac{1}{4}\mtrc_{3}(u_{4}-u_{4H})^{2}=-\mtrc_{1}+\co(r-r_H)^{2}\,,
\ee
and hence black holes within the SU(2)$\times$U(1) invariant sector satisfy $\mtrc_{1H}=0$. The formula \eqref{Hawking-general} for the Hawking temperature then gives
\bbxd
\vskip.4cm
\be\label{eq:Hawk-temp-1}
T=\frac{1}{4\pi}\sqrt{\pa_r(N^{-2})\dot{\mtrc}_{1}|_H}=\frac{1}{4\pi}\sqrt{\pa_r(N^{-2})\pa_{r}(e^{-u_{1}-3u_{2}})|_{H}} \,. 
\ee
\ebxd
However, combining the two expressions for the entropy in \eqref{entropy-on-ansatz} and \eqref{entropy-on-ansatz-U} leads to an alternative formula for the Hawking temperature in terms of the superpotential $\cu$, namely
\bbxd
\vskip.4cm
\be\label{T-superpotential}
T=-\frac{1}{12\p}e^{3u/2}\pa_{u_1}\cu|_H\,.
\ee
\ebxd
This expression for the temperature applies to both extremal and non-extremal black holes.

\subsubsection{Free energy}

The last thermodynamic variable we need to compute is the free energy -- specifically the grand canonical potential -- which corresponds to the renormalized Euclidean on-shell action. As we will show, for any solution in the SU(2)$\times$U(1) invariant sector, the on-shell action is determined in terms of the superpotential $\cu$, without explicit knowledge of the solution.       

Adding the Gibbons-Hawking term to the action \eqref{S_B-ansatz}, the total bulk action for solutions described by the SU(2)$\times$U(1) invariant ansatz \eqref{ansatz} takes the form
\be\label{total-bulk-action-ansatz}
\actn=\int dt \int_{r_H}^{r_0} dr\Big[L_{\rm 1D}+\frac{4\p^2}{\k_5^2}\pa_r\Big(\frac{8}{N}e^{-2u_1}\dot u_1-\frac13C_{IJK}a^Iv^Jv^K\Big)\Big]+\frac{1}{2\k_5^2}\int_{r=r_0} d^4x\sqrt{-\g}\,2K\,,
\ee
where $r_0$ is a radial cutoff and $r_H$ is the location of the horizon. Evaluating the Gibbons-Hawking term on the ansatz gives
\be
\frac{1}{\k_{5}^{2}}\int d^{4}x\,\sqrt{-\g}K=-\frac{32\pi^{2}}{\k_{5}^{2}N}e^{-2u_{1}}\dot{u}_{1}\int dt\,,
\ee
while Hamilton-Jacobi theory tells us that
\be
\int_{r_{H}}^{r_{0}}drL_{\text{1D}}=\cs_{\text{1D}}|_{r_0}-\cs_{\text{1D}}|_H \,,
\ee
where $\cs\sbtx{1D}$ is Hamilton's principal function given in \eqref{HJ1d}. Combining these results we can express the regularized on-shell action $\actn$ in terms of the superpotential $\cu$ as
\bbxd
\be\label{total-reg-action-ansatz}
\actn=\frac{4\pi^{2}}{\k_{5}^{2}}\int dt\Big(\Big[\cu+\Big(q_{I}-\frac{1}{3}C_{IJK}v^{J}v^{K}\Big)\big(a^{I}-u_{4}v^{I}\big)+ju_{4}\Big]_{r_{H}}^{r_{0}}-\frac{8}{N}e^{-2u_{1}}\dot{u}_{1}|_{r_H}\Big)\,.
\ee
\ebxd

The renormalized on-shell action for any solution in the SU(2)$\times$U(1) invariant sector is obtained from \eqref{total-reg-action-ansatz} by adding the boundary counterterms \eqref{counterterms-ansatz} and removing the radial cutoff, i.e. evaluating the limit $r_0\to\infty$. In particular, the renormalized on-shell action is
\be\label{total-ren-action-ansatz}
\actn\sbtx{ren}=\frac{4\pi^{2}}{\k_{5}^{2}}\lim_{r_0\to\infty}\int dt\Big(\cu\sbtx{ren}-\cu_H+\Big[\Big(q_{I}-\frac{1}{3}C_{IJK}v^{J}v^{K}\Big)\big(a^{I}-u_{4}v^{I}\big)+ju_{4}\Big]_{r_{H}}^{r_{0}}-\frac{8}{N}e^{-2u_{1}}\dot{u}_{1}|_{r_H}\Big)\,.
\ee
Using the formula \eqref{entropy-on-ansatz-U} for the entropy and Wick rotating to imaginary time, this determines the renormalized Euclidean on-shell action, $I=-i\actn_{\text{ren}}$, with
\bbxd
\vskip.0cm
\be\label{Euclidean-on-shell-action}
I=-S -\frac{4\pi^{2}\b}{\k_{5}^{2}}\lim_{r_{0}\to\infty}\Big(\cu_{\text{ren}}-\cu_H+\Big[(q_{I}-\frac{1}{3}C_{IJK}v^{J}v^{K})(a^{I}-u_{4}v^{I})+ju_{4}\Big]_{r_{H}}^{r_{0}}\Big)\,.
\ee
\ebxd

Notice that this expression for the renormalized Euclidean on-shell action is manifestly invariant under the transformations $u_4\to u_4+c$, $a^I\to a^I+a_0^I$ discussed in section \ref{sec:ansatz}, while under the constant shift transformations $a^I\to a^I+a_0^I$ it transforms as
\be
\label{I-shift-transformation}
I\to I+\frac{4\pi^{2}\b}{3\k_{5}^{2}}C_{IJK}a_0^{I}\big(v_{(0)}^{J}v_{(0)}^{K}-v_{H}^{J}v_{H}^{K}\big)\,.
\ee
This nonzero shift is due to the Bardeen-Zumino 3-form $\cx_I$ that appears in the transformation of the Chern-Simons form in \eqref{BZ-current}, evaluated at the boundary and on the horizon. 

Another noteworthy observation is that the Euclidean on-shell action depends on the difference $\cu\sbtx{ren}-\cu_H$, and so any constant shift of the superpotential $\cu$, such as the constant $\cu_0$ in \eqref{separable-U}, cancels out.

\subsubsection{Quantum statistical relation and the first law}

Black hole thermodynamics relates the Euclidean on-shell action \eqref{Euclidean-on-shell-action} with the grand canonical potential (Gibbs free energy) $G(T,\O,\F^I)\equiv M-TS-\O J-\F^{I}Q_{I}$ through a quantum statistical relation of the form
\be\label{QSR} 
I\sim\b G\,.
\ee
In fact, the Gibbs free energy corresponds to the {\em dimensionally reduced} on-shell action on the Euclidean time circle. This distinction is crucial in the presence of Chern-Simons terms. 

Using the above results for the thermodynamic variables and the fact that \eqref{separable-U} and \eqref{eq:regularity-generic} (which hold for both extremal and non-extremal black holes) imply that $\cu_H-\cu_0=0$, the renormalized Euclidean on-shell action \eqref{Euclidean-on-shell-action} takes the form
\bal\label{QSR1}
I = &\; \b\big(M^{\anom}-TS-\O J^{\anom}-\F^{I}Q_{I}^{\anom}\big)\NO \\
&\hskip-.3cm+\frac{4\pi^{2}\b}{3\k_{5}^{2}}C_{IJK}\big(\anom v_{(0)}^{J}v_{(0)}^{K}(\c^{\m}A_{\m}^{I})|_{H}^{\infty}+(v_{(0)}^{J}v_{(0)}^{K}-v_{H}^{J}v_{H}^{K})\c^{\m}A_{\m}^{I}|_{H}\big)\,,
\eal
where $\c^{\m}A^I_{\m}=a^{I}-u_{4H}v^{I}$. This result holds for any value of the parameter $\anom$. Notice that all thermodynamic variables on the r.h.s of \eqref{QSR1} are invariant under the constant shift transformations $a^I\to a^I+a_0^I$, since they involve the invariant combinations $\x^\m A^I_\m-\L^I_\x$, but the second line of \eqref{QSR1} provides the nontrivial transformation \eqref{I-shift-transformation}. However, the shift transformations are frozen by demanding that $\c^\m A^I_\m|_H=0$ so that the gauge field is regular at the tip of the Euclidean geometry. This puts to zero all terms in the second line of \eqref{QSR1} except for the one proportional to $\anom v_{(0)}^{J}v_{(0)}^{K}\c^{\m}A_{\m}^{I}|_{\infty}$. As we now explain, this term can be eliminated through a local counterterm that renders the on-shell action invariant under boundary gauge transformations and setting $\anom=1$, which corresponds to the Page charges. 

Of course, there exists no local counterterm that removes the R-symmetry and flavor anomalies in four dimensions that could be used to render the on-shell action gauge invariant. However, such a counterterm does exist for the dimensionally reduced on-shell action and involves the Bardeen-Zumino polynomial that relates the consistent and covariant currents as in \eqref{cov-currents}. The relevant counterterm takes the form \cite{Cheng:2021zjh}
\bbxd
\vskip.4cm
\be\label{thermal-ct}
I'\equiv I+\frac{\b}{24\k_{5}^{2}}\int_\infty d^3x \sqrt{-\g}\, \c^{i}A_{i}^{I}C_{IJK}\e^{jkl}F^J_{jk}A^K_l\,,
\ee
\ebxd
and ensures that $I'$ is invariant under gauge transformations that preserve the timelike isometry on the boundary. Notice that the timelike Killing vector $\c$ is the generator of the time circle on which the theory is reduced -- see footnote \footref{periodicity}. We should emphasize that in order for the Euclidean on-shell action to be rendered gauge invariant through the counterterm \eqref{thermal-ct} it is essential that the parameters $\anom$ defining the electric and Killing charges coincide. Evaluating the counterterm \eqref{thermal-ct} on our ansatz \eqref{ansatz} and imposing the regularity condition $\c^\m A^I_\m|_H=0$ we determine that the gauge invariant Euclidean on-shell action $I'$ is
\bbxd
\vskip.4cm
\be\label{QSR2}
I' = \b\big(M^{\anom}-TS-\O J^{\anom}-\F^{I}Q_{I}^{\anom}\big)+(\anom-1)\frac{4\pi^{2}\b}{3\k_{5}^{2}}C_{IJK} v_{(0)}^{J}v_{(0)}^{K}\c^{i}A_{(0)i}^{I}\,.
\ee
\ebxd
It follows that the desired relation between the Euclidean on-shell action and the Gibbs free energy holds only for the Page charges, corresponding to $\anom=1$. This is, therefore, the value of $\anom$ that we will adopt for most part of the subsequent analysis. 

The symplectic structure associated with the radial Hamiltonian description of the dynamics leads to an elegant and immediate derivation of the first law of thermodynamics for black holes in the SU(2)$\times$U(1) invariant sector. As we have seen in section \ref{sec:superpotential}, the Hamilton-Jacobi description separates the integration constants into those that parameterize a complete integral of the Hamilton-Jacobi equation, namely the mass and charge parameters, and those that arise from the integration of the flow equations. This separation is very natural in the context of holography, since the first set of integration constants maps to normalizable modes related to expectation values of local operators, while the latter correspond to non-normalizable modes that map to sources, i.e. background fields. This separation of integration constants is also ideally suited for describing the thermodynamics of AdS black holes, with the first set corresponding to conserved charges and the latter to chemical potentials. 

In particular, the Dirichlet variational principle for AdS black holes leads to the grand canonical ensemble version of the first law     
\bbxd
\vskip.2cm
\be\label{FirstLaw}
0=\d I' = \b\big(\d M^{\anom}-T\d S-\O \d J^{\anom}-\F^{I}\d Q_{I}^{\anom}\big)\,,
\ee
\ebxd
where the variations of all thermodynamic variables are evaluated by varying the integration constants parameterizing a complete integral of the Hamilton-Jacobi equation, that is the mass and charge parameters, while keeping fixed all integration constants arising from integrating the flow equations, which determine the chemical potentials.

\subsection{BPS black hole thermodynamics}

We are now ready to study the thermodynamics of 1/4 BPS black holes within the SU(2)$\times$U(1) invariant sector of the STU model, which are described by the exact superpotential \eqref{BPS-superpotential-F}. As we will see, a complete description of the thermodynamics requires that we consider complexified versions of such black holes, which we discuss in subsection \ref{sec:complexification}. As a prelude to that discussion, in the current subsection we apply the general results we obtained above to real 1/4 BPS solutions and determine the implications of supersymmetry.  Our analysis builds on extensive earlier work, most notably \cite{Hosseini:2017mds,Cabo-Bizet:2018ehj,Cassani:2019mms,Larsen:2020lhg} (see also \cite{Silva:2006xv,Kim:2006he}), but the 1/4 BPS superpotential \eqref{BPS-superpotential-F} provides a significant new perspective that leads to a first principles derivation of the entropy extremization functional for BPS black holes. Moreover, we will show that the conserved charges obey the generalized BPS relation derived in \cite{Papadimitriou:2017kzw}, which contains the supersymmetric Casimir energy as a consequence of the anomalous supersymmetry transformation of the $\cn=1$ supercurrent in the dual field theory.   

\subsubsection{Chemical potentials, conserved charges and BPS relation}

The angular velocity \eqref{eq:ang-vel-explicit} and electric chemical potentials \eqref{eq:el-potential-explicit} can be evaluated straightforwardly thanks to the supersymmetric integrals of motion $\mathscr{C}_{4}$ and $\mathscr{C}^{I}$ in \eqref{constants-of-motion-main}. From the definition of $g$ in \eqref{g-def} and the leading asymptotic form of the ansatz functions in \eqref{ansatz-asymptotics-u} follows that $\cos2g|_\infty=1$ at the boundary, while at the horizon we have $e^{-2u_{2}}|_H=0$. The integral of motion $\mathscr{C}_{4}$ then determines that
\be\label{u4H-BPS}
u_{4H}=u_{4(0)}-\projsp e^{-2u_{2(0)}-u_{3(0)}}\,,
\ee 
and so from \eqref{eq:ang-vel-explicit} we conclude that the supersymmetric angular velocity is
\bbxd
\vskip.2cm
\be\label{BPS-angular-velocity}
\O_{\text{BPS}}=\projsp e^{-2u_{2(0)}-u_{3(0)}}\,.
\ee
\ebxd
Importantly, this value of the BPS angular velocity implies that the Killing generator of the horizon \eqref{Killing-hor-general} coincides with the Killing spinor bilinear \eqref{KV}, namely  
\bbxd
\vskip.2cm
\be\label{KV=HG}
\c_{\text{BPS}}=\ck_{(0)}\,.
\ee
\ebxd 

Similarly, using the fact that 
\be\label{sing-limit}
\lim_{r\to\infty}e^{-\frac{1}{2}u}\sin2g=-\frac{\projsp\ell}{2}e^{2u_{3(0)}}\,,
\ee
and the integral of motion $\mathscr{C}^{I}$ in \eqref{constants-of-motion-main} we obtain
\bal\label{aIH-BPS}
a_H^{I}-u_{4H}v_H^{I}=&\;a_{(0)}^{I}-\big(u_{4(0)}-\projsp e^{-u_{3(0)}-2u_{2(0)}}\big)v_{(0)}^{I}+\frac{\ell}{2}e^{2u_{3(0)}-2u_{2(0)}}\NO\\
=&\;a_{(0)}^{I}-u_{4H}v_{(0)}^{I}+\frac{\ell}{2}e^{2u_{3(0)}-2u_{2(0)}}\,,
\eal 
where we used \eqref{u4H-BPS}. Hence, the electric chemical potentials \eqref{eq:el-potential-explicit} are given by
\bbxd
\vskip.3cm
\be\label{BPS-el-chemical-potentials}
\F^{I}_{\text{BPS}}=-\frac{\ell}{2}e^{-2u_{2(0)}+2u_{3(0)}}-\c^{\m}A_{\m}^{I}|_{(0)}+\L^I_\c\,,
\ee
\ebxd
or, separating the gravity and vector multiplets using the decomposition \eqref{gauge-field-decomposition},
\be
\F_{\text{BPS}}=\frac{\ell\sqrt{3}}{2}e^{-2u_{2(0)}+2u_{3(0)}}-\c^{\m}A_{\m}|_{(0)}+\L_\c\,,\qquad\wt{\F}^{a}_{\text{BPS}}=-\c^{\m}\wt A_{\m}^{a}|_{(0)}+\wt \L^a_\c\,.
\ee
Using \eqref{h-to-u}, the first equation in \eqref{susy-from-the-bulk}, \eqref{u4H-BPS} and \eqref{KV=HG}, one may verify that these expressions for the chemical potentials coincide with those in  \eqref{BPS-chemical-potentials-boundary} in appendix \ref{sec:boundary} which were determined using pure field theoretic arguments and boundary supersymmetry. 

As emphasized above, the gauge compensators $\L^I$ render the chemical potentials and conserved charges manifestly gauge invariant. However, at least within the SU(2)$\times$U(1) invariant sector, it is possible to fix $\L^I$ in a way  that preserves manifest invariance under the residual large gauge transformations. Moreover, it is crucial that such a choice does not depend on horizon data since the same gauge compensators are used in the definition of the field theory/holographic charges. With these conditions in mind, in the following we set 
\be\label{Lambda-fixing}
\L^I_\c=a_{(0)}^I-u_{4(0)}v_{(0)}^I=\c^{\m}A_{\m}^{I}|_{(0)}-\O_{\rm BPS} v_{(0)}^I\,,\qquad \L_{\pa_{\bar{\j}}}^{I}=0\,,\qquad\L_{\pa_{\bar{\f}}}^{I}=0\,.
\ee
With this choice of gauge compensators, the electric chemical potentials \eqref{BPS-el-chemical-potentials} simplify to
\bbxd
\vskip.3cm
\be\label{BPS-el-chemical-potentials-fixed}
\F^{I}_{\text{BPS}}=-\frac{\ell}{2}e^{-2u_{2(0)}+2u_{3(0)}}\big(1+2\projsp\ell^{-1}e^{-3u_{3(0)}} v_{(0)}^I\big)\,.
\ee
\ebxd
For the Gutowksi-Reall solutions \cite{Gutowski:2004yv,Kunduri:2006ek} these take the form (see \eqref{GR-FG-coeffs})
\be
\O_{\text{AdS}}=2\projsp\ell^{-1}\,,\qquad \F^{I}_{\text{AdS}}=-1\,.
\ee

Let us now turn to the conserved charges of 1/4 BPS black holes. The electric charges \eqref{el-charges-ansatz} and angular momentum \eqref{ang-mom-ansatz} for black holes in the SU(2)$\times$U(1) invariant sector remain unchanged when applied to supersymmetric black holes. However, the superpotential \eqref{BPS-superpotential-F} allows us to evaluate explicitly the mass \eqref{mass-ansatz} for BPS black holes. Inserting the expression \eqref{F-ren-BPS-derivatives} for $\cf_{\text{ren}}$ in the mass \eqref{mass-ansatz}, using \eqref{u4H-BPS} and fixing the gauge compensators as in \eqref{Lambda-fixing}, we obtain the BPS mass
\bbxd
\be
\label{BPS-mass}
\hskip-.3cm M_{\text{BPS}} =  \frac{4\pi^{2}}{\k_{5}^{2}}e^{-2u_{2(0)}}\Big[\frac{\ell^{3}}{4}\Big(e^{5u_{3(0)}}-\frac{5}{8}e^{8u_{3(0)}}\Big)+\frac{\ell}{2}e^{3u_{3(0)}}\sum_{I}\pa_{v^I_{(0)}}\cf_0(v_{(0)})+\projsp\cf_0(v_{(0)})\Big],\hskip-.1cm
\ee
\ebxd
where $\cf_0$ is defined in \eqref{F0-def}. Notice that with the choice \eqref{Lambda-fixing} for the gauge compensators the BPS mass is independent of the parameter $\anom$.\footnote{Recall that the quantum statistical relation \eqref{QSR2} requires that we set $\anom=1$ and $\c^\m A^I_\m|_H=0$. However, we keep the current discussion general and only impose these conditions when strictly necessary.}

From the BPS mass \eqref{BPS-mass}, electric charges \eqref{el-charges-ansatz}, angular momentum \eqref{ang-mom-ansatz} and velocity \eqref{BPS-angular-velocity}, as well as the electric chemical potentials \eqref{BPS-el-chemical-potentials} we obtain the BPS relation\footnote{This relation holds for any value of the gauge compensators $\L^I$ -- see the field theory result \eqref{BPS-relation} in appendix \ref{sec:boundary}.} 
\bbxd  
\vskip.3cm   
\be\label{BPS-relation-bulk}
M_{\text{BPS}}^{\anom}-\O_{\text{BPS}}J^{\anom}-\F_{\text{BPS}}^{I}Q_{I}^{\anom}=M_{\text{Casimir}}^{\anom}\,,
\ee
\ebxd
where $M_{\text{Casimir}}^{\anom}$ is a local expression of the field theory background that takes the form
\bbxd
\be\label{anomaly-charge-anstatz-bulk}
M_{\text{Casimir}}^{\anom} =  \frac{\pi^{2}\ell^3}{\k_{5}^{2}}e^{-2u_{2(0)}+2u_{3(0)}}\Big[e^{3u_{3(0)}}\Big(1-\frac{5}{8}e^{3u_{3(0)}}\Big)+\frac{2(\anom+2)}{3\ell^2}\sum_IC_{IJK}v_{(0)}^{J}v_{(0)}^{K}\Big]\,.
\ee
\ebxd
Using the first relation in \eqref{susy-from-the-bulk}, \eqref{h-to-u}, and the identity \eqref{decID} it is straightforward to confirm that this expression coincides with the supersymmetric Casimir energy obtained from the anomalous supersymmetry transformation of the supercurrent \cite{Papadimitriou:2017kzw}, as is shown explicitly in appendix \ref{sec:boundary} using an entirely field theoretic argument -- see eq.~\eqref{anomaly-charge-anstatz}. For purely electric asymptotically AdS solutions,  \eqref{BPS-relation-bulk} reduces to
\be
M_{\text{BPS}}^{\anom}-2\projsp\ell^{-1}J^{\anom}+\sum_{I}Q_{I}^{\anom}=\frac{3\pi^{2}\ell^{2}}{4\k_{5}^{2}}\,,
\ee
with $M_{\text{Casimir}}^{\anom}$ given by the Casimir mass of global AdS$_5$ \cite{Balasubramanian:1999re,Awad:1999xx,Papadimitriou:2005ii}.

\subsubsection{Supersymmetric attractor and black hole entropy}

Near the horizon, 1/4 BPS black holes described by the superpotential \eqref{BPS-superpotential-F} approach the general attractor solution \eqref{attractor}. As we showed in section \ref{sec:superpotential}, the entropy function that determines the values of the fields at the attractor point is closely related with the superpotential $\cf$ -- see eq.~\eqref{EF-F}. Using the parameterization \eqref{BPS-superpotential-F-complex} of the BPS superpotential in terms the complex variables $V^I_\pm$ introduced in \eqref{V-variables}, the attractor equations \eqref{attractor-eqs} can be solved exactly at the supersymmetric point. This allows us to express the BPS entropy in terms of the conserved charges, and to identify the nonlinear constraint among the charges that characterizes the supersymmetric attractor point.  

As we saw earlier in the discussion of extremal black holes in the SU(2)$\times$U(1) invariant sector, the relation \eqref{EF-F} between the entropy function and the effective superpotential $\cu$ allows us to formulate the attractor equations \eqref{attractor-eqs} in terms of the superpotential $\cu$ or $\cf$. In terms of $\cu$, these conditions take the form \eqref{U-horizon-extremization} (with $\pa_u\cu|_H=0$ in the extremal case), while in terms of $\cf$ they become \eqref{eq:regularity-generic} and \eqref{extremal-entropy-generic}. For generic extremal black holes these are nonlinear algebraic equations that are not easily solvable analytically. However, for supersymmetric ones we can solve them explicitly thanks to the BPS superpotential \eqref{BPS-superpotential-F} and its parameterization in terms of the complex superpotential \eqref{complex-real}, which is holomorphic in the variables $V^{I}\equiv V_{+}^{I}$ introduced in \eqref{V-variables-real}. Changing variables from $v^I$, $\vf^a$ to $V^I$, $\lbar V^I$, the derivative transformations
in \eqref{V-derivatives} allow us to write the attractor equations \eqref{eq:regularity-generic} as
\bbxd
\vskip.2cm
\be\label{Z-extremisation}
\cz|_{H}=\pa_{V^{I}}\cz|_{H}=0\,.
\ee
\ebxd
A key simplification in this parameterization is that the attractor equations $\pa_{u_{3}}\cz|_{H}=0$ and $\pa_{\lbar V^I}\cz|_H=0$ are automatically satisfied due to the form of the complex superpotential $\cz$ in \eqref{Zpm}. Namely, $\cz$ is holomorphic in $V^I$ and homogeneous in $e^{-u_3}$ and so $\pa_{u_{3}}\cz|_{H}=-\cz|_{H}$.

The BPS attractor equations \eqref{Z-extremisation} comprise eight real independent equations for the eight real components of the variables $u_{H}, u_{3H}, V^{I}_{H}$. However, only seven of these are independent due to the scalar constraint \eqref{LI-constraint}. These seven real parameters can be taken as $u_{H}, u_{3H}, \vf^{a}_{H}$ and $v^{I}_{H}$. It follows that the supersymmetric attractor equations \eqref{Z-extremisation}, besides determining the values of these seven independent parameters, impose a constraint on the parameters of the BPS superpotential, namely the charge parameters $j$ and $q_{I}$. We now solve the attractor equations \eqref{Z-extremisation} explicitly in order to determine the values of the fields on the horizon and the constraint among the charge parameters. As a consequence, we also obtain the BPS entropy in terms of the charge parameters.

The superpotential \eqref{Zpm} implies that the attractor equation $\pa_{V^{I}}\cz|_{H}=0$ takes the form 
\be\label{real-dV-Z}
q_{I}-4i\ell^{-1}e^{-\frac{3}{2}u_{H}}=C_{IJK}V_{H}^{J}V_{H}^{K}\,.
\ee
This can be solved to obtain $V_{H}^{I}$ in terms of $u_{H}$ and the charge parameters $q_I$, namely
\bbxd
\vskip.5cm
\be\label{VH-sol}
V_{H}^{I}=\pm\frac{1}{2}\sum_{J,K}C_{IJK}\sqrt{\frac{(q_{J}-4i\ell^{-1}e^{-\frac{3}{2}u_{H}})(q_{K}-4i\ell^{-1}e^{-\frac{3}{2}u_{H}})}{2(q_{I}-4i\ell^{-1}e^{-\frac{3}{2}u_{H}})}}\,,
\ee
\ebxd
where the same sign should be picked for all values of the index $I$. Using the definition of $V^I$ in \eqref{V-variables}, this determines $v_H^I$, $u_{3H}$ and $L_H^I$ (and hence $\vf^a_H$) via
\bbxd
\be\label{real-variables}
v_{H}^{I}=\textrm{Re}V^{I}_{H}\,,\quad L_{H}^{I}=\frac{\textrm{Im}V^{I}_{H}}{\textrm{Im}V^{1}_{H}\textrm{Im}V^{2}_{H}\textrm{Im}V^{3}_{H}}\,,\quad e^{3u_{3H}}=\projsp e^{\frac{3}{2}u_{H}}
\textrm{Im}V^{1}_{H}\textrm{Im}V^{2}_{H}\textrm{Im}V^{3}_{H}\,.
\ee
\ebxd
Recall that $e^{3u_{3H}}=b_H^2$ is the $S^3$ squashing parameter at the horizon - see eq.~\eqref{squashedS3}. 

In order to determine $u_H$ in terms of the charge parameters $j$ and $q_I$ we use the first attractor equation in \eqref{Z-extremisation}, which  -- due to \eqref{real-dV-Z} -- takes the form
\be\label{real-Z}
j-4i\projsp e^{-\frac{3}{2}u_{H}}=4V_{H}^{1}V_{H}^{2}V_{H}^{3}\,.
\ee
However, \eqref{real-dV-Z} also implies that
\be
\prod_{I=1}^{3}(q_{I}-4i\ell^{-1}e^{-\frac{3}{2}u_{H}})=8\big(V_{H}^{1}V_{H}^{2}V_{H}^{3}\big)^2\,.
\ee
Combining these two relations leads to the complex equation for $u_H$
\be\label{cubic}
\big(j-4i\projsp e^{-\frac{3}{2}u_{H}}\big)^{2}=2\prod_{I=1}^{3}(q_{I}-4i\ell^{-1}e^{-\frac{3}{2}u_{H}})\,.
\ee
The real and imaginary parts of this equation give the two alternative expressions for $u_H$
\bbxd
\vskip.5cm
\be\label{u-horizon}
e^{-\frac{3}{2}u_{H}}=\frac{\ell}{4}\Big(\frac{1}{2}\sum_{I}C^{IJK}q_{J}q_{K}-\projsp\ell j\Big)^{1/2}=\frac{\ell}{4}\Big(\frac{j^{2}-2q_{1}q_{2}q_{3}}{\ell^{2}-2\sum_{I}q_{I}}\Big)^{1/2}\,.
\ee
\ebxd

Together, the results \eqref{VH-sol}, \eqref{real-variables} and \eqref{u-horizon} completely determine the values of all fields on the horizon in terms of the charge parameters $j$ and $q_I$, therefore providing a complete solution to the supersymmetric attractor equations \eqref{Z-extremisation}. However, the two expressions for $u_H$ in \eqref{u-horizon} further imply that at the supersymmetric point of extremal solutions the charge parameters $j$ and $q_I$ satisfy the constraint
\bbxd
\vskip.5cm
\be\label{nonlinear-constraint}
j^{2}-2q_{1}q_{2}q_{3}=\Big(\frac{1}{2}\sum_{I}C^{IJK}q_{J}q_{K}-\projsp\ell j\Big)\Big(\ell^{2}-2\sum_{J}q_{J}\Big)\,,
\ee
\ebxd
together with the inequality
\be\label{entropy-reality}
\frac{1}{2}\sum_{I}C^{IJK}q_{J}q_{K}-\projsp\ell j\geq 0\,,
\ee
that ensures that $u_H$ -- and hence the entropy -- is real. Moreover, in order for the constraint \eqref{nonlinear-constraint} to have real solutions for $j$, the charge parameters $q_{I}$ must satisfy
\be\label{j-reality}
\big(1-2\ell^{-2}(q_{1}+q_{2})\big)\big(1-2\ell^{-2}(q_{2}+q_{3})\big)\big(1-2\ell^{-2}(q_{1}+q_{3})\big)\geq0\,.
\ee

Finally, the expressions \eqref{u-horizon} for $u_H$ allow us to evaluate the BPS entropy in terms of $j$ and $q_I$ using either the entropy function \eqref{EF} or the result \eqref{entropy-on-ansatz}. Namely,
\bbxd
\vskip.2cm
\be\label{BPS-entropy}
S_{\rm BPS}=2\p\ce_{\rm BPS}=\frac{8\pi^{3}\ell}{\k_{5}^{2}}\Big(\frac{1}{2}\sum_{I}C^{IJK}q_{J}q_{K}-\projsp\ell j\Big)^{\frac12}=\frac{8\pi^{3}\ell}{\k_{5}^{2}}\Big(\frac{j^{2}-2q_{1}q_{2}q_{3}}{\ell^{2}-2\sum_{I}q_{I}}\Big)^{\frac12}\,.
\ee
\ebxd
Either of these expressions may be used to write the entropy in terms of the conserved electric Page charges \eqref{el-charges-ansatz} and angular momentum \eqref{ang-mom-ansatz}. The first relation in \eqref{BPS-entropy} gives
\bal
\label{entropy-Page}
S_{\rm BPS} = &\; 2\pi\ell\Big[\frac12\sum_{I}C^{IJK}Q_{J}^{\anom=1}Q_{K}^{\anom=1} -\frac{\projsp\k_{5}^{2}\ell}{4\pi^{2}}\Big(J^{\anom=1}-\frac{4\pi^{2}}{3\k_{5}^{2}} C_{IJK}v_{(0)}^{I}v_{(0)}^{J}v_{(0)}^{K}\Big)\Big]^{1/2}\,,
\eal
which generalizes the formula obtained in \cite{Kim:2006he} (see also \cite{Cassani:2018mlh,Bombini:2019jhp}) to 1/4 BPS black holes with non-trivial magnetic flux on the boundary.\footnote{The expression obtained is \cite{Cassani:2018mlh} amounts to using different values for the parameter $\anom$ in the definition of the electric charges and angular momentum. However, we have seen above that such a choice is inconsistent with a well defined free energy.}

\subsection{Complex Euclidean saddles, free energy and entropy extremization}
\label{sec:complexification}

Let us now consider the free energy and associated on-shell action of 1/4 BPS black holes described by the superpotential \eqref{BPS-superpotential-F}. In the subsequent analysis we implicitly set $\anom=1$, as is required by the quantum statistical relation between the Euclidean on-shell action and the Gibbs free energy. The BPS relation \eqref{BPS-relation-bulk} implies that the Gibbs free energy of {\em real} BPS solutions is given by the supersymmetric Casimir mass \eqref{anomaly-charge-anstatz-bulk}, namely
\bbxd
\vskip.2cm
\be\label{Gibbs-BPS}
G_{\rm BPS}(T=0,\O_{\text{BPS}},\F_{\text{BPS}}^I)= M_{\text{BPS}}-\O_{\text{BPS}} J-\F_{\text{BPS}}^{I}Q_{I}=M_{\text{Casimir}}\,.
\ee
\ebxd
Hence, contrary to AdS$_4$ BPS black holes for which there is no Casimir energy \cite{Halmagyi:2017hmw,Cabo-Bizet:2017xdr}, the Euclidean on-shell action for AdS$_5$ BPS black holes is dominated by the Casimir energy and is infinite in the strict $\b\to\infty$ limit. In particular, the Euclidean on-shell action \eqref{QSR2} admits an expansion of the form 
\bbxd
\vskip.2cm
\be\label{on-shell-BPS}
I'=\b G=\b M_{\text{Casimir}}+I_0+\co(\b^{-1})\,.
\ee
\ebxd
We are now interested in the $\co(\b^0)$ term, $I_0$, which is related with the BPS entropy.  

\begin{figure}[h]
\begin{center}
\captionsetup{
singlelinecheck=false
}
\scalebox{.6}{\includegraphics{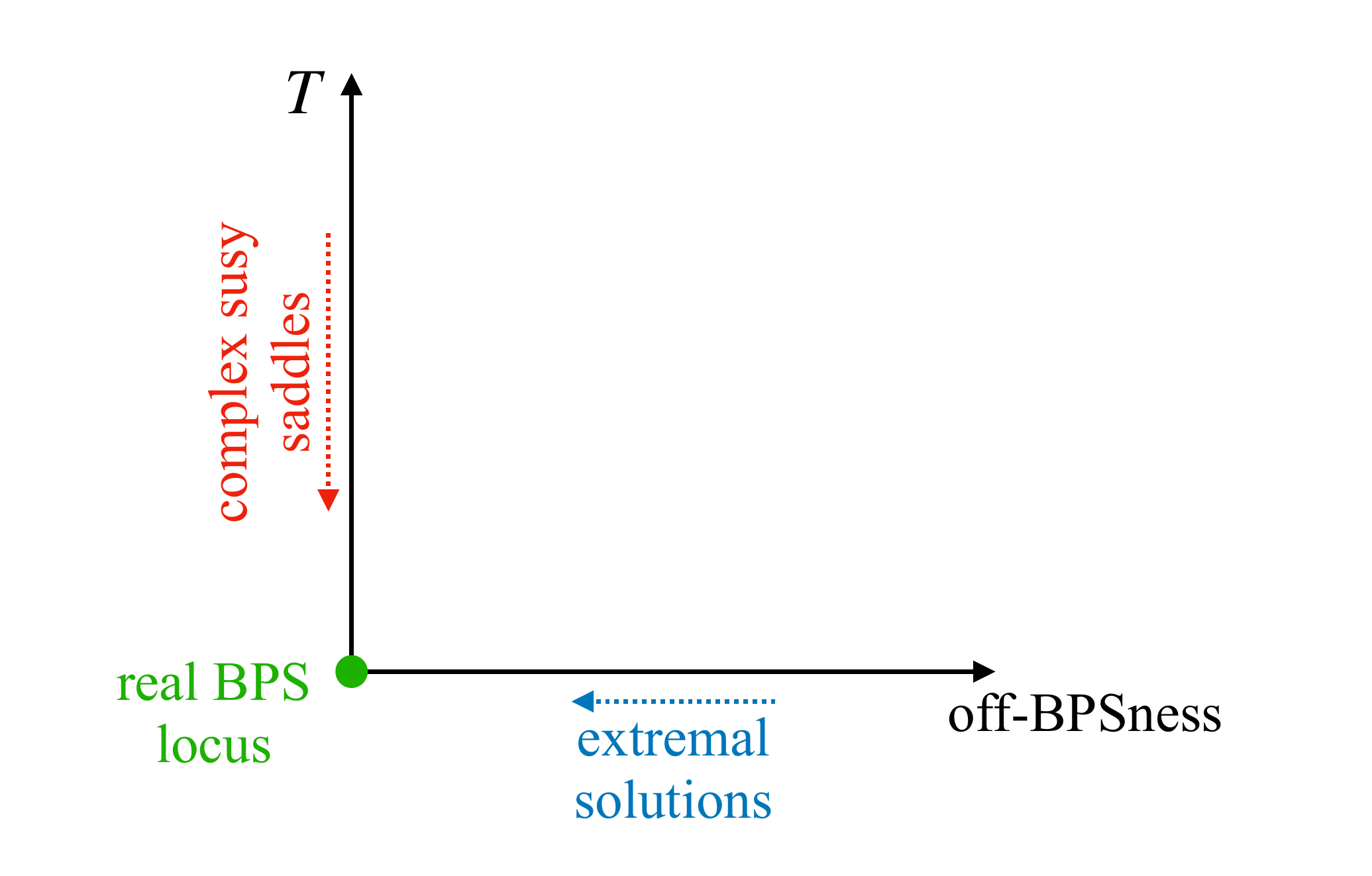}}
\vskip-.2cm
\caption{Schematic depiction of the space of regular black saddles. Real BPS solutions correspond to a point on the line of extremal, i.e. $T=\b^{-1}=0$ solutions. Moving away from the zero temperature line while preserving supersymmetry requires complexification.}
\label{fig:complex}
\end{center}
\end{figure}

The form of $I_0$ and higher order terms is not universal because it depends on the path along which one approaches the (real) BPS limit \cite{Silva:2006xv,Castro:2018ffi}  -- see fig.~\ref{fig:complex}. In order to match the field theory result obtained from supersymmetric localization, it is necessary to move away from the zero temperature line along a path that preserves supersymmetry \cite{Cabo-Bizet:2018ehj}. Such BPS solutions at finite (complex) temperature correspond to complex Euclidean saddles that cannot be analytically continued to real Lorentzian solutions. As we now show, these Euclidean solutions are described by the same BPS superpotential \eqref{BPS-superpotential-F}. 

\subsubsection{Complex supersymmetric saddles}

The BPS superpotential \eqref{BPS-superpotential-F} enables us to describe complex Euclidean saddles without having to revisit the entire problem in Euclidean supergravity. Complexifying the functions parameterizing the SU(2)$\times$U(1) invariant ansatz \eqref{ansatz}, as well as the charge parameters $j$ and $q_I$, 1/4 BPS solutions are still governed by the superpotential \eqref{BPS-superpotential-F} or \eqref{BPS-superpotential-F-complex}, except that the variables $V_\pm^I$ in \eqref{V-variables} and the functions $\cz_\pm$ in \eqref{Zpm} are now independent -- not complex conjugates. In particular, while for real BPS solutions both $\cz_+$ and $\cz_-$ vanish on the horizon, for complex BPS saddles only {\em one} of these functions is zero on the horizon. As we will see, this relaxes the constraint \eqref{nonlinear-constraint} among the charge parameters and leads to a nonzero -- albeit complex -- temperature.  

The fact that only one of $\cz_\pm$ vanishes at the horizon of complex BPS solutions implies that these solutions have a non-degenerate horizon. In particular, in the coordinate system \eqref{BH-metric}, the near horizon behavior of these functions is  
\be
\cz_+=\co(r-r_H)\,,\quad \cz_-=\co(1)\,,\qquad \text{or}\qquad \cz_+=\co(1)\,,\quad \cz_-=\co(r-r_H)\,.
\ee 
In either case, the results \eqref{NHE-ansatz} and \eqref{non-degenerate-scaling} for general non-extremal horizons apply, namely
\be
e^{-2u_{2}}=\co\big((r-r_{H})^{1/2}\big)\,,
\ee
and
\be\label{Zpm-horizon}
\cf_{\rm BPS}=\co((r-r_{H})^{1/2})\,,\qquad \pa_{u}\cf_{\rm BPS}=\co((r-r_{H})^{-1/2})\,,
\ee
so that the limit  
\be
\lim_{r\to r_H} e^{-2u_{2}}\pa_u\cf_{\rm BPS}\,,
\ee
exists. \eqref{U-horizon-extremization} implies that this limit is given by the product of the temperature and entropy. 

Using the expressions for $\cz_\pm$ in \eqref{Zpm} we find that for complex BPS solutions, depending on which of the two cases in \eqref{Zpm-horizon} is realized,
\bal
-12\pi T_\pm e^{-3u_{H}/2} =&\; \pa_{u}\cu_{\rm BPS}|_{H}= \lim_{r\to r_{H}}(e^{-2u_{2}}\pa_{u}\cf_{\rm BPS})\NO \\
= &\; 4\lim_{r\to r_{H}}e^{-2u_{2}}\Big(\sqrt{\cz_+/\cz_-}\,\pa_{u}\cz_{-}+\sqrt{\cz_{-}/\cz_{+}}\,\pa_{u}\cz_{+}\Big)\NO \\
= &\; -3i e^{-\frac{3}{2}u_{H}-u_{3H}}\bigg[\Big(\projsp\ell^{-1}\sum_{I}V_{-H}^{I}-1\Big)\lim_{r\to r_{H}}e^{-2u_{2}}\sqrt{\cz_+/\cz_-}\NO\\
&\;-\Big(\projsp\ell^{-1}\sum_{I}V_{+H}^{I}-1\Big)\lim_{r\to r_{H}}e^{-2u_{2}}\sqrt{\cz_{-}/\cz_{+}}\bigg]\,.
\eal
One of these terms vanishes, while the other gives the corresponding temperature. Namely,
\vskip-.1cm
\bbxd
\vskip.4cm
\be\label{T-complex}
\lim_{r\to r_{H}}e^{-2u_{2}}\sqrt{\cz_{\mp}/\cz_{\pm}}=\lim_{r\to r_{H}}e^{-2u_{2}}e^{\pm2i\projsp g}=\mp  \frac{4\pi i T_\pm e^{u_{3H}}}{1-\projsp \ell^{-1}\sum_{I}V_{\pm H}^{I}}\,.
\ee
\ebxd

Next, solving the horizon equations \eqref{eq:regularity-generic} we determine the complex BPS entropy in terms of the charge parameters $j$ and $q_I$. The complexified superpotential \eqref{BPS-superpotential-F-complex} is given by
\be
\cf_{\text{BPS}}(u,u_{3},V_{+}^{I},V_{-}^{I})=8\sqrt{\cz_{+}(u,u_{3},V_{+}^{I})\cz_{-}(u,u_{3},V_{-}^{I})}\,,
\ee
where $\cz_\pm$ are shown in \eqref{Zpm}. Using the change of variables in \eqref{V-derivatives} and the fact that $\cz_\pm$ are homogeneous in $e^{-u_3}$, we determine that the only independent attractor equations are 
\be
\cf_{\text{BPS}}|_H=0\,,\qquad \pa_{V^I_\pm}\cf_{\text{BPS}}|_H=0\,.
\ee
Since 
\be
\pa_{V^I_\pm}\cf_{\text{BPS}}=4\sqrt{\cz_\mp/\cz_\pm}\,\pa_{V^I_\pm}\cz_\pm\,,
\ee
these equations are equivalent to the two alternative sets of conditions on $\cz_\pm$
\bbxd
\be\label{complex-d-Z}
\cz_{+}|_{H}=\pa_{V_{+}^{I}}\cz_{+}|_{H}=0\,,\;\;\cz_{-}|_{H}\;\;\text{any}\,,\quad \text{\em or}\quad \cz_{-}|_{H}=\pa_{V_{-}^{I}}\cz_{-}|_{H}=0\,,\;\;\cz_{+}|_{H}\;\;\text{any}\,.
\ee
\ebxd
Each of these two sets of conditions is analogous to the attractor equations \eqref{Z-extremisation} for real BPS solutions, except that now one of the functions $\cz_\pm$ remains unconstrained. 

Proceeding as with the real BPS solutions, using the form of $\cz_\pm$ specified in \eqref{Zpm}, the horizon equations $\pa_{V_{+}^{I}}\cz_{+}|_{H}=0$ or $\pa_{V_{-}^{I}}\cz_{-}|_{H}=0$ read
\be\label{complex-dV-Z}
q_{I}\mp4i\ell^{-1}e^{-\frac{3}{2}u_{H}}=C_{IJK}V_{\pm H}^{J}V_{\pm H}^{K}\,,
\ee
and lead to the two alternative solutions
\bbxd
\bal
\label{VH-sol-complex}
&V_{+ H}^{I}=\pm\frac{1}{2}\sum_{J,K}C_{IJK}\sqrt{\frac{(q_{J}-4i\ell^{-1}e^{-\frac{3}{2}u_{H}})(q_{K}-4i\ell^{-1}e^{-\frac{3}{2}u_{H}})}{2(q_{I}-4i\ell^{-1}e^{-\frac{3}{2}u_{H}})}}\,,\quad V^I_{-H}\;\; \text{any}\,,\NO\\
&V_{- H}^{I}=\pm\frac{1}{2}\sum_{J,K}C_{IJK}\sqrt{\frac{(q_{J}+4i\ell^{-1}e^{-\frac{3}{2}u_{H}})(q_{K}+4i\ell^{-1}e^{-\frac{3}{2}u_{H}})}{2(q_{I}+4i\ell^{-1}e^{-\frac{3}{2}u_{H}})}}\,,\quad V^I_{+H}\;\; \text{any}\,,
\eal
\ebxd
where the overall sign is unrelated with the subscript of the variables $V_\pm^I$. 

Moreover, using \eqref{complex-dV-Z}, the horizon equations $\cz_{\pm}|_{H}=0$ reduce respectively to  
\be\label{complex-Z}
j\mp4i\projsp e^{-\frac{3}{2}u_{H}}=4V_{\pm H}^{1}V_{\pm H}^{2}V_{\pm H}^{3}\,.
\ee
However, \eqref{complex-dV-Z} also implies that
\be
\prod_{I=1}^{3}(q_{I}\mp4i\ell^{-1}e^{-\frac{3}{2}u_{H}})=8\big(V_{\pm H}^{1}V_{\pm H}^{2}V_{\pm H}^{3}\big)^2\,,
\ee
and hence
\bbxd
\vskip.4cm
\be\label{cubic-complex}
\big(j\mp4i\projsp e^{-\frac{3}{2}u_{H}}\big)^{2}=2\prod_{I=1}^{3}(q_{I}\mp4i\ell^{-1}e^{-\frac{3}{2}u_{H}})\,.
\ee
\ebxd
Since both the charge parameters $q_{I}$ and $j$, as well as $e^{-\frac{3}{2}u_{H}}$ are now complex, this is a cubic equation for $e^{-\frac{3}{2}u_{H}}$, which determines the entropy through \eqref{EF} or \eqref{entropy-on-ansatz}.

We should emphasize that although the value of $e^{-\frac{3}{2}u_H}$ is determined in terms of the charge parameters by \eqref{cubic-complex}, the value of the remaining variables on the horizon is not fully determined. Using the form of $V_\pm^I$ in \eqref{V-variables}, either of the two alternative solutions \eqref{VH-sol-complex} provides a set of three complex equations for a total of six complex parameters, $u_{3H}$, $v_H^I$ and $\vf^a_H$. We will now see that the undetermined parameters correspond to four complex chemical potentials subject to one complex constraint.

\subsubsection{Complex BPS thermodynamics}

Let us now consider the thermodynamics of these complex BPS black saddles. Starting with the chemical potentials, the nonzero temperature \eqref{T-complex} implies that 
\bal
\lim_{r\to r_{H}}e^{-2u_{2}}\cos 2g =&\;\mp  \frac{2\pi i T_\pm e^{u_{3H}}}{1-\projsp\ell^{-1}\sum_{I}V_{\pm H}^{I}}\,,\NO\\
\lim_{r\to r_{H}}e^{-2u_{2}}\sin 2g =&\; i\projsp\frac{2\pi i T_\pm e^{u_{3H}}}{1-\projsp\ell^{-1}\sum_{I}V_{\pm H}^{I}}\,.
\eal 
These limits, together with the supersymmetric integrals of motion $\mathscr C_{4}$ and $\mathscr C^{I}$ in \eqref{constants-of-motion-main} and the asymptotic values $\cos2g|_\infty=1$ and \eqref{sing-limit}, imply that
\bal
&u_{4H}\pm \frac{\projsp 2\pi i T_\pm }{1-\projsp \ell^{-1}\sum_{I}V_{\pm H}^{I}}=u_{4(0)}-\projsp e^{-2u_{2(0)}-u_{3(0)}}\,,\\
&a_H^{I}-u_{4H}v_H^{I}\mp \frac{\projsp 2\pi i T_\pm V^I_{\pm H}}{1-\projsp\ell^{-1}\sum_{I}V_{\pm H}^{I}}=a_{(0)}^{I}-\big(u_{4(0)}-\projsp e^{-u_{3(0)}-2u_{2(0)}}\big)v_{(0)}^{I}+\frac{\ell}{2}e^{2u_{3(0)}-2u_{2(0)}}\,.\NO
\eal

Inserting these relations in the expressions \eqref{angular-velocities-general} and \eqref{el-potential-general} for the angular velocity and electric chemical potentials and fixing $\L_\c$ as in \eqref{Lambda-fixing} leads to the complex potentials
\bbxd
\vskip.2cm
\be\label{Delta-def}
\O=\O_{\text{BPS}}+\D^\pm_{\O}T_\pm\,,\qquad\F^{I}=\F_{\text{BPS}}^{I}+ \D_\pm^{I}T_\pm\,,
\ee
\ebxd
where $\O_{\text{BPS}}$ and $\F_{\text{BPS}}^{I}$ denote the values of the chemical potentials for {\em real} BPS solutions given in \eqref{BPS-angular-velocity} and \eqref{BPS-el-chemical-potentials-fixed}. The values of the coefficients $\D^\pm_{\O}$, $\D_\pm^{I}$ are
\bbxd
\vskip.4cm
\be\label{deltas}
\D^\pm_{\O}=\frac{\pm 2\p i\projsp}{1-\projsp\ell^{-1}\sum_{I}V_{\pm H}^{I}}\,,\qquad
\D_\pm^{I} = -V_{\pm H}^{I}\D^\pm_{\O}\,,
\ee
\ebxd
and satisfy the constraint \cite{Hosseini:2017mds,Cabo-Bizet:2018ehj}
\bbxd
\vskip.4cm
\be\label{Delta-constraint}
\projsp\D^\pm_{\O}+\ell^{-1}\sum_{I}\D_\pm^{I}=\pm2\pi i\,.
\ee
\ebxd
This constraint implies that the complex solutions do not admit a real Lorentzian limit and therefore only make sense as complex saddles of the Euclidean path integral \cite{Bobev:2020pjk}.  

The conserved charges associated with complex BPS solutions coincide with their counterparts for real BPS black holes, except that all parameters are now complex. In particular, the mass for complex BPS solutions takes the form \eqref{BPS-mass}, while their angular momentum and electric charges are respectively given by \eqref{ang-mom-ansatz} and \eqref{el-charges-ansatz}. The only variables that get modified in the BPS relation \eqref{BPS-relation-bulk}, therefore, are the angular velocity and electric chemical potentials \eqref{Delta-def}, leading to the BPS relation for complex black saddles
\bbxd
\vskip.2cm
\be\label{BPS-complex}
M-\O J-\F^{I}Q_{I}=M_{\text{Casimir}}-(\D^\pm_{\O}J+\D_\pm^{I}Q_{I})T_\pm\,.
\ee
\ebxd

We now have all the ingredients to compute the $\co(\b^0)$ term in the on-shell action \eqref{on-shell-BPS}. The BPS relation \eqref{BPS-complex} implies that the Gibbs free energy for complex BPS black holes is
\be
G= M_{\text{Casimir}}-S T_\pm-(\D^\pm_{\O}J+\D_\pm^{I}Q_{I})T_\pm\,,
\ee  
and, therefore, the $\co(\b^0)$ term in the on-shell action \eqref{on-shell-BPS} is given by 
\bbxd
\vskip.2cm
\be\label{I0-BPS-1}
I_0=-S-(\D^\pm_{\O}J+\D_\pm^{I}Q_{I})\,.
\ee
\ebxd
Notice that all higher order terms in the temperature expansion of the on-shell action \eqref{on-shell-BPS} vanish identically when we approach real BPS black holes through complex non-extremal BPS solutions. This is not true for non-supersymmetric non-extremal solutions. 

The on-shell action term $I_0$ in \eqref{I0-BPS-1} can be expressed in terms of the chemical potentials \eqref{deltas} and the conserved charges using the solutions \eqref{complex-d-Z} of the horizon equations. Namely, contracting \eqref{complex-dV-Z} with $V_{\pm H}^{I}$ and subtracting the result from \eqref{complex-Z} leads to  
\bal
j-V_{\pm H}^{I}q_{I} = &\; -2V_{\pm H}^{1}V_{\pm H}^{2}V_{\pm H}^{3}\pm4i\projsp e^{-\frac{3}{2}u_{H}}\Big(1-\projsp\ell^{-1}\sum_{I}V_{\pm H}^{I}\Big)\NO \\
= &\; 2\frac{\D_\pm^{1}\D_\pm^{2}\D_\pm^{3}}{(\D^\pm_{\O})^{3}}\pm4ie^{-\frac{3}{2}u_{H}}\Big(\pm \frac{2\p i}{\D^\pm_{\O}}\Big)\,.
\eal
The entropy \eqref{entropy-on-ansatz} of complex BPS solutions, therefore, takes the form
\bbxd
\vskip.4cm
\be\label{S-BPS}
S^\pm_{\rm BPS}=\frac{8\p^2}{\k_5^2}\frac{\D_\pm^{1}\D_\pm^{2}\D_\pm^{3}}{(\D^\pm_{\O})^{2}}-\frac{4\p^2}{\k_5^2}\big(\D^\pm_{\O}j+\D_\pm^{I}q_{I}\big)\,.
\ee
\ebxd

Substituting this expression for the entropy in \eqref{I0-BPS-1} and using the relations \eqref{ang-mom-ansatz} and \eqref{el-charges-ansatz} respectively for the angular momentum and electric charges with $\anom=1$, we obtain
\bbxd
\vskip.4cm
\be
\label{c-susy-on-shell action}
I_0=-\frac{8\pi^{2}}{\k_{5}^{2}}\frac{\D_\pm^{1}\D_\pm^{2}\D_\pm^{3}}{(\D^\pm_{\O})^{2}}-\frac{4\pi^{2}}{3\k_{5}^{2}} \D^\pm_{\O}C_{IJK}v_{(0)}^{I}v_{(0)}^{J}v_{(0)}^{K}\,.
\ee
\ebxd
This generalizes the result for the finite part of the on-shell action \cite{Hosseini:2017mds} to the case $v^{I}_{(0)}\neq 0$.

\subsubsection{Entropy extremization}

As was first pointed out in \cite{Hosseini:2017mds}, the value \eqref{S-BPS} of the entropy of complex BPS solutions can be obtained from an extremization principle akin to Sen's entropy function approach for extremal black holes \cite{Sen:2007qy,Sen:2008vm}. The extremization functional proposed in \cite{Hosseini:2017mds} is the Legendre transform of the Euclidean on-shell action \eqref{c-susy-on-shell action} with respect to the angular momentum and electric charges. One then extremizes this functional with respect to the chemical potentials $\D^\pm_{\O}$ and $\D_\pm^{I}$, subject to the constraint \eqref{Delta-constraint}. We will now provide a derivation of this extremization principle and demonstrate that it is equivalent to the extremization of the complex superpotential $\cz_\pm$ on the horizon through equations \eqref{complex-d-Z}.   

Treating the values $u_H$, $u_{3H}$ and $V^I_{\pm H}$ of the ansatz functions on the horizon as arbitrary variables -- i.e. not imposing the conditions that result from the horizon equations \eqref{complex-d-Z} -- we can replace the variables $V^I_{\pm H}$ with $\D_\pm^{I}$ through the second equation in \eqref{deltas}. The value of the complex superpotential $\cz_{\pm}$ on the horizon can then be expressed in terms of the chemical potentials $\D_\pm^{I}$ as  
\bal\label{Zpm-Deltas}
\cz_{\pm} =&\;\frac{\projsp}{8\D^\pm_{\O}}e^{-u_{3H}}\Big(\frac{2\D_\pm^{1}\D_\pm^{2}\D_\pm^{3}}{(\D^\pm_{\O})^{2}}-\D^\pm_{\O}j-\D_{\pm}^{I}q_{I}\pm4ie^{-\frac{3}{2}u_H}\big(\projsp\D^\pm_{\O}+\sum_{I}\ell^{-1}\D_{\pm}^{I}\big)\Big)\,.
\eal

The last term in this expression coincides with the constraint \eqref{Delta-constraint}, except for the constant term $\pm 2\pi i$ that is missing. Moreover, we know that $\cz_\pm$ vanishes at the horizon\footnote{As we discussed earlier, only one of $\cz_\pm$ vanishes at the horizon, but here we consider the two possibilities simultaneously.} and so it cannot coincide with the entropy. These mismatches are addressed by considering
\bbxd
\vskip.4cm
\be\label{EF-F-S-BPS}
S^\pm_{\rm BPS}=2\p\ce^\pm_{\rm BPS}=\frac{32\p^3}{\k_5^2}\big(e^{-\frac{3}{2}u_H}+\projsp\p^{-1}e^{u_{3H}}\D^\pm_{\O}\cz_{\pm}(u_H,u_{3H},V_{\pm H}^{I})\big)\,.
\ee
\ebxd
By construction, this expression coincides on the horizon with \eqref{entropy-on-ansatz} and therefore gives the correct value for the entropy at the extremum. The same term also provides the constant $\pm 2\pi i$ that completes the constraint \eqref{Delta-constraint}. 

The expression \eqref{EF-F-S-BPS} for the entropy is analogous to the relation \eqref{EF-F} between Sen's entropy function for extremal black holes and the superpotential $\cf$. The expression for the temperature in \eqref{T-complex} and the definition of $\D_\O^\pm$ in \eqref{deltas} imply that the second term in \eqref{EF-F-S-BPS} can be written in terms of the real BPS superpotential $\cf_{\rm BPS}$ since
\be
\frac{1}{\p}e^{u_{3H}}\projsp\D^\pm_{\O}\cz_{\pm}=-\frac{1}{2\p T_\pm} e^{-2u_2}\sqrt{\cz_+\cz_-}|_H\,,
\ee
so that
\bbxd
\vskip.4cm
\be\label{EF-F-BPS}
\ce^\pm_{\rm BPS}=\frac{4\p^2}{\k_5^2}\Big(4e^{-\frac{3}{2}u}-\frac{1}{4\p T_\pm} e^{-2u_2}\cf_{\rm BPS}\Big)_H\,.
\ee
\ebxd
This coincides with the relation between Sen's entropy function and the superpotential $\cf$ for extremal black holes in \eqref{EF-F}, with the complex temperature acting effectively as the inverse AdS$_2$ radial coordinate $4\p T_\pm \sim r$.

The result \eqref{EF-F-S-BPS} provides also a derivation of the extremization functional of \cite{Hosseini:2017mds}. In particular, substituting the form of $\cz_\pm$ in terms of the chemical potentials $\D^\pm_{\O}$ and $\D_\pm^{I}$ given in \eqref{Zpm-Deltas}, we see that \eqref{EF-F-S-BPS} is a functional of $u_H$, $\D_\pm^{I}$ and $\D^\pm_{\O}$ only and takes the form
\bbxd
\be\label{EF-S-BPS}
S^\pm_{\rm BPS}=
\frac{4\p^2}{\k_5^2}\Big(\frac{2\D_\pm^{1}\D_\pm^{2}\D_\pm^{3}}{(\D^\pm_{\O})^{2}}-\D^\pm_{\O}j-\D_{\pm}^{I}q_{I}\pm 4ie^{-\frac{3}{2}u_H}\big(\projsp\D^\pm_{\O}+\sum_{I}\ell^{-1}\D_{\pm}^{I}\mp2\p i\big)\Big)\,.
\ee
\ebxd
This coincides with the extremization functional of \cite{Hosseini:2017mds} and is the Legendre transform of $I_0$ in \eqref{c-susy-on-shell action} with respect to the angular momentum and electric charges. The variable $u_H$ acts as a Lagrange multiplier imposing the constraint \eqref{Delta-constraint}. Extremizing \eqref{EF-S-BPS} with respect to all variables $u_H$, $\D_\pm^{I}$ and $\D^\pm_{\O}$ reproduces the horizon equations \eqref{complex-dV-Z} and \eqref{complex-Z}, as well as the constraint \eqref{Delta-constraint}, while it gives the correct value for the entropy at the extremum.

\section{Discussion and future directions}
\label{sec:discussion}

We reformulated the BPS equations of the SU(2)$\times$U(1) invariant sector of the five dimensional STU model as a set of gradient flow equations -- given in \eqref{flow-eqs-F} -- stemming from the 1/4 BPS superpotential \eqref{BPS-superpotential-F}. This superpotential enabled us to compute the holographic observables for all 1/4 BPS solutions in this sector and to study their thermodynamics. In particular, we showed that for complex BPS saddles, the 1/4 BPS superpotential reproduces the entropy extremization functional put forward in \cite{Hosseini:2017mds}. 

As part of our analysis, we obtained the Gutowski-Reall black holes \cite{Gutowski:2004yv} by integrating the gradient flow equations corresponding to the 1/4 BPS superpotential \eqref{BPS-superpotential-F}. However, these equations admit a much broader class of solutions that would be very interesting to explore. For example, we anticipate that the solutions found recently in \cite{Blazquez-Salcedo:2017kig,Blazquez-Salcedo:2017ghg} and \cite{Cassani:2018mlh,Bombini:2019jhp} can be further elucidated using their description in terms of the 1/4 BPS superpotential \eqref{BPS-superpotential-F}. Moreover, this superpotential could be a useful tool in the proof of uniqueness results such as the one of \cite{Lucietti:2021bbh} or in the classification of BPS solutions within the SU(2)$\times$U(1) invariant sector of the STU model.   

Another interesting prospect is to determine the effective superpotential describing non-supersymmetric solutions within the SU(2)$\times$U(1) invariant sector by solving the reduced Hamilton-Jacobi equation \eqref{HJ-F}. We obtained two exact non-supersymmetric solutions of this equation: \eqref{F_RN}, describing the Reissner-Nordstr\"om-AdS black hole, and \eqref{F-attractor} that describes the non-supersymmetric attractor solution \eqref{attractor}. However, it would be desirable to identify the effective superpotentials describing other important solutions, such as the Kerr-AdS$_5$ black hole or the general CLP solution \eqref{sol:CLP}. 

Allowing the ansatz functions in the SU(2)$\times$U(1) invariant ansatz \eqref{ansatz} to depend on time $t$ in addition to the radial coordinate $r$ leads to an effective two dimensional dilaton supergravity theory that is a consistent truncation of the 5D STU model. This is a generalization of the pure 5D gravity truncation obtained in \cite{Castro:2018ffi} and it would be interesting to extend the analysis there to this more general case. This effective 2D dilaton supergravity may also help probe the dual supersymmetric quantum mechanics whose index captures the BPS black hole entropy.

Throughout this paper we have focused exclusively on cohomogeneity one solutions of the STU model, i.e. solutions with an enhanced SU(2)$\times$U(1) isometry. This restriction was based solely on technical reasons and we would like to relax it in future work. In particular, only cohomogeneity one solutions can be described in terms of an effective superpotential. Higher cohomogeneity solutions can be described in terms of an effective {\em action} rather than a potential. Generic time-independent U(1)$\times$U(1) invariant solutions of the STU model require an ansatz that involves arbitrary functions of both $r$ and $\th$ and the corresponding effective action involves $\th$-derivatives. We anticipate that the effective action for supersymmetric solutions of this form -- in particular the general BPS black holes with two independent angular momenta obtained in \cite{Kunduri:2006ek} -- is closely related with the BPS superpotential \eqref{BPS-superpotential-F}. 

Higher cohomogeneity solutions also exist within the SU(2)$\times$U(1) invariant sector if one allows for time dependence. Examples of exact effective actions in that case are already known in the context of two dimensional dilaton gravity -- e.g. eq.~(A.2) in \cite{Castro:2018ffi} and eq.~(3.22) in \cite{Chaturvedi:2020jyy}. In fact, the necessary framework for describing higher cohomogeneity solutions is already available and -- in the case of the STU model -- it is reviewed in appendix \ref{sec:holography}. The effective action is always given by the functional $\cs[\g,A^I,\vf^a]$ that satisfies the Hamilton-Jacobi equation \eqref{HJ-eqn}, while the corresponding gradient flow equations are given by \eqref{5D-flow-generic}. Specializing this general framework to solutions of a specific cohomogeneity results in a gradient flow description in terms of the relevant effective action. 

It should be emphasized that although the focus of the present work is on solutions of the STU model in five dimensions, the effective superpotential approach is applicable to any supergravity theory in any dimension and has already been used in several different contexts to describe domain wall \cite{Freedman:2003ax,Papadimitriou:2006dr,Skenderis:2006rr,Papadimitriou:2007sj,Janssen:2007rc,HoyosBadajoz:2008fw,Lindgren:2015lia,Cremonini:2020rdx,Kim:2020dqx} and static  black holes \cite{Andrianopoli:2009je,DallAgata:2010ejj,Hyun:2012bc,Lindgren:2015lia,Dorronsoro:2016pin,Klemm:2016kxw,Klemm:2017pxv,Cabo-Bizet:2017xdr}. To our knowledge, the exact superpotential \eqref{BPS-superpotential-F} is the first example for a rotating solution. In general, the superpotential approach relies on a radial Hamiltonian formulation of the dynamics as described in appendix \ref{sec:hamiltonian}. An example of this formalism in the presence of hypermultiplets can be found in \cite{An:2017ihs}. 

Finally, our analysis provided glimpses of the direct relation between the supersymmetric Casimir energy obtained from the anomalous supersymmetry transformation of the $\cn=1$ supercurrent  \cite{Papadimitriou:2017kzw,Papadimitriou:2019gel,Papadimitriou:2019yug} with the result obtained from an analysis of supersymmetric partition functions on $S^1\times S^3$ \cite{Assel:2014paa,Lorenzen:2014pna,Assel:2015nca,Bobev:2015kza,Martelli:2015kuk,BenettiGenolini:2016qwm,Brunner:2016nyk}. We will report on this connection elsewhere.

\section*{Acknowledgments}

We are grateful to Alejandro Cabo-Bizet, Davide Cassani, Mirjam Cveti\v c, Morteza Hosseini, Seok Kim, Lampros Lamprou, Finn Larsen, James Lucietti, Dario Martelli, Sameer Murthy, Leo Pando Zayas, Maria Jos\'e Rodriguez, Kostas Skenderis, Chiara Toldo, Oscar Varela and especially Alberto Zaffaroni for valuable discussions. IP also thanks the Aspen Center for Physics and King's College London for hospitality, and the Utah State University for hospitality and partial financial support during the early stages of this work. PN is supported by a Leverhulme Trust Research Project Grant. 

\appendix

\renewcommand{\theequation}{\Alph{section}.\arabic{equation}}

\setcounter{section}{0}

\section*{Appendices}
\setcounter{section}{0}

\section{Holographic dictionary for the 5D STU model}
\label{sec:holography}

In this appendix, we summarize the radial Hamiltonian formulation of the bosonic sector of $\cn=2$ FI gauged supergravity in five dimensions and we use it in order to determine the boundary counterterms for the STU model and to derive the anomalous Ward identities of the dual field theory (see also \cite{Bombini:2019jhp}). A more general analysis that includes the fermionic fields can be found in \cite{Papadimitriou:2017kzw,An:2017ihs}. In the last part of this appendix we provide explicit expressions for the renormalized 1-point functions for the SU(2)$\times$U(1) sector described by the ansatz \eqref{ansatz} in terms of the effective superpotential, $\cu$, introduced in section \ref{sec:superpotential}.

\subsection{Radial Hamiltonian dynamics}
\label{sec:hamiltonian}

The starting point for the Hamiltonian formulation of the bosonic sector of $\cn=2$ FI gauged supergravity is the ADM decomposition of the five dimensional metric and gauge fields \cite{Arnowitt:1960es} 
\be\label{ADM}
ds^2=(N^2+N_iN^i)dr^2+2N_idrdx^i+\g_{ij}dx^idx^j,\qquad A^I=\elfg^I dr+A^I_idx^i,
\ee
where $i,j,$ run over the transverse coordinates, the radial coordinate, $r$, is identified with the Hamiltonian time, and the conformal boundary is located at $r\to\infty$. This decomposition amounts to a field redefinition, replacing the bulk metric, $g_{\m\n}$, with the induced metric on the radial slices, $\g_{ij}$, as well as the lapse and shift functions, respectively $N$ and $N_i$. Moreover, the bulk gauge fields, $A_\m^I$, are traded for the components along the radial slices, $A_i^I$, and the radial components, $\elfg^I$.

Inserting the ADM decomposition of the bulk metric, the bulk Ricci scalar becomes 
\be
R[g]=\car[\g]+K^2-K_{ij}K^{ij}+\nabla_{\m}\left(-2Kn^{\m}+2n^{\nu}\nabla_{\nu}n^{\m}\right),
\ee
where $\car[\g]$ is the Ricci scalar of the induced metric, the extrinsic curvature, $K_{ij}$, is given by 
\be
K_{ij}=\frac{1}{2N}\left(\dot{\g}_{ij}-D_iN_j-D_jN_i\right),
\ee
and $n^{\m}=(1/N,-N^i/N)$ is the unit normal vector to the constant-$r$ hypersurfaces. We use a dot to denote a derivative with respect to the radial coordinate, $\dot{}\equiv \pa_r$, while $D_i$ stands for the covariant derivative with respect to the induced metric $\g_{ij}$. Moreover, the volume element and the inverse metric take respectively the form
\be
\sqrt{-g}=N\sqrt{-\g},\qquad g^{\m\nu}=\begin{pmatrix}\frac{1}{N^2}&-\frac{N^i}{N^2}\\-\frac{N^i}{N^2}&\g^{ij}+\frac{N^iN^j}{N^2}\end{pmatrix}\,.
\ee
Finally, we follow the conventions of \cite{Papadimitriou:2017kzw} and define the four dimensional Levi-Civita symbol as $\ve^{(4)}_{ijkl}\equiv\ve^{(5)}_{rijkl}$ (see footnote \footref{foot-orientation} for our convention for the five dimensional Levi-Civita symbol). Accordingly, the four dimensional Levi-Civita tensor is defined as $\e_{(4)}^{ijkl}\equiv N\e_{(5)}^{rijkl}$ (see eq.~(A.20) in \cite{Papadimitriou:2017kzw}). Since the dimensionality of the Levi-Civita tensor can be deduced from the number of indices, we will not indicate the dimension explicitly in the following.
 
With the help of these identities, we find that the bosonic action \eqref{full-action} can be written as 
\be
\actn =\int \!\!dr  L\,,
\ee 
where the radial Lagrangian takes the form
\bal\label{5D-r-lagrangian}
L&=\frac{1}{2\k_5^2}\int d^4 x\; N\sqrt{-\g}\Big(K^2-K_{ij}K^{ij}-\frac{1}{2N^2}(\dot \vf^a-N^i\pa_i \vf^a)^2 \NO\\
&\hskip0.5in-\frac{1}{2N^{2}}G_{IJ}\g^{ij}(\dot A^I_{i}-\pa_{i}\elfg^I-N^{k}F^I_{ki})(\dot A^J_{j}-\pa_{j}\elfg^J-N^{l}F^J_{lj})\NO\\
&\hskip0.5in-\frac{1}{24N}C_{IJK}\big(4(\dot A^I_i-\pa_i \elfg^I)\e^{ijkl}F^J_{jk}A^K_l+\elfg^I\e^{ijkl}F^J_{ij}F^K_{kl}\big)\NO\\
&\hskip0.5in+\car[\g]-\frac{1}{4}G_{IJ}F^I_{ij}F^{J ij}-\frac12\pa_i \vf^a\pa^i \vf^a-V\Big)\,.
\eal
The nontrivial canonical momenta following from this Lagrangian are 
\bal\label{5Dmomenta}
\pi^{ij}=&\;\frac{\d L}{\d\dot{\g}_{ij}}=\frac{1}{2\k_5^2}\sqrt{-\g}\big(K\g^{ij}-K^{ij}\big),\NO \\
\pi_a=&\;\frac{\d L}{\d\dot \vf^a}=-\frac{1}{2\k_5^2}N^{-1}\sqrt{-\g}(\dot \vf^a-N^i\pa_i \vf^a), \NO\\
\pi^{i}_I=&\;\frac{\d L}{\d\dot A^I_{i}}=\frac{1}{2\k_5^2}\sqrt{-\g}\Big(-\frac1NG_{IJ}\g^{ij}(\dot A^J_{j}-\pa_j \elfg^J-N^{l}F^J_{lj})-\frac{1}{6}C_{IJK}\e^{ijkl}F^J_{jk}A^K_l\Big)\,,
\eal
while those conjugate to the variables $N, N_i$ and $\elfg^I$ vanish identically. Consequently, 
these variables are non-dynamical Lagrange multipliers imposing a number of first class constraints.  

These constraints can be determined from the radial Hamiltonian, which is obtained by the Legendre transform of the Lagrangian \eqref{5D-r-lagrangian}, namely 
\be\label{5D-H}
H=\int d^4 x\;(\pi^{ij}\dot{\g}_{ij}+\pi_a\dot \vf^a+\pi_I^{i}\dot A^I_{i})-L=\int d^4 x\;(N\ch+N_i\ch^i+\elfg^I\cf_I),
\ee
where
\bal\label{5D-constraints}
\ch\equiv&-\frac{\k^2_5}{\sqrt{-\g}}\Big(2\Big(\g_{ik}\g_{jl}-\frac{1}{3}\g_{ij}\g_{kl}\Big)\pi^{ij}\pi^{kl}+\p_a\p^a+G^{IJ}\hat\P_{I i}\hat\P^{i}_J\Big) \NO \\
&+\frac{\sqrt{-\g}}{2\k_5^2}\Big(-\car[\g]+\frac12\pa_i \vf^a\pa^i \vf^a+\frac{1}{4}G_{IJ}F^I_{ij}F^{J ij}+V\Big)\,, \NO\\
\ch_i\equiv&-2D_j\pi^j_i+\pi_a\pa_i \vf^a+F^{I}_{ij}\hat\P_I^{j}\,, \\
\cf_I\equiv&
-D_i\check\P^i_I\,,\NO
\eal
and we have defined the shifted gauge momenta 
\be\label{CS-momenta}
\hat\P^i_{I}\equiv\p^i_I+\frac{1}{12\k_5^2}C_{IJK}\sqrt{-\g}\,\e^{ijkl}F^J_{jk}A^K_l,\qquad \check\P^i_{I}\equiv\p^i_I-\frac{1}{24\k_5^2}C_{IJK}\sqrt{-\g}\,\e^{ijkl}F^J_{jk}A^K_l.
\ee
Since the canonical momenta conjugate to the variables $N, N_i$ and $\elfg^I$ vanish identically, Hamilton's equations for these variables lead to the first class constraints
\be\label{5D-1st-class-constraints}
\ch=\ch^i=\cf_I=0\,.
\ee
It follows that the Hamiltonian \eqref{5D-H} vanishes identically on the constraint submanifold, 
which reflects the invariance of the bulk action \eqref{5Daction} under radial and transverse diffeomorphisms, as well as U(1)$^{n_V+1}$ gauge transformations (up to boundary terms).

Notice that the form of the gauge momentum in \eqref{5Dmomenta} implies that the quantities $\hat\P^i_I$ are gauge invariant. As we discuss in section \ref{sec:thermodynamics}, they determine the Maxwell charges of the $n_V+1$ gauge fields. We will also see below that $\hat\P^i_I$ correspond to the covariant currents in the dual field theory. They differ from the consistent currents obtained from $\p_I^i$ by local Bardeen-Zumino terms \cite{Bardeen:1984pm}. In contrast, the quantities $\check\P^i_I$ are not gauge invariant, but they are conserved by virtue of the constraints \eqref{5D-1st-class-constraints}. They define the electric Page charges \cite{Marolf:2000cb}.

\subsubsection{Hamilton's equations}

Hamilton's equations follow from the Hamiltonian \eqref{5D-H}. Half of these determine the radial evolution of the generalized coordinates, $\g_{ij}$, $A_i^I$, $\vf^a$, and are equivalent to the expressions \eqref{5Dmomenta} for the canonical momenta. In the gauge 
\be\label{GF}
N_i=\elfg^I=0,\quad N\;\text{arbitrary}\,,
\ee
these take the form
\bal\label{Heqs1}
\dot\g_{ij}=&\;\frac{\d H}{\d\p_{ij}}=-\frac{4\k^2_5N}{\sqrt{-\g}}\Big(\g_{ik}\g_{jl}-\frac{1}{3}\g_{ij}\g_{kl}\Big)\pi^{kl}\,,\NO\\
\dot\vf^a=&\;\frac{\d H}{\d\p_a}=-\frac{2\k_5^2N}{\sqrt{-\g}}\p_a\,,\NO\\
\dot A_i^I=&\;\frac{\d H}{\d\p^i_I}=-\frac{2\k_5^2N}{\sqrt{-\g}}G^{IJ}\hat\P_{Ji}\,.
\eal

The other half of Hamilton's equations determine the radial evolution of the canonical momenta. In the gauge \eqref{GF}, they read  
\bal\label{Heqs2}
\dot\p^{ij}=&\;-\frac{\d H}{\d\g_{ij}}=\frac{\k_5^2N}{\sqrt{-\g}}\Big[4\Big(\p^{ik}\p_k{}^j-\frac13\p^k_k\p^{ij}\Big)+G^{IJ}\hat\P^i_I\hat\P^j_J\Big]\NO\\
&\hskip0.72in+\frac{\sqrt{-\g}}{2\k_5^2}\Big[-N\car^{ij}+D^iD^jN-\g^{ij}\square_\g N+\frac12 N\pa^i\vf^a\pa^j\vf^a+\frac12 NG_{IJ}F^{Iik}F^{Jj}{}_k\NO\\
&\hskip0.72in-N\g^{ij}\Big(-\car[\g]+\frac12\pa_k \vf^a\pa^k \vf^a+\frac{1}{4}G_{IJ}F^I_{kl}F^{J kl}+V\Big)\Big]\,,\\
\rule{.0cm}{.8cm}\dot\p_a=&\;-\frac{\d H}{\d\vf^a}=\frac{\sqrt{-\g}}{2\k_5^2}\Big(\pa_iN\pa^i\vf^a+N\square_\g\vf^a-N\pa_a V-\frac{N}{4}\pa_a G_{IJ}F^I_{ij}F^{J ij}\Big)+\frac{\k_5^2N}{\sqrt{-\g}}\pa_aG^{IJ}\hat\P_{Ii}\hat\P^i_J\,,\NO\\
\rule{.0cm}{.8cm}\dot\p^i_I=&\;-\frac{\d H}{\d A_i^I}=\frac{\sqrt{-\g}}{2\k_5^2}D_j(NG_{IJ}F^{Jji})+\frac{1}{6}C_{IJK}\e^{ijkl}\Big(NG^{LJ}\hat\P_{Lk}F^K_{jl}+2D_j\big(NG^{LJ}\hat\P_{Lk}A^K_l\big)\Big)\, .\NO
\eal

In the case of the STU model, the two sets of Hamilton's equations, \eqref{Heqs1} and \eqref{Heqs2}, are equivalent to the second order field equations \eqref{eoms}. As an illustration that is useful for our analysis in the main text, we can combine the last equations in \eqref{Heqs1} and \eqref{Heqs2} in order to obtain the Maxwell equation in \eqref{eoms} in terms of Hamiltonian variables. Namely, 
\bal
\pa_r\big(C_{IJK}\sqrt{-\g}\,\e^{ijkl}F^J_{jk}A^K_l\big)=&\;C_{IJK}\sqrt{-\g}\,\e^{ijkl}\pa_r\big(F^J_{jk}A^K_l\big)\\
=&\;C_{IJK}\sqrt{-\g}\,\e^{ijkl}\big(2D_j\dot A^J_kA^K_l+F^J_{jk}\dot A^K_l\big)\NO\\
=&\;C_{IJK}\sqrt{-\g}\,\e^{ijkl}\big(2D_j(\dot A^J_kA^K_l)-\dot A^J_kF_{jl}^K+F^J_{jk}\dot A^K_l\big)\NO\\
=&\;2C_{IJK}\sqrt{-\g}\,\e^{ijkl}D_j(\dot A^J_kA^K_l)+2C_{IJK}\sqrt{-\g}\,\e^{ijkl}F^J_{jk}\dot A^K_l\NO\\
=&\,-4\k_5^2C_{IJK}\e^{ijkl}D_j(NG^{JL}\hat\P_{Lk} A^K_l)+4\k_5^2C_{IJK}\e^{ijkl}NG^{KL}\hat\P_{Lk}F^J_{jl}\,,\NO
\eal
where we have used the last equation in \eqref{Heqs1} in the last step. From the definition of $\check\P_I^i$ in \eqref{CS-momenta} and the last equation in \eqref{Heqs2} follows that 
\be\label{HMaxwell}
\dot{\check\P}^i_I=\frac{\sqrt{-\g}}{2\k_5^2}D_j\Big(NG_{IJ}F^{Jji}+\frac{\k_5^2}{\sqrt{-\g}}C_{IJK}\e^{ijkl}NG^{LJ}\hat\P_{Lk}A^K_l\Big)\,.
\ee
This equation is equivalent to the Maxwell equation in \eqref{eoms}. However, it makes manifest the fact that, for homogeneous backgrounds, the Page charge, obtained by integrating the Page current $\check\P_I^i$ over a Cauchy surface, is radially conserved.

\subsubsection{Hamilton-Jacobi equation}

In the Hamilton-Jacobi formulation of the dynamics, the canonical momenta are expressed as gradients of the Hamilton-Jacobi functional $\cs[\g,A^I,\vf^a]$, namely 
\be\label{5D-HJ-momenta}
\p^{ij}=\frac{\d\cs}{\d\g_{ij}}\,,\qquad \p_I^{i}=\frac{\d\cs}{\d A^I_{i}}\,,\qquad \p_a=\frac{\d\cs}{\d\vf^a}\,.
\ee
Notice that, since the canonical momenta conjugate to the variables $N, N_i$ and $\elfg^I$ vanish identically, the Hamilton-Jacobi functional is independent of these variables.

Inserting the expressions \eqref{5D-HJ-momenta} for the canonical momenta in the constraints \eqref{5D-1st-class-constraints} leads to a set of functional partial differential equations for $\cs[\g,A^I,\vf^a]$ -- the Hamilton-Jacobi equations. The constraints $\ch^i=0$ and $\cf_I=0$ imply respectively that $\cs[\g,A^I,\vf^a]$ is invariant under diffeomorphisms tangent to the radial slices and U(1)$^{n_V+1}$ gauge transformations. The Hamiltonian constraint, $\ch=0$, however, leads to the nonlinear equation
\bal
\label{HJ-eqn}
&\dfrac{\k^2_5}{\sqrt{-\g}}\bigg[2\Big(\g_{ik}\g_{jl}-\dfrac{1}{3}\g_{ij}\g_{kl}\Big)\dfrac{\d\cs}{\d \g_{ij}} \dfrac{\d\cs}{\d \g_{kl}}
+\Big(\dfrac{\d\cs}{\d \vf^a}\Big)^2
\NO \\
&+G^{IJ}\g_{ij}
\Big(\dfrac{\d\cs}{\d A^I_i}+\dfrac{1}{12\k_5^2}C_{IJK}\sqrt{-\g}\,\e^{iklm}F^J_{kl}A^K_m\Big)
\Big(\dfrac{\d\cs}{\d A^J_j}+\dfrac{1}{12\k_5^2}C_{JPQ}\sqrt{-\g}\,\e^{jpqs}F^P_{pq}A^Q_s\Big)
\bigg]\NO\\ 
&=\dfrac{\sqrt{-\g}}{2\k_5^2}\Big(-\car[\g]+\frac12\pa_i \vf^a\pa^i \vf^a+\frac{1}{4}G_{IJ}F^I_{ij}F^{J ij}+V\Big)\,.
\eal
It is this equation that determines the dependence of $\cs[\g,A^I,\vf^a]$ on the generalized coordinates and so we often refer to this single equation as {\em the} Hamilton-Jacobi equation. 

The significance of the Hamilton-Jacobi functional is twofold. Firstly, evaluated at some radial cutoff $r_0$, $\cs[\g,A^I,\vf^a]$ coincides with the bulk on-shell action, $\actn$, evaluated again with an upper radial cutoff $r_0$, up to a constant. In particular, solving the Hamilton-Jacobi equation \eqref{HJ-eqn} asymptotically determines the local divergent terms of the on-shell action in covariant form, and hence the boundary counterterms required to renormalize the theory. In the next subsection of this appendix, we will briefly review the procedure for asymptotically solving the Hamilton-Jacobi equation for the STU model.     

The second key role that a solution of the Hamilton-Jacobi equation plays is that it leads to first order gradient flow equations for the radial dependence of the generalized coordinates. In particular, given a solution, $\cs[\g,A^I,\vf^a]$, of \eqref{HJ-eqn}, inserting the expressions \eqref{5D-HJ-momenta} for the canonical momenta in the Hamilton equations \eqref{Heqs1} results in the gradient flow equations
\bal\label{5D-flow-generic}
\dot{\g}_{ij}=&\;-\frac{4\k_5^2N}{\sqrt{-\g}}\Big(\g_{ik}\g_{jl}-\frac{1}{3}\g_{ij}\g_{kl}\Big)\frac{\d \cs}{\d\g_{kl}}\,,\NO\\ 
\dot\vf^a=&\;-\frac{2\k_5^2N}{\sqrt{-\g}}\frac{\d \cs}{\d \vf^a}\,,\NO\\
\dot{A}_{i}^I=&\;-\frac{2\k_5^2N}{\sqrt{-\g}}G^{IJ}\g_{ij}\Big(\frac{\d \cs}{\d A_{j}^J}+\frac{1}{12\k_5^2}C_{JKL}\sqrt{-\g}\,\e^{jklm}F^K_{kl}A^L_m\Big)\, .
\eal
The equivalence between Hamilton's equations and the second order field equations implies that the solutions of these first order equations solve the second order equations as well. It is precisely this observation that we utilize in section \ref{sec:superpotential}.

\subsection{Holographic renormalization}
\label{sec:counterterms}

As we have just discussed, the asymptotic solution of the Hamilton-Jacobi equation \eqref{HJ-eqn} determines the local and covariant boundary counterterms that are required to renormalize the on-shell action and the 1-point functions of the operators dual to the bulk supergravity fields \cite{Balasubramanian:1999re,deBoer:1999tgo,deHaro:2000vlm,Bianchi:2001kw,Martelli:2002sp,Papadimitriou:2004ap}. More fundamentally, the boundary counterterms are required in order for the variational principle to be well defined \cite{Papadimitriou:2005ii}. 

We now briefly review how the Hamilton-Jacobi equation \eqref{HJ-eqn} can be solved asymptotically using a systematic recursive procedure, following the approach to holographic renormalization developed in   
\cite{Papadimitriou:2004ap, Papadimitriou:2011qb} (see \cite{Papadimitriou:2016yit} for a review). A more detailed analysis for $\cn=2$ gauged supergravity in five dimensions, including the fermion fields, can be found in \cite{Papadimitriou:2017kzw,An:2017ihs}.

\subsubsection{Boundary counterterms}

The general asymptotic solution of the Hamilton-Jacobi equation \eqref{HJ-eqn} can be sought in the form of a covariant expansion in eigenfunctions of a suitable linear functional operator. This could be the dilatation operator \cite{Papadimitriou:2004ap}, $\d_D$, or more generally the operator \cite{Papadimitriou:2011qb}
\be\label{gamma-derivative}
\d_{\g} =\int d^4 x \,\,2\g_{ij} \dfrac{\d}{\d \g_{ij}} \,,
\ee
which enables one to determine the counterterms without specifying the scalar functions that parameterize the bulk action \eqref{5Daction} in advance. 

Using $\d_\g$, one proceeds by formally expanding the Hamilton-Jacobi functional, $\cs$, in eigenfunctions of decreasing eigenvalues as   
\be\label{HJ-expansion}
\cs = \cs_{(0)} +  \cs_{(2)} +  \cs_{(4)} + \cdots \,,
\ee
where $\d_{\g} \cs_{(2n)}=(4-2n)\cs_{(2n)}$ and $n$ counts the number of inverse metrics, $\g^{ij}$. Since the induced metric diverges as we approach the conformal boundary, a covariant expansion in powers of the inverse metric, $\g^{ij}$, is compatible with an asymptotic expansion. Inserting the formal expansion 
\eqref{HJ-expansion} in equation \eqref{HJ-eqn} and collecting terms of equal $\d_{\g}$-weight results in a set of recursive relations that determine $\cs_{(2n)}$. This recursive procedure is described in detail in \cite{Papadimitriou:2011qb}. Here, we only provide a summary, specifically for the STU model.  

Diffeomorphism and gauge covariance -- i.e. the constraints $\ch^i=0$ and $\cf_I=0$ in \eqref{5D-1st-class-constraints} -- imply that the asymptotically leading term (highest $\d_{\g}$-weight) is of the form 
\be\label{S0}
\cs_{(0)}=\frac{1}{\k_5^2}\int d^4 x\sqrt{-\g}\, U(\vf) \, ,
\ee
for some function $U(\vf)$ that depends only on the scalars. The Hamilton-Jacobi equation requires that this function is determined by the scalar potential in \eqref{scalar-potential-STU} through the equation
\be\label{U-eq}
(\pa_{a}U)^2-\frac{2}{3}U^2=\frac12V \,.
\ee
Notice that this equation coincides with the relation \eqref{V-W-eq} between the scalar potential and the STU model superpotential, $W$. It follows that $U=W$ is a solution of \eqref{U-eq}, but, in contrast to the STU model in four dimensions \cite{Cabo-Bizet:2017xdr}, it is {\em not} the correct solution that determines the boundary counterterms.   

To see why, recall that, near the conformal boundary where $\vf^a=0$, the scalar potential admits a Taylor expansion of the form
\be
V=-\frac{12}{\ell^2}+ \frac12 m_\vf^2\vec\vf^2+\cdots.
\ee
Consequently, equation \eqref{U-eq} admits two distinct branches of asymptotic solutions, whose form depends on the value of the scalar mass $m_\vf^2$. For $-4< m_\vf^2\ell^2<0$, the two branches of asymptotic solutions are 
\be\label{U-exp}
U_\pm(\vf)=\frac{3}{\ell}+\frac{1}{4\ell}\D_\pm\vec\vf^2+\cdots,
\ee  
where
\be
m_\vf^2\ell^2=\D_\pm(\D_\pm-4),\qquad \D_++\D_-=4,\quad \D_+>\D_-.
\ee
However, when the scalar mass saturates the Breitenlohner-Freedman (BF) bound \cite{Breitenlohner:1982bm}, i.e.  $m_\vf^2\ell^2=-4$, as the scalar potential \eqref{scalar-potential-STU} implies is the case for the STU model, then the two asymptotic solutions of \eqref{U-eq} are
\be\label{U-exp-BF}
U_-(\vf)=\frac{3}{\ell}+\frac{1}{2\ell}\vec\vf^2\Big(1+\frac{2}{\log \vec\vf^2}\Big)+\cdots\,,\qquad U_+(\vf)=\frac{3}{\ell}+\frac{1}{2\ell}\vec\vf^2+\cdots\,.
\ee

In either case, the asymptotic solution relevant for the boundary counterterms is $U_-(\vf)$, which corresponds to the slower asymptotic falloff of the scalar fields \cite{Papadimitriou:2004rz}. The reason why the superpotential \eqref{superpotential-STU} cannot be used as a counterterm for the 5D STU model is that it belongs to the $U_+$ branch of asymptotic solutions, which describe a vacuum expectation value for the dual dimension 2 scalar operators. The relation between $\cs\sub{0}$ and the leading asymptotic behavior of the induced fields follows from the flow equations \eqref{5D-flow-generic}. In particular, setting the lapse function to $N=1$ and inserting \eqref{S0} with $U_-$ in the flow equations, one finds that the leading asymptotic behavior of the induced fields for the STU model is
\be\label{leading-asymptotics}
\g_{ij} \sim e^{2r/\ell} g_{(0)ij}(x) \, , \qquad
A^I_i \sim A^I_{(0)i}(x) \, , \qquad
\vf^a \sim e^{-2r/\ell} \frac{r}{\ell} \vf^a_{(0)}(x) \, ,
\ee
where $g_{(0)ij}(x)$, $A^I_{(0)i}(x)$ and $\vf^a_{(0)}(x)$ are arbitrary functions of the transverse coordinates and are interpreted as sources of local operators in the dual field theory. In fact, the full asymptotic expansions of the induced fields can be obtained by inserting the covariant expansion \eqref{HJ-expansion} of the Hamilton-Jacobi functional in the flow equations \eqref{5D-flow-generic} \cite{Papadimitriou:2011qb}. 

The next-to-highest $\d_{\g}$-weight term in the covariant expansion \eqref{HJ-expansion} takes the form
\be
\cs_{(2)}=\frac{1}{\k_5^2}\int d^4 x\sqrt{-\g}\, \big(M(\vf)\pa_i \vf^a \pa^i \vf^a+\X(\vf)\car[\g]\big) \, ,
\ee
where the functions $M(\vf)$ and $\X(\vf)$ satisfy the system of equations
\bal
&2\pa_{a}U \pa_{a}\X-\frac23 U\X+\frac12=0 \, ,\NO\\
&2M\pa_a U-U\pa_a\X=0\,,\NO\\
&\frac23U(M\d_{ab}-3\pa_a\pa_b\X)+2\pa_c U\pa^c M \d_{ab}+\frac14=0\,.
\eal
Inserting the asymptotic form of $U_-(\vf)$ in \eqref{U-exp} or \eqref{U-exp-BF} we find that 
\be
M_-(\vf)=\frac{\ell}{8(\D_--1)}+\co(\vec\vf^2),\qquad \X_-(\vf)=\frac{\ell}{4}+\frac{\ell\D_-}{48(\D_--1)}\vec\vf^2+\co(\vec\vf^4)\,, 
\ee
where $\D_-=2$ when the BF bound is saturated, as is the case for the STU model. The asymptotic behavior \eqref{leading-asymptotics} of the induced fields in this case implies that the only term in $\cs\sub{2}$ that does not vanish asymptotically is the leading term in $\X$. Hence, for the purpose of determining the counterterms, there is no loss of generality in ignoring all asymptotically vanishing terms and setting
\be
\cs_{(2)}=\frac{1}{\k_5^2}\int d^4 x\sqrt{-\g}\, \frac{\ell}{4}\car[\g] \, .
\ee

The next order term, $\cs\sub{4}$, can be determined following the same procedure. A subtlety that arises at this order is that the $\d_\g$ eigenvalue of $\cs\sub{4}$ is zero in five dimensions, which leads to an explicit dependence on the radial cutoff, $r_0$, reflecting the holographic Weyl anomaly \cite{Henningson:1998gx}. However, as for $\cs\sub{2}$, all scalar dependence drops out asymptotically and so the result for the STU model coincides with that for pure gravity coupled to three Maxwell fields \cite{Bianchi:2001de,Bianchi:2001kw,Papadimitriou:2017kzw}. Putting everything together, the boundary counterterms, $\actn_{\text{ct}}\equiv-\sum_{n\leq 2}\cs\sub{2n}$, for the 5D STU model are
\be
\label{counterterms}
\actn_{\text{ct}}= -\frac{1}{\k_5^2}\int
d^4x\sqrt{-\g}\bigg(\frac{3}{\ell} +\frac{1}{2\ell} \vec\vf^2\Big(1-\frac{\ell}{2r_0}\Big) + \frac{\ell}{4} \car +\frac{\ell^2 r_0}{8}\Big(\car_{ij}\car^{ij}-\frac13\car^2-\frac{1}{\ell^2}F^I_{ij}F^{I ij}\Big)\bigg) \, ,
\ee
and the renormalized on-shell action is defined as the limit
\bbxd
\vskip.3cm
\be\label{Sren}
\actn_{\text{ren}}=\lim_{r_{0}\to\infty} (\actn + \actn_{\text{ct}})\,.
\ee
\ebxd

Notice that, besides the linear dependence on the radial cutoff, $r_0$, which is related to the conformal anomaly, the counterterms \eqref{counterterms} for the STU model contain an additional explicit cutoff dependence proportional to $\vec\vf^2$. This term arises whenever the scalar mass saturates the BF bound \cite{Bianchi:2001de,Bianchi:2001kw,Papadimitriou:2004rz} and can be deduced directly from the form of the $U_-(\vf)$ solution in \eqref{U-exp-BF}. In particular, locality of the counterterms requires that they be analytic in the induced fields. This implies that the $\log\vec\vf^2$ term in $U_-(\vf)$ must be decomposed as $-4r_0+\cdots$, where we have used the leading asymptotic behavior of the scalar fields in \eqref{leading-asymptotics} and the ellipses stand for subleading terms. It can be easily seen that these subleading terms contribute only asymptotically vanishing terms in $\cs\sub{0}$ and can therefore be eliminated, ensuring the locality of the counterterms.

\subsubsection{Supersymmetric scheme}

Although the boundary counterterms \eqref{counterterms} are uniquely and unambiguously determined, it is possible to include an arbitrary linear combination of conformal invariants to these. Such terms are local and covariant, remain finite as the radial cutoff is removed, and they correspond to a choice of renormalization scheme.       

The number of possible conformal invariants is severely restricted by the requirement that they preserve supersymmetry. In particular, this means that any candidate supersymmetric counterterm must not contain mixed terms between the gravity and vector multiplets. There exist five parity even such conformal invariants that can be constructed out of the field content of the 5D STU model, namely
\bal
\label{invariants}
&\ci_1=\int d^4x\sqrt{-\g}\,\mathscr{E}\,,\qquad \ci_2=\int d^4x\sqrt{-\g}\,\cw^2\,,\qquad \ci_3=\int d^4x\sqrt{-\g}\,\frac{1}{r_0^2}\vec\vf^2\,,\NO\\
&\ci_4=\int d^4x\sqrt{-\g}\,F^I_{ij}F^{Iij}\,,\qquad \ci_5=\sum_I\int d^4x\sqrt{-\g}\,C_{IJK}F^J_{ij}F^{Kij}\,,
\eal
where $\mathscr{E}$ is the Euler density of the induced metric and $\cw^2$ is the square of its Weyl tensor: 
\be
\label{Euler-Weyl}
\mathscr{E}=\car^{ijkl}\car_{ijkl}-4\car^{ij}\car_{ij}+\car^2\,,\qquad
\cw^2=\car^{ijkl}\car_{ijkl}-2\car^{ij}\car_{ij}+\frac13\car^2\,.
\ee
In addition, there are three parity odd invariants that do not involve mixed terms between the gravity and vector multiplets   
\be
\label{invariants-odd}
\ci_6=\int d^4x\sqrt{-\g}\,\mathscr{P},\quad \ci_7=\int d^4x\sqrt{-\g}\,\e^{ijkl}F^I_{ij}F^I_{kl}\,,\quad \ci_8=\sum_I\int d^4x\sqrt{-\g}\,\e^{ijkl}C_{IJK}F^J_{ij}F^K_{kl}\,,
\ee
where $\mathscr{P}$ is the Pontryagin density of the induced metric
\be
\mathscr{P}=\frac12\e^{ijkl}\car_{ijmn}\car_{kl}{}^{mn}\,.
\ee

Nevertheless, only certain linear combinations of these invariants are supersymmetric. Focusing on terms that preserve parity, these are the Euler density $\ci_1$ and
\bbxd
\bal
\label{susy-invariants}
\ci_2'=&\;\ci_2-\frac{2}{3\ell^2}\big(\ci_4+\ci_5\big)=\int d^4x\sqrt{-\g}\,\Big[\cw^2-\frac{2}{3\ell^2}\Big(\d_{JK}+\sum_IC_{IJK}\Big)F^J_{ij}F^{Kij}\Big]\,,\\
\ci_3'=&\;\frac{1}{3\ell^2}(6\ci_3-2\ci_4+\ci_5)=\frac{1}{3\ell^2}\int d^4x\sqrt{-\g}\,\Big[\frac{6}{r_0^2}\vec\vf^2-\Big(2\d_{JK}-\sum_IC_{IJK}\Big)F^J_{ij}F^{Kij}\Big]\,.\NO
\eal 
\ebxd
Using the decomposition \eqref{gauge-field-decomposition} of the gauge fields, it is straightforward to verify that $\ci_2'$ involves only the graviphoton, while $\ci_3'$ depends only on the vector multiplet fields. A parity preserving supersymmetric scheme, therefore, is parameterized by an arbitrary linear combination of $\ci_1$, $\ci_2'$, and $\ci_3'$. However, $\ci_1$ does not affect any correlation functions and is numerically zero for the backgrounds we consider. This leaves the local superconformal invariants $\ci_2'$ and $\ci_3'$ that parameterize the supersymmetric scheme relevant to our analysis.

\subsubsection{One-point functions}
\label{sec:1pt}

The radial Hamiltonian formulation of the bulk theory allows us to compute the 1-point functions of the dual operators by evaluating the on-shell canonical momenta directly from the supergravity solution. In particular, one need not evaluate the on-shell action in order to determine the 1-point functions.

The renormalized 1-point functions are obtained from the limits
\bal
\label{1pt-generic}
\<\ct^i{}_j\> \equiv &\; -\lim_{r_{0}\to\infty}2\frac{e^{4r_{0}/\ell}}{\sqrt{-\g}}\g_{kj}\frac{\d}{\d\g_{ik}}(\actn +\actn_{\text{ct}})=-\lim_{r_{0}\to\infty}2\frac{e^{4r_{0}/\ell}}{\sqrt{-\g}}(\pi^i{}_j+\pi^i_{\text{ct}\,j})\,,\NO \\
\<\cj_{I}^{i}\> \equiv &\; \lim_{r_{0}\to\infty}\frac{e^{4r_{0}/\ell}}{\sqrt{-\g}}\frac{\d}{\d A_{i}^{I}}(\actn +\actn_{\text{ct}})=\lim_{r_{0}\to\infty}\frac{e^{4r_{0}/\ell}}{\sqrt{-\g}}(\pi_{I}^{i}+\pi_{\text{ct}\,I}^i)\,,\NO \\
\<\co_{a}\>\equiv &\; \lim_{r_{0}\to\infty}\frac{(r_{0}/\ell)e^{2r_{0}/\ell}}{\sqrt{-\g}}\frac{\d}{\d\vf^{a}}(\actn +\actn_{\text{ct}})=\lim_{r_{0}\to\infty}\frac{(r_{0}/\ell)e^{2r_{0}/\ell}}{\sqrt{-\g}}(\pi_{a}+\pi_{\text{ct}\,a})\, ,
\eal
which are computed as follows. Firstly, the contribution of the counterterms to the canonical momenta is determined by differentiating the boundary counterterms \eqref{counterterms} with respect to the relevant field. Allowing for the supersymmetric renormalization scheme dependence corresponding to adding a linear combination of superconformal invariants \eqref{susy-invariants}, namely
\bal
\label{susy-invariants-combo}
&-\frac{\ell^2}{16\k_5^2}\big(\a_2\,\ci_2'+2\a_3\,\ci_3'\big)=\\
&-\frac{\ell^2}{16\k_5^2}\int d^4x\sqrt{-\g}\,\Big[\a_2\cw^2-\frac{2}{3\ell^2}\Big((\a_2+2\a_3)\d_{JK}+(\a_2-
\a_3)\sum_IC_{IJK}\Big)F^J_{ij}F^{Kij}+\frac{4\a_3}{\ell^2r_0^2}\vec\vf^2\Big]\,,\NO
\eal
where $\a_2$ and $\a_3$ are arbitrary constants that specify the supersymmetric scheme, we get
\bal
\label{counterterms-momenta}
\p_{\rm ct}^{ij}\equiv&\;\frac{\d\actn\sbtx{ct}}{\d\g_{ij}}=-\frac{1}{2\k_5^2}\sqrt{-\g}\,\bigg[\frac{3}{\ell} +\frac{1}{2\ell} \vec\vf^2\Big(1-\frac{\ell}{2r_0}+\frac{\ell\a_3}{2r_0^2}\Big) + \frac{\ell}{4} \car+\frac{\ell^2 r_0}{8}\Big(\car_{kl}\car^{kl}-\frac13\car^2\Big)\NO\\
&-\frac{1}{24}\Big((3r_0+\a_2+2\a_3)\d_{JK}+(\a_2-
\a_3)\sum_IC_{IJK}\Big)F^J_{kl}F^{Kkl}\bigg]\g^{ij}+\frac{1}{\k_5^2}\sqrt{-\g}\,\bigg[\frac{\ell}{4}\car^{ij}\NO\\
&+\frac{\ell^2 }{4}(r_0+\a_2)\cb^{ij}-\frac{1}{12}\Big((3r_0+\a_2+2\a_3)\d_{JK}+(\a_2-
\a_3)\sum_IC_{IJK}\Big)F^{Ji}{}_{l}F^{Kjl}\bigg]\,,\NO\\
\p_{\text{ct}\,I}^{i}\equiv&\;\frac{\d\actn\sbtx{ct}}{\d A^I_i}=-\frac{1}{6\k_5^2}\sqrt{-\g}\Big((3r_0+\a_2+2\a_3)\d_{IJ}+(\a_2-
\a_3)\sum_KC_{KIJ}\Big)D_jF^{J ji}\,,\NO\\
\p_{\text{ct}\,a}\equiv&\;\frac{\d\actn\sbtx{ct}}{\d\vf^a}=-\frac{1}{\k_5^2\ell}\sqrt{-\g}\,\Big(1-\frac{\ell}{2r_0^2}(r_0-\a_3)\Big)\vf_a\,,
\eal 
where the Bach tensor in 4 dimensions, $\cb_{ij}$, can be expressed in terms of the Weyl tensor as
\be\label{bBach}
\cb^{ij}=\Big(D_k D_l+\frac12 \car_{kl}\Big)\cw^{ikjl}\,.
\ee
Recall that the Weyl tensor corresponds to the traceless part of the Riemann tensor and can be compactly expressed in terms of the Schouten tensor
\be\label{Schouten}
\cp_{ij}=\frac{1}{2}\Big(\car_{ij}-\frac{1}{6}\car \g_{ij}\Big)\,,
\ee
as
\be
\label{Weyl}
\cw_{ikjl}=\car_{ikjl}+\g_{il}\cp_{kj}+\g_{kj}\cp_{il}-\g_{ij}\cp_{kl}-\g_{kl}\cp_{ij}\,.
\ee

Given these expressions for the contribution of the counterterms to the canonical momenta, the 1-point functions \eqref{1pt-generic} are computed by evaluating the canonical momenta \eqref{5Dmomenta} on a given supergravity background.

\subsubsection{Ward identities}
\label{sec:WardIDs}

The radial Hamiltonian formulation of the bulk theory also provides a straightforward derivation of the anomalous Ward identities in the dual theory. In particular, the first class constraints \eqref{5D-constraints} are in one-to-one correspondence with the holographic Ward identities.  

Using the boundary counterterms \eqref{counterterms} and the definition of the renormalized 1-point functions in \eqref{1pt-generic}, we obtain the identities
\bbxd
\bal\label{WardIDs}
&D_{(0)j}\<\ct^{j}{}_{i}\>+\<\co_{a}\> D_{(0)i}\vf_{(0)}^{a}+F_{(0)ij}^{I}\Big(\<\cj_{I}^{j}\>+\frac{1}{12\k_{5}^{2}}C_{IJK}\epsilon_{(0)}^{jklp}F_{(0)kl}^{J}A_{(0)p}^{K}\Big) = 0\,,\NO\\
&\<\ct^i{}_i\>+2\vf^a_{(0)}\<\co_a\>=\ca_{(0)}^W\,,\NO\\
&D_{(0)i}\<\cj_{I}^{i}\> = \ca_{(0)I}\,,
\eal
\ebxd
corresponding respectively to boundary diffeomorphisms, local Weyl rescalings, and U(1)$^{n_V+1}$ gauge transformations. The quantities $\ca_{(0)}^W$  and $\ca_{I(0)}$ respectively in the trace and U(1)$^{n_V+1}$ Ward identities denote the local functions of the sources 
\bbxd
\bal\label{anomalies}
\ca_{(0)}^W=&\;\frac{\ell^3}{8\k_5^2}\Big(\frac13\car^2[g\sub{0}]-\car_{ij}[g\sub{0}]\car^{ij}[g\sub{0}]+\frac{1}{\ell^2}F^I_{(0)ij}F_{(0)}^{I ij}-\frac{2}{\ell^4}\vec\vf_{(0)}^2\Big)\,,\NO\\
\ca_{(0)I}=&\;\frac{1}{48\k_{5}^{2}}C_{IJK}\epsilon_{(0)}^{ijkl}F_{(0)ij}^{J}F_{(0)kl}^{K}\,,
\eal
\ebxd
and represent the conformal, R-symmetry, and U(1)$^{n_V}$ flavor symmetry anomalies of the dual quantum field theory. Note that the integrated conformal anomaly can be expressed in terms of the invariants \eqref{invariants} as
\be
\int d^4x\sqrt{-g_{(0)}}\ca_{(0)}^W=\frac{\ell^3}{16\k_5^2}\lim_{r_0\to\infty}(\ci_1-\ci_2'-2\ci_3')\,.
\ee

We should note that the U(1) current operators, $\cj_I^i$, defined in \eqref{1pt-generic} are the {\em consistent} currents and the anomalies in \eqref{WardIDs} are therefore the corresponding consistent anomalies, which satisfy the Wess-Zumino consistency conditions. The gauge invariant currents, are the so called {\em covariant} currents, which differ from the consistent ones by local Bardeen-Zumino terms \cite{Bardeen:1984pm}, namely 
\be\label{cov-currents}
\<\cj^i_{\text{cov}\,I}\>=\<\cj^i_I\>+\frac{1}{12\k_5^2}C_{IJK}\,\e_{(0)}^{ijkl}F^J_{(0)jk}A^K_{(0)l}\,.
\ee
These currents correspond to the gauge invariant momenta $\hat\P_I^i$ in \eqref{CS-momenta}. Although the anomalies \eqref{anomalies} -- and hence the divergence of both the consistent and covariant currents --  vanish numerically on supersymmetric backgrounds, the Bardeen-Zumino terms do not, as is clear from eq.~\eqref{cov-currents-STU} below.

\subsection{One-point functions for the SU(2)$\times$U(1) sector}
\label{STU-1pt-fns}

So far, the holographic analysis in this appendix applies to the full STU model. In this final subsection we compute the renormalized 1-point functions \eqref{1pt-generic} specifically for solutions within the ansatz \eqref{ansatz}. These 1-point functions are used in section \ref{sec:thermodynamics} in order to evaluate the conserved charges of black hole solutions.  

Within the ansatz \eqref{ansatz}, the boundary counterterms \eqref{counterterms} can be expressed in terms of an effective point particle superpotential (cf. \eqref{HJ1d}) as  
\be\label{counterterms-ansatz}
\actn_{\text{ct}}=\frac{4\pi^2}{\k_5^2} \int dt \,\, \cu_{\text{ct}}= \frac{4\pi^2}{\k_5^2} \int dt \,\, e^{-2u_2}\cf_{\text{ct}}\, ,
\ee
where
\bbxd
\bal\label{counterterms-ansatz-F}
&-\cf_{\text{ct}} =
\frac{12}{\ell}e^{-2u}
-\frac{\ell}{2}e^{-u+u_{3}}(e^{3u_{3}}-4)
+\frac{2}{\ell}e^{-2u}\vec{\vf}^{2}\Big(1-\frac{\ell}{2r_0^2}(r_0-\a_3)\Big)\\
&+\frac{\ell^2}{3} (r_0+\a_2)\,e^{2u_3}(e^{3u_3}-1)^2-\frac13\,e^{2u_3}\Big((3r_0+\a_2+2\a_3)\d_{JK}+(\a_2-
\a_3)\sum_IC_{IJK}\Big)v^Jv^K\,,\NO
\eal
\ebxd
and as in \eqref{counterterms-momenta} above the constants $\a_2$ and $\a_3$ correspond to the linear combination of superconformal invariants \eqref{susy-invariants-combo} and characterize the supersymmetric renormalization scheme. 

For solutions within the ansatz \eqref{ansatz}, the canonical momenta \eqref{5Dmomenta} can be expressed in terms of the point particle momenta \eqref{1dmomenta} as follows: 
\bal
\label{5D1D-momenta}
\p^{tt}=&\;\frac{1}{64\p^2}\sin\th\, e^{u_1+3u_2}(\p_{u_1}+\p_{u_2}),\NO\\
\p^{t\j}=&\;\frac{1}{64\p^2}\sin\th\,\big(-e^{u_1+3u_2}u_4(\p_{u_1}+\p_{u_2})+2e^{u_1-u_2-2u_3}\p_{u_4}\big),\NO\\
\p^{\th\th}=&\;\frac{1}{3\times64\p^2}\sin\th\, e^{u_1-u_2+u_3}(-3\p_{u_1}+\p_{u_2}-2\p_{u_3}),\NO\\
\p^{\j\j}=&\;\frac{1}{3\times64\p^2}\Big(\cos\th\cot\th\, e^{u_1-u_2+u_3}(-3\p_{u_1}+\p_{u_2}-2\p_{u_3})\NO\\
&+\sin\th\, e^{u_1-u_2-2u_3}(-3\p_{u_1}+\p_{u_2}+4\p_{u_3})\NO\\
&+3\sin\th\,u_4\big(e^{u_1+3u_2}u_4(\p_{u_1}+\p_{u_2})-4e^{u_1-u_2-2u_3}\p_{u_4}\big)\Big),\NO\\
\p^{\j\f}=&\;-\frac{1}{3\times64\p^2}\cot\th\, e^{u_1-u_2+u_3}(-3\p_{u_1}+\p_{u_2}-2\p_{u_3}),\NO\\
\p^{\f\f}=&\;\frac{1}{3\times64\p^2}\text{cosec}\,\th\, e^{u_1-u_2+u_3}(-3\p_{u_1}+\p_{u_2}-2\p_{u_3}),\NO\\
\p_I^t=&\;\frac{1}{16\p^2}\sin\th\,\Big(\p^a_I-\frac{4\p^2}{3\k_5^2}C_{IJK}v^Jv^K\Big),\quad
\p_I^\j=\;\frac{1}{16\p^2}\sin\th\,\Big(\p^v_I-\frac{8\p^2}{3\k_5^2}C_{IJK}a^Jv^K\Big),\NO\\
\p_1=&\;\frac{1}{16\p^2}\sin\th\,\p_{\vf_1},\quad \p_2=\;\frac{1}{16\p^2}\sin\th\,\p_{\vf_2}.
\eal

Using these relations, one can evaluate the renormalized 1-point functions \eqref{1pt-generic} in terms of the renormalized superpotential
\be
\cf_{\text{ren}}=\cf + \cf_{\text{ct}}\,,
\ee
by invoking the first order equations \eqref{flow-eqs-U}. For generic solutions within the ansatz \eqref{ansatz}, we find that the non-vanishing components of the stress tensor 1-point functions are
\bbxd
\bal\label{T-vevs}
\<\ct^{t}{}_{\j}\> = &\;  \sec\th\,\<\ct^t{}_\f\>= -\frac{1}{4\k_{5}^{2}}e^{2u_{1(0)}}\Big(j-v_{(0)}^Iq_I+\frac13C_{IJK}v_{(0)}^Iv_{(0)}^Jv_{(0)}^K\Big)\,,\NO \\
\<\ct^{t}{}_{t}\> = &\;u_{4(0)}\<\ct^t{}_\j\> -\frac{1}{4\k_{5}^{2}}\lim_{r_{0}\to\infty}e^{4r_{0}/\ell}e^{2u}\cf_{\rm ren}\,,\NO \\
\<\ct^{\j}{}_{\f}\> = &\; -\cos\th\Big(\frac{1}{4\k_{5}^{2}}\lim_{r_{0}\to\infty}e^{4r_{0}/\ell}e^{2u}\pa_{u_3}\cf_{\rm ren}+u_{4(0)}\<\ct^t{}_\j\>\Big)\,,\NO\\
\<\ct^{\th}{}_{\th}\> = &\; \<\ct^\f{}_\f\>=\frac{1}{6\k_{5}^{2}}\lim_{r_{0}\to\infty}e^{4r_{0}/\ell}e^{2u}\pa_{u}\cf_{\rm ren}-\frac13\big(\<\ct^t{}_t\>+\sec\th\<\ct^\j{}_\f\>\big)\,,\NO \\
\<\ct^{\j}{}_{\j}\> = &\; \<\ct^\f{}_\f\>+\sec\th\<\ct^\j{}_\f\>\,,\NO\\
\<\ct^{\j}{}_{t}\> = &\;u_{4(0)}\big(\<\ct^\f{}_\f\>-\<\ct^t{}_t\>+\sec\th \<\ct^\j{}_\f\>\big)+\big(u_{4(0)}^2-e^{-4u_{2(0)}-2u_{3(0)}}\big)\<\ct^t{}_\j\>\,.
\eal
\ebxd
Similarly, the nonzero components of the covariant U(1)$^3$ currents \eqref{cov-currents} take the form
\bbxd
\bal\label{J-vevs}
\<\cj_{\text{cov}\,I}^{t}\> = &\; \frac{1}{4\k_{5}^{2}}e^{2u_{1(0)}}\big(q_{I}-C_{IJK}v_{(0)}^{J}v_{(0)}^{K}\big)\,, \\
\<\cj_{\text{cov}\,I}^{\j}\> = &\;\frac{1}{4\k_{5}^{2}}\lim_{r_{0}\to\infty}e^{4r_{0}/\ell}e^{2u}\pa_{v^I}\cf_{\rm ren}-u_{4(0)}\<\cj_{\text{cov}\,I}^{t}\>\,,\NO
\eal
\ebxd
where we have used the fact that for SU(2)$\times$U(1) invariant solutions 
\be\label{cov-currents-STU}
\<\cj^i_{\text{cov}\,I}\>=\<\cj^i_I\>-\frac{1}{6\k_5^2}\d^i_te^{2u_{1(0)}} C_{IJK}v^J_{(0)}v^K_{(0)}+\frac{1}{6\k_5^2}\d^i_\j e^{2u_{1(0)}}C_{IJK}\,v^J_{(0)}a^K_{(0)}\,.
\ee
Finally, the scalar 1-point functions take the form
\bbxd
\vskip.4cm
\be\label{scalar-vevs}
\<\co_{a}\> =  \frac{1}{4\k_{5}^{2}}\lim_{r_{0}\to\infty}(r_{0}/\ell)e^{2r_{0}/\ell}e^{2u}\pa_{\vf^a}\cf_{\rm ren}\,.
\ee
\ebxd

\subsubsection{Ward identities}
\label{sec:WardIDs-ansatz}

It is straightforward to verify that these 1-point functions satisfy the diffeomorphism and U(1)$^3$  Ward identities in \eqref{WardIDs} identically. However, the trace Ward identity leads to a general condition on $\cf\sbtx{ren}$, without reference to any specific solution. In particular, the expressions \eqref{T-vevs} determine that the 1-point function of the trace of the stress tensor is
\be
\<\ct^i{}_i\>=\frac{1}{2\k_{5}^{2}}\lim_{r_{0}\to\infty}e^{4r_{0}/\ell}e^{2u}\pa_u\cf_{\rm ren}\,.
\ee
Evaluating the conformal anomaly on the ansatz \eqref{ansatz}, the trace Ward identity in \eqref{WardIDs} implies that $\cf\sbtx{ren}$ satisfies the condition
\be\label{trace-ansatz}
\lim_{r_{0}\to\infty}e^{4r_{0}/\ell}e^{2u}\Big(\pa_u\cf_{\rm ren}+\vf^a\pa_a\cf_{\rm ren}+\frac{\ell^3}{2}e^{2u_3}\Big(\frac13 (e^{3u_3}-1)^2-\frac{1}{\ell^2}v^Iv^I\Big)+\frac{\ell}{4r_0^2}e^{-2u}\vec\vf^2\Big)=0\,.
\ee

\section{Field theory backgrounds and BPS relations}
\label{sec:boundary}

In this appendix we summarize the structure of the field theory backgrounds on the boundary induced by bulk supergravity solutions within the SU(2)$\times$U(1) invariant ansatz \eqref{ansatz}. We first discuss the most general backgrounds compatible with the ansatz, before deriving the constraints imposed by boundary supersymmetry.

\subsection{Parameterizations of the round and squashed $S^3$}

The round $S^3$ can be thought of as a surface in $\bb C^2$. Introducing complex coordinates $w_{1,2}$ on $\bb C^2$, the round $S^3$ of unit radius is the surface
\be
|w_{1}|^{2}+|w_{2}|^{2}=1.
\ee
For the purposes of this paper it is particularly useful to view $S^3$ as the Hopf vibration $S^1 \hookrightarrow S^3 \rightarrow S^2$, which leads to the following three coordinate parameterizations:\footnote{Recall that any odd dimensional sphere $S^{2n+1}$ can be written as a $U(1)$ fibration over $\bb C\bb P^{n}$ -- see e.g. appendix B of \cite{Larios:2019kbw}.}
\bbxd
\begin{tabular}{lll}
\\
{\bf Hopf} &  & \\
$0\leq \f_a,\,\f_b,\,4\vth\leq 2\p$ & $w_1=\sin\vth e^{i\f_a}$ & $ds^2=d\vth^2+\sin^2\vth d\f_a^2+\cos^2\vth d\f_b^2$  \\ 
& $w_2=\cos\vth e^{i\f_b}$ & \\ 
{\bf Euler} &  & \\
$0\leq \j,\,2\f,\,4\th\leq 4\p$ & $w_1=\sin(\frac{\th}{2}) e^{\frac{i(\j-\f)}{2}}$ & $ds^2=\frac14((d\j+\cos\th d\f)^2+d\th^2+\sin^2\th d\f^2)$  \\ 
&  $w_2=\cos(\frac{\th}{2}) e^{\frac{i(\j+\f)}{2}}$ &  \\ 
{\bf Projective} &  & \\
$0\leq \J\leq 2\p$ & $w_1=ze^{i\Psi}/\sqrt{1+z\bar{z}}$ & $ds^2=(d\Psi+\bs)^{2}+dzd\bar{z}/(1+z\bar{z})^{2}$  \\ 
& $w_2=e^{i\Psi}/\sqrt{1+z\bar{z}}$ & \\   
\end{tabular}
\ebxd
where the one-form
\be
\bs= \frac{i}{2(1+z\bar{z})}(zd\bar{z}-\bar{z}dz)\,,
\ee
is the potential of the K\"ahler form on $\bb C\bb P^1$, i.e. 
\be
\text{J}_{\fatc\fatp^{1}}=\frac{1}{2}d\bs=\frac{i}{2}\frac{dz\wedge d\bar{z}}{(1+z\bar{z})^{2}}=-\frac{1}{4}\sin\th d\th\wedge d\f\,.
\ee

The left-invariant one-forms $\s_{1,2,3}$ satisfy the relations $d\s_1=\s_2\wedge \s_3$ (and cyclic permutations) and in the above coordinate systems take the form 
\bbxd
\begin{tabular}{ll}
\\
{\bf Hopf} 
  & $\s_1=-2\sin(\f_a+\f_b)d\vth+\cos(\f_a+\f_b)\sin(2\vth)(d\f_b-d\f_a)$\\
  & $\s_2=2\cos(\f_a+\f_b)d\vth+\sin(\f_a+\f_b)\sin(2\vth)(d\f_b-d\f_a)$\\ 
  & $\s_3=2\sin^2\vth d\f_a+2\cos^2\vth d\f_b$\\&\\
{\bf Euler}
  & $\s_1=-\sin\j d\th+\cos\j\sin\th d\f$\\
  & $\s_2=\cos\j d\th+\sin\j \sin\th d\f$\\ 
  & $\s_3=d\j+\cos\th d\f$\\&\\
{\bf Projective} 
  & $\s_1=\frac{i}{(1+z\bar{z})}(e^{2i\J}dz-e^{-2i\J}d\bar{z})$\\
  & $\s_2=\frac{1}{(1+z\bar{z})}(e^{-2i\J}d\bar{z}+e^{2i\J}dz)$\\ 
  & $\s_3=2d\J+\frac{i}{(1+z\bar{z})}(zd\bar{z}-\bar{z}dz)=2(d\J+\bs)$\\
\end{tabular}
\ebxd

The metric on the round $S^3$ can be expressed in terms of the left-invariant one-forms as 
\be\label{S3-metric-appendix}
d\O_3^2=\frac14(\s_1^2+\s_2^2+\s_3^2)=\frac14 \s_3^2+\frac14d\Om_2^2\,,
\ee
where $d\O_{2}^{2}$ is the canonical metric on $S^2$ 
\be
d\O_{2}^{2} = \s_{1}^{2}+\s_{2}^{2}=d\th^2+\sin^2\th d\f^2\,.
\ee
When $S^2$ is viewed as $\bb C\bb P^1$, this is related to the Fubini-Study metric on $\bb C\bb P^1$ as
\be
ds_{\rm FS}^{2}(\fatc\fatp^{1}) = \frac{dzd\bar{z}}{(1+z\bar{z})^{2}}=\frac{1}{4}d\O_{2}^{2}\,.
\ee

\subsubsection{Squashed $S^{3}$}

The squashed (Berger) sphere can also be parameterized in terms of the above coordinate systems and the corresponding left-invariant one-forms. In particular, the metric on the squashed $S^3$ with squashing parameter $b^2>0$ is given by (see e.g. \cite{Blazquez-Salcedo:2017kig,Blazquez-Salcedo:2017ghg}) 
\be\label{squashedS3}
d\O_{3}^{2}(b) = \frac14(b^{2}\s_3^2+d\O_{2}^{2})=b^{2}(d\Psi+\bs)^{2}+ds_{\rm FS}^{2}(\fatc\fatp^{1})\,.
\ee

\subsubsection{Lens spaces}

Another generalization of $S^3$ that is covered by the SU(2)$\times$U(1) ansatz \eqref{ansatz} is the family of lens spaces, $L(p,q)$. Parameterized by a pair of coprime integers $p,q\in \bb N$, these are quotients of $S^3$ (viewed as a surface in $\bb C^2$ as above) by the $\bb Z/p$ action $(w_1,w_2)\mapsto (e^{2\p ip}w_1,e^{2\p iq/p}w_2)$. Although we do not discuss explicit examples of this type, our analysis applies equally to this case, albeit with some minor modifications, such as a different overall factor in the 1D effective Lagrangian \eqref{1DLagrangian}.

\subsection{Field theory backgrounds induced from the bulk} 

Given the parameterizations of $S^3$ and its squashing discussed above, we can now describe the structure of the field theory backgrounds induced on the boundary by generic bulk solutions within the SU(2)$\times$U(1) ansatz \eqref{ansatz}. The field theory background is specified by the leading coefficients in the asymptotic expansions \eqref{leading-asymptotics} of the supergravity fields, $g_{(0)ij}$,  $A_{(0)i}^I$ and $\vf^a_{(0)}$. In general, these are arbitrary functions of the boundary coordinates, but for solutions within the ansatz \eqref{ansatz} their form is constrained. Field theory backgrounds induced on the boundary by solutions within the SU(2)$\times$ U(1) ansatz are parameterized by the twelve constants $u_{1,2,3,4(0)}$, $a^I_{(0)}$, $v^I_{(0)}$, $\vf^a_{(0)}$, or equivalently $h_{1,2,3(0)}$, $u_{4(0)}$, $a^I_{(0)}$, $v^I_{(0)}$, $\vf^a_{(0)}$,  where $h_{1,2,3(0)}$ correspond to the asymptotic values of the functions introduced in \eqref{h-to-u}.

The boundary metric for SU(2)$\times$ U(1) invariant AlAdS solutions takes the form
\be\label{b-metric-ansatz}
ds_{(0)}^2=g_{(0)ij}dx^{i}dx^{j}=-\mtrc_{1(0)}dt^{2}+\frac{\mtrc_{2(0)}}{4}d\O_{2}^{2}+\frac{\mtrc_{3(0)}}{4}(\s_{3}+u_{4(0)}dt)^{2}\,.
\ee
For example, global AdS$_5$ induces the boundary metric (up to an arbitrary Weyl factor)
\be
ds^2_{(0)}=-dt^{2}+\ell^{2}d\O_{3}^{2}\,,
\ee
which corresponds to 
\be
\mtrc_{1(0)}=1\,,\qquad\mtrc_{2(0)}=\mtrc_{3(0)}=\ell^2\,,\qquad u_{4(0)}=0\,.
\ee
Similarly, the solutions discussed in \cite{Cassani:2018mlh} induce a boundary metric with a squashed $S^3$, 
\be
ds^2_{(0)}=(2\acaso)^{2}\Big(-\frac{1}{\vcas^{2}}dt^{2}+\frac{\ell^{2}\vcas^{2}}{4}\s_{3}^{2}+\frac{\ell^{2}}{4}d\O_{2}^{2}\Big)\,,
\ee
which corresponds to the values
\be
\mtrc_{1(0)}=\frac{4\acaso^{2}}{\vcas^{2}}\,,\qquad\mtrc_{2(0)}=4\ell^{2}\acaso^{2}\,,\qquad\mtrc_{3(0)}=4\ell^{2}\acaso^{2}\vcas^{2}\,,\qquad u_{4(0)}=0\,.
\ee

The general form of the boundary gauge fields for SU(2)$\times$U(1) invariant solutions is 
\be\label{b-gauge-ansatz}
A_{(0)}^{I}=A_{(0)i}^{I}dx^{i}=a_{(0)}^{I}dt+v_{(0)}^{I}\s_{3}=a_{(0)}^{I}dt+2v_{(0)}^{I}d\J+\frac{iv_{(0)}^{I}}{(1+z\bar{z})}(zd\bar{z}-\bar{z}dz)\,.
\ee
The corresponding field strengths can be easily evaluated, namely 
\be
F_{(0)}^{I}=4v_{(0)}^{I}\text{J}_{\fatc\fatp^{1}}=2v_{(0)}^{I}\frac{idz\wedge d\bar{z}}{(1+z\bar{z})^{2}}=-v_{(0)}^{I}\sin\th d\th\wedge d\f\,,
\ee
and so their only non-vanishing component is
\be\label{b-magnetic-field}
F_{(0)\th\f}^{I}=-v_{(0)}^{I}\sin\th\,.
\ee

\subsubsection{Symmetry transformations and non-rotating frame}

The field theory background induced on the boundary by supergravity solutions within the SU(2)$\times$U(1) ansatz can be simplified by means of the residual symmetry transformations $u_4\to u_4+c$, $a^I\to a^I +a_0^I+ c v^I$, discussed in section \ref{sec:superpotential}. In particular, these transformations can be used to set to zero the background parameters $a^I_{(0)}$ and $u_{4(0)}$, at least locally. Doing so may be desirable in some situations, but one should pay attention to global considerations, as is evident from the discussion in section \ref{sec:thermodynamics}. Notice that $-u_{4(0)}\equiv \O_{\infty}$ corresponds to the angular velocity on the boundary and so setting $u_{4(0)}=0$ amounts to going to a non-rotating frame at the boundary by means of the coordinate transformation $\j \to \j+\O_{\infty}t$.

\subsection{Constraints from boundary supersymmetry}

So far we discussed field theory backgrounds induced on the boundary by SU(2)$\times$U(1) invariant bulk supergravity solutions, without demanding boundary supersymmetry. As we now show, imposing supersymmetry on the boundary leads to certain constraints on the parameters $(h_{1,2,3(0)}, u_{4(0)},a^I_{(0)}, v^I_{(0)},\vf^a_{(0)})$ that describe the field theory backgrounds.  

Recall that 5D $\cn=2$ gauged supergravity coupled to $n_V$ vector multiplets induces 4D $\cn=1$ off-shell conformal supergravity coupled to $n_V$ vector multiplets on the boundary \cite{Balasubramanian:2000pq}. It follows that supersymmetric field theory backgrounds on the conformal boundary correspond to solutions of the $\cn=1$ conformal supergravity Killing spinor equations \cite{Festuccia:2011ws}. The general structure of such backgrounds has been discussed extensively in the literature \cite{Klare:2012gn,Dumitrescu:2012ha,Genolini:2016ecx,Papadimitriou:2017kzw}. Here we are interested in the subclass of such backgrounds that are also compatible with the SU(2)$\times$U(1) invariant ansatz \eqref{ansatz}. 

\subsubsection{Gravity multiplet}

The gravity multiplet backgrounds compatible with the SU(2)$\times$U(1) symmetry are described by the parameters $(h_{1,2,3(0)}, u_{4(0)},a_{(0)}, v_{(0)})$, where the boundary graviphoton components $a_{(0)}$, $v_{(0)}$ are determined by the relation \eqref{gauge-field-decomposition}, namely 
\be
a_{(0)}=-\frac{1}{\sqrt{3}}(a_{(0)}^1+a_{(0)}^2+a_{(0)}^3),\qquad v_{(0)}=-\frac{1}{\sqrt{3}}(v_{(0)}^1+v_{(0)}^2+v_{(0)}^3)\,.
\ee

Euclidean $\cn=1$ supersymmetric backgrounds of the Killing spinor equations of conformal supergravity are $T^2$ fibrations over a Riemann surface $\S$. In the special case when one of the $T^2$ cycles is trivially fibered, the Lorentzian version of such backgrounds takes the form (we follow the conventions of \cite{Papadimitriou:2017kzw} except that $A_{\rm here}=-2 A_{\rm there}$, $\wt t_{\rm here}=t_{\rm there}$, $\wt\j_{\rm here}=\j_{\rm there}$)       
\bal\label{susy-background-mu}
ds^2_{(0)} =&\; \Om^2(z,\bar z)\Big[-\ell^{-2}d\wt t^2+\Big( d\wt\j+\frac{i}{2}\pa_{\bar z}\m d\bar z-\frac{i}{2}\pa_z\m dz\Big)^2+4e^w dz d\bar z\Big],\NO\\
A_{(0)} =&\; \frac{2\ell}{\sqrt 3}\Big[-\frac{1}{8\ell} e^{-w}\pa_z\pa_{\bar z}\m\, d \wt t+\frac 14 e^{-w}\pa_z\pa_{\bar z}\m\Big( d\wt\j+\frac{i}{2}\pa_{\bar z}\m  d\bar z-\frac{ i}{2}\pa_z\m dz\Big)\NO\\
&\hskip1.4cm+\frac{i}{4}(\pa_{\bar z}w d\bar z -\pa_z w dz )+\g'\ell^{-1} d \wt t+\g d\wt\j +d\l(z,\bar z)\Big]\,,
\eal 
where $\O(z,\bar z)$, $w(z,\bar z)$, $\m(z,\bar z)$ and $\l(z,\bar z)$ are arbitrary functions on the Riemann surface $\S$, while $\g$ and $\g'$ are constants. We refer the reader to \cite{Genolini:2016ecx} for a discussion of the geometric significance of these functions and constants, as well as of the generic case when both $T^2$ cycles are non-trivially fibered. 

The most general choice of $\O(z,\bar z)$, $w(z,\bar z)$, $\m(z,\bar z)$, $\l(z,\bar z)$, $\g$ and $\g'$ such that the background \eqref{susy-background-mu} is compatible with the SU(2)$\times$U(1) invariant ansatz is
\bal\label{boundary-bgd}
&\O^2=\frac{\ell^2}{4}\,,\qquad \m=2sb\log(1+z\bar{z})\,,\qquad  e^{w}=s^2e^{-\m/sb}=\frac{s^2}{(1+z\bar{z})^{2}}\,,\NO\\
& \g=-\frac{1}{2sb}\,,\qquad \g'=\text{arbitrary}\,,\qquad \l=\text{const.}\,,
\eal
where $s$ and $b$ are arbitrary nonzero constants. Noticing that these satisfy
\be
 e^{-w}\pa_{z}\pa_{\bar{z}}\m=\frac{2b}{s}\,,
\ee
and rescaling the $T^2$ coordinates as
\be\label{coordinate-rescaling}
\wt t =2h_{1(0)}^{1/2} t,\qquad \wt\j=2sb\J,
\ee
where $h_{1(0)}>0$ is another arbitrary constant, we find that the background \eqref{susy-background-mu} becomes
\bal
ds^2_{(0)} =&\; -h_{1(0)}dt^2+\frac{\ell^2s^2}{4}d\O_2^2+\frac{\ell^2s^2b^2}{4}\s_3^2,\NO\\
A_{(0)} =&\; \frac{2}{\sqrt 3}\Big[2h_{1(0)}^{1/2}\Big(\g'-\frac{b}{4s} \Big) d t+\frac{\ell}{2}(b^2-1)\s_3\Big]\,.
\eal 

Comparing with \eqref{b-metric-ansatz} and \eqref{b-gauge-ansatz}, we read off 
\be
h_{2(0)}=\ell^2s^2\,,\qquad h_{3(0)}=\ell^2s^2b^2\,,\qquad u_{4(0)}=0\,,\qquad a_{(0)}= \frac{4h_{1(0)}^{1/2}}{\sqrt{3}}\Big(\g'-\frac{b}{4s} \Big)\,,
\ee
as well as
\bbxd
\vskip.4cm
\be\label{b-susy-gravity}
v_{(0)}=\frac{\ell}{\sqrt 3}\(h_{3(0)}/h_{2(0)}-1\)\,.
\ee
\ebxd
These expressions allow for arbitrary values for $h_{1,2,3(0)}$ and $a_{(0)}$. The apparent restriction to $u_{4(0)}=0$ is an artifact due the trivial fibering of the time coordinate we started from. Starting instead from the most general $\cn=1$ background, one finds that any value of $u_{4(0)}$ is compatible with supersymmetry. Alternatively, as we have seen, a nonzero value for $u_{4(0)}$ can be generated by a coordinate transformation, which provides an alternative argument showing that any value of $u_{4(0)}$ preserves supersymmetry. However, the relation between the magnetic charge parameter $v_{(0)}$ and the squashing of the $S^3$ in \eqref{b-susy-gravity} is a genuine constraint imposed by boundary supersymmetry. As we show in section \ref{sec:superpotential}, the same constraint follows from the bulk BPS equations.    

It is instructive to present an alternative derivation of the supersymmetry constraint \eqref{b-susy-gravity}, by solving the $\cn=1$ conformal supergravity Killing spinor equations directly within the $\sutwo\times\uone$ ansatz. This is a warm up for the more complicated analysis of the bulk Killing spinor equations in appendix \ref{sec:KS} and allows us to obtain the explicit form of the conformal Killing spinor, as well as to easily generalize the above result to nonzero $u_{4(0)}$. In terms of chiral spinors, the conformal supergravity Killing spinor equation for the gravity multiplet takes the form       
\be
\label{eq:CKSBoundary}
\cd_{(0)i}\e_{(0)}^+\equiv\Big(D_{(0)i}-\frac{i\sqrt{3}}{2\ell}A_{(0)i}\Big)\e_{(0)}^+=\frac{1}{\ell}\Gamma_{(0)i}\e_{(0)}^-\,,
\ee
where $\e^\pm_{(0)}$ are chiral (commuting) spinors satisfying $\Gamma_{5}\epsilon^\pm_{(0)}=\pm\epsilon^\pm_{(0)}$, with $\Gamma_{5}=i\Gamma_{\tflat}\Gamma_{1}\Gamma_{2}\Gamma_{3}$. Notice that contracting with $\Gamma_{(0)}^{i}$ determines
\be
\e_{(0)}^-=\frac{\ell}{4}\Gamma_{(0)}^{i}\cd_{(0)i}\e_{(0)}^+\,,\label{eq:CKSBoundary-Minus}
\ee
and so \eqref{eq:CKSBoundary} can be viewed as an equation for the spinor $\e_{(0)}^+$. 

In order to solve the conformal Killing spinor equation \eqref{eq:CKSBoundary}, the spinor $\e_{(0)}^+$ must satisfy the additional projection condition
\be\label{b-spinor-projection}
\Gamma_{12}\e_{(0)}^+=-i \projsp \e_{(0)}^+\,,
\ee
where $\projsp=\pm 1$, since $\G_{12}^2=-\bb 1$. (Further details on our spinor conventions may be found in appendix \ref{sec:KS} and appendix A of \cite{Papadimitriou:2017kzw}.) Since $\e_{(0)}^+$ is also chiral, the corresponding backgrounds will preserve two real supercharges. Using this spinor projection and the vielbein basis
\bbxd
\vskip.4cm
\be\label{b-vielbein}
 e^{\underline t}_{(0)}=h_{1(0)}^{1/2} dt\,,\qquad
e^{1,2}_{(0)}=\frac12h_{2(0)}^{1/2}\s_{1,2}\,,\qquad e^3_{(0)}=\frac12h_{3(0)}^{1/2}(\s_3+u_{4(0)} dt)\,,
\ee
\ebxd
the conformal Killing spinor equation \eqref{eq:CKSBoundary} is equivalent to the set of equations 
\bal\label{eq:CKSBoundary-basis}
\pa_{\th}\e_{(0)}^+ = &\; 0\,,\NO \\
\pa_{\f}\e_{(0)}^+ = &\; \cos\th\,\pa_{\j}\e_{(0)}^+ + i \projsp \sin\th\,\pa_{\th}\e_{(0)}^+\,,\NO \\
\pa_{\j}\e_{(0)}^+ = &\; \frac{i\projsp}{2\mtrc_{2(0)}}\big(\mtrc_{2(0)}(1+\projsp\sqrt{3}\ell^{-1}v_{(0)})-\mtrc_{3(0)}\big)\e_{(0)}^+\,,\NO \\
\pa_{t}\e_{(0)}^+ = &\; \frac{u_{4(0)}-2\projsp\mtrc_{1(0)}^{1/2}}{\mtrc_{3(0)}^{1/2}}\pa_{\j}\e_{(0)}^++\frac{i}{2}\Big(2\mtrc_{1(0)}^{1/2}\mtrc_{3(0)}^{-1/2}(1+\projsp\sqrt{3}\ell^{-1}v_{(0)})\NO\\
&+\sqrt{3}\ell^{-1}(a_{(0)}-u_{4(0)}v_{(0)})-\projsp u_{4(0)}-3\mtrc_{1(0)}^{1/2}\mtrc_{3(0)}^{1/2}\mtrc_{2(0)}^{-1}\Big)\e_{(0)}^+\,.
\eal

The first two equations in \eqref{eq:CKSBoundary-basis} imply that $\e_{(0)}^+$ is independent of $\th$, $\j$ and $\f$. As a result, the third equation leads to the algebraic constraint
\bbxd
\vskip.4cm
\be\label{eq:v-squashing}
v_{(0)}=\frac{\projsp\ell}{\sqrt 3}\(h_{3(0)}/h_{2(0)}-1\)\,,
\ee
\ebxd
which coincides with \eqref{b-susy-gravity} for the spinor projection choice $\projsp=1$, which was implicit in the earlier discussion. Finally, the $t$-dependence of $\e_{(0)}^+$ is determined by the last equation in \eqref{eq:CKSBoundary-basis}, which, using the constraint \eqref{eq:v-squashing}, can be simplified as
\be
\pa_{t}\e_{(0)}^+ =  i\o_0\e_{(0)}^+\,,
\ee
where the constant $\om_0$ is given by
\bbxd
\vskip.4cm
\be\label{BKS-frequency}
\o_0 =  \frac{1}{2}\Big(\sqrt{3}\,\ell^{-1}(a_{(0)}-u_{4(0)}v_{(0)})-\projsp u_{4(0)}-\mtrc_{1(0)}^{1/2}\mtrc_{3(0)}^{1/2}\mtrc_{2(0)}^{-1}\Big)\,.
\ee
\ebxd

It follows that the conformal Killing spinor $\e^+_{(0)}$ takes the form
\bbxd
\vskip.3cm
\be\label{epsilon-plus}
\e_{(0)}^+(t) =  \mtrc_{1(0)}^{1/4}e^{i\o_0t}\epsilon_{0}\,,
\ee
\ebxd
where $\epsilon_{0}$ is a constant (commuting) spinor satisfying the projections $\Gamma_{5}\epsilon_{0}=\epsilon_{0}$,  $\Gamma_{12}\epsilon_{0}=-i \projsp\epsilon_{0}$, and normalized
such that $\epsilon_{0}^{\dagger}\epsilon_{0}=1$. The factor of $\mtrc_{1(0)}^{1/4}$
has been introduced to ensure the proper normalization ($\ck_{(0)}^{t}=1$)
of the associated Killing vector
\be\label{KV}
\ck_{(0)}=-i\bar\e_{(0)}^{\,+}\Gamma_{(0)}^{i}\e_{(0)}^+\pa_{i}=\pa_{t}-\big(u_{4(0)}-2\projsp\mtrc_{1(0)}^{1/2}\mtrc_{3(0)}^{-1/2}\big)\pa_{\j}\,.
\ee
Finally, the $\e_{(0)}^-$ component of the conformal Killing spinor can be evaluated using \eqref{eq:CKSBoundary-Minus}:\footnote{Notice that the conformal Killing spinor \eqref{epsilon-plus}-\eqref{epsilon-minus} differs from the one in \cite{Genolini:2016ecx,Papadimitriou:2017kzw} by a local Lorentz rotation due to the fact that the vielbein basis \eqref{b-vielbein} is rotated relative to that used in those references. More specifically,  $e^{\underline{a}}_{(0)\text{there}}=(e^{-L})^{\underline{a}}{}_{\underline{b}}\,e^{\underline{b}}_{(0)}$, and so   $\epsilon^+_{(0)\text{there}}=e^{-\frac{1}{4}L_{\underline{a}\underline{b}}\Gamma^{\underline{a}\underline{b}}}\epsilon^+_{(0)}=e^{i\projsp\g\wt{\j}}\epsilon^+_{(0)}$, where $\g$ and $\wt\j$ are defined in \eqref{boundary-bgd} and \eqref{coordinate-rescaling}.}
\vskip-.2cm
\bbxd
\vskip.4cm
\be\label{epsilon-minus}
\e_{(0)}^-=-i\frac{\ell \projsp}{2}\mtrc_{3(0)}^{1/2}\mtrc^{-1}_{2(0)}\Gamma_{3}\e_{(0)}^+=-i\frac{\ell\projsp}{2}\mtrc_{1(0)}^{1/4}\mtrc_{3(0)}^{1/2}\mtrc^{-1}_{2(0)} e^{i\o_0t} \Gamma_{3}\epsilon_{0}\,.
\ee
\ebxd

\subsubsection{Vector multiplets}

The vector multiplet backgrounds compatible with the SU(2)$\times$U(1) symmetry are described by the parameters $(\wt a^a_{(0)}, \wt v^a_{(0)},\vf^a_{(0)})$, where again the boundary components of the vector multiplet gauge fields $\wt a^a_{(0)}$, $\wt v^a_{(0)}$ are determined by the relation \eqref{gauge-field-decomposition}, namely 
\bal
&\wt a^1_{(0)}=\frac{1}{\sqrt{6}}(a_{(0)}^1+a_{(0)}^2-2a_{(0)}^3),\qquad \wt v^1_{(0)}=\frac{1}{\sqrt{6}}(v_{(0)}^1+v_{(0)}^2-2v_{(0)}^3)\,,\NO\\
&\wt a^2_{(0)}=\frac{1}{\sqrt{2}}(a_{(0)}^1-a_{(0)}^2),\qquad \wt v^2_{(0)}=\frac{1}{\sqrt{2}}(v_{(0)}^1-v_{(0)}^2)\,.
\eal

A supersymmetric vector multiplet background satisfies the Killing spinor equations following from setting to zero the supersymmetry transformations of the gauginos. In our conventions these take the form (see e.g. (6.49) in \cite{Freedman:2012zz}. Following \cite{Papadimitriou:2017kzw}, we use conventions such that the boundary spinors are Weyl.)
\be
\d\l_{(0)}^a=-\frac{i}{4}\Big(\frac{i}{2}\G_{(0)}^{ij}\wt F^a_{(0)ij}+\frac{1}{\ell}\vf^a_{(0)}\Big)\e^+_{(0)} =0\, .
\ee
\eqref{b-magnetic-field} implies that the only nonzero component of the vector multiplet field strengths is  
\be
\wt F_{(0)\th\f}^{a}=-\wt v_{(0)}^{a}\sin\th\,.
\ee
Using the vielbein basis \eqref{b-vielbein} and the spinor projection \eqref{b-spinor-projection}, setting the supersymmetry transformation of the gauginos to zero leads to the vector multiplet constraints  
\bbxd
\vskip.3cm
\be
\label{b-susy-vector}
\vf^a_{(0)}=-4\projsp\ell \wt v^a_{(0)}h^{-1}_{2(0)}\,,
\ee
\ebxd
relating the scalar sources in the vector multiplets to the corresponding magnetic fluxes. As we verify in section \ref{sec:superpotential}, these constraints follow also from the bulk BPS equations.

\subsection{Anomalous fermionic currents and BPS states}

In the previous subsection we saw that supersymmetry imposes certain constraints on the field theory background, i.e. on the possible sources for local operators. However, the expectation values of local operators in a supersymmetric state are also restricted by supersymmetry and satisfy further conditions. As we will review in this subsection, these conditions lead to the BPS relations among the conserved charges. A key observation for determining the correct form of the BPS relations is the fact that the fermionic operators in the theory, namely the supercurrent $\cs^i$ and the vector multiplet currents $j^\l_a$, possess an anomalous transformation under rigid supersymmetry \cite{Papadimitriou:2017kzw,An:2017ihs}.    

The anomalous rigid supersymmetry transformations of the fermionic operators in the conformal current multiplet and in the two vector multiplets can be read off the results of \cite{Papadimitriou:2017kzw,Papadimitriou:2019gel,Papadimitriou:2019yug}. To do so, we decompose the current operators $\cj^I$ into the R-current and vector multiplet Noether currents using the relations \eqref{gauge-field-decomposition} for the background gauge fields, namely    
\be\label{R-vector-currents}
\cj_R^i=-\frac{1}{\sqrt{3}}(\cj_1^i+\cj_2^i+\cj_3^i),\quad \wt \cj_1^i =\frac{1}{\sqrt 6}(\cj_1^i+\cj_2^i-2 \cj_3^i),\quad \wt \cj_2^i=\frac{1}{\sqrt 2}(\cj_1^i-\cj_2^i)\,.
\ee
From the conservation law of the currents $\cj^I$ in \eqref{WardIDs} and the anomalies \eqref{anomalies} follows that the covariant divergence of the currents \eqref{R-vector-currents} are given respectively by the anomalies 
\bal
\ca_{(0)R}=&\;-\frac{1}{24\sqrt{3}\k_{5}^{2}}\epsilon_{(0)}^{ijkl}\Big(F_{(0)ij}F_{(0)kl}-\frac12\wt F_{(0)ij}^{1}\wt F_{(0)kl}^{1}-\frac12\wt F_{(0)ij}^{2}\wt F_{(0)kl}^{2}\Big)\,,\NO\\
\wt\ca_{(0)1}=&\;\frac{1}{24\sqrt{6}\k_{5}^{2}}\epsilon_{(0)}^{ijkl}\Big(\sqrt 2F_{(0)ij}\wt F_{(0)kl}^{1}-\wt F_{(0)ij}^{1}\wt F_{(0)kl}^{1}+\wt F_{(0)ij}^{2}\wt F_{(0)kl}^{2}\Big)\,,\NO\\
\wt\ca_{(0)2}=&\;\frac{1}{24\sqrt{6}\k_{5}^{2}}\epsilon_{(0)}^{ijkl}\Big(\sqrt{2}F_{(0)ij}\wt F_{(0)kl}^{2}+2\wt F_{(0)ij}^{1}\wt F_{(0)kl}^{2}\Big)\,.
\eal
The corresponding covariant currents \eqref{cov-currents} take the form
\bbxd
\bal\label{R-vector-cov-currents}
\hspace{-.1cm}\<\cj_{R\,{\rm cov}}^i\>=&\,\<\cj_{R}^i\>-\frac{1}{6\sqrt{3}\k_{5}^{2}}\e_{(0)}^{ijkl}\Big(F_{(0)kl} A_{(0)j}-\frac12\wt F^1_{(0)kl}\wt A^1_{(0)j}-\frac12\wt F^2_{(0)kl}\wt A^2_{(0)j}\Big),\\
\hspace{-.1cm}\rule{.0cm}{.7cm}\<\wt\cj^i_{1\,{\rm cov}}\>=&\,\<\wt\cj^i_1\>+\frac{1}{12\sqrt{6}\k_{5}^{2}}\e_{(0)}^{ijkl}\big(\sqrt{2}F_{(0)kl}\wt A_{(0)j}^{1}+\sqrt{2}\wt F^{1}_{(0)kl}A_{(0)j}-2 \wt F_{(0)kl}^{1}\wt A^{1}_{(0)j}+2\wt F_{(0)kl}^{2}\wt A^{2}_{(0)j}\big),\hspace{-.1cm}\NO\\
\hspace{-.1cm}\rule{.0cm}{.7cm}\<\wt\cj^i_{2\,{\rm cov}}\>=&\,\<\wt\cj^i_2\>+\frac{1}{12\sqrt{6}\k_{5}^{2}}\e_{(0)}^{ijkl}\big(\sqrt{2}F_{(0)kl}\wt A_{(0)j}^{2}+\sqrt{2}\wt F^{2}_{(0)kl}A_{(0)j}+2\wt F_{(0)kl}^{1}\wt A^{2}_{(0)j}+2\wt F_{(0)kl}^{2}\wt A^{1}_{(0)j}\big).\hspace{-.1cm}\NO
\eal
\ebxd

Using the above expressions for the R-current and two flavor currents, the anomalous transformations of the fermionic operators under the rigid supersymmetry associated with the conformal Killing spinor \eqref{epsilon-plus}-\eqref{epsilon-minus} can be read off 
\cite{Papadimitriou:2017kzw,Papadimitriou:2019gel,Papadimitriou:2019yug}. Accounting for the minor differences in conventions, we obtain 
\bbxd
\bal\label{fermionic-current-transformations}
\hspace{-.1cm}\d_\e\<\cs^i\>
=&\,-\frac{1}{2}\G_{(0)}^j\e^+_{(0)}\<\ct^i_j\>-\frac{i\ell}{4\sqrt{3}}\big(4\d^{[i}_j\d^{k]}_l-i \e_{(0)}{}^i{}_{j}{}^{k}{}_l\big) \G_{(0)}^j\cd_{(0)k}\big(\e^+_{(0)}\<\cj_{R\,{\rm cov}}^l\>\big)+\frac{\sqrt{3}i}{2}\e^-_{(0)}\<\cj_{R\,{\rm cov}}^i\>\hspace{-.2cm}\NO\\
&-\frac{i\ell}{4\sqrt{3}\k_{5}^{2}}\big(\G_{(0)}^{[i}{}_k\d^{j]}_l-\d^{[i}_k\d^{j]}_l-i\e_{(0)}^{ij}{}_{kl}\big)\cd_{(0)j}\big(F_{(0)}^{kl}\e^-_{(0)}\big)\NO\\
&-\frac{3\ell^2}{8\k_5^2}g_{(0)}^{k[l}\G_{(0)}^{ij]}\cd_{(0)j}\Big(\big(R_{(0)kl}-\tfrac16 R_{(0)}g_{(0)kl}\big)\e^-_{(0)}\Big),\NO\\
\hspace{-.1cm}\rule{.0cm}{.7cm}\d_\e\<j^\l_a\>
=&\,-\frac{1}{\ell}\G_{(0)i}\e^+_{(0)}\<\wt\cj^i_{a\,{\rm cov}}\>+i\G_{(0)}^i\cd_{(0)i}\big(\e^+_{(0)}\<\co_a\>\big),
\eal
\ebxd
where $\e^\pm_{(0)}$ are the (commuting) components \eqref{epsilon-plus}-\eqref{epsilon-minus} of the conformal Killing spinor. Notice that only the covariant currents enter in these transformations, as follows from general properties of the supersymmetry algebra and of the corresponding Ward identities \cite{Minasian:2021png}.

The relations among the expectation values of the bosonic operators in supersymmetric states follow from the fact that the left hand side of the expressions \eqref{fermionic-current-transformations} must vanish in a supersymmetric state. In particular, multiplying the identities \eqref{fermionic-current-transformations} from the left with $-i\bar\e_{(0)}^{\,+}$ and using the conformal Killing spinor properties  
\be
\cd_{(0)i}\e_{(0)}^+=\frac{1}{\ell}\Gamma_{(0)i}\e_{(0)}^-\,,\qquad \bar\e_{(0)}^{\,+}\stackrel{\leftarrow}{\cd}_{(0)i}=-\frac{1}{\ell}\bar\e_{(0)}^{\,-}\Gamma_{(0)i}\,,\qquad \ck_{(0)}^i=-i\bar\e_{(0)}^{\,+}\Gamma_{(0)}^{i}\e_{(0)}^+\,,
\ee
results in the following two identities for the expectation values in supersymmetric states:
\vskip-.1cm
\bbxd
\bal\label{BPS-relations-VEVs}
\hspace{-.25cm}0=&\,-i\bar\e_{(0)}^{\,+}\d_\e\<\cs^i\>\sbtx{BPS}
=-\frac{1}{2}\ck_{(0)}^j\<\ct^i_j\>\sbtx{BPS}+\frac{\sqrt{3}}{2}\big(\bar\e_{(0)}^{\,+}\e^-_{(0)}+\bar\e_{(0)}^{\,-}\e^+_{(0)}\big)\<\cj_{R\,{\rm cov}}^i\>\sbtx{BPS}\NO\\
&+\frac{i\sqrt{3}}{8\k_{5}^{2}}\big(\bar\e_{(0)}^{\,-}\G_{(0)}^j\e^-_{(0)}\big)\e_{(0)}{}^{i}{}_{jkl}F_{(0)}^{kl}-\frac{i\ell}{2\k_5^2}\big(\bar\e_{(0)}^{\,-}\G_{(0)}^{[i}\e^-_{(0)}\big)g_{(0)}^{k]l}\big(R_{(0)kl}-\tfrac16 R_{(0)}g_{(0)kl}\big)+\cd_{(0)j}\cv^{ij},\NO\\
\hspace{-.25cm}\rule{.0cm}{.7cm}0=&\,-i\bar\e_{(0)}^{\,+}\d_\e\<j^\l_a\>\sbtx{BPS}
=-\frac{1}{\ell}\ck_{(0)i}\<\wt\cj^i_{a\,{\rm cov}}\>\sbtx{BPS}+i\ck_{(0)}^i\pa_i\<\co_a\>\sbtx{BPS}+\frac{4}{\ell}\bar\e_{(0)}^{\,+}\e^-_{(0)}\<\co_a\>\sbtx{BPS},\hspace{-.05cm}
\eal
\ebxd
where $\cv^{ij}$ is the antisymmetric tensor
\bal
\cv^{ij}\equiv&\; -\frac{i\ell}{4\sqrt{3}}\big(4\d^{[i}_k\d^{j]}_l-i \e_{(0)}{}^i{}_{k}{}^{j}{}_l\big) \ck_{(0)}^k\<\cj_{R\,{\rm cov}}^l\>
-\frac{\ell}{4\sqrt{3}\k_{5}^{2}}\bar\e_{(0)}^{\,+}\big(\G_{(0)}^{[i}{}_k\d^{j]}_l-\d^{[i}_k\d^{j]}_l-i\e_{(0)}^{ij}{}_{kl}\big)\e^-_{(0)}F_{(0)}^{kl}\NO\\
&+\frac{3i\ell^2}{8\k_5^2}\bar\e_{(0)}^{\,+}g_{(0)}^{k[l}\G_{(0)}^{ij]}\e^-_{(0)}\big(R_{(0)kl}-\tfrac16 R_{(0)}g_{(0)kl}\big)\,.
\eal

The BPS relations \eqref{BPS-relations-VEVs} can be written in a more useful form using the properties of the supersymmetric background. Firstly, using the form of the Killing vector \eqref{KV} we get
\bal
\ck_{(0)}^iA_{(0)i}\stackrel{\eqref{eq:v-squashing}}{=}&\;a_{(0)}-u_{4(0)}v_{(0)}+\frac{2}{\sqrt{3}}\ell\mtrc_{1(0)}^{1/2}\mtrc_{3(0)}^{1/2}\mtrc^{-1}_{2(0)}-\frac{2\ell}{\sqrt{3}}\mtrc_{1(0)}^{1/2}\mtrc_{3(0)}^{-1/2}\NO\\
\stackrel{\eqref{BKS-frequency}}{=}&\;\frac{\ell}{\sqrt{3}}\big(2\o_0 +\projsp u_{4(0)}-2\mtrc_{1(0)}^{1/2}\mtrc_{3(0)}^{-1/2}\big)+\sqrt{3}\ell\mtrc_{1(0)}^{1/2}\mtrc_{3(0)}^{1/2}\mtrc^{-1}_{2(0)}\,.
\eal
Moreover, the expressions \eqref{epsilon-plus}-\eqref{epsilon-minus} for the conformal Killing spinor imply that
\bal
\bar\e_{(0)}^{\,+}\e^-_{(0)}=\bar\e_{(0)}^{\,-}\e^+_{(0)}=&\;
-i\frac{\ell\projsp}{2}\mtrc_{1(0)}^{1/2}\mtrc_{3(0)}^{1/2}\mtrc^{-1}_{2(0)} i\e_0^\dag\G_{{\underline t}3}\e_0=\frac{\ell}{2}\mtrc_{1(0)}^{1/2}\mtrc_{3(0)}^{1/2}\mtrc^{-1}_{2(0)}\,,
\eal
where we have used the identity $i\Gamma_{5}\Gamma_{12}=\Gamma_{\tflat 3}$ in the last step. Combining these determines 
\be
\sqrt{3}\big(\bar\e_{(0)}^{\,+}\e^-_{(0)}+\bar\e_{(0)}^{\,-}\e^+_{(0)}\big)=\ck_{(0)}^iA_{(0)i}-\frac{\ell}{\sqrt{3}}\big(2\o_0 +\projsp u_{4(0)}-2\mtrc_{1(0)}^{1/2}\mtrc_{3(0)}^{-1/2}\big)\,.
\ee

These results allow us to write the integral of the first identity in \eqref{BPS-relations-VEVs} over a Cauchy surface $\cc_{(0)}$ in the form
\bal
0=&\;\int_{\cc_{(0)}}d\s_{i}\Big[\ck_{(0)}^j\<\ct^i_j\>\sbtx{BPS}-\big(\ck_{(0)}^jA^I_{(0)j}-\L^I_{\ck_{(0)}}\big)\<\cj_{I}^{i\anom}\>\sbtx{BPS}\NO\\
&+\Big(\frac{\ell}{\sqrt{3}}\big(2\o_0 +\projsp u_{4(0)}-2\mtrc_{1(0)}^{1/2}\mtrc_{3(0)}^{-1/2}\big)-\L_{\ck_{(0)}}\Big)\<\cj_{R}^{i\anom}\>\sbtx{BPS}+\big(\ck_{(0)}^j{\wt A}^a_{(0)j}-\wt\L^a_{\ck_{(0)}}\big)\<\wt\cj^{i\anom}_{a}\>\sbtx{BPS}\NO\\
&-\sqrt{3}\big(\bar\e_{(0)}^{\,+}\e^-_{(0)}+\bar\e_{(0)}^{\,-}\e^+_{(0)}\big)\big(\<\cj_{R\,{\rm cov}}^i\>\sbtx{BPS}-\<\cj_{R}^{i\anom}\>\sbtx{BPS}\big)\NO\\
&-\frac{i\sqrt{3}}{4\k_{5}^{2}}\big(\bar\e_{(0)}^{\,-}\G_{(0)}^j\e^-_{(0)}\big)\e_{(0)}{}^{i}{}_{jkl}F_{(0)}^{kl}+\frac{i\ell}{\k_5^2}\big(\bar\e_{(0)}^{\,-}\G_{(0)}^{[i}\e^-_{(0)}\big)g_{(0)}^{k]l}\big(R_{(0)kl}-\tfrac16 R_{(0)}g_{(0)kl}\big)\Big]\,,
\eal
where the one-parameter family of currents $\<\cj_{I}^{i\anom}\>$ was defined in \eqref{U(1)-current-family} and the gauge compensators $\L^I_{\ck_{(0)}}$ associated with a conformal Killing vector were introduced in \eqref{conformal-Killing}. Using the definition of the U(1) charges in \eqref{el-charges-omega} and of the Killing charges in \eqref{conformal-Killing-charges}, this identity leads to the BPS relation
\bbxd
\vskip.3cm
\be
\label{BPS-relation}
-Q^{\anom}[\ck_{(0)}]-\F_{\text{BPS}}\, Q_R^{\anom}-\wt{\F}_{\text{BPS}}^{a}\,\wt Q^{\anom}_{a}=Q^\anom_{\text{anomaly}}\,,
\ee
\ebxd
where we have assumed that -- as is the case for the backgrounds here -- the quantities  
\bbxd
\bal\label{BPS-chemical-potentials-boundary}
\F_{\text{BPS}}\equiv &\;-\frac{\ell}{\sqrt{3}}\big(2\o_0 +\projsp u_{4(0)}-2\mtrc_{1(0)}^{1/2}\mtrc_{3(0)}^{-1/2}\big)+\L_{\ck_{(0)}}\,,\NO\\
\wt{\F}_{\text{BPS}}^{a}\equiv &\;-\wt a_{(0)}^a+\big(u_{4(0)}-2\projsp\mtrc_{1(0)}^{1/2}\mtrc_{3(0)}^{-1/2}\big)\wt v^a_{(0)}+\wt\L^a_{\ck_{(0)}}\,,
\eal
\ebxd
are constant, while the background anomaly charge, $Q^\anom_{\text{anomaly}}$, corresponds to the supersymmetric Casimir energy and is given by \cite{Papadimitriou:2017kzw} 
\bbxd
\bal\label{anomaly-charge}
Q^\anom_{\text{anomaly}}\equiv &\;\int_{\cc_{(0)}}d\s_{i}\Big(-\frac{(\anom+2)}{12\k_{5}^{2}}\big(\bar\e_{(0)}^{\,+}\e^-_{(0)}+\bar\e_{(0)}^{\,-}\e^+_{(0)}\big)\e_{(0)}^{ijkl}\big(F_{(0)kl} A_{(0)j}-\tfrac12\wt F^a_{(0)kl}\wt A^a_{(0)j}\big)\NO\\
&\hspace{2.cm}+\frac{i\ell}{2\k_5^2}\big(\bar\e_{(0)}^{\,-}\G_{(0)j}\e^-_{(0)}\big)\big(R_{(0)}^{ij}-\tfrac12 R_{(0)}g_{(0)}^{ij}+\tfrac{\sqrt{3}}{2\ell}\e_{(0)}^{ijkl}F_{(0)kl}\big)\Big)\,.
\eal
\ebxd
On supersymmetric backgrounds of the form \eqref{b-metric-ansatz}-\eqref{b-gauge-ansatz} this becomes 
\bbxd
\vskip.05cm
\be\label{anomaly-charge-anstatz}
Q^\anom_{\text{anomaly}}= \frac{2\p^2\ell^3}{\k_5^2}h_{1(0)}^{1/2}h_{3(0)}^{-1/2}b^2\Big(b^2\big(1-\tfrac{5b^2}{8}\big)+\tfrac{4(\anom+2)}{9}(b^2-1)^2-\tfrac{2(\anom+2)}{3\ell^2}\wt v_{(0)}^a\wt v_{(0)}^a\Big)\,,
\ee
\ebxd
where recall that $b^2=h_{3(0)}h_{2(0)}^{-1}$ is the $S^3$ squashing parameter and we have used the relation \eqref{eq:v-squashing}. Notice that this background is nonzero even for global AdS, in which case it coincides with the Casimir mass of global AdS$_5$ \cite{Balasubramanian:1999re,Awad:1999xx,Papadimitriou:2005ii}
\be\label{anomaly-charge-AdS}
Q^{\rm AdS}_{\text{anomaly}}= \frac{3\p^2\ell^2}{4\k_5^2}\,.
\ee
In section \ref{sec:thermodynamics} we show that the quantities \eqref{BPS-chemical-potentials-boundary} are indeed the electric chemical potentials of the BPS black holes and provide a bulk derivation of the BPS relation \eqref{BPS-relation}.

As we have just seen, the BPS relation \eqref{BPS-relation} among the conserved charges follows from the first constraint in \eqref{BPS-relations-VEVs}. However, provided $\bar\e_{(0)}^{\,+}\e^-_{(0)}=\bar\e_{(0)}^{\,-}\e^+_{(0)}\neq 0$, the second relation in \eqref{BPS-relations-VEVs} determines the expectation values of the scalar operators in the vector multiplets in terms of those of the flavor currents, namely 
\bbxd
\vskip.3cm
\be\label{BPS-scalar-VEVs}
\<\co_a\>\sbtx{BPS}=\frac{1}{2\ell}\mtrc_{1(0)}^{-1/2}\mtrc_{3(0)}^{-1/2}\mtrc_{2(0)}\ck_{(0)i}\<\wt\cj^i_{a\,{\rm cov}}\>\sbtx{BPS}\,.
\ee
\ebxd
These scalar VEVs parameterize a sector of the Coulomb branch of the dual theory, which is lifted when $\bar\e_{(0)}^{\,+}\e^-_{(0)}=\bar\e_{(0)}^{\,-}\e^+_{(0)}\neq 0$, as is the case in all backgrounds we consider here.

\section{Bulk Killing spinors and BPS equations}
\label{sec:KS}

In this appendix we present the derivation of the superpotential for $1/4$-BPS solutions within the SU(2)$\times$U(1) invariant ansatz \eqref{ansatz} and determine the explicit form of the corresponding bulk Killing spinor. We follow the spinor conventions of \cite{Freedman:2012zz} and the radial decomposition described in appendix A of \cite{Papadimitriou:2017kzw}. Moreover, our convention is that the flat gamma matrices in five dimensions satisfy $i\G_{\underline{t}12}=\G_{\underline{r}3}$, as has been assumed in writing the bulk action \eqref{5Daction} and supersymmetry transformations \eqref{susy-transformations}. This leads to the following relations that we use repeatedly in this appendix:  
\bbxd
\vskip.2cm
\be\label{gamma-relations}
i\G_{\underline{t}12}=\G_{\underline{r}3}\,,\qquad
i\G_{123}=\G_{\underline{rt}}\,,\qquad 
-i\G_{\underline{t}3}=\G_{\underline{r}12}\,.
\ee
\ebxd

\subsection{Vielbein and spin connection}

In order to determine the form of the bulk BPS equations for SU(2)$\times$U(1) invariant solutions we need to specify a vielbein basis compatible with the ansatz \eqref{ansatz} and compute the corresponding spin connection. A vielbein basis compatible with \eqref{ansatz} is      
\bbxd
\bal\label{vielbein}
&e^{\underline r}=Ndr\,,\quad e^{\underline t}=e^{-\frac12 u_1-\frac32 u_2} dt\,,\NO\\
&e^{1,2}=e^{-\frac12 u_1+\frac12 u_2-\frac12 u_3}\s_{1,2}\,,\quad e^3=e^{-\frac12 u_1+\frac12 u_2+ u_3}(\s_3+u_4 dt)\,,
\eal
\ebxd
where $(\underline{r}, \underline{t}, 1, 2, 3)$ refer to frame bundle indices. The nonzero components of the corresponding spin connection are
\bbxd
\bal\label{spin-connection}
\begin{aligned}
\hspace{-.3cm}\o_{t}\,^{\underline{r}\underline{t}}  = &\, \frac{1}{2N}e^{-\frac{u_{1}}{2}-\frac{3u_{2}}{2}}\left(\dot{u}_{1}+3\dot{u}_{2}+e^{4u_{2}+2u_{3}}u_{4}\dot{u}_{4}\right),\\
\hspace{-.3cm}\o_{\j}\,^{\underline{r}\underline{t}} = &\, \frac{1}{2N}e^{-\frac{u_{1}}{2}+\frac{5u_{2}}{2}+2u_{3}}\dot{u}_{4}, \\
\hspace{-.3cm}\o_{\f}\,^{\underline{r}\underline{t}}  = &\, \frac{1}{2N}e^{-\frac{u_{1}}{2}+\frac{5u_{2}}{2}+2u_{3}}\dot{u}_{4}\cos\th, \\
\hspace{-.3cm}\o_{t}\,^{\underline{r}3}  = &\, \frac{1}{2N}e^{\frac{1}{2}\left(-u_{1}+u_{2}+2u_{3}\right)}\left(u_{4}\left(\dot{u}_{1}-\dot{u}_{2}-2\dot{u}_{3}\right)-\dot{u}_{4}\right),\\
\hspace{-.3cm}\o_{\j}\,^{\underline{r}3}  = &\, \frac{1}{2N}e^{\frac{1}{2}\left(-u_{1}+u_{2}+2u_{3}\right)}\left(\dot{u}_{1}-\dot{u}_{2}-2\dot{u}_{3}\right), \\
\hspace{-.3cm}\o_{\f}\,^{\underline{r}3}  = &\, \frac{1}{2N}e^{\frac{1}{2}(-u_{1}+u_{2}+2u_{3})}\left(\dot{u}_{1}-\dot{u}_{2}-2\dot{u}_{3}\right)\cos\th,\\
\hspace{-.3cm}\o_{\th}\,^{\underline{r}1} = &\, -\frac{1}{2N}e^{\frac{1}{2}\left(-u_{1}+u_{2}-u_{3}\right)}\left(\dot{u}_{1}-\dot{u}_{2}+\dot{u}_{3}\right)\sin\j, \\
\hspace{-.3cm}\o_{\th}\,^{\underline{r}2} = &\, \frac{1}{2N}e^{\frac{1}{2}\left(-u_{1}+u_{2}-u_{3}\right)}\left(\dot{u}_{1}-\dot{u}_{2}+\dot{u}_{3}\right)\cos\j,\\
\hspace{-.3cm}\o_{\f}\,^{\underline{r}1} = &\, \frac{1}{2N}e^{\frac{1}{2}\left(-u_{1}+u_{2}-u_{3}\right)}\left(\dot{u}_{1}-\dot{u}_{2}+\dot{u}_{3}\right)\cos\j\sin\th, \\
\hspace{-.3cm}\o_{\f}\,^{\underline{r}2}  = &\, \frac{1}{2N}e^{\frac{1}{2}\left(-u_{1}+u_{2}-u_{3}\right)}\left(\dot{u}_{1}-\dot{u}_{2}+\dot{u}_{3}\right)\sin\j\sin\th,
\end{aligned}
\hspace{.4cm}
\begin{aligned}
\o_{r}\,^{\underline{t}3} =&\, -\frac{1}{2}e^{2u_{2}+u_{3}}\dot{u}_{4},\\
\o_{t}\,^{12} =&\, -\frac{1}{2}e^{3u_{3}}u_{4},\\
\o_{\j}\,^{12} = &\, -\frac{1}{2}(e^{3u_{3}}-2),\\
\o_{\f}\,^{12} = &\, -\frac{1}{2}(e^{3u_{3}}-2)\cos\th,\\
\o_{\th}\,^{13} =&\,  -\frac{1}{2}e^{\frac{3u_{3}}{2}}\cos\j,\\
\o_{\f}\,^{13} = &\,-\frac{1}{2}e^{\frac{3u_{3}}{2}}\sin\j\sin\th,\\
\o_{\th}\,^{23} =&\,  -\frac{1}{2}e^{\frac{3u_{3}}{2}}\sin\j,\\
\o_{\f}\,^{23} = &\,\frac{1}{2}e^{\frac{3u_{3}}{2}}\cos\j\sin\th.\\
\rule{.0cm}{1.6cm}
\end{aligned}
\hspace{-.4cm}
\eal
\ebxd

\subsection{BPS equations}

Given the form of the vielbein \eqref{vielbein} and spin connection \eqref{spin-connection}, we can evaluate the gravitino and gaugino supersymmetry transformations \eqref{susy-transformations} in order to determine the bulk Killing spinor equations. After some algebra we find that these take the form 
\bbxd
\bal\label{BPSeqs}
\hspace{-.1cm}&(-N^{-1}e^{-2u}\pa_r+\Th_r)\e=0,\quad
(\pa_t+\Th_t)\e=0,\quad
(\pa_\j+\Th_\j)\e=0,\quad
\big(\G_{2}\pa_\th+e^{\G_{12}\j}\G_{\underline{r}}\Th\big)\e=0,\hspace{-.3cm}\NO\\
\hspace{-.1cm}&(\pa_\f-\cos\th\;\pa_\j-\G_{12}\sin\th\pa_\th)\e=0,\quad
(N^{-1}\dot\vf^a+\Th^a_{\vf})\e=0,
\eal
\ebxd
where we have defined 
\bal\label{thetas}
\Th_0=&\;\Big(\frac{\k_5^2}{4\p^2}\p_I^v-u_4e^{u_3}\pa_{v^I}\cf_0\Big)L^I,\NO\\
\rule{0cm}{.5cm}
\Th_r=&\;\frac16 e^{-2u}W\G_{\underline r}+\frac18(\cf_0- 2e^{-u+u_3}L_Iv^I)\G_{\underline{t}3}-\frac{i}{12}e^{-\frac{1}{2}u+2u_2+u_3}\Th_0\G_{3}+\frac{i}{12}e^{-\frac12 u+u_3}L^I\pa_{v^I}\cf_0\G_{\underline{t}},\NO\\
\rule{0cm}{.5cm}
\Th_\j=&\;e^{\frac32u+u_3}\G_{\underline{r}3}\Big(\frac{\k_5^2}{32\times 6\p^2}e^{2u_2}(\p_{u_1}+\p_{u_2}+4\p_{u_3})-\frac{i}{24}e^{-\frac12u+u_3}L^I\pa_{v^I}\cf_0\G_{\underline{t}}\NO\\
&+\frac16e^{-2u}W\G_{\underline{r}}-\frac{i}{4}e^{-\frac32u-u_3}(e^{3u_3}-2)\G_{\underline{t}}-\frac18(\cf_0+ 2e^{-u+u_3}L_Iv^I)\G_{\underline{t}3}\NO\\
&-\frac{i}{2\ell}e^{-\frac32u-u_3}\sum_Iv^I\G_{\underline{r}3}-\frac{i}{12}e^{-\frac12u+2u_2+u_3}\Th_0\G_{3}\Big),\NO\\
\rule{0cm}{.5cm}
\Th_t=&\;\frac{i}{2\ell} \sum_I a^I+u_4\Big(\Th_\j-\frac{i}{2\ell}\sum_I v^I-\frac12\G_{12}\Big)\NO\\
&+e^{\frac32u-2u_2}\Big(-\frac16e^{-2u}W\G_{\underline t}-\frac18(\cf_0- 2e^{-u+u_3}L_Iv^I)\G_{\underline{r}3}-\frac{1}{24}e^{-\frac12u+2u_2+u_3}\Th_0\G_{12}\NO\\
&+\frac{\k_5^2}{32\times 6\p^2}e^{2u_2}(\p_{u_1}-3\p_{u_2})\G_{\underline{rt}}-\frac{i}{12}e^{-\frac12u+u_3}L^I\pa_{v^I}\cf_0\G_{\underline r}\Big),\NO\\
\rule{0cm}{.5cm}
\Th=&\;-e^{\frac32u-\frac12u_3}\Big(\frac{\k_5^2}{32\times 6\p^2}e^{2u_2}(\p_{u_1}+\p_{u_2}-2\p_{u_3})-\frac{i}{24}e^{-\frac12u+u_3}L^I\pa_{v^I}\cf_0\G_{\underline{t}}\NO\\
&+\frac16e^{-2u}W\G_{\underline r}+\frac{i}{4}e^{-\frac32u+2u_3}\G_{\underline{t}}+\frac{1}{2}e^{-u+u_3}L_Iv^I\G_{\underline{t}3}+\frac{i}{24}e^{-\frac12u+2u_2+u_3}\Th_0\G_{3}\Big),\NO\\
\rule{0cm}{.5cm}
\Th^a_\vf=&\;\pa_{\vf^a}\Big(2W\G_{\underline r}+3ie^{u+u_3}L_I v^I\G_{\underline{r}12}-\frac{i}{2}e^{\frac32u+u_3}L^I\pa_{v^I}\cf_0\G_{\underline{t}}\Big)\NO\\
&+\frac{i}{2}e^{\frac32u+2u_2+u_3}\Big(\frac{\k_5^2}{4\p^2}\p_I^v-u_4e^{u_3}\pa_{v^I}\cf_0\Big)\pa_{\vf^a}L^I\G_{3}\,.
\eal
Recall that $\cf_0$ was defined in \eqref{F0-def}. 

\subsection{BPS superpotential and solution of the Killing spinor equations}

We are now ready to solve the Killing spinor equations \eqref{BPSeqs}. To do so we demand that the Killing spinor $\e(r,t)$ only depends on the coordinates $r$ and $t$ and is a simultaneous eigenvector of two commuting matrices, namely
\bbxd
\vskip.25cm
\be\label{eigenspinor}
\G_{12}\e=-i \projsp\e\,,\qquad P\e\equiv\big(c_1(r)\G_{\underline{r}}+ic_2(r)\G_{\underline{t}}\big)\e=\l_P(r)\e\,,
\ee  
\ebxd
where the functions $c_1(r)$ and $c_2(r)$ are functions of the ansatz variables and their conjugate momenta that will be determined shortly. Since
\be
\G_{12}^2=-\bb 1\,, \qquad P^2=(c_1^2+c_2^2)\bb 1\,,
\ee 
it follows that the eigenvalues of these matrices are respectively
\be\label{eigenvalues}
\projsp=\pm 1\,,\qquad \l_P=\pm \sqrt{c_1^2+c_2^2}\,.
\ee

The property \eqref{eigenspinor} of the Killing spinor allows us to simplify the Killing spinor equations further. In particular, we find that
\bal\label{simplified-BPS}
&\Th_r\e=-\Big(P+\frac23\pa_uP+\frac{i}{12}e^{-\frac{1}{2}u+2u_2+u_3}\Th_0\G_{3}\Big)\e\,,\NO\\
\rule{0cm}{.5cm}
&\big(\Th_\j-2e^{\frac32u_3}\G_{\underline{r}3}\Th\big)\e=e^{\frac32u+u_3}\G_{\underline{r}3}\Big(\frac{\k_5^2}{64\p^2}e^{2u_2}(\p_{u_1}+\p_{u_2})+P\Big)\e\,,\NO\\
\rule{0cm}{.5cm}
&\big(\Th_\j+e^{\frac32u_3}\G_{\underline{r}3}\Th\big)\e=e^{\frac32u+u_3}\G_{\underline{r}3}\Big(\frac{\k_5^2}{32\p^2}e^{2u_2}\p_{u_3}-\pa_{u_3}P-\frac{i}{8}e^{-\frac12u+2u_2+u_3}\Th_0\G_{3}\Big)\e\,,\NO\\
\rule{0cm}{.5cm}
&\Th_t\e=\Big[\frac{i}{2\ell}\sum_I a^I+u_4\Big(\Th_\j-\frac{i}{2\ell}\sum_I v^I+\frac{i\projsp}{2}\Big)+\frac{i\projsp}{24}e^{u+u_3}\Th_0\Big]\e\NO\\
&\hskip.7cm+e^{\frac32u-2u_2}\G_{\underline{rt}}\Big[-\Big(\frac{\k_5^2}{64\p^2}e^{2u_2}(\p_{u_1}+\p_{u_2})+P\Big)+\Big(\frac{\k_5^2}{48\p^2}e^{2u_2}\p_{u_1}-\frac23\pa_uP\Big)\Big]\e\,,\NO\\
\rule{0cm}{.5cm}
&\Th^a_\vf\e=4e^{2u}\Big(\pa_{\vf^a}P+\frac{i}{8}e^{\frac32u+2u_2+u_3}\Big(\frac{\k_5^2}{4\p^2}\p_I^v-u_4e^{u_3}\pa_{v^I}\cf_0\Big)\pa_{\vf^a}L^I\G_{3}\Big)\e\,,
\eal
where we identified 
\bbxd
\bal\label{c1&2}
c_1=&\;\frac12e^{-2u}W-\frac{\projsp}{8}\big(\cf_0-6e^{-u+u_3}L_Iv^I\big)\,,\NO\\
c_2=&\;\frac{1}{4}e^{-\frac32u-u_3}\Big(e^{3u_3}+2-2\projsp\ell^{-1}\sum_Iv^I\Big)-\frac{1}{8}e^{-\frac12u+u_3}L^I\pa_{v^I}\cf_0\,,
\eal
\ebxd
and so \eqref{eigenvalues} determines that (the sign is fixed by requiring AdS asymptotics -- see \eqref{U-exp-BF})
\bbxd
\bal\label{eigenvalue-superpotential}
\l_{P}=&\;\Big[\Big(\frac12e^{-2u}W-\frac{\projsp}{8}\big(\cf_0-6e^{-u+u_3}L_Iv^I\big)\Big)^2\NO\\
&+\Big(\frac{1}{4}e^{-\frac32u-u_3}\Big(e^{3u_3}+2-2\projsp\ell^{-1}\sum_Iv^I\Big)-\frac{1}{8}e^{-\frac12u+u_3}L^I\pa_{v^I}\cf_0\Big)^2\Big]^{1/2}\,.
\eal
\ebxd

The functions $c_1$ and $c_2$ in \eqref{c1&2} satisfy a number of relations that are useful for proving several of the subsequent results in this appendix. By direct computation one verifies that
\bal\label{c-equations}
c_{2}+\frac{2}{3}\pa_{u}c_{2}+\frac{1}{12}e^{-\frac{1}{2}u+u_{3}}L^{I}\pa_{v^{I}}\cf_{0} = &\; 0\,,\NO \\
c_{1}+\pa_{u_{3}}c_{1}+\projsp e^{-\frac{1}{2}u+u_{3}}L^{I}\pa_{v^{I}}c_{2} = &\; 0\,,\NO \\
c_{2}+\pa_{u_{3}}c_{2}-\projsp e^{-\frac{1}{2}u+u_{3}}L^{I}\pa_{v^{I}}c_{1} = &\; 0\,,\NO \\
\pa_{L^{I}}c_{1}+\projsp e^{-\frac{1}{2}u+u_{3}}\pa_{v^{I}}c_{2} = &\; 0\,,\NO \\
\pa_{L^{I}}c_{2}-\projsp e^{-\frac{1}{2}u+u_{3}}\pa_{v^{I}}c_{1}+\frac{3}{4}e^{-\frac{3}{2}u+2u_{3}}L_{I} = &\; 0\,,
\eal
where the derivatives $\pa_{L^{I}}c_{1}$, $\pa_{L^{I}}c_{2}$ treat $L^{I}$ as independent variables.

\subsubsection{BPS superpotential}

Remarkably, the function \eqref{eigenvalue-superpotential} can be identified with the superpotential for 1/4 BPS solutions of the STU model within the SU(2)$\times$U(1) invariant ansatz. To see this, notice that since the Killing spinor only depends on $r$ and $t$, the Killing spinor equations \eqref{BPSeqs} require that $\Th\e=0$ and $\Th_\j\e=0$. Hence, the second equation in \eqref{simplified-BPS} implies that 
\be
\l_P=-\frac{\k_5^2}{64\p^2}e^{2u_2}(\p_{u_1}+\p_{u_2})\,.
\ee     
Comparing this with the flow equations \eqref{flow-eqs-F} we conclude that  
\bbxd
\vskip.2cm
\be\label{BPS-superpotential-appendix}
\cf_{\rm BPS}=8\l_{P}\,,
\ee
\ebxd
where $\l_{P}$ is given in \eqref{eigenvalue-superpotential}. 
As an independent check, one can verify that \eqref{BPS-superpotential-appendix} is an exact solution of the Hamilton-Jacobi equation \eqref{HJ-F}.

\subsubsection{Bulk Killing spinor}

Having identified the BPS superpotential, let us now proceed to solve the remaining BPS equations in order to determine the explicit form of the Killing spinor. Using the form of the BPS superpotential in \eqref{BPS-superpotential-appendix} and the flow equations \eqref{flow-eqs-F}, the remaining Killing spinor equations can be further simplified to the following system of equations: 
\bbxd
\bal\label{further-simplified-BPS}
&(P-\l_P)\pa_{u_3}\e=i e^{-\frac12u+u_3}L^I\pa_{v^I}\l_{P}\G_{3}\e\,,\NO\\
&(P-\l_P)\pa_{\vf^a}\e=i e^{-\frac12u+u_3}\pa_{\vf^a}L^I\pa_{v^I}\l_{P}\G_{3}\e\,,\NO\\
&N^{-1}e^{-2u}\dot\e+\Big(\l_P+\frac23\pa_u\l_P\Big)\e+\frac23(P-\l_P)(\pa_{u_3}-\pa_u)\e=0\,,\\
&-i\projsp\pa_t\e-\frac{i}{3}e^{\frac32u-2u_2}\G_{3}(P-\l_P)(2\pa_{u}+\pa_{u_3})\e+\frac12\Big(\frac{\projsp}{\ell}\sum_I (a^I-u_4v^I)+u_4\Big)\e=0\,.\NO
\eal
\ebxd
In writing these we have utilized that
\be
\Th_0=e^{-2u_2}L^I\pa_{v^I}\cf_{\rm BPS}=8e^{-2u_2}L^I\pa_{v^I}\l_{P}\,.
\ee

To proceed further, we look for a solution of the form 
\bbxd
\vskip.3cm
\be\label{spinor-ansatz}
\e=e^{-\frac14 u-u_2+ig\G_3}e^{i\o_0 t}\e_0\,,
\ee
\ebxd
where $g$ is a function of the ansatz variables to be determined, $\o_0$ is a constant and $\e_0$ is a constant spinor. The first part of the exponent that is proportional to the identity matrix follows from the third equation in \eqref{further-simplified-BPS} and the flow equations \eqref{flow-eqs-F}, which imply that 
\be
N^{-1}e^{-2u}\Big(-\frac14\dot u-\dot u_2\Big) +\l_P+\frac23\pa_u\l_P=0\,.
\ee
As a result, the spinor ansatz \eqref{spinor-ansatz} eliminates all the $\G_3$ dependence from the Killing spinor equations \eqref{further-simplified-BPS}. Inserting this form of the Killing spinor in \eqref{further-simplified-BPS} leads to a set of scalar equations for the function $g$. Namely,
\bbxd 
\bal\label{g-equations}
& e^{-2u}g'-\frac43\l_P(\pa_{u_3}-\pa_u)g=0,\NO\\
&\pa_{u_3}g=-\frac{1}{2}e^{-\frac12u+u_3}\l_{P}^{-1}L^I\pa_{v^I}\l_{P},\NO\\
&\pa_{\vf_a}g=-\frac{1}{2}e^{-\frac12u+u_3}\l_{P}^{-1}\pa_{\vf^a}L^I\pa_{ v^I}\l_{P},\NO\\
&(2\pa_{u}+\pa_{u_3})g=\frac{3}{4}e^{-\frac32u+2u_2}\l_P^{-1}\Big(u_4+\projsp\ell^{-1}\sum_I (a^I-u_4v^I)+2\projsp\o_0\Big).
\eal
\ebxd

However, the spinor ansatz \eqref{spinor-ansatz} is not automatically compatible with the diagonalization of the projectors $\G_{12}$ and $P$ in \eqref{eigenspinor}. In fact, demanding compatibility determines the unknown function $g$ algebraically. To see this, notice that inserting the spinor ansatz \eqref{spinor-ansatz} in the second relation in \eqref{eigenspinor} and using the relations \eqref{gamma-relations} leads to the constraint
\be\label{spinor-constraint}
\big(c_1-ic_2\projsp \G_{3}\big)\G_{\underline{r}}\e_0=\l_{P}e^{2ig\G_3}\e_0=\l_{P}(\cos 2g+i\G_3\sin 2g)\e_0\,.
\ee
Since $\G_{12}$ commutes with $\G_3$, the first relation in \eqref{eigenspinor} is automatically ensured by requiring that the constant spinor $\e_0$ itself satisfies $\G_{12}\e_0=-i\projsp \e_0$. However, the second relation in \eqref{eigenspinor} cannot be imposed on the constant spinor $\e_0$. Demanding instead that $\e_0$ satisfies the radiality condition 
\bbxd
\vskip.2cm
\be
\G_{\underline{r}}\e_0=\l_r\e_0\,,\qquad \l_r=\pm 1\,,
\ee
\ebxd
the constraint \eqref{spinor-constraint} implies that the Killing spinor \eqref{spinor-ansatz} satisfies the second relation in \eqref{eigenspinor} provided the function $g$ is given by
\bbxd
\vskip.4cm
\be\label{g-function}
\cos 2g = \l_r\frac{c_1}{\l_{P}}\,,\qquad \sin2g=-\projsp\l_r\frac{c_2}{\l_{P}}\,.
\ee
\ebxd
In particular, with this choice of $g$ we have
\bal
P\e = &\;\big(c_1\G_{\underline{r}}+ic_2\G_{\underline{t}}\big)e^{-\frac14 u-u_2+ig\G_3}e^{i\o_0 t}\e_0\NO\\
=&\;e^{-\frac14 u-u_2}e^{i\o_0 t}\big(c_1\G_{\underline{r}}+ic_2\G_{\underline{t}}\big)e^{i g\G_3}\e_0\NO\\
=&\; e^{-\frac14 u-u_2}e^{i\o_0 t}\big(c_1-\projsp ic_2\G_3\big)e^{-i g\G_3}\l_r\e_0\NO\\
=&\; \l_{P}e^{-\frac14 u-u_2}e^{i\o_0 t}\big(\cos 2g+i \G_3\sin 2g\big)e^{-i g\G_3}\e_0=\l_{P}\e\,,
\eal
as required. In the following and throughout the paper we set $\l_r=1$.

Given the expressions \eqref{g-function} for the function $g=g(u,u_3,v^I,\vf^a)$, we can now verify that the reduced Killing spinor equations \eqref{g-equations} are satisfied. To this end, it is useful to observe that any derivative of the functions $g$ and $\l_{P}$ can be expressed in terms of the corresponding derivatives of $c_1$ and $c_2$ through the identities
\bal\label{dc1-dc2}
\pa\l_{P}  = &\; \cos2g\,\pa c_{1}-\projsp \sin2g\,\pa c_{2}\,,\NO \\
-2\l_{P}\pa g = &\; \sin2g\,\pa c_{1}+\projsp \cos2g\,\pa c_{2}\,.
\eal
Applying these to the derivatives $\pa_{u_3} g$ and $\pa_{\vf^a}g$, the second and third equations in \eqref{g-equations} can be verified straightforwardly. In particular, one can easily check that the second and third equations of \eqref{c-equations} imply the second equation in \eqref{g-equations} while the fourth and the fifth equations of \eqref{c-equations}  imply the third of \eqref{g-equations}. Moreover, utilizing the flow equations \eqref{flow-eqs-F} and the superpotential \eqref{BPS-superpotential-appendix}, the first equation in \eqref{g-equations} can be expressed in the form 
\be
\frac{1}{3}(\l_P+\pa_{u_3}\l_P)\pa_{u_3}g+\pa_{\vf^a}\l_P\pa_{\vf^a}g+e^{-u+2u_3}G^{IJ}\pa_{v^I}\l_P\pa_{v^J}g=0\,,
\ee
which can also be easily verified. A weaker version of the last equation in \eqref{g-equations} is
\be
\Big[e^{\frac32u-2u_2}\l_P(2\pa_{u}+\pa_{u_3})g-\frac{3}{4}\Big(u_4+\projsp\ell^{-1}\sum_I (a^I-u_4v^I)\Big)\Big]'=0\,,
\ee
i.e. the quantity inside the square brackets is constant. This may be verified explicitly using again the flow equations \eqref{flow-eqs-F} and the superpotential \eqref{BPS-superpotential-appendix}. The last equation in \eqref{g-equations} identifies the value of this constant with the frequency of the Killing spinor.

\subsubsection{Supersymmetric integrals of motion}

The integral of motion we have just derived is in fact a special case of a set of integrals of motion that supersymmetry implies. As we will now show, the quantities 
\bbxd
\bal\label{constants-of-motion}
\mathscr C_{4} \equiv &\; u_{4}-\projsp e^{-2u_{2}-u_{3}}\cos2g\,,\NO \\
\mathscr C^{I} \equiv &\; a^{I}-u_{4}v^{I}+\projsp e^{-2u_{2}}(e^{-u_{3}}\cos2g\,v^{I}-e^{-\frac{1}{2}u}\sin2g\,L^{I})\,,
\eal
\ebxd
are all constant along supersymmetric flows, defining a hyperplane on which supersymmetric point particles move. The fact that $\mathscr C^{I}$ are constant is related to the consistency of the gauge choice in eq.~(2.46) of \cite{Gutowski:2004yv}.

Showing the conservation of $\mathscr C_4$ is straightforward. We have
\be\label{u4-constraint}
\mathscr C'_4=u_{4}'+\frac{4\projsp}{3}e^{2u-2u_{2}-u_{3}}(2-\pa_{u_{3}}+\pa_{u})c_{1}
=u_{4}'-\frac{1}{2}e^{2u-2u_{2}-u_{3}}\cf_{0}=0\,,
\ee
where in the first equality we eliminated $g'$ using the first equation of
\eqref{g-equations} and $u_{2}'$, $u_{3}'$ using the flow equations \eqref{flow-eqs-F} 
with $\cf=\cf_{\text{BPS}}$. In the second equality we used the explicit
form of $c_{1}$ in \eqref{c1&2} and in the third equality the flow equation
for $u_{4}$ in \eqref{flow-eqs-F}. 

In order to demonstrate the conservation of $\mathscr C^I$, we begin by evaluating
\bal
\mathscr C'^I = &\; a'^{I}-u_{4}v'^{I}+(\projsp e^{-2u_{2}-u_{3}}\cos2g)v'^{I}-(\projsp e^{-2u_{2}-\frac{1}{2}u}\sin2g\,L^{I})'\NO \\
= &\; \projsp \Big(-(e^{-2u_{2}-\frac{1}{2}u}\sin2g)'L^{I}+\frac{1}{2}e^{\frac{3}{2}u-2u_{2}}\sin2g\,(\pa_{a}L^{I}\pa_{a}L^{J})\pa_{L^{J}}\cf\Big)\NO \\
&-\frac{1}{2}e^{u-2u_{2}+u_{3}}G^{IJ}\big(\projsp \cos2g\,\pa_{v^{J}}\cf+\pa_{v^{J}}\cf_{0}\big)\,,
\eal
where in the first equality we rearranged terms and used
that $\mathscr C'_4=0$, while in the second equality we used the
flow equations for $a'^{I}-u_{4}v'^{I}$ and for the scalars. Moreover, from the flow equations for $u$ and $u_{2}$, as well as the first equation in \eqref{g-equations}, we have
\bal
&(e^{-2u_{2}-\frac{1}{2}u}\sin2g)' =  \frac23e^{\frac{3}{2}u-2u_{2}}\big(4\l_{P}\cos2g\,(\pa_{u_{3}}-\pa_{u})g-\sin2g\,(3+2\pa_{u})\l_{P}\big)\NO \\
&\quad= -\frac16e^{\frac{3}{2}u-2u_{2}}\Big(8\cos2g\big(\sin2g\,\pa_{u_{3}}c_{1}+\projsp \cos2g\,\pa_{u_{3}}c_{2}\big)
+\projsp e^{-\frac{1}{2}u+u_{3}}L^{J}\pa_{v^{J}}\cf_{0}\Big)\NO \\
&\quad=e^{u-2u_{2}+u_{3}}L^{J}\Big(\frac{4}{3}\cos2g\big(\projsp \sin2g\,\pa_{v^{J}}c_{2}-\cos2g\,\pa_{v^{J}}c_{1}\big)-\frac{\projsp}{6}\pa_{v^{J}}\cf_{0}\Big) \,, 
\eal
where in the second equality we used \eqref{dc1-dc2} and the first equation in \eqref{c-equations}. In the third equality, we eliminated the $\pa_{u_{3}}$ derivatives using the second and the third equations in \eqref{c-equations}, as well as the definition of $g$ in \eqref{g-function}. Finally, from \eqref{dc1-dc2} we also have
\bal
\pa_{L^{I}}\cf = &\; 8(\cos2g\,\pa_{L^{I}}c_{1}-\projsp \sin2g\,\pa_{L^{I}}c_{2})\,,\NO \\
\pa_{v^{I}}\cf = &\; 8(\cos2g\,\pa_{v^{I}}c_{1}-\projsp \sin2g\,\pa_{v^{I}}c_{2})\,.
\eal

Collecting the above expressions and using \eqref{inverse-GIJ}, we obtain
\bal
e^{2u_{2}}\mathscr C'^{I} = &\; \pa_{a}L^{I}\pa_{a}L^{J}\Big(-\frac{1}{2}e^{u+u_{3}}\pa_{v^{J}}\cf_{0}
- 4e^{u+u_{3}}\projsp\cos2g\big(\cos2g\,\pa_{v^{J}}c_{1}-\projsp\sin2g\,\pa_{v^{J}}c_{2}\big)\NO \\
&+4e^{\frac{3}{2}u}\sin2g\big(\cos2g\,\pa_{L^{I}}c_{1}-\projsp\sin2g\,\pa_{L^{I}}c_{2}\big)\Big)\NO \\
= &\; -\frac{1}{2}\pa_{a}L^{I}\pa_{a}L^{J}\big(e^{u+u_{3}}\pa_{v^{J}}\cf_{0}+8\projsp e^{\frac{3}{2}u}\pa_{L^{I}}c_{2}\big)= 0\,,
\eal
where we used the fourth and fifth equations in \eqref{c-equations}, the explicit form of $c_{2}$ in \eqref{c1&2}, and the identity \eqref{special-geometry-id}. This concludes the proof that $\mathscr C_{4}$ and $\mathscr C^{I}$ defined in \eqref{constants-of-motion} are conserved. These supersymmetric integrals of motion  determine $u_{4}$ and the electric components of the gauge fields, $a^{I}$, in terms of the remaining ansatz variables. As we discuss in detail in section \ref{sec:thermodynamics}, they play an important role in the study of BPS black hole thermodynamics. 

We conclude this appendix with the observation that the Killing spinor frequency $\o_0$ in the last equation in \eqref{g-equations} can be expressed in terms of the constants $\mathscr C^I$ as 
\bbxd
\vskip.5cm
\be\label{KS-frequency}
\omega_{0}=-\frac{1}{2\ell}\sum_{I}\mathscr C^{I}\,,
\ee
\ebxd
which follows directly from the identities
\be
c_{1}+(2\pa_{u}+\pa_{u_{3}})c_{1}=-\frac{3}{2}e^{-2u}W\,,\qquad c_{2}+(2\pa_{u}+\pa_{u_{3}})c_{2}=\frac{3}{2}e^{-\frac{3}{2}u-u_{3}}\big(\projsp \ell^{-1}\sum_{I}v^{I}-1\big)\,.
\ee

\bibliographystyle{JHEP}
\bibliography{susythermo}

\end{document}

%% file: macros.tex
\def\){\right)}
\def\({\left( }
\def\]{\right] }
\def\[{\left[ }

\def\NO{\nonumber}

\newcommand{\be}{\begin{equation}}
\newcommand{\ee}{\end{equation}}

\def\bea{\begin{eqnarray}}
\def\eea{\end{eqnarray}}

\def\bal#1\eal{\begin{align}#1\end{align}}

\def\bald{\begin{aligned}}
\def\eald{\end{aligned}}

\def\bsub{\begin{subequations}}
\def\esub{\end{subequations}}

\def\beqx{\begin{displaymath}}
\def\eeqx{\end{displaymath}}

\newcommand{\bmat}{\left(\begin{array}}
\newcommand{\emat}{\end{array}\right)}

\def\a{\alpha}
\def\b{\beta}
\def\c{\chi}
\def\d{\delta}
\def\e{\epsilon}
\def\f{\phi}
\def\g{\gamma}

\def\j{\psi}
\def\k{\kappa}
\def\l{\lambda}
\def\m{\mu}
\def\n{\nu}
\def\o{\omega}
    \def\om{\omega}
\def\p{\pi}

    \def\th{\theta}
\def\r{\rho}
\def\s{\sigma}
\def\t{\tau}
\def\x{\xi}
\def\z{\zeta}
\def\D{\Delta}
\def\F{\Phi}
\def\G{\Gamma}
\def\J{\Psi}
\def\L{\Lambda}
\def\O{\Omega}
    \def\Om{\Omega}
\def\P{\Pi}

    \def\Th{\Theta}
\def\S{\Sigma}

\def\X{\Xi}

\def\ve{\varepsilon}

    \def\vth{\vartheta}

\def\vf{\varphi}

\def\bs{\bbsigma}

\def\ca{{\cal A}}
\def\cb{{\cal B}}
\def\cc{{\cal C}}
\def\cd{{\cal D}}
\def\ce{{\cal E}}
\def\cf{{\cal F}}

\def\ch{{\cal H}}
\def\ci{{\cal I}}
\def\cj{{\cal J}}
\def\ck{{\cal K}}
\def\cl{{\cal L}}
\def\cm{{\cal M}}
\def\cn{{\cal N}}
\def\co{{\cal O}}
\def\cp{{\cal P}}

\def\car{{\cal R}}
\def\cs{{\cal S}}
\def\ct{{\cal T}}
\def\cu{{\cal U}}
\def\cv{{\cal V}}
\def\cw{{\cal W}}
\def\cx{{\cal X}}

\def\cz{{\cal Z}}

\def\bb#1{\ensuremath{\mathbb{#1}}} 

\def\bo{{\raise-.3ex\hbox{\large$\Box$}}}               
\def\pa{\partial}                                       
\def\face{{\raise.2ex\hbox{$\displaystyle \bigodot$}\mskip-2.2mu \llap {$\ddot
        \smile$}}}                                   
\def\>{\rangle}                                      
\def\<{\langle}                                      

\def\sbtx#1{{}_{\rm #1}}                           
\newcommand{\sub}[1]{\phantom{}_{(#1)}\phantom{}}    
\def\wt#1{\widetilde{#1}}                            
\def\Hat#1{\widehat{#1}}                             
\def\lbar#1{\ensuremath{\overline{#1}}}              
\def\leftrightarrowfill{$\mathsurround=0pt \mathord\leftarrow \mkern-6mu
        \cleaders\hbox{$\mkern-2mu \mathord- \mkern-2mu$}\hfill
        \mkern-6mu \mathord\rightarrow$}        
\def\dvec#1{\vbox{\ialign{##\crcr
        \leftrightarrowfill\crcr\noalign{\kern-1pt\nointerlineskip}
        $\hfil\displaystyle{#1}\hfil$\crcr}}}           

\def\-{\hphantom{-}}

\global\long\def\ca{\mathcal{A}}
\global\long\def\cb{\mathcal{B}}
\global\long\def\cc{\mathcal{C}}
\global\long\def\cd{\mathcal{D}}
\global\long\def\ce{\mathcal{E}}
\global\long\def\cf{\mathcal{F}}

\global\long\def\ch{\mathcal{H}}
\global\long\def\ci{\mathcal{I}}
\global\long\def\cj{\mathcal{J}}
\global\long\def\ck{\mathcal{K}}
\global\long\def\cl{\mathcal{L}}
\global\long\def\cm{\mathcal{M}}
\global\long\def\cn{\mathcal{N}}
\global\long\def\co{\mathcal{O}}
\global\long\def\cp{\mathcal{P}}

\global\long\def\cs{\mathcal{S}}
\global\long\def\ct{\mathcal{T}}
\global\long\def\cu{\mathcal{U}}
\global\long\def\cv{\mathcal{V}}
\global\long\def\cw{\mathcal{W}}
\global\long\def\cx{\mathcal{X}}

\global\long\def\cz{\mathcal{Z}}

\global\long\def\fatc{\mathbb{C}}

\global\long\def\fatp{\mathbb{P}}

\global\long\def\uone{\text{U}(1)}

\global\long\def\sutwo{\text{SU}(2)}

\global\long\def\tflat{\underline{t}}

\global\long\def\anom{\varpi}
\global\long\def\kbulk{\xi}
\global\long\def\kbdy{\xi_{(0)}}
\global\long\def\actn{\mathtt{S}}

\global\long\def\vcas{v_{\text{c}}}

\global\long\def\acaso{a_{\text{c}0}}

\global\long\def\fatc{\mathbb{C}}

\global\long\def\fatp{\mathbb{P}}

\global\long\def\mtrc{h}

\global\long\def\vcas{v_{\text{c}}}

\global\long\def\acaso{a_{\text{c}0}}

\global\long\def\projsp{\lambda_{12}}

\def\elfg{\ensuremath{A_r}}
\def\actn{{\mathtt{S}}}

\def\anom{\varpi}

%% file: susybhs_arXiv2.bbl
\providecommand{\href}[2]{#2}\begingroup\raggedright\begin{thebibliography}{100}

\bibitem{Papadimitriou:2017kzw}
I.~Papadimitriou, \emph{{Supercurrent anomalies in 4d SCFTs}},
  \href{https://doi.org/10.1007/JHEP07(2017)038}{\emph{JHEP} {\bfseries 07}
  (2017) 038} [\href{https://arxiv.org/abs/1703.04299}{{\ttfamily
  1703.04299}}].

\bibitem{Strominger:1996sh}
A.~Strominger and C.~Vafa, \emph{{Microscopic origin of the Bekenstein-Hawking
  entropy}}, \href{https://doi.org/10.1016/0370-2693(96)00345-0}{\emph{Phys.
  Lett. B} {\bfseries 379} (1996) 99}
  [\href{https://arxiv.org/abs/hep-th/9601029}{{\ttfamily hep-th/9601029}}].

\bibitem{Benini:2015eyy}
F.~Benini, K.~Hristov and A.~Zaffaroni, \emph{{Black hole microstates in
  AdS$_{4}$ from supersymmetric localization}},
  \href{https://doi.org/10.1007/JHEP05(2016)054}{\emph{JHEP} {\bfseries 05}
  (2016) 054} [\href{https://arxiv.org/abs/1511.04085}{{\ttfamily
  1511.04085}}].

\bibitem{Benini:2016rke}
F.~Benini, K.~Hristov and A.~Zaffaroni, \emph{{Exact microstate counting for
  dyonic black holes in AdS4}},
  \href{https://doi.org/10.1016/j.physletb.2017.05.076}{\emph{Phys. Lett. B}
  {\bfseries 771} (2017) 462}
  [\href{https://arxiv.org/abs/1608.07294}{{\ttfamily 1608.07294}}].

\bibitem{Zaffaroni:2019dhb}
A.~Zaffaroni, \emph{{AdS black holes, holography and localization}},
  \href{https://doi.org/10.1007/s41114-020-00027-8}{\emph{Living Rev. Rel.}
  {\bfseries 23} (2020) 2} [\href{https://arxiv.org/abs/1902.07176}{{\ttfamily
  1902.07176}}].

\bibitem{Strominger:1997eq}
A.~Strominger, \emph{{Black hole entropy from near horizon microstates}},
  \href{https://doi.org/10.1088/1126-6708/1998/02/009}{\emph{JHEP} {\bfseries
  02} (1998) 009} [\href{https://arxiv.org/abs/hep-th/9712251}{{\ttfamily
  hep-th/9712251}}].

\bibitem{Sen:2007qy}
A.~Sen, \emph{{Black Hole Entropy Function, Attractors and Precision Counting
  of Microstates}}, \href{https://doi.org/10.1007/s10714-008-0626-4}{\emph{Gen.
  Rel. Grav.} {\bfseries 40} (2008) 2249}
  [\href{https://arxiv.org/abs/0708.1270}{{\ttfamily 0708.1270}}].

\bibitem{Sen:2008vm}
A.~Sen, \emph{{Quantum Entropy Function from AdS(2)/CFT(1) Correspondence}},
  \href{https://doi.org/10.1142/S0217751X09045893}{\emph{Int. J. Mod. Phys. A}
  {\bfseries 24} (2009) 4225}
  [\href{https://arxiv.org/abs/0809.3304}{{\ttfamily 0809.3304}}].

\bibitem{Maldacena:2016upp}
J.~Maldacena, D.~Stanford and Z.~Yang, \emph{{Conformal symmetry and its
  breaking in two dimensional Nearly Anti-de-Sitter space}},
  \href{https://doi.org/10.1093/ptep/ptw124}{\emph{PTEP} {\bfseries 2016}
  (2016) 12C104} [\href{https://arxiv.org/abs/1606.01857}{{\ttfamily
  1606.01857}}].

\bibitem{Kinney:2005ej}
J.~Kinney, J.~M. Maldacena, S.~Minwalla and S.~Raju, \emph{{An Index for 4
  dimensional super conformal theories}},
  \href{https://doi.org/10.1007/s00220-007-0258-7}{\emph{Commun. Math. Phys.}
  {\bfseries 275} (2007) 209}
  [\href{https://arxiv.org/abs/hep-th/0510251}{{\ttfamily hep-th/0510251}}].

\bibitem{Romelsberger:2005eg}
C.~Romelsberger, \emph{{Counting chiral primaries in N = 1, d=4 superconformal
  field theories}},
  \href{https://doi.org/10.1016/j.nuclphysb.2006.03.037}{\emph{Nucl. Phys. B}
  {\bfseries 747} (2006) 329}
  [\href{https://arxiv.org/abs/hep-th/0510060}{{\ttfamily hep-th/0510060}}].

\bibitem{Grant:2008sk}
L.~Grant, P.~A. Grassi, S.~Kim and S.~Minwalla, \emph{{Comments on 1/16 BPS
  Quantum States and Classical Configurations}},
  \href{https://doi.org/10.1088/1126-6708/2008/05/049}{\emph{JHEP} {\bfseries
  05} (2008) 049} [\href{https://arxiv.org/abs/0803.4183}{{\ttfamily
  0803.4183}}].

\bibitem{Chang:2013fba}
C.-M. Chang and X.~Yin, \emph{{1/16 BPS states in $\mathcal N=$ 4
  super-Yang-Mills theory}},
  \href{https://doi.org/10.1103/PhysRevD.88.106005}{\emph{Phys. Rev. D}
  {\bfseries 88} (2013) 106005}
  [\href{https://arxiv.org/abs/1305.6314}{{\ttfamily 1305.6314}}].

\bibitem{Cabo-Bizet:2018ehj}
A.~Cabo-Bizet, D.~Cassani, D.~Martelli and S.~Murthy, \emph{{Microscopic origin
  of the Bekenstein-Hawking entropy of supersymmetric AdS$_{5}$ black holes}},
  \href{https://doi.org/10.1007/JHEP10(2019)062}{\emph{JHEP} {\bfseries 10}
  (2019) 062} [\href{https://arxiv.org/abs/1810.11442}{{\ttfamily
  1810.11442}}].

\bibitem{Choi:2018hmj}
S.~Choi, J.~Kim, S.~Kim and J.~Nahmgoong, \emph{{Large AdS black holes from
  QFT}},  \href{https://arxiv.org/abs/1810.12067}{{\ttfamily 1810.12067}}.

\bibitem{Benini:2018ywd}
F.~Benini and P.~Milan, \emph{{Black Holes in 4D $\mathcal{N}$=4
  Super-Yang-Mills Field Theory}},
  \href{https://doi.org/10.1103/PhysRevX.10.021037}{\emph{Phys. Rev. X}
  {\bfseries 10} (2020) 021037}
  [\href{https://arxiv.org/abs/1812.09613}{{\ttfamily 1812.09613}}].

\bibitem{Honda:2019cio}
M.~Honda, \emph{{Quantum Black Hole Entropy from 4d Supersymmetric Cardy
  formula}}, \href{https://doi.org/10.1103/PhysRevD.100.026008}{\emph{Phys.
  Rev. D} {\bfseries 100} (2019) 026008}
  [\href{https://arxiv.org/abs/1901.08091}{{\ttfamily 1901.08091}}].

\bibitem{ArabiArdehali:2019tdm}
A.~Arabi~Ardehali, \emph{{Cardy-like asymptotics of the 4d $ \mathcal{N}=4 $
  index and AdS$_{5}$ blackholes}},
  \href{https://doi.org/10.1007/JHEP06(2019)134}{\emph{JHEP} {\bfseries 06}
  (2019) 134} [\href{https://arxiv.org/abs/1902.06619}{{\ttfamily
  1902.06619}}].

\bibitem{Kim:2019yrz}
J.~Kim, S.~Kim and J.~Song, \emph{{A 4d $ \mathcal{N} $ = 1 Cardy Formula}},
  \href{https://doi.org/10.1007/JHEP01(2021)025}{\emph{JHEP} {\bfseries 01}
  (2021) 025} [\href{https://arxiv.org/abs/1904.03455}{{\ttfamily
  1904.03455}}].

\bibitem{Cabo-Bizet:2019osg}
A.~Cabo-Bizet, D.~Cassani, D.~Martelli and S.~Murthy, \emph{{The asymptotic
  growth of states of the 4d $ \mathcal{N}=1 $ superconformal index}},
  \href{https://doi.org/10.1007/JHEP08(2019)120}{\emph{JHEP} {\bfseries 08}
  (2019) 120} [\href{https://arxiv.org/abs/1904.05865}{{\ttfamily
  1904.05865}}].

\bibitem{Hosseini:2019lkt}
S.~M. Hosseini, K.~Hristov and A.~Zaffaroni, \emph{{Microstates of rotating
  AdS$_{5}$ strings}},
  \href{https://doi.org/10.1007/JHEP11(2019)090}{\emph{JHEP} {\bfseries 11}
  (2019) 090} [\href{https://arxiv.org/abs/1909.08000}{{\ttfamily
  1909.08000}}].

\bibitem{Hosseini:2020mut}
S.~M. Hosseini and A.~Zaffaroni, \emph{{Universal AdS Black Holes in Theories
  with 16 Supercharges and Their Microstates}},
  \href{https://doi.org/10.1103/PhysRevLett.126.171604}{\emph{Phys. Rev. Lett.}
  {\bfseries 126} (2021) 171604}
  [\href{https://arxiv.org/abs/2011.01249}{{\ttfamily 2011.01249}}].

\bibitem{Hosseini:2017mds}
S.~M. Hosseini, K.~Hristov and A.~Zaffaroni, \emph{{An extremization principle
  for the entropy of rotating BPS black holes in AdS$_{5}$}},
  \href{https://doi.org/10.1007/JHEP07(2017)106}{\emph{JHEP} {\bfseries 07}
  (2017) 106} [\href{https://arxiv.org/abs/1705.05383}{{\ttfamily
  1705.05383}}].

\bibitem{Hosseini:2018dob}
S.~M. Hosseini, K.~Hristov and A.~Zaffaroni, \emph{{A note on the entropy of
  rotating BPS AdS$_7\times S^4$ black holes}},
  \href{https://doi.org/10.1007/JHEP05(2018)121}{\emph{JHEP} {\bfseries 05}
  (2018) 121} [\href{https://arxiv.org/abs/1803.07568}{{\ttfamily
  1803.07568}}].

\bibitem{Choi:2018fdc}
S.~Choi, C.~Hwang, S.~Kim and J.~Nahmgoong, \emph{{Entropy Functions of BPS
  Black Holes in AdS$_{4}$ and AdS$_{6}$}},
  \href{https://doi.org/10.3938/jkps.76.101}{\emph{J. Korean Phys. Soc.}
  {\bfseries 76} (2020) 101}
  [\href{https://arxiv.org/abs/1811.02158}{{\ttfamily 1811.02158}}].

\bibitem{Hosseini:2019ddy}
S.~M. Hosseini and A.~Zaffaroni, \emph{{Geometry of $\mathcal{I}$-extremization
  and black holes microstates}},
  \href{https://doi.org/10.1007/JHEP07(2019)174}{\emph{JHEP} {\bfseries 07}
  (2019) 174} [\href{https://arxiv.org/abs/1904.04269}{{\ttfamily
  1904.04269}}].

\bibitem{Assel:2014paa}
B.~Assel, D.~Cassani and D.~Martelli, \emph{{Localization on Hopf surfaces}},
  \href{https://doi.org/10.1007/JHEP08(2014)123}{\emph{JHEP} {\bfseries 08}
  (2014) 123} [\href{https://arxiv.org/abs/1405.5144}{{\ttfamily 1405.5144}}].

\bibitem{Lorenzen:2014pna}
J.~Lorenzen and D.~Martelli, \emph{{Comments on the Casimir energy in
  supersymmetric field theories}},
  \href{https://doi.org/10.1007/JHEP07(2015)001}{\emph{JHEP} {\bfseries 07}
  (2015) 001} [\href{https://arxiv.org/abs/1412.7463}{{\ttfamily 1412.7463}}].

\bibitem{Assel:2015nca}
B.~Assel, D.~Cassani, L.~Di~Pietro, Z.~Komargodski, J.~Lorenzen and
  D.~Martelli, \emph{{The Casimir Energy in Curved Space and its Supersymmetric
  Counterpart}}, \href{https://doi.org/10.1007/JHEP07(2015)043}{\emph{JHEP}
  {\bfseries 07} (2015) 043}
  [\href{https://arxiv.org/abs/1503.05537}{{\ttfamily 1503.05537}}].

\bibitem{Bobev:2015kza}
N.~Bobev, M.~Bullimore and H.-C. Kim, \emph{{Supersymmetric Casimir Energy and
  the Anomaly Polynomial}},
  \href{https://doi.org/10.1007/JHEP09(2015)142}{\emph{JHEP} {\bfseries 09}
  (2015) 142} [\href{https://arxiv.org/abs/1507.08553}{{\ttfamily
  1507.08553}}].

\bibitem{Martelli:2015kuk}
D.~Martelli and J.~Sparks, \emph{{The character of the supersymmetric Casimir
  energy}}, \href{https://doi.org/10.1007/JHEP08(2016)117}{\emph{JHEP}
  {\bfseries 08} (2016) 117}
  [\href{https://arxiv.org/abs/1512.02521}{{\ttfamily 1512.02521}}].

\bibitem{BenettiGenolini:2016qwm}
P.~Benetti~Genolini, D.~Cassani, D.~Martelli and J.~Sparks, \emph{{The
  holographic supersymmetric Casimir energy}},
  \href{https://doi.org/10.1103/PhysRevD.95.021902}{\emph{Phys. Rev. D}
  {\bfseries 95} (2017) 021902}
  [\href{https://arxiv.org/abs/1606.02724}{{\ttfamily 1606.02724}}].

\bibitem{Brunner:2016nyk}
F.~Brünner, D.~Regalado and V.~P. Spiridonov, \emph{{Supersymmetric Casimir
  Energy and $SL(3,\mathbb{Z})$ Transformations}},
  \href{https://arxiv.org/abs/1611.03831}{{\ttfamily 1611.03831}}.

\bibitem{Papadimitriou:2019gel}
I.~Papadimitriou, \emph{{Supersymmetry anomalies in $\mathcal{N}=1$ conformal
  supergravity}}, \href{https://doi.org/10.1007/JHEP04(2019)040}{\emph{JHEP}
  {\bfseries 04} (2019) 040}
  [\href{https://arxiv.org/abs/1902.06717}{{\ttfamily 1902.06717}}].

\bibitem{Papadimitriou:2019yug}
I.~Papadimitriou, \emph{{Supersymmetry anomalies in new minimal supergravity}},
  \href{https://doi.org/10.1007/JHEP09(2019)039}{\emph{JHEP} {\bfseries 09}
  (2019) 039} [\href{https://arxiv.org/abs/1904.00347}{{\ttfamily
  1904.00347}}].

\bibitem{Ferrara:1996dd}
S.~Ferrara and R.~Kallosh, \emph{{Supersymmetry and attractors}},
  \href{https://doi.org/10.1103/PhysRevD.54.1514}{\emph{Phys. Rev. D}
  {\bfseries 54} (1996) 1514}
  [\href{https://arxiv.org/abs/hep-th/9602136}{{\ttfamily hep-th/9602136}}].

\bibitem{Larsen:2006xm}
F.~Larsen, \emph{{The Attractor Mechanism in Five Dimensions}}, {\emph{Lect.
  Notes Phys.} {\bfseries 755} (2008) 249}
  [\href{https://arxiv.org/abs/hep-th/0608191}{{\ttfamily hep-th/0608191}}].

\bibitem{LopesCardoso:2007qid}
G.~Lopes~Cardoso, A.~Ceresole, G.~Dall'Agata, J.~M. Oberreuter and J.~Perz,
  \emph{{First-order flow equations for extremal black holes in very special
  geometry}}, \href{https://doi.org/10.1088/1126-6708/2007/10/063}{\emph{JHEP}
  {\bfseries 10} (2007) 063} [\href{https://arxiv.org/abs/0706.3373}{{\ttfamily
  0706.3373}}].

\bibitem{Cacciatori:2009iz}
S.~L. Cacciatori and D.~Klemm, \emph{{Supersymmetric AdS(4) black holes and
  attractors}}, \href{https://doi.org/10.1007/JHEP01(2010)085}{\emph{JHEP}
  {\bfseries 01} (2010) 085} [\href{https://arxiv.org/abs/0911.4926}{{\ttfamily
  0911.4926}}].

\bibitem{DallAgata:2010ejj}
G.~Dall'Agata and A.~Gnecchi, \emph{{Flow equations and attractors for black
  holes in N = 2 U(1) gauged supergravity}},
  \href{https://doi.org/10.1007/JHEP03(2011)037}{\emph{JHEP} {\bfseries 03}
  (2011) 037} [\href{https://arxiv.org/abs/1012.3756}{{\ttfamily 1012.3756}}].

\bibitem{Cabo-Bizet:2017xdr}
A.~Cabo-Bizet, U.~Kol, L.~A. Pando~Zayas, I.~Papadimitriou and V.~Rathee,
  \emph{{Entropy functional and the holographic attractor mechanism}},
  \href{https://doi.org/10.1007/JHEP05(2018)155}{\emph{JHEP} {\bfseries 05}
  (2018) 155} [\href{https://arxiv.org/abs/1712.01849}{{\ttfamily
  1712.01849}}].

\bibitem{Bobev:2020pjk}
N.~Bobev, A.~M. Charles and V.~S. Min, \emph{{Euclidean black saddles and
  AdS$_{4}$ black holes}},
  \href{https://doi.org/10.1007/JHEP10(2020)073}{\emph{JHEP} {\bfseries 10}
  (2020) 073} [\href{https://arxiv.org/abs/2006.01148}{{\ttfamily
  2006.01148}}].

\bibitem{Papadimitriou:2005ii}
I.~Papadimitriou and K.~Skenderis, \emph{{Thermodynamics of asymptotically
  locally AdS spacetimes}},
  \href{https://doi.org/10.1088/1126-6708/2005/08/004}{\emph{JHEP} {\bfseries
  08} (2005) 004} [\href{https://arxiv.org/abs/hep-th/0505190}{{\ttfamily
  hep-th/0505190}}].

\bibitem{Gunaydin:1983bi}
M.~Gunaydin, G.~Sierra and P.~Townsend, \emph{{The Geometry of N=2
  Maxwell-Einstein Supergravity and Jordan Algebras}},
  \href{https://doi.org/10.1016/0550-3213(84)90142-1}{\emph{Nucl. Phys. B}
  {\bfseries 242} (1984) 244}.

\bibitem{Freedman:2012zz}
D.~Z. Freedman and A.~Van~Proeyen, \emph{{Supergravity}}. Cambridge Univ.
  Press, Cambridge, UK, 2012.

\bibitem{Chong:2005hr}
Z.~W. Chong, M.~Cvetič, H.~Lü and C.~N. Pope, \emph{{General non-extremal
  rotating black holes in minimal five-dimensional gauged supergravity}},
  \href{https://doi.org/10.1103/PhysRevLett.95.161301}{\emph{Phys. Rev. Lett.}
  {\bfseries 95} (2005) 161301}
  [\href{https://arxiv.org/abs/hep-th/0506029}{{\ttfamily hep-th/0506029}}].

\bibitem{Looyestijn:2010pb}
H.~Looyestijn, E.~Plauschinn and S.~Vandoren, \emph{{New potentials from
  Scherk-Schwarz reductions}},
  \href{https://doi.org/10.1007/JHEP12(2010)016}{\emph{JHEP} {\bfseries 12}
  (2010) 016} [\href{https://arxiv.org/abs/1008.4286}{{\ttfamily 1008.4286}}].

\bibitem{Chong:2005da}
Z.~W. Chong, M.~Cvetič, H.~Lü and C.~N. Pope, \emph{{Five-dimensional gauged
  supergravity black holes with independent rotation parameters}},
  \href{https://doi.org/10.1103/PhysRevD.72.041901}{\emph{Phys. Rev.}
  {\bfseries D72} (2005) 041901}
  [\href{https://arxiv.org/abs/hep-th/0505112}{{\ttfamily hep-th/0505112}}].

\bibitem{Klemm:2000nj}
D.~Klemm and W.~A. Sabra, \emph{{Supersymmetry of black strings in D = 5 gauged
  supergravities}},
  \href{https://doi.org/10.1103/PhysRevD.62.024003}{\emph{Phys. Rev.}
  {\bfseries D62} (2000) 024003}
  [\href{https://arxiv.org/abs/hep-th/0001131}{{\ttfamily hep-th/0001131}}].

\bibitem{Azzola:2018sld}
M.~Azzola, D.~Klemm and M.~Rabbiosi, \emph{{AdS$_5$ black strings in the stu
  model of FI-gauged $N=2$ supergravity}},
  \href{https://doi.org/10.1007/JHEP10(2018)080}{\emph{JHEP} {\bfseries 10}
  (2018) 080} [\href{https://arxiv.org/abs/1803.03570}{{\ttfamily
  1803.03570}}].

\bibitem{Cvetic:1999xp}
M.~Cvetič, M.~J. Duff, P.~Hoxha, J.~T. Liu, H.~Lü, J.~X. Lu et~al.,
  \emph{{Embedding AdS black holes in ten-dimensions and eleven-dimensions}},
  \href{https://doi.org/10.1016/S0550-3213(99)00419-8}{\emph{Nucl. Phys.}
  {\bfseries B558} (1999) 96}
  [\href{https://arxiv.org/abs/hep-th/9903214}{{\ttfamily hep-th/9903214}}].

\bibitem{Romans:1991nq}
L.~J. Romans, \emph{{Supersymmetric, cold and lukewarm black holes in
  cosmological Einstein-Maxwell theory}},
  \href{https://doi.org/10.1016/0550-3213(92)90684-4}{\emph{Nucl. Phys. B}
  {\bfseries 383} (1992) 395}
  [\href{https://arxiv.org/abs/hep-th/9203018}{{\ttfamily hep-th/9203018}}].

\bibitem{London:1995ib}
L.~London, \emph{{Arbitrary dimensional cosmological multi - black holes}},
  \href{https://doi.org/10.1016/0550-3213(94)00511-C}{\emph{Nucl. Phys. B}
  {\bfseries 434} (1995) 709}.

\bibitem{Behrndt:1998jd}
K.~Behrndt, M.~Cvetič and W.~Sabra, \emph{{Nonextreme black holes of
  five-dimensional N=2 AdS supergravity}},
  \href{https://doi.org/10.1016/S0550-3213(99)00243-6}{\emph{Nucl. Phys. B}
  {\bfseries 553} (1999) 317}
  [\href{https://arxiv.org/abs/hep-th/9810227}{{\ttfamily hep-th/9810227}}].

\bibitem{Hawking:1998kw}
S.~W. Hawking, C.~J. Hunter and M.~Taylor, \emph{{Rotation and the AdS / CFT
  correspondence}},
  \href{https://doi.org/10.1103/PhysRevD.59.064005}{\emph{Phys. Rev.}
  {\bfseries D59} (1999) 064005}
  [\href{https://arxiv.org/abs/hep-th/9811056}{{\ttfamily hep-th/9811056}}].

\bibitem{Cvetic:2004hs}
M.~Cvetič, H.~Lü and C.~N. Pope, \emph{{Charged Kerr-de Sitter black holes in
  five dimensions}},
  \href{https://doi.org/10.1016/j.physletb.2004.08.011}{\emph{Phys. Lett. B}
  {\bfseries 598} (2004) 273}
  [\href{https://arxiv.org/abs/hep-th/0406196}{{\ttfamily hep-th/0406196}}].

\bibitem{Cvetic:2004ny}
M.~Cvetič, H.~Lü and C.~N. Pope, \emph{{Charged rotating black holes in five
  dimensional U(1)**3 gauged N=2 supergravity}},
  \href{https://doi.org/10.1103/PhysRevD.70.081502}{\emph{Phys. Rev. D}
  {\bfseries 70} (2004) 081502}
  [\href{https://arxiv.org/abs/hep-th/0407058}{{\ttfamily hep-th/0407058}}].

\bibitem{Wu:2011gq}
S.-Q. Wu, \emph{{General Nonextremal Rotating Charged AdS Black Holes in
  Five-dimensional $U(1)^3$ Gauged Supergravity: A Simple Construction
  Method}}, \href{https://doi.org/10.1016/j.physletb.2011.12.031}{\emph{Phys.
  Lett.} {\bfseries B707} (2012) 286}
  [\href{https://arxiv.org/abs/1108.4159}{{\ttfamily 1108.4159}}].

\bibitem{Cvetic:2005zi}
M.~Cvetič, G.~W. Gibbons, H.~Lü and C.~N. Pope, \emph{{Rotating black holes
  in gauged supergravities: Thermodynamics, supersymmetric limits, topological
  solitons and time machines}},
  \href{https://arxiv.org/abs/hep-th/0504080}{{\ttfamily hep-th/0504080}}.

\bibitem{Gutowski:2004yv}
J.~B. Gutowski and H.~S. Reall, \emph{{General supersymmetric AdS(5) black
  holes}}, \href{https://doi.org/10.1088/1126-6708/2004/04/048}{\emph{JHEP}
  {\bfseries 04} (2004) 048}
  [\href{https://arxiv.org/abs/hep-th/0401129}{{\ttfamily hep-th/0401129}}].

\bibitem{Behrndt:1998ns}
K.~Behrndt, A.~H. Chamseddine and W.~Sabra, \emph{{BPS black holes in N=2
  five-dimensional AdS supergravity}},
  \href{https://doi.org/10.1016/S0370-2693(98)01208-8}{\emph{Phys. Lett. B}
  {\bfseries 442} (1998) 97}
  [\href{https://arxiv.org/abs/hep-th/9807187}{{\ttfamily hep-th/9807187}}].

\bibitem{Klemm:2000gh}
D.~Klemm and W.~Sabra, \emph{{General (anti-)de Sitter black holes in
  five-dimensions}},
  \href{https://doi.org/10.1088/1126-6708/2001/02/031}{\emph{JHEP} {\bfseries
  02} (2001) 031} [\href{https://arxiv.org/abs/hep-th/0011016}{{\ttfamily
  hep-th/0011016}}].

\bibitem{Cacciatori:2003kv}
S.~L. Cacciatori, D.~Klemm and W.~A. Sabra, \emph{{Supersymmetric domain walls
  and strings in D = 5 gauged supergravity coupled to vector multiplets}},
  \href{https://doi.org/10.1088/1126-6708/2003/03/023}{\emph{JHEP} {\bfseries
  03} (2003) 023} [\href{https://arxiv.org/abs/hep-th/0302218}{{\ttfamily
  hep-th/0302218}}].

\bibitem{Gauntlett:2003fk}
J.~P. Gauntlett and J.~B. Gutowski, \emph{{All supersymmetric solutions of
  minimal gauged supergravity in five-dimensions}},
  \href{https://doi.org/10.1103/PhysRevD.70.089901}{\emph{Phys. Rev. D}
  {\bfseries 68} (2003) 105009}
  [\href{https://arxiv.org/abs/hep-th/0304064}{{\ttfamily hep-th/0304064}}].

\bibitem{Gutowski:2004ez}
J.~B. Gutowski and H.~S. Reall, \emph{{Supersymmetric AdS(5) black holes}},
  \href{https://doi.org/10.1088/1126-6708/2004/02/006}{\emph{JHEP} {\bfseries
  02} (2004) 006} [\href{https://arxiv.org/abs/hep-th/0401042}{{\ttfamily
  hep-th/0401042}}].

\bibitem{Kunduri:2006ek}
H.~K. Kunduri, J.~Lucietti and H.~S. Reall, \emph{{Supersymmetric multi-charge
  AdS(5) black holes}},
  \href{https://doi.org/10.1088/1126-6708/2006/04/036}{\emph{JHEP} {\bfseries
  04} (2006) 036} [\href{https://arxiv.org/abs/hep-th/0601156}{{\ttfamily
  hep-th/0601156}}].

\bibitem{Kunduri:2007qy}
H.~K. Kunduri and J.~Lucietti, \emph{{Near-horizon geometries of supersymmetric
  AdS(5) black holes}},
  \href{https://doi.org/10.1088/1126-6708/2007/12/015}{\emph{JHEP} {\bfseries
  12} (2007) 015} [\href{https://arxiv.org/abs/0708.3695}{{\ttfamily
  0708.3695}}].

\bibitem{Kunduri:2006uh}
H.~K. Kunduri, J.~Lucietti and H.~S. Reall, \emph{{Do supersymmetric anti-de
  Sitter black rings exist?}},
  \href{https://doi.org/10.1088/1126-6708/2007/02/026}{\emph{JHEP} {\bfseries
  02} (2007) 026} [\href{https://arxiv.org/abs/hep-th/0611351}{{\ttfamily
  hep-th/0611351}}].

\bibitem{Bhattacharyya:2010yg}
S.~Bhattacharyya, S.~Minwalla and K.~Papadodimas, \emph{{Small Hairy Black
  Holes in $AdS_5 x S^5$}},
  \href{https://doi.org/10.1007/JHEP11(2011)035}{\emph{JHEP} {\bfseries 11}
  (2011) 035} [\href{https://arxiv.org/abs/1005.1287}{{\ttfamily 1005.1287}}].

\bibitem{Markeviciute:2018yal}
J.~Markeviciute and J.~E. Santos, \emph{{Evidence for the existence of a novel
  class of supersymmetric black holes with AdS$_5\times$S$^5$ asymptotics}},
  \href{https://doi.org/10.1088/1361-6382/aaf680}{\emph{Class. Quant. Grav.}
  {\bfseries 36} (2019) 02LT01}
  [\href{https://arxiv.org/abs/1806.01849}{{\ttfamily 1806.01849}}].

\bibitem{Markeviciute:2018cqs}
J.~Markeviciute, \emph{{Rotating Hairy Black Holes in AdS$_5\times$S$^5$}},
  \href{https://doi.org/10.1007/JHEP03(2019)110}{\emph{JHEP} {\bfseries 03}
  (2019) 110} [\href{https://arxiv.org/abs/1809.04084}{{\ttfamily
  1809.04084}}].

\bibitem{Lucietti:2021bbh}
J.~Lucietti and S.~G. Ovchinnikov, \emph{{Uniqueness of supersymmetric AdS$_5$
  black holes with $SU(2)$ symmetry}},
  \href{https://doi.org/10.1088/1361-6382/ac13b7}{\emph{Class. Quant. Grav.}
  {\bfseries 38} (2021) 195019}
  [\href{https://arxiv.org/abs/2105.08542}{{\ttfamily 2105.08542}}].

\bibitem{Durgut:2021rma}
T.~Durgut and H.~K. Kunduri, \emph{{Supersymmetric multi-charge solitons in
  AdS$_5$}},  \href{https://arxiv.org/abs/2111.06831}{{\ttfamily 2111.06831}}.

\bibitem{Blazquez-Salcedo:2017kig}
J.~L. Bl\'azquez-Salcedo, J.~Kunz, F.~Navarro-L\'erida and E.~Radu, \emph{{New
  black holes in $D=5$ minimal gauged supergravity: Deformed boundaries and
  frozen horizons}},
  \href{https://doi.org/10.1103/PhysRevD.97.081502}{\emph{Phys. Rev. D}
  {\bfseries 97} (2018) 081502}
  [\href{https://arxiv.org/abs/1711.08292}{{\ttfamily 1711.08292}}].

\bibitem{Blazquez-Salcedo:2017ghg}
J.~L. Bl\'azquez-Salcedo, J.~Kunz, F.~Navarro-L\'erida and E.~Radu,
  \emph{{Squashed, magnetized black holes in $D=5$ minimal gauged
  supergravity}}, \href{https://doi.org/10.1007/JHEP02(2018)061}{\emph{JHEP}
  {\bfseries 02} (2018) 061}
  [\href{https://arxiv.org/abs/1711.10483}{{\ttfamily 1711.10483}}].

\bibitem{Cassani:2018mlh}
D.~Cassani and L.~Papini, \emph{{Squashing the boundary of supersymmetric
  AdS$_{5}$ black holes}},
  \href{https://doi.org/10.1007/JHEP12(2018)037}{\emph{JHEP} {\bfseries 12}
  (2018) 037} [\href{https://arxiv.org/abs/1809.02149}{{\ttfamily
  1809.02149}}].

\bibitem{Bombini:2019jhp}
A.~Bombini and L.~Papini, \emph{{General supersymmetric $\hbox {AdS}_5$ black
  holes with squashed boundary}},
  \href{https://doi.org/10.1140/epjc/s10052-019-7015-x}{\emph{Eur. Phys. J. C}
  {\bfseries 79} (2019) 515}
  [\href{https://arxiv.org/abs/1903.00021}{{\ttfamily 1903.00021}}].

\bibitem{Bardeen:1984pm}
W.~A. Bardeen and B.~Zumino, \emph{{Consistent and Covariant Anomalies in Gauge
  and Gravitational Theories}},
  \href{https://doi.org/10.1016/0550-3213(84)90322-5}{\emph{Nucl. Phys. B}
  {\bfseries 244} (1984) 421}.

\bibitem{Freedman:2003ax}
D.~Freedman, C.~Nunez, M.~Schnabl and K.~Skenderis, \emph{{Fake supergravity
  and domain wall stability}},
  \href{https://doi.org/10.1103/PhysRevD.69.104027}{\emph{Phys. Rev. D}
  {\bfseries 69} (2004) 104027}
  [\href{https://arxiv.org/abs/hep-th/0312055}{{\ttfamily hep-th/0312055}}].

\bibitem{Papadimitriou:2006dr}
I.~Papadimitriou, \emph{{Non-Supersymmetric Membrane Flows from Fake
  Supergravity and Multi-Trace Deformations}},
  \href{https://doi.org/10.1088/1126-6708/2007/02/008}{\emph{JHEP} {\bfseries
  02} (2007) 008} [\href{https://arxiv.org/abs/hep-th/0606038}{{\ttfamily
  hep-th/0606038}}].

\bibitem{Skenderis:2006rr}
K.~Skenderis and P.~K. Townsend, \emph{{Hamilton-Jacobi method for curved
  domain walls and cosmologies}},
  \href{https://doi.org/10.1103/PhysRevD.74.125008}{\emph{Phys. Rev. D}
  {\bfseries 74} (2006) 125008}
  [\href{https://arxiv.org/abs/hep-th/0609056}{{\ttfamily hep-th/0609056}}].

\bibitem{Papadimitriou:2007sj}
I.~Papadimitriou, \emph{{Multi-Trace Deformations in AdS/CFT: Exploring the
  Vacuum Structure of the Deformed CFT}},
  \href{https://doi.org/10.1088/1126-6708/2007/05/075}{\emph{JHEP} {\bfseries
  05} (2007) 075} [\href{https://arxiv.org/abs/hep-th/0703152}{{\ttfamily
  hep-th/0703152}}].

\bibitem{Janssen:2007rc}
B.~Janssen, P.~Smyth, T.~Van~Riet and B.~Vercnocke, \emph{{A First-order
  formalism for timelike and spacelike brane solutions}},
  \href{https://doi.org/10.1088/1126-6708/2008/04/007}{\emph{JHEP} {\bfseries
  04} (2008) 007} [\href{https://arxiv.org/abs/0712.2808}{{\ttfamily
  0712.2808}}].

\bibitem{HoyosBadajoz:2008fw}
C.~Hoyos-Badajoz, C.~Nunez and I.~Papadimitriou, \emph{{Comments on the String
  dual to N=1 SQCD}},
  \href{https://doi.org/10.1103/PhysRevD.78.086005}{\emph{Phys. Rev. D}
  {\bfseries 78} (2008) 086005}
  [\href{https://arxiv.org/abs/0807.3039}{{\ttfamily 0807.3039}}].

\bibitem{Lindgren:2015lia}
J.~Lindgren, I.~Papadimitriou, A.~Taliotis and J.~Vanhoof, \emph{{Holographic
  Hall conductivities from dyonic backgrounds}},
  \href{https://doi.org/10.1007/JHEP07(2015)094}{\emph{JHEP} {\bfseries 07}
  (2015) 094} [\href{https://arxiv.org/abs/1505.04131}{{\ttfamily
  1505.04131}}].

\bibitem{Cremonini:2020rdx}
S.~Cremonini, L.~Li, K.~Ritchie and Y.~Tang, \emph{{Constraining
  Non-Relativistic RG Flows with Holography}},
  \href{https://arxiv.org/abs/2006.10780}{{\ttfamily 2006.10780}}.

\bibitem{Kim:2020dqx}
N.~Kim and S.-J. Kim, \emph{{The Hamilton-Jacobi equation and holographic
  renormalization group flows on sphere}},
  \href{https://doi.org/10.1007/JHEP10(2020)068}{\emph{JHEP} {\bfseries 10}
  (2020) 068} [\href{https://arxiv.org/abs/2006.16727}{{\ttfamily
  2006.16727}}].

\bibitem{Andrianopoli:2009je}
L.~Andrianopoli, R.~D'Auria, E.~Orazi and M.~Trigiante, \emph{{First Order
  Description of D=4 static Black Holes and the Hamilton-Jacobi equation}},
  \href{https://doi.org/10.1016/j.nuclphysb.2010.02.020}{\emph{Nucl. Phys. B}
  {\bfseries 833} (2010) 1} [\href{https://arxiv.org/abs/0905.3938}{{\ttfamily
  0905.3938}}].

\bibitem{Hyun:2012bc}
S.~Hyun, J.~Jeong and S.-H. Yi, \emph{{Fake Supersymmetry and Extremal Black
  Holes}}, \href{https://doi.org/10.1007/JHEP03(2013)042}{\emph{JHEP}
  {\bfseries 03} (2013) 042} [\href{https://arxiv.org/abs/1210.6273}{{\ttfamily
  1210.6273}}].

\bibitem{Dorronsoro:2016pin}
J.~Diaz~Dorronsoro, B.~Truijen and T.~Van~Riet, \emph{{Comments on fake
  supersymmetry}}, \href{https://doi.org/10.1088/1361-6382/aa64b4}{\emph{Class.
  Quant. Grav.} {\bfseries 34} (2017) 095003}
  [\href{https://arxiv.org/abs/1606.07730}{{\ttfamily 1606.07730}}].

\bibitem{Klemm:2016kxw}
D.~Klemm, N.~Petri and M.~Rabbiosi, \emph{{Black string first order flow in $N
  = 2, d = 5$ abelian gauged supergravity}},
  \href{https://doi.org/10.1007/JHEP01(2017)106}{\emph{JHEP} {\bfseries 01}
  (2017) 106} [\href{https://arxiv.org/abs/1610.07367}{{\ttfamily
  1610.07367}}].

\bibitem{Klemm:2017pxv}
D.~Klemm and M.~Rabbiosi, \emph{{First order flow equations for nonextremal
  black holes in AdS (super)gravity}},
  \href{https://doi.org/10.1007/JHEP10(2017)149}{\emph{JHEP} {\bfseries 10}
  (2017) 149} [\href{https://arxiv.org/abs/1706.05862}{{\ttfamily
  1706.05862}}].

\bibitem{Andrianopoli:2007gt}
L.~Andrianopoli, R.~D'Auria, E.~Orazi and M.~Trigiante, \emph{{First order
  description of black holes in moduli space}},
  \href{https://doi.org/10.1088/1126-6708/2007/11/032}{\emph{JHEP} {\bfseries
  11} (2007) 032} [\href{https://arxiv.org/abs/0706.0712}{{\ttfamily
  0706.0712}}].

\bibitem{Perz:2008kh}
J.~Perz, P.~Smyth, T.~Van~Riet and B.~Vercnocke, \emph{{First-order flow
  equations for extremal and non-extremal black holes}},
  \href{https://doi.org/10.1088/1126-6708/2009/03/150}{\emph{JHEP} {\bfseries
  03} (2009) 150} [\href{https://arxiv.org/abs/0810.1528}{{\ttfamily
  0810.1528}}].

\bibitem{Ceresole:2009iy}
A.~Ceresole, G.~Dall'Agata, S.~Ferrara and A.~Yeranyan, \emph{{First order
  flows for N=2 extremal black holes and duality invariants}},
  \href{https://doi.org/10.1016/j.nuclphysb.2009.09.003}{\emph{Nucl. Phys. B}
  {\bfseries 824} (2010) 239}
  [\href{https://arxiv.org/abs/0908.1110}{{\ttfamily 0908.1110}}].

\bibitem{Kiritsis:2012ma}
E.~Kiritsis and V.~Niarchos, \emph{{The holographic quantum effective potential
  at finite temperature and density}},
  \href{https://doi.org/10.1007/JHEP08(2012)164}{\emph{JHEP} {\bfseries 08}
  (2012) 164} [\href{https://arxiv.org/abs/1205.6205}{{\ttfamily 1205.6205}}].

\bibitem{Gnecchi:2012kb}
A.~Gnecchi and C.~Toldo, \emph{{On the non-BPS first order flow in N=2
  U(1)-gauged Supergravity}},
  \href{https://doi.org/10.1007/JHEP03(2013)088}{\emph{JHEP} {\bfseries 03}
  (2013) 088} [\href{https://arxiv.org/abs/1211.1966}{{\ttfamily 1211.1966}}].

\bibitem{Chamblin:1999tk}
A.~Chamblin, R.~Emparan, C.~V. Johnson and R.~C. Myers, \emph{{Charged AdS
  black holes and catastrophic holography}},
  \href{https://doi.org/10.1103/PhysRevD.60.064018}{\emph{Phys. Rev. D}
  {\bfseries 60} (1999) 064018}
  [\href{https://arxiv.org/abs/hep-th/9902170}{{\ttfamily hep-th/9902170}}].

\bibitem{Choi:2008he}
J.~Choi, S.~Lee and S.~Lee, \emph{{Near Horizon Analysis of Extremal AdS(5)
  Black Holes}},
  \href{https://doi.org/10.1088/1126-6708/2008/05/002}{\emph{JHEP} {\bfseries
  05} (2008) 002} [\href{https://arxiv.org/abs/0802.3330}{{\ttfamily
  0802.3330}}].

\bibitem{Morales:2006gm}
J.~F. Morales and H.~Samtleben, \emph{{Entropy function and attractors for AdS
  black holes}},
  \href{https://doi.org/10.1088/1126-6708/2006/10/074}{\emph{JHEP} {\bfseries
  10} (2006) 074} [\href{https://arxiv.org/abs/hep-th/0608044}{{\ttfamily
  hep-th/0608044}}].

\bibitem{Castro:2018ffi}
A.~Castro, F.~Larsen and I.~Papadimitriou, \emph{{5D rotating black holes and
  the nAdS$_{2}$/nCFT$_{1}$ correspondence}},
  \href{https://doi.org/10.1007/JHEP10(2018)042}{\emph{JHEP} {\bfseries 10}
  (2018) 042} [\href{https://arxiv.org/abs/1807.06988}{{\ttfamily
  1807.06988}}].

\bibitem{Bianchi:2001kw}
M.~Bianchi, D.~Z. Freedman and K.~Skenderis, \emph{{Holographic
  renormalization}},
  \href{https://doi.org/10.1016/S0550-3213(02)00179-7}{\emph{Nucl. Phys. B}
  {\bfseries 631} (2002) 159}
  [\href{https://arxiv.org/abs/hep-th/0112119}{{\ttfamily hep-th/0112119}}].

\bibitem{Sahoo:2010sp}
B.~Sahoo and H.-U. Yee, \emph{{Electrified plasma in AdS/CFT correspondence}},
  \href{https://doi.org/10.1007/JHEP11(2010)095}{\emph{JHEP} {\bfseries 11}
  (2010) 095} [\href{https://arxiv.org/abs/1004.3541}{{\ttfamily 1004.3541}}].

\bibitem{Genolini:2016ecx}
P.~Benetti~Genolini, D.~Cassani, D.~Martelli and J.~Sparks, \emph{{Holographic
  renormalization and supersymmetry}},
  \href{https://doi.org/10.1007/JHEP02(2017)132}{\emph{JHEP} {\bfseries 02}
  (2017) 132} [\href{https://arxiv.org/abs/1612.06761}{{\ttfamily
  1612.06761}}].

\bibitem{Marolf:2000cb}
D.~Marolf, \emph{{Chern-Simons terms and the three notions of charge}},  in
  \emph{{Quantization, gauge theory, and strings. Proceedings, International
  Conference dedicated to the memory of Professor Efim Fradkin, Moscow, Russia,
  June 5-10, 2000. Vol. 1+2}}, pp.~312--320, 6, 2000,
  \href{https://arxiv.org/abs/hep-th/0006117}{{\ttfamily hep-th/0006117}}.

\bibitem{Witten:1998qj}
E.~Witten, \emph{{Anti-de Sitter space and holography}}, {\emph{Adv. Theor.
  Math. Phys.} {\bfseries 2} (1998) 253}
  [\href{https://arxiv.org/abs/hep-th/9802150}{{\ttfamily hep-th/9802150}}].

\bibitem{Page:1984qv}
D.~N. Page, \emph{{Classical Stability of Round and Squashed Seven Spheres in
  Eleven-dimensional Supergravity}},
  \href{https://doi.org/10.1103/PhysRevD.28.2976}{\emph{Phys. Rev.} {\bfseries
  D28} (1983) 2976}.

\bibitem{Copsey:2005se}
K.~Copsey and G.~T. Horowitz, \emph{{The Role of dipole charges in black hole
  thermodynamics}},
  \href{https://doi.org/10.1103/PhysRevD.73.024015}{\emph{Phys. Rev. D}
  {\bfseries 73} (2006) 024015}
  [\href{https://arxiv.org/abs/hep-th/0505278}{{\ttfamily hep-th/0505278}}].

\bibitem{Benini:2007gx}
F.~Benini, F.~Canoura, S.~Cremonesi, C.~Nunez and A.~V. Ramallo,
  \emph{{Backreacting flavors in the Klebanov-Strassler background}},
  \href{https://doi.org/10.1088/1126-6708/2007/09/109}{\emph{JHEP} {\bfseries
  09} (2007) 109} [\href{https://arxiv.org/abs/0706.1238}{{\ttfamily
  0706.1238}}].

\bibitem{Kunduri:2013vka}
H.~K. Kunduri and J.~Lucietti, \emph{{The first law of soliton and black hole
  mechanics in five dimensions}},
  \href{https://doi.org/10.1088/0264-9381/31/3/032001}{\emph{Class. Quant.
  Grav.} {\bfseries 31} (2014) 032001}
  [\href{https://arxiv.org/abs/1310.4810}{{\ttfamily 1310.4810}}].

\bibitem{Elgood:2020nls}
Z.~Elgood, T.~Ort\'\i{}n and D.~Pere\~n\'\i{}guez, \emph{{The first law and
  Wald entropy formula of heterotic stringy black holes at first order in
  $\alpha'$}}, \href{https://doi.org/10.1007/JHEP05(2021)110}{\emph{JHEP}
  {\bfseries 05} (2021) 110}
  [\href{https://arxiv.org/abs/2012.14892}{{\ttfamily 2012.14892}}].

\bibitem{Jensen:2012kj}
K.~Jensen, R.~Loganayagam and A.~Yarom, \emph{{Thermodynamics, gravitational
  anomalies and cones}},
  \href{https://doi.org/10.1007/JHEP02(2013)088}{\emph{JHEP} {\bfseries 02}
  (2013) 088} [\href{https://arxiv.org/abs/1207.5824}{{\ttfamily 1207.5824}}].

\bibitem{Minasian:2021png}
R.~Minasian, I.~Papadimitriou and P.~Yi, \emph{{Anomalies and Supersymmetry}},
  \href{https://arxiv.org/abs/2104.13391}{{\ttfamily 2104.13391}}.

\bibitem{Balasubramanian:1999re}
V.~Balasubramanian and P.~Kraus, \emph{{A Stress tensor for Anti-de Sitter
  gravity}}, \href{https://doi.org/10.1007/s002200050764}{\emph{Commun. Math.
  Phys.} {\bfseries 208} (1999) 413}
  [\href{https://arxiv.org/abs/hep-th/9902121}{{\ttfamily hep-th/9902121}}].

\bibitem{Awad:1999xx}
A.~M. Awad and C.~V. Johnson, \emph{{Holographic stress tensors for Kerr - AdS
  black holes}}, \href{https://doi.org/10.1103/PhysRevD.61.084025}{\emph{Phys.
  Rev. D} {\bfseries 61} (2000) 084025}
  [\href{https://arxiv.org/abs/hep-th/9910040}{{\ttfamily hep-th/9910040}}].

\bibitem{Cheng:2021zjh}
P.~Cheng, R.~Minasian and S.~Theisen, \emph{{Anomalies as Obstructions: from
  Dimensional Lifts to Swampland}},
  \href{https://arxiv.org/abs/2106.14912}{{\ttfamily 2106.14912}}.

\bibitem{Cassani:2019mms}
D.~Cassani and L.~Papini, \emph{{The BPS limit of rotating AdS black hole
  thermodynamics}}, \href{https://doi.org/10.1007/JHEP09(2019)079}{\emph{JHEP}
  {\bfseries 09} (2019) 079}
  [\href{https://arxiv.org/abs/1906.10148}{{\ttfamily 1906.10148}}].

\bibitem{Larsen:2020lhg}
F.~Larsen and S.~Paranjape, \emph{{Thermodynamics of Near BPS Black Holes in
  AdS$_4$ and AdS$_7$}},  \href{https://arxiv.org/abs/2010.04359}{{\ttfamily
  2010.04359}}.

\bibitem{Silva:2006xv}
P.~J. Silva, \emph{{Thermodynamics at the BPS bound for Black Holes in AdS}},
  \href{https://doi.org/10.1088/1126-6708/2006/10/022}{\emph{JHEP} {\bfseries
  10} (2006) 022} [\href{https://arxiv.org/abs/hep-th/0607056}{{\ttfamily
  hep-th/0607056}}].

\bibitem{Kim:2006he}
S.~Kim and K.-M. Lee, \emph{{1/16-BPS Black Holes and Giant Gravitons in the
  AdS(5) X S**5 Space}},
  \href{https://doi.org/10.1088/1126-6708/2006/12/077}{\emph{JHEP} {\bfseries
  12} (2006) 077} [\href{https://arxiv.org/abs/hep-th/0607085}{{\ttfamily
  hep-th/0607085}}].

\bibitem{Halmagyi:2017hmw}
N.~Halmagyi and S.~Lal, \emph{{On the On-Shell: The Action of AdS$_4$ Black
  Holes}},  \href{https://arxiv.org/abs/1710.09580}{{\ttfamily 1710.09580}}.

\bibitem{Chaturvedi:2020jyy}
P.~Chaturvedi, I.~Papadimitriou, W.~Song and B.~Yu, \emph{{AdS$_{3}$ gravity
  and the complex SYK models}},
  \href{https://doi.org/10.1007/JHEP05(2021)142}{\emph{JHEP} {\bfseries 05}
  (2021) 142} [\href{https://arxiv.org/abs/2011.10001}{{\ttfamily
  2011.10001}}].

\bibitem{An:2017ihs}
O.~S. An, \emph{{Anomaly-corrected supersymmetry algebra and supersymmetric
  holographic renormalization}},
  \href{https://doi.org/10.1007/JHEP12(2017)107}{\emph{JHEP} {\bfseries 12}
  (2017) 107} [\href{https://arxiv.org/abs/1703.09607}{{\ttfamily
  1703.09607}}].

\bibitem{Arnowitt:1960es}
R.~L. Arnowitt, S.~Deser and C.~W. Misner, \emph{{Canonical variables for
  general relativity}},
  \href{https://doi.org/10.1103/PhysRev.117.1595}{\emph{Phys. Rev.} {\bfseries
  117} (1960) 1595}.

\bibitem{deBoer:1999tgo}
J.~de~Boer, E.~P. Verlinde and H.~L. Verlinde, \emph{{On the holographic
  renormalization group}},
  \href{https://doi.org/10.1088/1126-6708/2000/08/003}{\emph{JHEP} {\bfseries
  08} (2000) 003} [\href{https://arxiv.org/abs/hep-th/9912012}{{\ttfamily
  hep-th/9912012}}].

\bibitem{deHaro:2000vlm}
S.~de~Haro, S.~N. Solodukhin and K.~Skenderis, \emph{{Holographic
  reconstruction of space-time and renormalization in the AdS / CFT
  correspondence}}, \href{https://doi.org/10.1007/s002200100381}{\emph{Commun.
  Math. Phys.} {\bfseries 217} (2001) 595}
  [\href{https://arxiv.org/abs/hep-th/0002230}{{\ttfamily hep-th/0002230}}].

\bibitem{Martelli:2002sp}
D.~Martelli and W.~Mueck, \emph{{Holographic renormalization and Ward
  identities with the Hamilton-Jacobi method}},
  \href{https://doi.org/10.1016/S0550-3213(03)00060-9}{\emph{Nucl. Phys.}
  {\bfseries B654} (2003) 248}
  [\href{https://arxiv.org/abs/hep-th/0205061}{{\ttfamily hep-th/0205061}}].

\bibitem{Papadimitriou:2004ap}
I.~Papadimitriou and K.~Skenderis, \emph{{AdS / CFT correspondence and
  geometry}}, \href{https://doi.org/10.4171/013-1/4}{\emph{IRMA Lect. Math.
  Theor. Phys.} {\bfseries 8} (2005) 73}
  [\href{https://arxiv.org/abs/hep-th/0404176}{{\ttfamily hep-th/0404176}}].

\bibitem{Papadimitriou:2011qb}
I.~Papadimitriou, \emph{{Holographic Renormalization of general dilaton-axion
  gravity}}, \href{https://doi.org/10.1007/JHEP08(2011)119}{\emph{JHEP}
  {\bfseries 08} (2011) 119} [\href{https://arxiv.org/abs/1106.4826}{{\ttfamily
  1106.4826}}].

\bibitem{Papadimitriou:2016yit}
I.~Papadimitriou, \emph{{Lectures on Holographic Renormalization}},
  \href{https://doi.org/10.1007/978-3-319-31352-8_4}{\emph{Springer Proc.
  Phys.} {\bfseries 176} (2016) 131}.

\bibitem{Breitenlohner:1982bm}
P.~Breitenlohner and D.~Z. Freedman, \emph{{Positive Energy in anti-De Sitter
  Backgrounds and Gauged Extended Supergravity}},
  \href{https://doi.org/10.1016/0370-2693(82)90643-8}{\emph{Phys. Lett. B}
  {\bfseries 115} (1982) 197}.

\bibitem{Papadimitriou:2004rz}
I.~Papadimitriou and K.~Skenderis, \emph{{Correlation functions in holographic
  RG flows}}, \href{https://doi.org/10.1088/1126-6708/2004/10/075}{\emph{JHEP}
  {\bfseries 10} (2004) 075}
  [\href{https://arxiv.org/abs/hep-th/0407071}{{\ttfamily hep-th/0407071}}].

\bibitem{Henningson:1998gx}
M.~Henningson and K.~Skenderis, \emph{{The Holographic Weyl anomaly}},
  \href{https://doi.org/10.1088/1126-6708/1998/07/023}{\emph{JHEP} {\bfseries
  07} (1998) 023} [\href{https://arxiv.org/abs/hep-th/9806087}{{\ttfamily
  hep-th/9806087}}].

\bibitem{Bianchi:2001de}
M.~Bianchi, D.~Z. Freedman and K.~Skenderis, \emph{{How to go with an RG
  flow}}, \href{https://doi.org/10.1088/1126-6708/2001/08/041}{\emph{JHEP}
  {\bfseries 08} (2001) 041}
  [\href{https://arxiv.org/abs/hep-th/0105276}{{\ttfamily hep-th/0105276}}].

\bibitem{Larios:2019kbw}
G.~Larios, P.~Ntokos and O.~Varela, \emph{{Embedding the SU(3) sector of SO(8)
  supergravity in $D=11$}},
  \href{https://doi.org/10.1103/PhysRevD.100.086021}{\emph{Phys. Rev. D}
  {\bfseries 100} (2019) 086021}
  [\href{https://arxiv.org/abs/1907.02087}{{\ttfamily 1907.02087}}].

\bibitem{Balasubramanian:2000pq}
V.~Balasubramanian, E.~G. Gimon, D.~Minic and J.~Rahmfeld,
  \emph{{Four-dimensional conformal supergravity from AdS space}},
  \href{https://doi.org/10.1103/PhysRevD.63.104009}{\emph{Phys. Rev. D}
  {\bfseries 63} (2001) 104009}
  [\href{https://arxiv.org/abs/hep-th/0007211}{{\ttfamily hep-th/0007211}}].

\bibitem{Festuccia:2011ws}
G.~Festuccia and N.~Seiberg, \emph{{Rigid Supersymmetric Theories in Curved
  Superspace}}, \href{https://doi.org/10.1007/JHEP06(2011)114}{\emph{JHEP}
  {\bfseries 06} (2011) 114} [\href{https://arxiv.org/abs/1105.0689}{{\ttfamily
  1105.0689}}].

\bibitem{Klare:2012gn}
C.~Klare, A.~Tomasiello and A.~Zaffaroni, \emph{{Supersymmetry on Curved Spaces
  and Holography}}, \href{https://doi.org/10.1007/JHEP08(2012)061}{\emph{JHEP}
  {\bfseries 08} (2012) 061} [\href{https://arxiv.org/abs/1205.1062}{{\ttfamily
  1205.1062}}].

\bibitem{Dumitrescu:2012ha}
T.~T. Dumitrescu, G.~Festuccia and N.~Seiberg, \emph{{Exploring Curved
  Superspace}}, \href{https://doi.org/10.1007/JHEP08(2012)141}{\emph{JHEP}
  {\bfseries 08} (2012) 141} [\href{https://arxiv.org/abs/1205.1115}{{\ttfamily
  1205.1115}}].

\end{thebibliography}\endgroup
